\documentclass[epj]{svjour}
\usepackage{graphicx}
\usepackage{color}

\def\mb#1{\setbox0=\hbox{$#1$}\kern-.025em\copy0\kern-\wd0
\kern-0.05em\copy0\kern-\wd0\kern-.025em\raise.0233em\box0}

\begin{document}
   \title{Dynamical stability of systems with long-range interactions:
   application of the Nyquist method to the HMF model}

 \author{P.H. Chavanis and L. Delfini}

\institute{$^1$ Universit\'e de Toulouse, UPS, Laboratoire de
Physique Th\'eorique (IRSAMC), F-31062 Toulouse, France\\
$^2$ CNRS,  Laboratoire de
Physique Th\'eorique (IRSAMC), F-31062 Toulouse, France\\
\email{chavanis@irsamc.ups-tlse.fr; Luca.Delfini@irsamc.ups-tlse.fr}
}

\titlerunning{Application of the Nyquist method to the HMF model}

   \date{To be included later }

   \abstract{We apply the Nyquist method to the Hamiltonian Mean Field
   (HMF) model in order to settle the linear dynamical stability of a
   spatially homogeneous distribution function with respect to the
   Vlasov equation. We consider the case of Maxwell (isothermal) and
   Tsallis (polytropic) distributions and show that the system is
   stable above a critical kinetic temperature $T_c$ and unstable
   below it. Then, we consider a symmetric double-humped distribution,
   made of the superposition of two decentered Maxwellians, and show
   the existence of a re-entrant phase in the stability diagram. When
   we consider an asymmetric double-humped distribution, the
   re-entrant phase disappears above a critical value of the asymmetry
   factor $\Delta>1.09$. We also consider the HMF model with a
   repulsive interaction. In that case, single-humped distributions
   are always stable. For asymmetric double-humped distributions,
   there is a re-entrant phase for $1<\Delta<25.6$, a double
   re-entrant phase for $25.6<\Delta<43.9$ and no re-entrant phase for
   $\Delta>43.9$. Finally, we extend our results to arbitrary
   potentials of interaction and mention the connexion between the HMF
   model, Coulombian plasmas and gravitational systems. We discuss the
   relation between linear dynamical stability and formal nonlinear
   dynamical stability and show their equivalence for spatially
   homogeneous distributions. We also provide a criterion of dynamical
   stability for inhomogeneous systems.
 }

   \maketitle
%

\section{Introduction}
\label{energy}

Systems with long-range interactions have recently been the object of
considerable interest \cite{houches,assise}.  These systems have
applications in different areas of physics, astrophysics,
hydrodynamics and biology \footnote{For example, there exists beautiful analogies between self-gravitating systems, two-dimensional vortices and bacterial populations investigated by one of the authors [3-5].}.  Furthermore, they display very peculiar
phenomena with respect to ordinary systems with short-range
interactions such as negative specific heats
[6-9], inequivalence of statistical
ensembles [10-16], numerous types of phase
transitions persisting at the thermodynamic limit
\cite{bb,ijmpb}, slow
collisional relaxation
[19-25], kinetic
blocking due to the absence of resonances
[26-30,25], non-trivial
dependence of the collisional relaxation time with the number $N$ of
particles [31-33], long-lived metastable states whose
lifetimes increase exponentially with $N$ \cite{rm,lifetime}, algebraic
decay of the correlation functions \cite{corr,rp,bd,kinvortex,ybd}, front
structure and slow relaxation of the velocity tails
\cite{cl,kinvortex},  long-lived quasi stationary states (QSS) [20-23,31,32,40,41,24,25,42], violent
collisionless relaxation of the Vlasov equation
[43-47,29,48-51], out-of-equilibrium phase
transitions [52-59],
re-entrant phases \cite{epjb,reentrant}, dynamical phase transitions
\cite{cc}, curious effects due to spatial inhomogeneity \cite{curious},
non-ergodic behaviors \cite{ymiller,latora,prt,break}, glassy dynamics
\cite{glass}, non-ideal effects due to the influence of a thermal bath
[29,66-68] etc... Certainly, the physical richness of these
systems (Pandora box) will lead to further investigations and
applications.

For systems with long-range interactions, the mean field approximation
is exact in a proper thermodynamic limit where the number of particles
$N\rightarrow +\infty$ while the strength of the interaction $k\sim 1/N$
(coupling constant) goes to zero and the volume $V\sim 1$ remains fixed
\cite{houches,assise,assiseph}. For a fixed interval of times $[0,T]$ and $N\rightarrow
+\infty$, the evolution of the distribution function $f({\bf r},{\bf
v},t)$ is governed by the (mean field) Vlasov equation.  The Vlasov
equation describes the collisionless dynamics of the system for
sufficiently ``short" times with respect to the collisional relaxation
time: $t\ll t_R(N)$. In fact, the relaxation time $t_R(N)$ increases
rapidly with the number of particles $N$ so that, in practice, the
domain of validity of the Vlasov equation is huge
\cite{assiseph}. Since the Vlasov equation admits an
infinite number of steady states, the system can be found in quasi
stationary states (QSS) that differ from the Boltzmann
distribution. These are long-lived stable steady states of the Vlasov
equation. For example, if we place initially the system in a stable
steady state of the Vlasov equation $f_0({\bf r},{\bf v})$, it will
remain in that state for a very long time until ``collisions" (more
precisely correlations due to finite $N$ effects) make it slowly
evolve \cite{assiseph}. On the other hand, if we start from an
unstable steady state of the Vlasov equation, or from an unsteady
initial condition, the system will spontaneously evolve through phase
mixing and violent relaxation towards a QSS on the coarse-grained
scale (see, e.g.,
\cite{bt,ch}). This distribution function $\overline{f}_{QSS}({\bf
r},{\bf v})$ is a stable steady state of the Vlasov equation that can
be predicted by the statistical theory of Lynden-Bell \cite{lb} if the
evolution is ergodic (i.e. if the system ``mixes" efficiently
\cite{assiseph}). As a result, it is important on general grounds to
study the dynamical stability of stationary solutions of the Vlasov
equation and devise general stability criteria.

This problem has been first considered in plasma physics
[70-72] and astrophysics [20-23]. In plasma physics, the interaction between like-sign
charges is repulsive. However, in neutral plasmas, the presence of
opposite charges screens the interaction between like-sign charges
(Debye shielding) so that the system is spatially
homogeneous. Several methods have been developed to study the
dynamical stability of spatially homogeneous distribution functions $f({\bf v})$
with respect to the Vlasov equation. One approach is to determine
criteria of nonlinear dynamical stability or formal nonlinear
stability by minimizing an energy-Casimir functional
\cite{holm}. Another approach is to determine criteria of linear stability by linearizing the Vlasov equation around a steady
state, taking the Laplace-Fourier transform to obtain the dispersion relation and studying the sign of the
imaginary part of the complex pulsation
[70-72]. In some cases, the dispersion
relation can be solved analytically and the pulsation can be obtained
explicitly. In more complicated cases, the dispersion relation must be
solved numerically. We can also have recourse to other methods to
settle the linear stability of the system without being
required to solve the dispersion relation. In this respect, Nyquist
\cite{nyquist} has introduced a powerful method to investigate the linear
stability of a spatially homogeneous distribution of the Vlasov
equation. This is based on a graphical construction that can be easily
implemented in practice. This method tells whether the distribution is linearly stable (or
unstable) but it does not give the value of the decay rate (or growth
rate) of the perturbation. Using this criterion, it can be shown with
almost no calculation that the Maxwellian distribution is always
dynamically stable in a plasma for perturbations with arbitrary
wavenumbers.

For gravitational systems, the interaction between masses is
attractive leading to possible collapse (Jeans instability). We could
imagine using the Nyquist method to settle the dynamical stability of
a galaxy described by the Vlasov equation. However, we are rapidly confronted to the difficulty that a
realistic stellar system is spatially inhomogeneous
[20-23]. This considerably complicates the
stability analysis and precludes the direct application of the Nyquist
method. In that case, we must use other methods like the Antonov
[75-77] criterion of linear
dynamical stability or methods of nonlinear dynamical stability
\cite{lemougrav}. Using these methods, it can be shown that any
distribution function of the form $f=f(\epsilon)$ with
$f'(\epsilon)<0$ where $\epsilon=v^2/2+\Phi({\bf r})$ is the
individual energy of a star is linearly \cite{bt}, and even
nonlinearly \cite{lemougrav}, dynamically stable with respect to the
Vlasov equation.  The Nyquist method can be used (see Appendix
\ref{sec_ana}) to investigate the linear dynamical stability of a
rather unrealistic infinite and homogeneous distribution of stars
making the Jeans swindle \cite{bt}. For a Maxwellian distribution, it
leads to a graphical illustration of the Jeans instability showing
that the system is unstable with respect to perturbations of
sufficiently low wavenumbers $k<k_J$, where $k_J=(4\pi G\beta
n)^{1/2}$ is the Jeans wavenumber.

In this paper, we consider the Hamiltonian Mean Field (HMF) model
which is a very much studied toy model of systems with long-range
interactions \cite{ar,cvb}. In this model, the interaction between
particles is attractive like gravity but, unlike gravity, the system
can be spatially homogeneous. Therefore, for the HMF model, it is
possible to use the Nyquist method of plasma physics to investigate
the linear stability of a homogeneous distribution. However, since the
interaction is attractive, a crucial sign changes in the dispersion
relation and this leads to new results with respect to plasma physics.
As a direct application of this method, we can recover by a graphical
construction the fact that the Maxwellian distribution is linearly
stable with respect to the Vlasov equation if $T>T_c$ and linearly
unstable if $T<T_c$ where $T_c=\frac{kM}{4\pi}$ is a critical
temperature. This instability is similar to the Jeans instability in
astrophysics where the temperature $T$ plays the role of the
wavenumber $k$.

The paper is organized as follows. In Sec. \ref{sec_dyn}, we recall basic results
concerning the Vlasov equation, the dispersion relation and the
Nyquist method in relation to the HMF model. We first consider the HMF
model with an attractive interaction (ferromagnetic). Using the
Nyquist method, we establish general linear stability criteria for
single-humped and double-humped distributions.  In Secs. \ref{sec_maxwell} and \ref{sec_tsallis},
we apply the Nyquist method to the Maxwell (isothermal) and the
Tsallis (polytropic) distributions. We show that the system is
linearly stable if $T>T_c$ and linearly unstable if $T<T_c$ where
$T_c=\frac{n}{n+1}\frac{kM}{4\pi}$ is a critical kinetic temperature depending
on the polytropic index $n$ (the Maxwellian distribution is recovered
for $n\rightarrow +\infty$). In Sec. \ref{sec_sdh}, we consider a symmetric
double-humped distribution made of the superposition of two decentered
Maxwellians (with separation $v_a$). We show that this distribution has a rich
stability diagram in the $(v_a,T)$ plane exhibiting a re-entrant
phase. In Sec. \ref{sec_adh}, we generalize our study to the case of an
asymmetric double-humped distribution where the amplitude of one
maximum is larger than the other by a factor
$\Delta>1$. We determine how the stability diagram in
the $(v_a,T)$ plane changes as a function of $\Delta$ and show that the
re-entrant phase disappears above a very small asymmetry
$\Delta>\Delta_c=1.09$. In Sec. \ref{sec_rep}, we consider the HMF model with a
repulsive interaction (antiferromagnetic). We show that single-humped
distributions (like the Maxwell and the Tsallis distributions) are
always linearly stable. We also establish general linear stability
criteria for double-humped distributions. More specifically, we
investigate the linear stability of a double-humped distribution made
of two Maxwellian distributions. For $\Delta=1$ (symmetric case), we
show the existence of a re-entrant phase. A double re-entrant phase
appears for $\Delta>\Delta_{c}^{(1)}=25.6$ and they both disappear for
$\Delta>\Delta_{c}^{(2)}=43.9$. In Sec. \ref{sec_genr} we discuss the extension of
our results to arbitrary attractive or repulsive potentials of
interaction and make the connexion between the HMF model, Coulombian
plasmas and gravitational systems.  We also discuss the relation
between linear and formal nonlinear stability and show their
equivalence for spatially homogeneous systems.

\section{Dynamical stability of the HMF model}
\label{sec_dyn}

\subsection{The Vlasov equation}
\label{sec_vlasov}

The HMF model consists in $N$ particles of unit mass $m=1$ moving on a
circle and interacting via a cosine potential \cite{ar,cvb}. The
microscopic dynamics of this system is governed by the Hamiltonian
equations
\begin{eqnarray}
\frac{d\theta_{i}}{dt}=\frac{\partial H}{\partial v_{i}}, \qquad \frac{dv_{i}}{dt}=-\frac{\partial H}{\partial \theta_{i}},
\label{v1}
\end{eqnarray}
\begin{eqnarray}
H=\sum_{i=1}^{N}\frac{v_{i}^{2}}{2}-\frac{k}{2\pi}\sum_{i<j}\cos(\theta_{i}-\theta_{j}),
\label{v2}
\end{eqnarray}
where $\theta_{i}$ is the angle that particle $i$ makes with an axis
of reference, $v_{i}=d\theta_i/dt$ is its velocity and $k>0$ is the
coupling constant. This system conserves the energy $E=H$ and the total mass $M=Nm=N$.

The evolution of the $N$-body distribution $P_{N}(\theta_{1},v_{1},...,\theta_{N},v_{N},t)$ is governed by the
Liouville equation
\begin{eqnarray}
{\partial P_{N}\over\partial t}+\sum_{i=1}^{N}\biggl (v_{i}{\partial P_{N}\over\partial\theta_{i}}+F_{i}{\partial P_{N}\over\partial v_{i}}\biggr )=0
\label{v3}
\end{eqnarray}
where
\begin{eqnarray}
F_{i}=-\frac{\partial \Phi}{\partial\theta_{i}}=-{k\over 2\pi}\sum_{j\neq i}\sin(\theta_{i}-\theta_{j})=\sum_{j\neq i}F(j\rightarrow i)
\label{v4}
\end{eqnarray}
is the total force experienced by particle $i$ due to the interaction
with the other particles. From the Liouville equation, we can
construct the BBGKY hierarchy for the reduced distributions
$P_{j}(\theta_{1},v_{1},...,\theta_{j},v_{j},t)=\int
P_{N}(\theta_{1},v_{1},...,\theta_{N},v_{N},t)\prod_{k=j+1}^{N}d\theta_{k}dv_{k}$ \cite{hb3}.

We consider the thermodynamic limit $N\rightarrow +\infty$ in such a
way that the dimensionless energy $\epsilon=8\pi E/kM^2$ and the
dimensionless temperature $\eta=kM/4\pi T$ are of order unity (these
scalings are obtained by comparing the kinetic and the potential
energies in Eq. (\ref{v2}) yielding $E\sim N\langle
v^2\rangle\sim NT\sim kN^2$). We can renormalize the parameters so that the
coupling constant $k\sim 1/N$ while $\beta\sim 1$, $E/N\sim 1$ and
$V=2\pi\sim 1$. The dynamical time $t_{D}\sim 2\pi/\sqrt{\langle
v^2\rangle}\sim 1/\sqrt{k\rho}$ is also of order unity \footnote{In
this paper, we use the notations of
\cite{cvb} that are similar to those introduced in astrophysics. The
link with the notations of \cite{yamaguchi} is made by taking
$k=2\pi/N$, $\eta=\beta/2$ and $\epsilon=4(U-1/2)$.}. In the
thermodynamic limit $N\rightarrow +\infty$, it can be shown from
scaling arguments \cite{hb3} that the $N$-body distribution factorizes
in a product of $N$ one-body distributions
\begin{eqnarray}
P_{N}(\theta_{1},v_{1},...,\theta_{N},v_{N},t)=\prod_{i=1}^{N} P_{1}(\theta_{i},v_{i},t).
\label{v5}
\end{eqnarray}
Substituting this result in the first equation  of the BBGKY hierarchy, we find that the one-body distribution $P_{1}(\theta,v,t)$, or equivalently the smooth distribution function $f(\theta,v,t)=NP_{1}(\theta,v,t)$, satisfies the Vlasov equation \footnote{A rigorous derivation of the Vlasov equation is given by Braun \& Hepp \cite{bh}.}
\begin{equation}
\label{v6} {\partial f\over\partial t}+v{\partial
f\over\partial\theta}- {\partial\Phi\over\partial \theta}{\partial
f\over\partial v}=0,
\end{equation}
where
\begin{equation}
\label{v7} \Phi(\theta,t)=-{k\over
2\pi}\int_{0}^{2\pi}\cos(\theta-\theta')\rho(\theta',t)d\theta',
\end{equation}
is the mean field potential and $\rho(\theta,t)=\int f(\theta,v,t) \,
dv$ is the spatial density. The mean field average force experienced by a
particle located in $\theta$ is $\langle F\rangle=-\Phi'(\theta)$.

The Vlasov equation admits an infinite number of stationary
solutions. A spatially homogeneous system ($\Phi=0$) with distribution
function (DF)  $f=f(v)$ is always a steady state of the Vlasov equation. If
the system is spatially inhomogeneous ($\Phi\neq 0$), the general form
of steady distributions of the Vlasov equation is
$f=f(\epsilon)$ where $\epsilon=v^2/2+\Phi(\theta)$ is the
individual energy.  In this paper, we shall only consider spatially
homogeneous distributions so that $f=f(v)$.

\subsection{The dispersion relation}
\label{sec_dr}

We want to study the linear dynamical stability \footnote{In fact, we
consider here the {\it spectral} stability of a distribution function
$f=f(v)$. For infinite dimensional systems, spectral stability does
not necessarily imply {\it linear} stability \cite{holm}. However, by
an abuse of language, we shall often say ``linearly stable'' instead of
``spectrally stable''.} of a spatially homogeneous stationary solution
of the Vlasov equation described by a distribution function
$f=f(v)$. Linearizing the Vlasov equation around the steady state and
taking the Laplace-Fourier transform (writing the perturbation as $\delta
f_{n\omega}\sim\delta\Phi_{n\omega}\sim e^{i(n\theta-\omega t)}$), we
obtain the dispersion relation
\cite{inagaki,choi,cvb}:
\begin{eqnarray}
\epsilon(n,\omega)\equiv 1+{k\over 2}(\delta_{n,1}-\delta_{n,-1})\int_{C} {f'(v)\over nv-\omega}dv=0,
\label{dr1}
\end{eqnarray}
where $\epsilon(n,\omega)$ is the dielectric function and the integral
must be performed along the Landau contour $(C)$
\cite{nicholson}. There exists solutions only for $n=\pm 1$ so that
only these modes can propagate \footnote{This differs from the case of
gaseous systems described by the dispersion relation
(\ref{tard2}).}. For the mode $n=+1$ ($n=-1$ gives the same result),
Eq. (\ref{dr1}) reduces to
\begin{eqnarray}
\epsilon(\omega)\equiv 1+{k\over 2}\int_{C} {{f'(v)}\over v-{\omega}}dv=0,
\label{dr2}
\end{eqnarray}
which is the fundamental equation studied in this paper.  For a given
distribution $f(v)$, this equation determines the complex pulsation(s)
$\omega=\omega_{r}+i\omega_{i}$ of the linearized perturbation $\delta
f$. Since $\delta f\sim e^{\omega_i t}$, the system is linearly stable
if $\omega_{i}<0$ and linearly unstable if $\omega_{i}>0$. For future reference, let us recall that
\begin{eqnarray}
\int_{C} {{f'(v)}\over v-{\omega}}dv=\int_{-\infty}^{+\infty} {{f'(v)}\over v-{\omega}}dv, \quad (\omega_i>0),
\label{dr3}
\end{eqnarray}
\begin{eqnarray}
\int_{C} {{f'(v)}\over v-{\omega}}dv=P\int_{-\infty}^{+\infty} {{f'(v)}\over v-{\omega}}dv+i\pi  f'(\omega), \quad (\omega_i=0),\nonumber\\
\label{dr4}
\end{eqnarray}
\begin{eqnarray}
\int_{C} {{f'(v)}\over v-{\omega}}dv=\int_{-\infty}^{+\infty} {{f'(v)}\over v-{\omega}}dv+ 2\pi i f'(\omega), \quad (\omega_i<0).\nonumber\\
\label{dr5}
\end{eqnarray}

\subsection{The condition of marginal stability}
\label{sec_marginal}

The condition of marginal
stability corresponds to
\begin{equation}
\label{m1}
\omega_{i}=0,
\end{equation}
where $\omega_{i}$ is the imaginary part of $\omega$. In that case,
$\omega=\omega_r$ is real and the integral in Eq. (\ref{dr2}) can be evaluated
along the real axis by using the Plemelj formula \cite{nicholson}:
\begin{equation}
\label{m2}
\frac{1}{u-a}=P\frac{1}{u-a} + i \pi \delta(u-a),
\end{equation}
where $P$ denotes the principal value. We obtain
\begin{eqnarray}
1+{k\over 2}{P}\int_{-\infty}^{+\infty} {{f'(v)}\over
v-\omega_{r}}dv+i\pi {k\over 2} f'(\omega_{r})=0.
\label{m3}
\end{eqnarray}
Identifying the real and the
imaginary parts, if follows that the marginal mode is determined by the equations \cite{cvb}:
\begin{eqnarray}
1+{k\over 2}\int_{-\infty}^{+\infty} {{f'(v)}\over v-\omega_{r}}dv=0,
\label{m4}
\end{eqnarray}
\begin{eqnarray}
f'(\omega_{r})=0.
\label{m5}
\end{eqnarray}
The second relation fixes the pulsation $\omega_{r}$ of the
perturbation and the first relation determines the point of marginal
stability in the series of equilibria \footnote{When condition
(\ref{m5}) is fulfilled the integral is convergent in $v=\omega_r$ so
that the principal part is not necessary.}.  Note that the
distribution $f(v)$ can be  relatively arbitrary. However, if $f(v)$ has a
single maximum at $v=0$, then $\omega_{r}=0$ (implying $\omega=0$) and
the condition of marginal stability becomes
\begin{eqnarray}
1+{k\over 2}\int_{-\infty}^{+\infty} {{f'(v)}\over v}dv=0.
\label{m6}
\end{eqnarray}

\subsection{The Nyquist Method}
\label{sec_nyquist}

To determine whether the distribution $f=f(v)$ is stable or unstable,
one possibility is to solve the dispersion relation (\ref{dr2}) and
determine the sign of the imaginary part of the complex
pulsation. This can be done analytically in some simple cases
\cite{cvb}. In this paper, we use another strategy. We shall
apply the Nyquist method introduced in plasma physics. This is a
graphical method that does not require to solve the dispersion
relation. The details of the method are explained in \cite{nicholson}
and we just recall how it works in practice. In the
$\epsilon$-plane, we plot the Nyquist curve \footnote{This curve is
also called an hodograph.}
$(\epsilon_{r}(\omega_{r}),\epsilon_{i}(\omega_{r}))$ parameterized by
$\omega_{r}$ going from $-\infty$ to $+\infty$. This curve is closed
and always rotates in the counterclockwise sense. If the Nyquist curve
does not encircle the origin, the system is stable. If the Nyquist
curve encircles the origin one or more times, the system is
unstable. The number $N$ of tours around the origin gives the number
of zeros of $\epsilon(\omega)$ in the upper half plane, i.e. the
number of unstable modes with $\omega_{i}>0$.  The Nyquist method by
itself does not give the growth rate of the instability.

\subsection{General properties}
\label{sec_gp}

Let us determine some general properties of the Nyquist curve for
the HMF model. Taking $\omega_{i}=0$, we have
\begin{eqnarray}
\epsilon(\omega_{r})=1+{k\over 2}{P}\int_{-\infty}^{+\infty} {{f'(v)}\over v-\omega_{r}}dv+i\pi {k\over 2} f'(\omega_{r}).
\label{gp1}
\end{eqnarray}
Therefore, the real and imaginary parts of the dielectric function  $\epsilon(\omega_{r})=\epsilon_{r}(\omega_{r})+i\epsilon_{i}(\omega_{r})$ are
\begin{eqnarray}
\epsilon_{r}(\omega_{r})=1+{k\over 2}{P}\int_{-\infty}^{+\infty} {{f'(v)}\over v-\omega_{r}}dv,
\label{gp2}
\end{eqnarray}
\begin{eqnarray}
\epsilon_{i}(\omega_{r})=\pi {k\over 2} f'(\omega_{r}).
\label{gp3}
\end{eqnarray}
To apply the Nyquist method, we have to plot the curve
$(\epsilon_{r}(\omega_{r}),\epsilon_{i}(\omega_{r}))$ parameterized by
$\omega_{r}$ going from $-\infty$ to $+\infty$. Let us consider the
asymptotic behavior for $\omega_{r}\rightarrow \pm\infty$. Since $f(v)$
tends to zero for $v\rightarrow \pm\infty$, we conclude that
$\epsilon_{i}(\omega_{r})\rightarrow 0$ for $\omega_{r}\rightarrow
\pm\infty$ and that  $\epsilon_{i}(\omega_{r})>0$ for
$\omega_{r}\rightarrow -\infty$ while $\epsilon_{i}(\omega_{r})<0$ for
$\omega_{r}\rightarrow +\infty$.  On the other hand, integrating by
parts in Eq. (\ref{gp2}), we obtain
\begin{eqnarray}
\epsilon_{r}(\omega_{r})=1+{k\over 2}{P}\int_{-\infty}^{+\infty} {{f(v)}\over (v-\omega_{r})^{2}}dv,
\label{gp4}
\end{eqnarray}
provided that $f(v)$ decreases sufficiently rapidly. Therefore, for $\omega_{r}\rightarrow \pm\infty$, we obtain at leading order
\begin{eqnarray}
\epsilon_{r}(\omega_{r})\simeq 1+{k\over 2}\frac{\rho}{\omega_{r}^{2}}, \qquad (\omega_{r}\rightarrow \pm\infty).
\label{gp5}
\end{eqnarray}
In particular, $\epsilon_{r}(\omega_{r})\rightarrow 1$ for $\omega_{r}\rightarrow
\pm\infty$.
From these results, we conclude that the behavior of the Nyquist curve
close to the limit point $(1,0)$ is the one represented in
Fig. \ref{maxwell}. In addition, according to Eq. (\ref{gp3}), the
Nyquist curve crosses the $x$-axis at each value of $\omega_{r}$
corresponding to an extremum of $f(v)$. For $\omega_r=v_{ext}$, where
$v_{ext}$ is a velocity at which the distribution is extremum
$(f'(v_{ext})=0)$, the imaginary part of the dielectric function
$\epsilon_i(v_{ext})=0$ and the real part of the dielectric function
\begin{equation}
\epsilon_r(v_{ext})= 1+\frac{k}{2} \int_{-\infty}^{\infty}\frac{f'(v)}{v-v_{ext}}\, dv.
\end{equation}
Subtracting the value $f'(v_{ext})=0$ in the numerator of the integrand, and  integrating by parts, we obtain
\begin{equation}
\epsilon_r(v_{ext})= 1- \frac{k}{2}\int_{-\infty}^{\infty}\frac{f(v_{ext})-f(v)}{(v-v_{ext})^2}\, dv.
\label{gp6}
\end{equation}
If $v_{Max}$ denotes the velocity corresponding to the global maximum of
the distribution, we clearly have
\begin{eqnarray}
\epsilon_{r}(v_{Max}) <1.
\label{gp7}
\end{eqnarray}
Finally, for an attractive interaction, we note that if the zero of
$f'(v)$ corresponds to a maximum (resp. minimum) of $f$ then the
hodograph crosses the real axis downward (resp. upward). On the other
hand, if $f(v)$ is symmetric with respect to $v=0$, then
$\epsilon_{i}(\omega_r=0)=\pi\frac{k}{2}f'(0)=0$ and
$\epsilon_{r}'(\omega_r=0)=\frac{k}{2}P\int_{-\infty}^{+\infty}\frac{f'(v)}{v^2}dv=0$
so that the Nyquist curve has a vertical tangent at
$(\epsilon_{r}(0),0)$ when $\omega_{r}=0$ provided that
$\epsilon_{i}'(\omega_r=0)\propto f''(0)\neq 0$.

\subsection{Single-humped distributions}
\label{sec_sh}

Let us assume that the distribution $f(v)$ has a single maximum at
$v=v_{0}$ (so that $f'(v_0)=0$) and tends to zero for $v\rightarrow
\pm \infty$. Then, the Nyquist curve cuts the $x$-axis  ($\epsilon_{i}(\omega_{r})$ vanishes) at the limit point
$(1,0)$ when $\omega_{r}\rightarrow \pm \infty$ and at the point
$(\epsilon_{r}(v_0),0)$ when $\omega_{r}=v_{0}$. Due to its behavior
close to the limit point $(1,0)$, the fact that it rotates in the
counterclockwise sense, and the property that $\epsilon_r(v_0)<1$ according to Eq. (\ref{gp7}),
the Nyquist curve must necessarily behave like in
Fig. \ref{maxwell}.  Therefore, the Nyquist curve starts on the real
axis at $\epsilon_r(\omega_r) =1$ for $\omega_r \rightarrow -\infty$,
then going in counterclockwise sense it crosses the real axis at the
point $\epsilon_r(v_0)<1$ and returns on the real axis at
$\epsilon_r(\omega_r) =1$ for $\omega_r \rightarrow +
\infty$. According to the Nyquist criterion exposed in Sec. \ref{sec_nyquist}, we
conclude that a single-humped distribution is linearly
stable if
\begin{eqnarray}
\epsilon_{r}(v_0)=1+{k\over 2}\int_{-\infty}^{+\infty} {{f'(v)}\over v-v_0}dv>0,
\label{sh1}
\end{eqnarray}
and linearly unstable if $\epsilon_{r}(v_0)<0$. The equality
corresponds to the marginal stability condition
(\ref{m4})-(\ref{m5}). When the system is unstable, there is only one
mode of instability since the Nyquist curve rotates only one time
around the origin. This stability criterion, which is valid for
single-humped distributions, was stated without proof in
\cite{cvb}. It is here {\it justified} from the Nyquist method.
In particular, a symmetric distribution $f=f(v)$ with a single maximum at
$v_{0}=0$ is linearly dynamically stable iff
\begin{eqnarray}
1+{k\over 2}\int_{-\infty}^{+\infty} {{f'(v)}\over v}dv>0.
\label{sh2}
\end{eqnarray}
This coincides with the criterion of formal stability given in
\cite{yamaguchi,cvb}. Therefore, for spatially homogeneous distributions,
the criterion of linear dynamical stability coincides with
the criterion of formal nonlinear dynamical stability (see Sec. \ref{sec_genr}).

\subsection{Double-humped distributions}
\label{sec_doubh}

Let us consider a double-humped distribution with a global maximum at
$v_{Max}$, a minimum at $v_{min}$ and a local maximum at $v_{max}$. We
assume that $v_{Max}<v_{min}<v_{max}$. The Nyquist curve will cut the
$x$-axis at the limit point $(1,0)$ and at three other points
$(\epsilon_{r}(v_{Max}),0)$, $(\epsilon_{r}(v_{min}),0)$ and
$(\epsilon_{r}(v_{max}),0)$. We also know that the Nyquist curve can only
rotate in the counterclockwise sense and that $\epsilon_{r}(v_{Max}) <1$ according to Eq.
(\ref{gp7}). Then, we can convince ourselves by making drawings of the following results. If

$(+++)$: $\epsilon_{r}(v_{Max})>0$, $\epsilon_{r}(v_{min})>0$, $\epsilon_{r}(v_{max})>0$,

$(+--)$: $\epsilon_{r}(v_{Max})>0$, $\epsilon_{r}(v_{min})<0$, $\epsilon_{r}(v_{max})<0$,

$(--+)$: $\epsilon_{r}(v_{Max})<0$, $\epsilon_{r}(v_{min})<0$, $\epsilon_{r}(v_{max})>0$,

$(+-+)$ \footnote{This case is relatively tricky. It is obtained by
making two loops in the S-E and N-W quadrants of the $\epsilon$-plane
(assuming that this is possible for some double-humped
distributions). The origin is then encircled by a sort of triangle but
since the curve rotates clockwise along this triangle, the origin is
in fact exterior to the Nyquist curve. Thus the system is stable.}:
$\epsilon_{r}(v_{Max})>0$, $\epsilon_{r}(v_{min})<0$,
$\epsilon_{r}(v_{max})>0$,

\noindent the Nyquist curve does not encircle the origin so the
system is stable. If

$(---)$: $\epsilon_{r}(v_{Max})<0$, $\epsilon_{r}(v_{min})<0$, $\epsilon_{r}(v_{max})<0$,

$(-++)$: $\epsilon_{r}(v_{Max})<0$, $\epsilon_{r}(v_{min})>0$, $\epsilon_{r}(v_{max})>0$,

$(++-)$:  $\epsilon_{r}(v_{Max})>0$, $\epsilon_{r}(v_{min})>0$, $\epsilon_{r}(v_{max})<0$,

\noindent the Nyquist curve rotates one time
around the origin so that there is one mode of instability. Finally, if

$(-+-)$: $\epsilon_{r}(v_{Max})<0$, $\epsilon_{r}(v_{min})>0$, $\epsilon_{r}(v_{max})<0$,

\noindent the Nyquist curve rotates two times around the origin
so that there are two modes of instability. Cases   $(+++)$, $(---)$, $(-++)$ and  $(-+-)$ are observed in
Sec. \ref{sec_adh} for an asymmetric double-humped distribution
made of two Maxwellians. The other cases cannot be obtained from this
distribution but they may be obtained from other distributions.

If the double-humped distribution is symmetric with respect to the
origin with two maxima at $\pm v_{*}$ and a minimum at $v=0$, only
three cases can arise.  If

$(+++)$: $\epsilon_{r}(v_{*})>0$, $\epsilon_{r}(0)>0$,

$(+-+)$: $\epsilon_{r}(v_{*})>0$, $\epsilon_{r}(0)<0$,

\noindent the Nyquist curve does not encircle the origin so the
system is stable. If

$(---)$: $\epsilon_{r}(v_{*})<0$, $\epsilon_{r}(0)<0$,

\noindent the Nyquist curve rotates one time
around the origin so that there is one mode of instability. Finally, if

$(-+-)$: $\epsilon_{r}(v_{*})<0$, $\epsilon_{r}(0)>0$,

\noindent the Nyquist curve rotates two times around the origin
so that there are two modes of instability. Cases  $(+++)$, $(---)$ and  $(-+-)$ are
observed in Sec. \ref{sec_sdh} for a symmetric double-humped
distribution made of two Maxwellians.

\subsection{A particular solution of $\epsilon(\omega)=0$}
\label{sec_exp}

For the attractive HMF model  we can
look for a solution of the equation $\epsilon(\omega)=0$ in the form
$\omega=i\omega_{i}$ corresponding to $\omega_{r}=0$. In that case,
the perturbation grows ($\omega_{i}>0$) or decays ($\omega_{i}<0$)
without oscillating. For $\omega_{i}>0$, the equation
$\epsilon(\omega)=0$ becomes
\begin{eqnarray}
1+\frac{k}{2}\int_{-\infty}^{+\infty}\frac{f'(v)}{v-i\omega_{i}}\, dv=0.
\label{mon1}
\end{eqnarray}
Multiplying the numerator by $v+i\omega_{i}$ and separating real and imaginary parts, we obtain
\begin{eqnarray}
1+\frac{k}{2}\int_{-\infty}^{+\infty}\frac{v f'(v)}{v^2+\omega_{i}^2}\, dv=0,
\label{mon2}
\end{eqnarray}
\begin{eqnarray}
\int_{-\infty}^{+\infty}\frac{f'(v)}{v^2+\omega_{i}^2}\, dv=0.
\label{mon3}
\end{eqnarray}
If we consider distribution functions $f(v)$ that are symmetric with
respect to $v=0$, Eq. (\ref{mon3}) is always satisfied. Then, the
growth rate $\omega_i$ is given by Eq. (\ref{mon2}).

For $\omega_{i}<0$, the equation $\epsilon(\omega)=0$ becomes
\begin{eqnarray}
1+\frac{k}{2}\int_{-\infty}^{+\infty}\frac{f'(v)}{v-i\omega_{i}}\, dv+ik\pi f'(i\omega_{i})=0.
\label{mon4}
\end{eqnarray}
Multiplying the numerator by $v+i\omega_{i}$, assuming that $f(v)$ is even, and separating real and imaginary parts, we obtain
\begin{eqnarray}
1+\frac{k}{2}\int_{-\infty}^{+\infty}\frac{v f'(v)}{v^2+\omega_{i}^2}\, dv+ik\pi f'(i\omega_{i})=0,
\label{mon5}
\end{eqnarray}
and Eq. (\ref{mon3}). The equation for the imaginary part is always satisfied.
Then, the damping rate $\omega_i$ is given by Eq. (\ref{mon5}). 

Note that the solution $\omega=i\omega_i$ exists only for attractive
interactions. Note also that there may exist other solutions to the
equation $\epsilon(\omega)=0$. However, for unstable single-humped
distributions, the Nyquist curve encircles the origin only once (see
Sec. \ref{sec_sh}) so that $\omega=i\omega_i$ with $\omega_{i}>0$ is
the only solution of $\epsilon(\omega)=0$ (for stable single-humped
distributions there may be other solutions of $\epsilon(\omega)=0$
with $\omega_{r}\neq 0$ and $\omega_{i}<0$). Explicit solutions of
$\epsilon(\omega)=0$ with $\omega=i\omega_{i}$ are given in \cite{cvb}
for the Maxwell and Tsallis distributions.

\section{The Maxwell distribution}
\label{sec_maxwell}

We consider the Maxwell (or isothermal) distribution
\begin{equation}
\label{mm1}
f(v)=\left (\frac{\beta}{2 \pi}\right )^{1/2}\rho\, e^{-\beta \frac{v^2}{2}},
\end{equation}
where $\rho=M/(2\pi)$ is the spatial density and $T\equiv
1/\beta=\langle v^2\rangle$ is the kinetic temperature \footnote{For
any distribution function $f({\bf v})$, we define the kinetic
temperature by the relation $\frac{1}{2}m\langle ({\bf v}-\langle {\bf
v}\rangle)^2\rangle=\frac{d}{2}T_{kin}$ where $d$ is the dimension of
space. The energy (conserved control parameter) is then
$E=\frac{1}{2}Nm\langle v^2\rangle=\frac{1}{2}Nm\langle
v\rangle^{2}+\frac{d}{2}NT_{kin}$.  For one dimensional systems
($d=1$) and particles of unit mass ($m=1$), the kinetic temperature is
equal to the velocity dispersion $T_{kin}=\langle v^2\rangle$ when
$\langle v\rangle =0$.}. We justify here the Maxwell distribution as a
particular stationary solution of the Vlasov equation
\cite{cvb,cstsallis}.  The Maxwell distribution represents also the
statistical equilibrium state of the system. In that case, $T$ can be
interpreted as the thermodynamical temperature.

The Maxwellian distribution has a single maximum at $v=0$. Therefore,
the condition of marginal stability (\ref{m5}) implies $\omega_{r}=0$. From
Eq. (\ref{m4}), we find that the Maxwellian distribution is marginally stable
for $T=T_c$ where we have introduced the critical temperature
\begin{equation}
\label{mm2}
T_{c}=\frac{kM}{4\pi}.
\end{equation}
According to the criterion (\ref{sh2}), the Maxwell distribution is linearly dynamically  stable if $T>T_{c}$ and linearly dynamically unstable if $T<T_{c}$.

\begin{figure}
\begin{center}
\includegraphics[clip,scale=0.3]{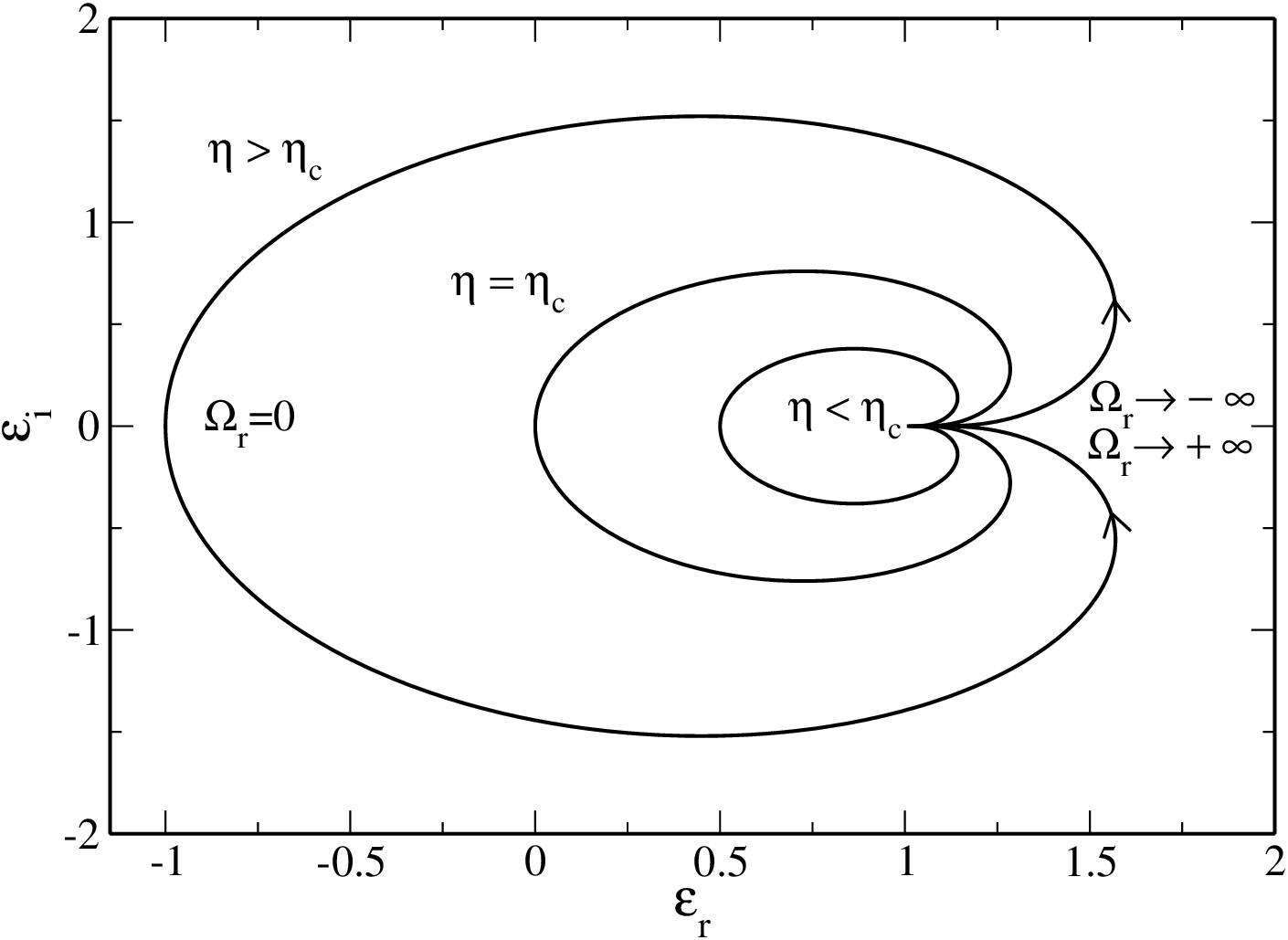}
\caption{Nyquist curve for the Maxwellian distribution (\ref{mm1}). The DF is stable for $\eta<\eta_c$ ($T>T_c$), marginally stable for $\eta=\eta_c$ ($T=T_c$) and unstable for $\eta>\eta_c$ ($T<T_c$). We have taken $\eta= 2, 1, 0.5$ from the outer to the inner curve. }
\label{maxwell}
\end{center}
\end{figure}

The dielectric function (\ref{dr2}) associated to the Maxwellian
distribution is
\begin{equation}
\label{mm3}
\epsilon(\omega)=1- \frac{k}{2}\left (\frac{\beta}{2 \pi}\right )^{1/2}\rho  \int_{C} \frac{\beta v}{v-\omega}e^{-\beta \frac{v^2}{2}}\, dv.
\end{equation}
Introducing the dimensionless temperature
\begin{equation}
\label{mm4}
\eta=\frac{\beta kM}{4\pi},
\end{equation}
and the dimensionless pulsation and velocity
\begin{equation}
\label{mm5}
\Omega=\left (\frac{4\pi}{kM}\right )^{1/2}\omega,\qquad  V=\left (\frac{4\pi}{kM}\right )^{1/2} v,
\end{equation}
the dielectric function can be rewritten
\begin{equation}
\label{mm6}
\epsilon(\Omega)=1-\eta W(\sqrt{\eta}\Omega),
\end{equation}
where
\begin{eqnarray}
W(z)={1\over\sqrt{2\pi}}\int_{C}{x\over x-z}e^{-{x^{2}\over 2}}\, dx
 \label{mm7}
\end{eqnarray}
is the $W$-function of plasma physics \cite{balescubook,ichimaru}. We note
that $W(0)=1$. For any complex number $z$, we have the analytical
expression
\begin{equation}
\label{mm8}
W(z)=1-z e^{-\frac{z^2}{2}}\int_{0}^{z}e^{\frac{x^2}{2}}\, dx + i
\sqrt{\frac{\pi}{2}}z e^{-\frac{z^2}{2}}.
\end{equation}

When $\Omega_{i}=0$, the real and imaginary parts of the dielectric function  $\epsilon(\Omega_{r})=\epsilon_{r}(\Omega_{r})+i\epsilon_{i}(\Omega_{r})$ can be written
\begin{equation}
\label{mm9}
\epsilon_{r}(\Omega_{r})=1-\eta W_{r}(\sqrt{\eta}\Omega_r),
\end{equation}
\begin{equation}
\label{mm10}
\epsilon_{i}(\Omega_{r})=-\eta W_{i}(\sqrt{\eta}\Omega_r),
\end{equation}
with
\begin{equation}
\label{mm11}
W_{r}(z)=1-z e^{-\frac{z^2}{2}}\int_{0}^{z}e^{\frac{x^2}{2}}\, dx,
\end{equation}
\begin{equation}
\label{mm12}
W_{i}(z)=\sqrt{\frac{\pi}{2}}z e^{-\frac{z^2}{2}},
\end{equation}
where $z$ is here a real number. The condition of marginal stability corresponds to $\epsilon_{r}(\Omega_{r})=\epsilon_{i}(\Omega_{r})=0$. The condition $\epsilon_{i}(\Omega_{r})=0$, which is equivalent to $f'(\Omega_{r})=0$, implies $\Omega_{r}=0$. Then, the condition $\epsilon_{r}(\Omega_{r})=0$ leads to $\eta=\eta_{c}$ with
\begin{equation}
\label{mm13}
\eta_{c}=1.
\end{equation}

To apply the Nyquist method, we need to plot the curve
$(\epsilon_{r}(\Omega_{r}),\epsilon_{i}(\Omega_{r}))$ in the
$\epsilon$-plane. For $\Omega_{r}\rightarrow \pm \infty$, this curve
tends to the point $(1,0)$ in the manner described in Sec. \ref{sec_sh}. On the
other hand, for $\Omega_r=0$, it crosses the $x$-axis at
$(\epsilon_r(0)=1-\eta,0)$.  The Nyquist curve is represented in
Fig. \ref{maxwell} for several values of the inverse temperature $\eta$. For
$\eta < 1$ (i.e. $T > T_c$), the Nyquist curve does not encircle the
origin so that the Maxwellian distribution is stable. For $\eta > 1$
(i.e. $T < T_c$) the Nyquist curve encircles the origin so that the
Maxwellian distribution is unstable. For $\eta = 1$
(i.e. $T = T_c$) the Nyquist curve passes through the origin  so that the
Maxwellian distribution is marginally stable.

The stability limits obtained with the Nyquist method for isothermal
and polytropic (see next section) distributions coincide with those
found in \cite{cvb}. However, the Nyquist method does not give the
value of the growth rate (in the unstable regime) or decay rate (in
the stable regime) of the perturbation. These values have been
obtained in \cite{cvb} by solving the dispersion relation (\ref{dr2}).

\section{The Tsallis  distributions}
\label{sec_tsallis}

\begin{figure}
\begin{center}
\includegraphics[clip,scale=0.3]{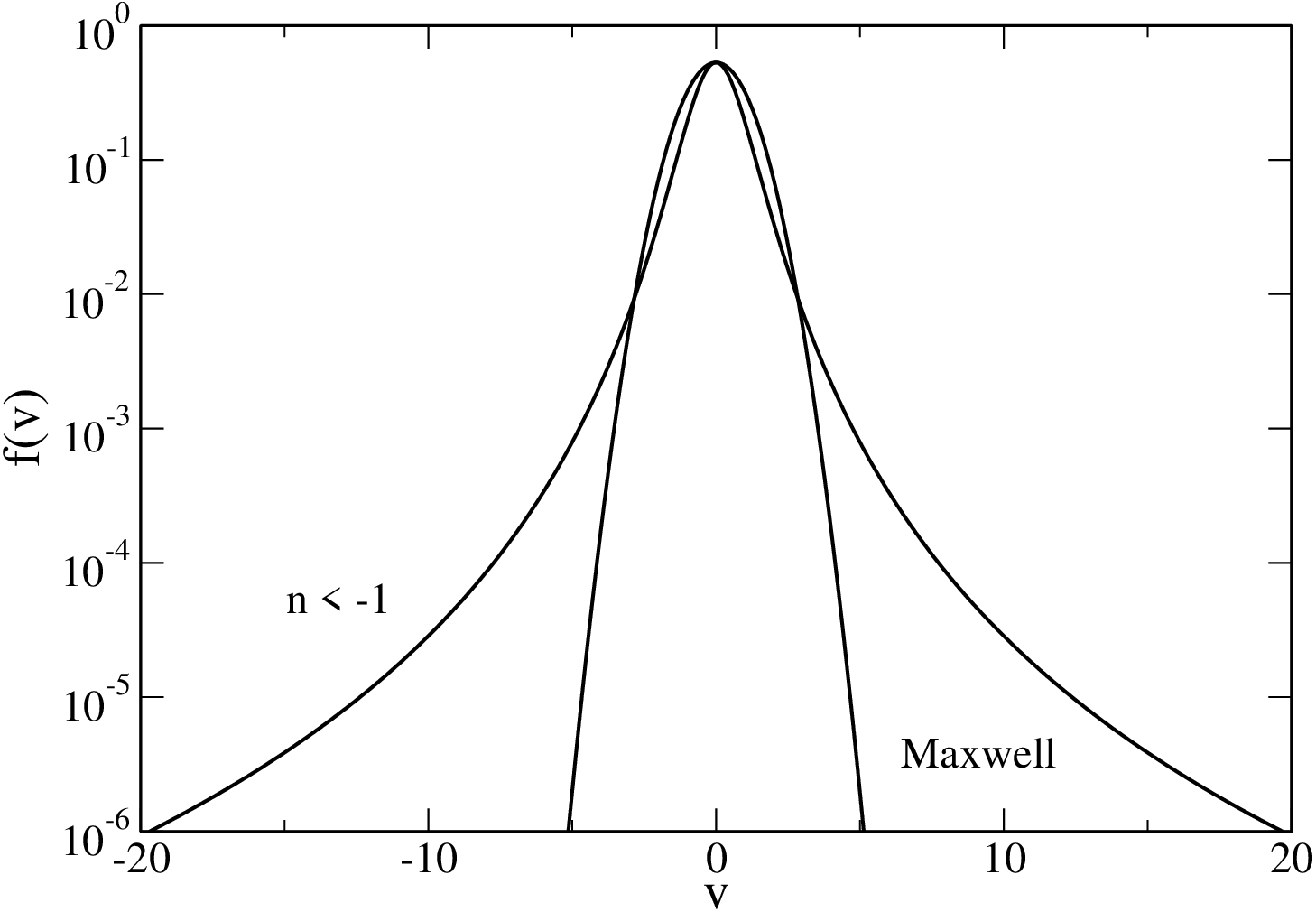}
\caption{Tsallis distribution with $n<-1$ (power-law tail) and Maxwell distribution (corresponding to $n=\pm\infty$).}
\label{dneg}
\end{center}
\end{figure}

\begin{figure}
\begin{center}
\includegraphics[clip,scale=0.3]{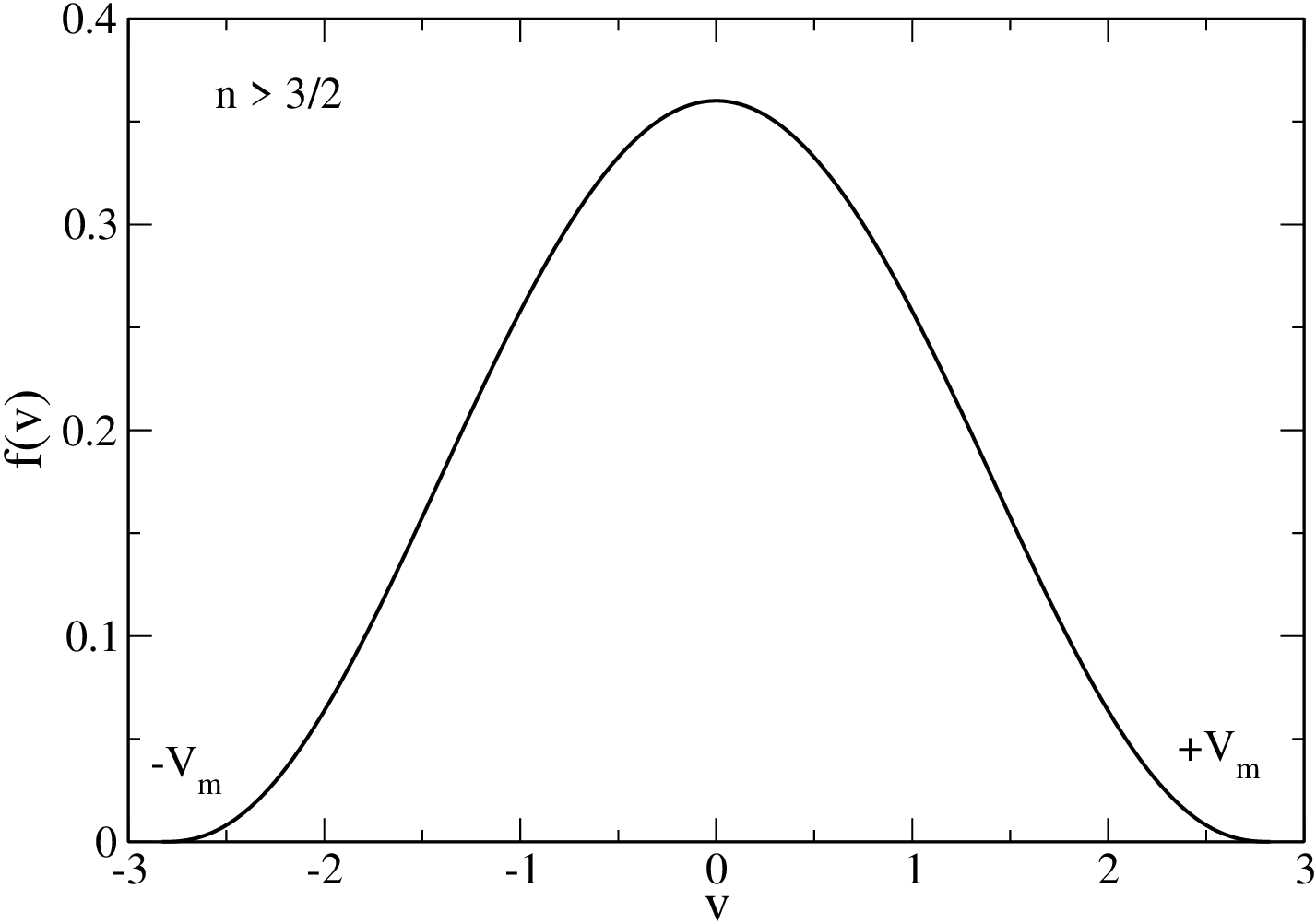}
\caption{Tsallis distribution with $n>3/2$ (specifically $n=3$). It has a compact support with a horizontal tangent at $v=\pm v_{m}$.}
\label{tsallisH}
\end{center}
\end{figure}

\begin{figure}
\begin{center}
\includegraphics[clip,scale=0.3]{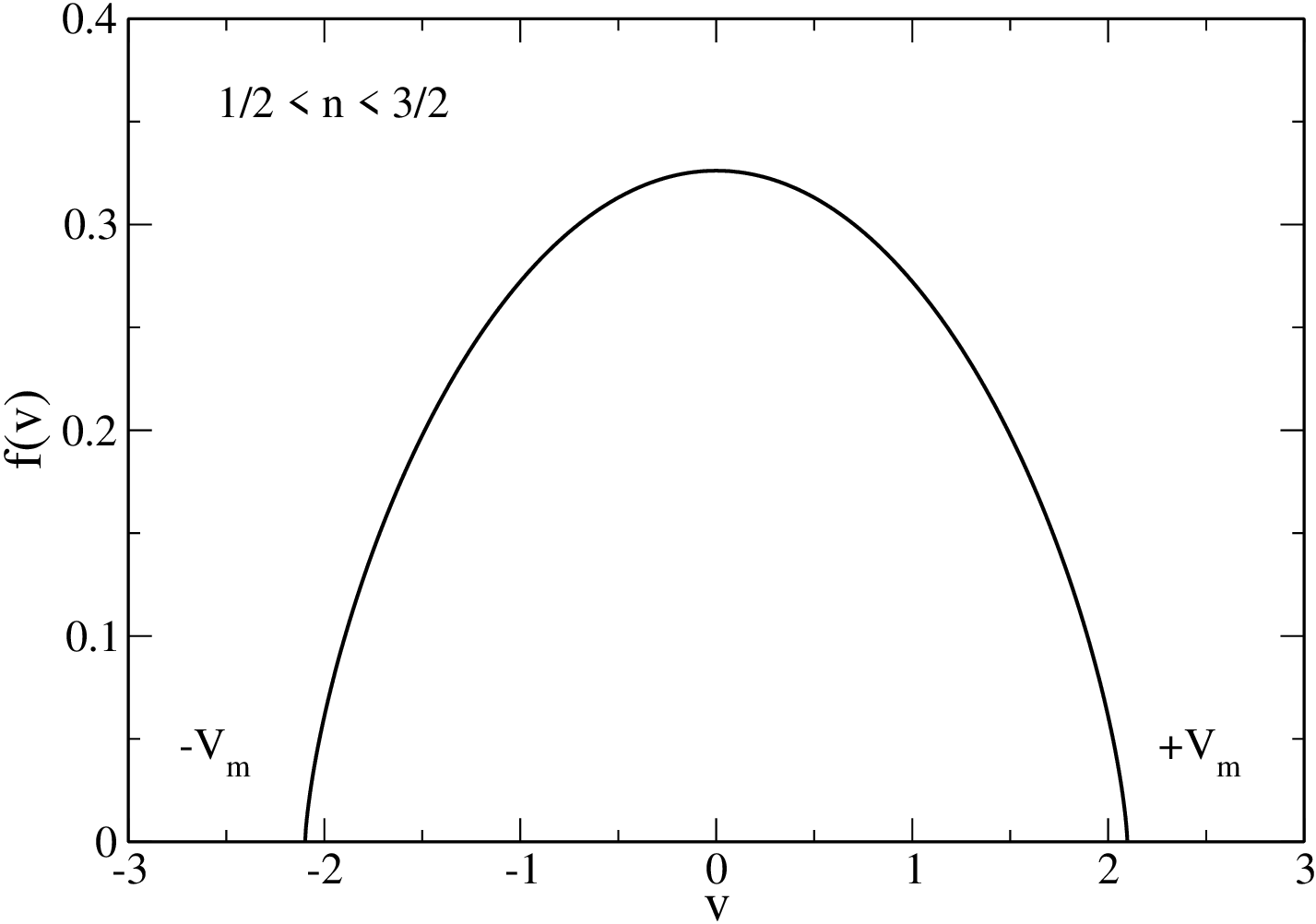}
\caption{Tsallis distribution with $1/2<n<3/2$ (specifically $n=1.2$). It has a compact support with a vertical tangent at $v=\pm v_{m}$.}
\label{tsallisV}
\end{center}
\end{figure}

We consider the Tsallis (or polytropic) distributions written in the form (see \cite{cvb} and Appendix \ref{sec_tc}):
\begin{equation}
\label{t1}
f(v)=B_n \frac{\rho}{\sqrt{2\pi T}}\left [1-\frac{v^2}{2(n+1)T} \right ]_+^{n-1/2},
\end{equation}
where $\rho=\frac{M}{2\pi}$ is the density, $T\equiv 1/\beta=\langle v^2\rangle$ is the kinetic temperature and $B_n$ is a normalization constant given by
\begin{eqnarray}
\label{t2}
B_n&=&\frac{\Gamma(n+1)}{\Gamma(n+1/2)(n+1)^{1/2}}, \qquad n>{1\over 2},\\
B_n&=&\frac{\Gamma(1/2-n)}{\Gamma(-n)[-(n+1)]^{1/2}}, \qquad n<{-1}.
\end{eqnarray}
We limit ourselves to these indices so that the distributions are
normalizable and their second moments $\langle v^2\rangle$ exist. Some
representative distributions are plotted in Figs. \ref{dneg},
\ref{tsallisH} and \ref{tsallisV}. For $n > 1/2 $ the distributions
have a compact support since they vanish at $v=\pm v_{m}$ where
\begin{equation}
\label{t3}
v_{m}=\sqrt{2(n+1)T}.
\end{equation}
For $v > v_{m}$, we set $f=0$. Therefore, the notation $[x]_+=x$ for
$x>0$ and $[x]_+=0$ for $x<0$. For $n<-1$, the distribution functions
remain strictly positive for all $v$ and they decrease algebraically
like $|v|^{-(1-2n)}$ for $v\rightarrow \pm\infty$. For $n\rightarrow
\pm\infty$, we recover the Maxwell distribution (\ref{mm1}). We justify
here the Tsallis distributions (\ref{t1}) as particular stationary
solutions of the Vlasov equation \cite{cvb,cstsallis}. These
distributions possess a lot of interesting mathematical properties so
that their study is interesting in its own right
\footnote{Such distributions have been extensively studied in
astrophysics because they represent an important class of stationary
solutions of the Vlasov-Poisson system called stellar polytropes
\cite{bt}. Stellar polytropes have been introduced by
Plummer \cite{plummer} and Eddington \cite{edd}. However, these
distributions were written in a form that is less convenient than the
``Tsallis form'' (\ref{tc1}) or (\ref{t1}). The nice forms (\ref{tc1}) or (\ref{t1})
allow us to make a smooth connexion between polytropic and isothermal
distributions for $q\rightarrow 1$ or $n\rightarrow\infty$. See
\cite{cstsallis} for further discussion on these historical issues.}. We also
note that the Tsallis distributions (\ref{t1}) have been observed in
numerical simulations of the HMF model either in case of incomplete
relaxation \cite{latora,campa} or during the collisional evolution of
the system \cite{cc} (see discussion in Sec. 3.4 of \cite{hb3}).

The Tsallis distributions have a single maximum at $v=0$. Therefore,
the condition of marginal stability (\ref{m5}) implies
$\omega_{r}=0$ (it is shown below that the minimum at $v=\pm v_m$ for $n>3/2$ does not lead to
marginal stability). From Eq. (\ref{m4}), we find that the Tsallis
distribution of index $n$ is marginally stable for $T=T_c$ where we
have introduced the critical temperature
\begin{equation}
\label{t4}
T_{c}=\frac{n}{n+1}\frac{kM}{4\pi}.
\end{equation}
For $n\rightarrow \pm\infty$, we recover the critical temperature
(\ref{mm2}) corresponding to the Maxwell distribution. On the other
hand, according to the criterion (\ref{sh2}), the Tsallis distribution of
index $n$ is linearly dynamically stable if $T>T_{c}$ and linearly
dynamically unstable if $T<T_{c}$.

The dielectric function associated to the Tsallis distribution of index $n$ is
\begin{eqnarray}
\label{t5}
\epsilon(\omega)=1- \frac{k}{2}\frac{\rho}{\sqrt{2 \pi T}} B_n \left (n-\frac{1}{2}\right )\frac{1}{(n+1)T} \nonumber\\
\times \int_{C} \frac{v\left [1-\frac{v^2}{2(n+1)T} \right ]_+^{n-3/2}}{v-\omega}\, dv.
\end{eqnarray}
Introducing the dimensionless temperature (\ref{mm4})
and the dimensionless pulsation (\ref{mm5}) it can be rewritten
\begin{equation}
\label{t6}
\epsilon(\Omega)=1-\frac{n}{n+1}\eta W^{(n)}(\sqrt{\eta}\Omega),
\end{equation}
where
\begin{eqnarray}
\label{t7}
W^{(n)}(z)= \frac{1}{\sqrt{2\pi}}\frac{B_n}{n}\left (n-\frac{1}{2}\right )\int_{C}\frac{x[1-\frac{x^2}{2(n+1)}]_+^{n-3/2}}{x-z}dx,\nonumber\\
\end{eqnarray}
generalizes the $W$ function of plasma physics to the case of Tsallis distributions \cite{cvb}. We note that $W^{(n)}(0)=1$.

When $\Omega_{i}=0$, the real and imaginary parts of the dielectric function $\epsilon(\Omega_{r})=\epsilon_{r}(\Omega_{r})+i\epsilon_{i}(\Omega_{r})$ can be written
\begin{equation}
\label{t8}
\epsilon_{r}(\Omega_{r})=1-\frac{n}{n+1}\eta W_{r}^{(n)}(\sqrt{\eta}\Omega_r),
\end{equation}
\begin{equation}
\label{t9}
\epsilon_{i}(\Omega_{r})=-\frac{n}{n+1}\eta W_{i}^{(n)}(\sqrt{\eta}\Omega_r),
\end{equation}
with
\begin{eqnarray}
\label{t10}
W_{r}^{(n)}(z)=\frac{1}{\sqrt{2\pi}}\frac{B_n}{n}\left (n-\frac{1}{2}\right )P\int_{-\infty}^{+\infty}\frac{x[1-\frac{x^2}{2(n+1)}]_{+}^{n-3/2}}{x-z}dx,\nonumber\\
\end{eqnarray}
\begin{equation}
\label{t11}
W_{i}^{(n)}(z)=\sqrt{\frac{\pi}{2}}\frac{B_n}{n}\left (n-\frac{1}{2}\right )z\left [1-\frac{z^2}{2(n+1)}\right ]_+^{n-3/2},
\end{equation}
where $z$ is here a real number. The condition of marginal stability
corresponds to
$\epsilon_{r}(\Omega_{r})=\epsilon_{i}(\Omega_{r})=0$. The condition
$\epsilon_{i}(\Omega_{r})=0$, which is equivalent to
$f'(\Omega_{r})=0$, implies $\Omega_r=0$. Then, the relation $\epsilon_{r}(\Omega_{r})=0$ leads to $\eta=\eta_{c}$ with
\begin{equation}
\label{t12}
\eta_{c}=\frac{n+1}{n}.
\end{equation}

To apply the Nyquist method, we need to plot the curve
$(\epsilon_{r}(\Omega_{r}),\epsilon_{i}(\Omega_{r}))$  in the $\epsilon$-plane.
We have to distinguish different cases:

\begin{figure}
\begin{center}
\includegraphics[clip,scale=0.3]{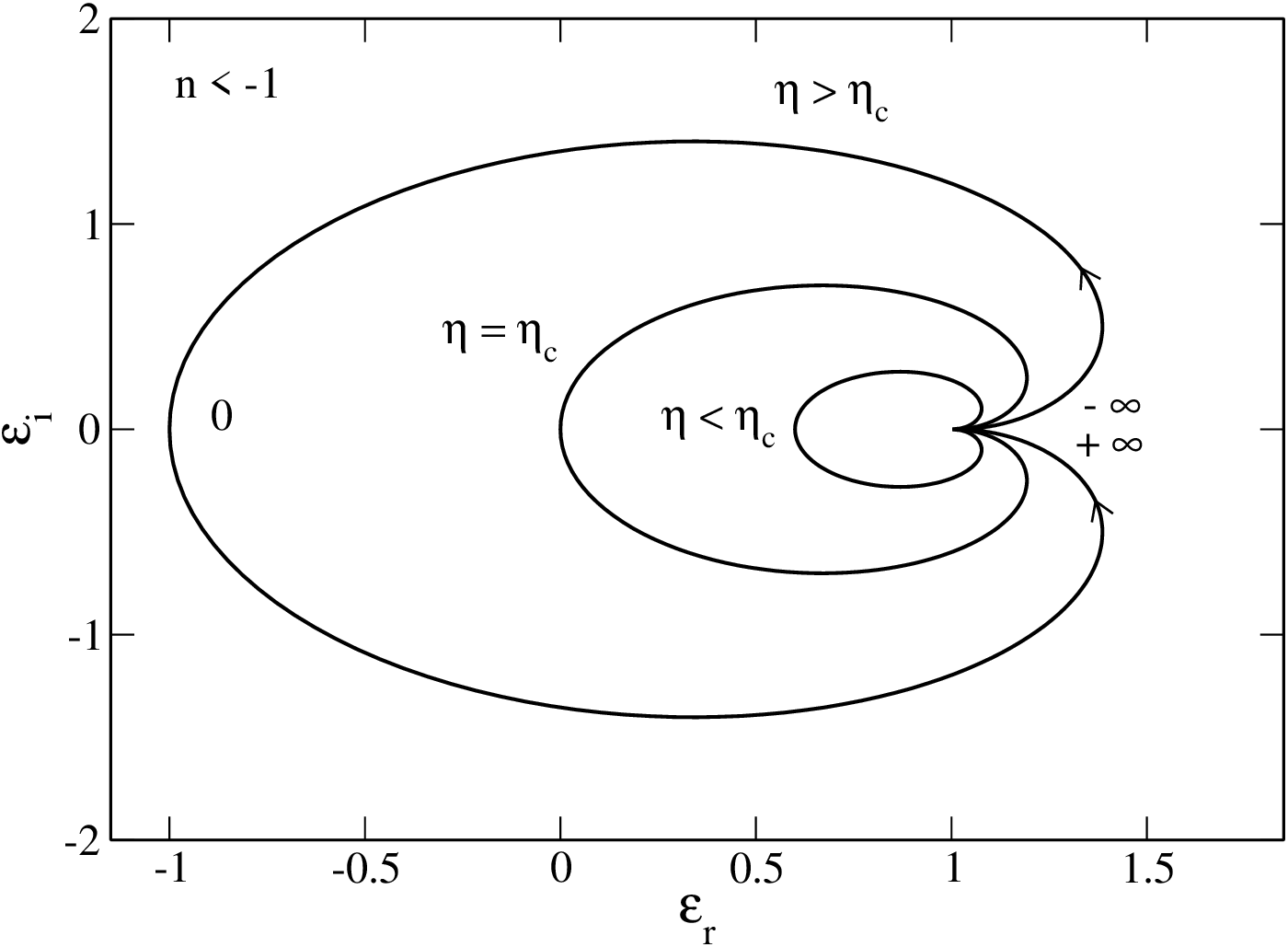}
\caption{Nyquist curve for the Tsallis distributions with $n<-1$ (specifically $n=-2$ yielding $\eta_{c}=1/2$). The DF is stable for $\eta<\eta_c$, marginally stable for $\eta=\eta_c$ and unstable for $\eta>\eta_c$. We have taken $\eta= 1, 0.5, 0.4$ from the outer to the inner curve.}
\label{tsallisneg}
\end{center}
\end{figure}

$\bullet$ For $n<-1$, the Tsallis distributions have a single maximum
at $v=0$ and they decrease algebraically at infinity.  For
$\Omega_{r}\rightarrow \pm \infty$, the curve
$(\epsilon_{r}(\Omega_{r}),\epsilon_{i}(\Omega_{r}))$ tends to the
point $(1,0)$ in the manner described in Sec. \ref{sec_sh}. On the other hand,
for $\Omega_r=0$, it crosses the $x$-axis at
$(\epsilon_r(0)=1-\eta/\eta_c,0)$.  The Nyquist curve is represented
in Fig. \ref{tsallisneg} for several values of the inverse temperature
$\eta$. For $\eta < \eta_c$ (i.e. $T > T_c$), the Nyquist curve does
not encircle the origin so that the Tsallis distribution is
stable. For $\eta > \eta_c$ (i.e. $T < T_c$) the Nyquist curve
encircles the origin so that the Tsallis distribution is unstable. For
$\eta = \eta_c$ (i.e. $T = T_c$) the Nyquist curve passes through the
origin so that the Tsallis distribution is marginally
stable. Therefore, the Nyquist curve of polytropic
distributions with index $n<-1$ is very similar to the Nyquist curve
of Maxwellian distributions (recovered for $n=-\infty$).
They essentially differ  in the value of
the critical temperature $\eta_{c}$.

\begin{figure}
\begin{center}
\includegraphics[clip,scale=0.3]{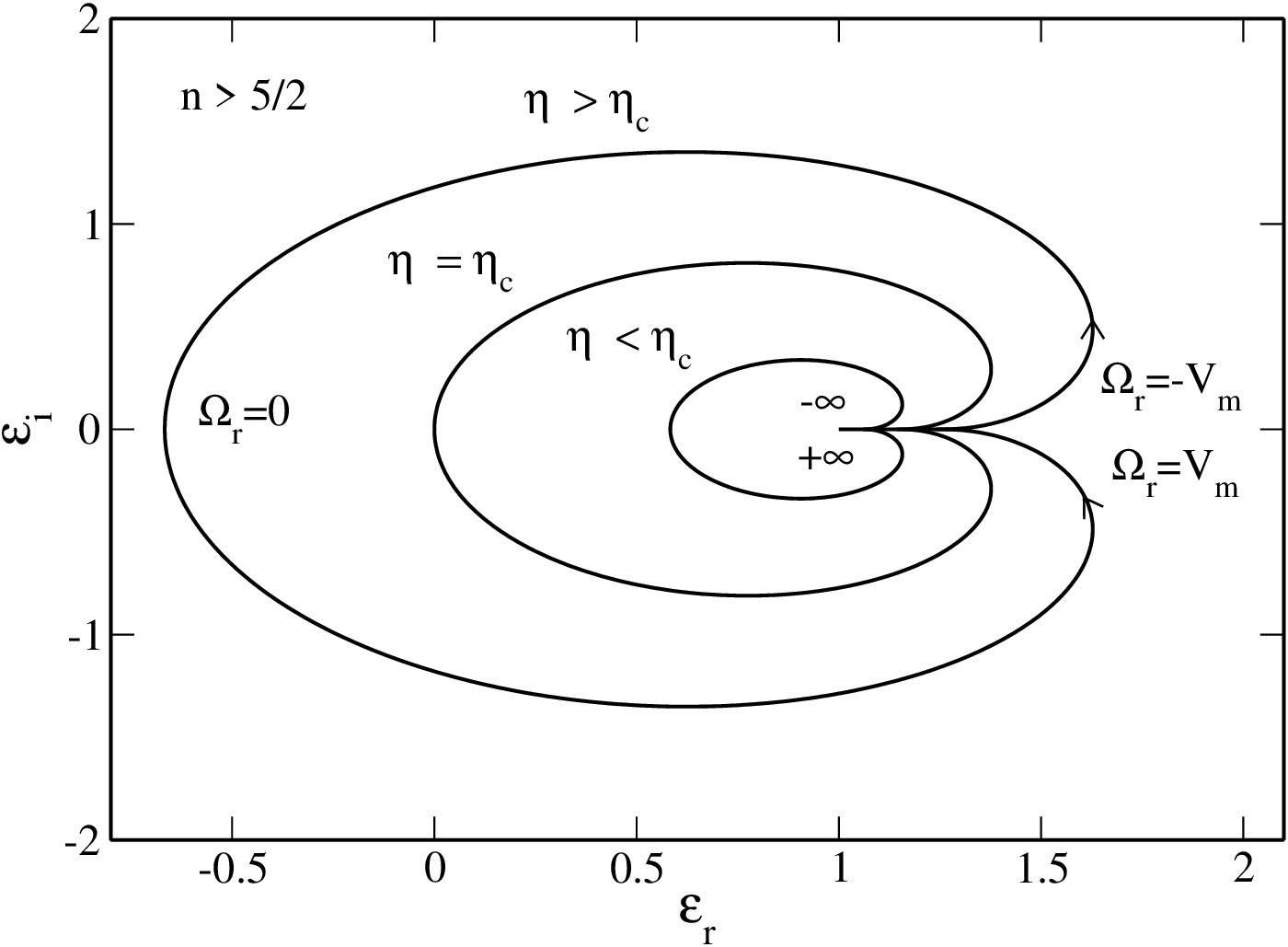}
\caption{Nyquist curve for the Tsallis distribution with $n>5/2$ (specifically $n=5$ yielding $\eta_{c}=6/5$). The DF is stable for $\eta<\eta_c$, marginally stable for $\eta=\eta_c$ and unstable for $\eta>\eta_c$. We have taken $\eta= 2, 6/5, 0.5$ from the outer to the inner curve.  The curve has  a horizontal tangent at $(\epsilon_{r}(V_{m}),0)$.}
\label{tsallis5}
\end{center}
\end{figure}

\begin{figure}
\begin{center}
\includegraphics[clip,scale=0.3]{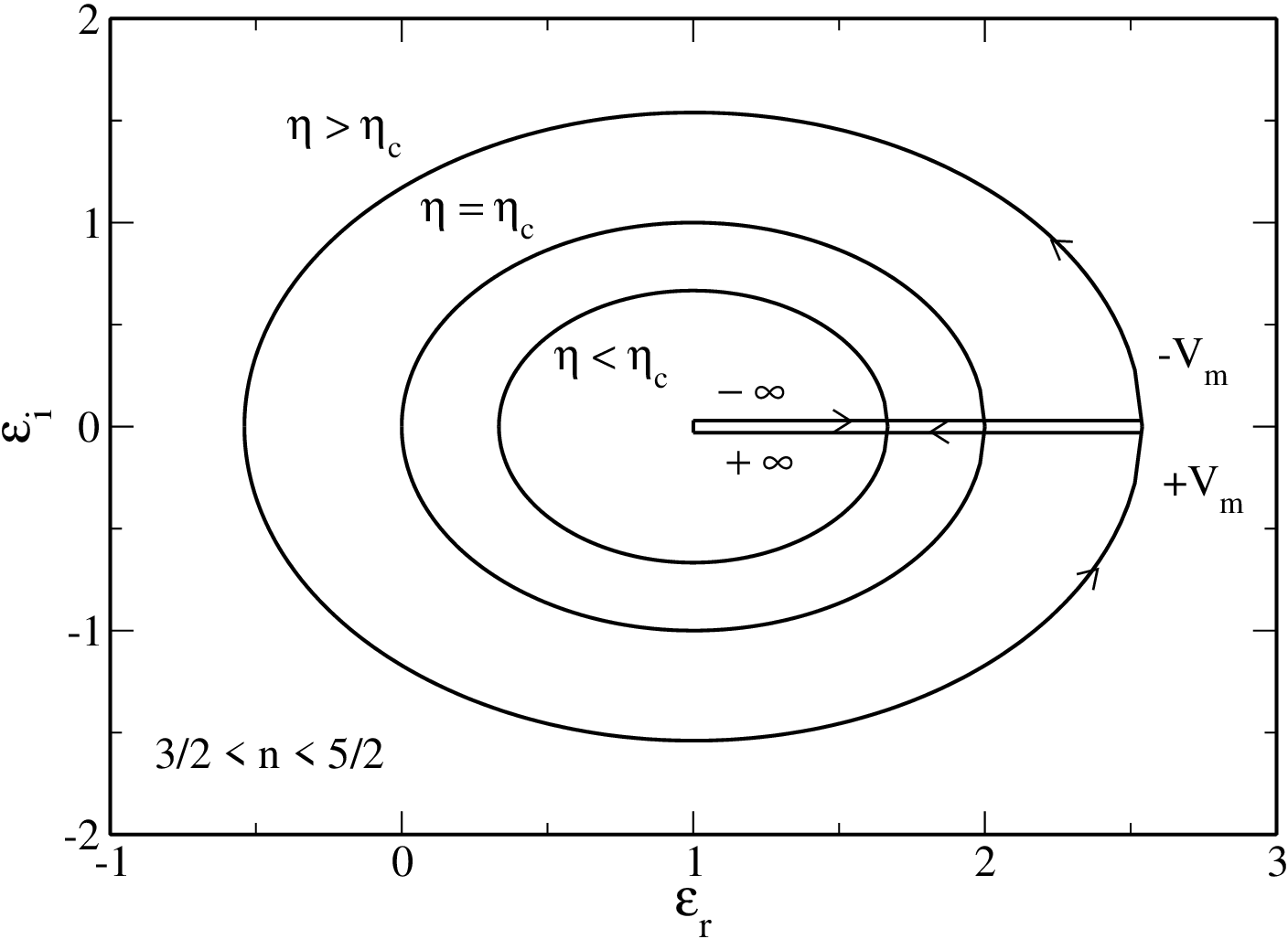}
\caption{Nyquist curve for the Tsallis distribution with $3/2<n<5/2$ (specifically $n=2$ yielding $\eta_{c}=3/2$). The DF is stable for $\eta<\eta_c$, marginally stable for $\eta=\eta_c$ and unstable for $\eta>\eta_c$. We have taken $\eta= 2, 3/2, 0.5$ from the outer to the inner curve. The curve has a vertical tangent at $(\epsilon_{r}(V_{m}),0)$.}
\label{tsallis2}
\end{center}
\end{figure}

\begin{figure}
\begin{center}
\includegraphics[clip,scale=0.3]{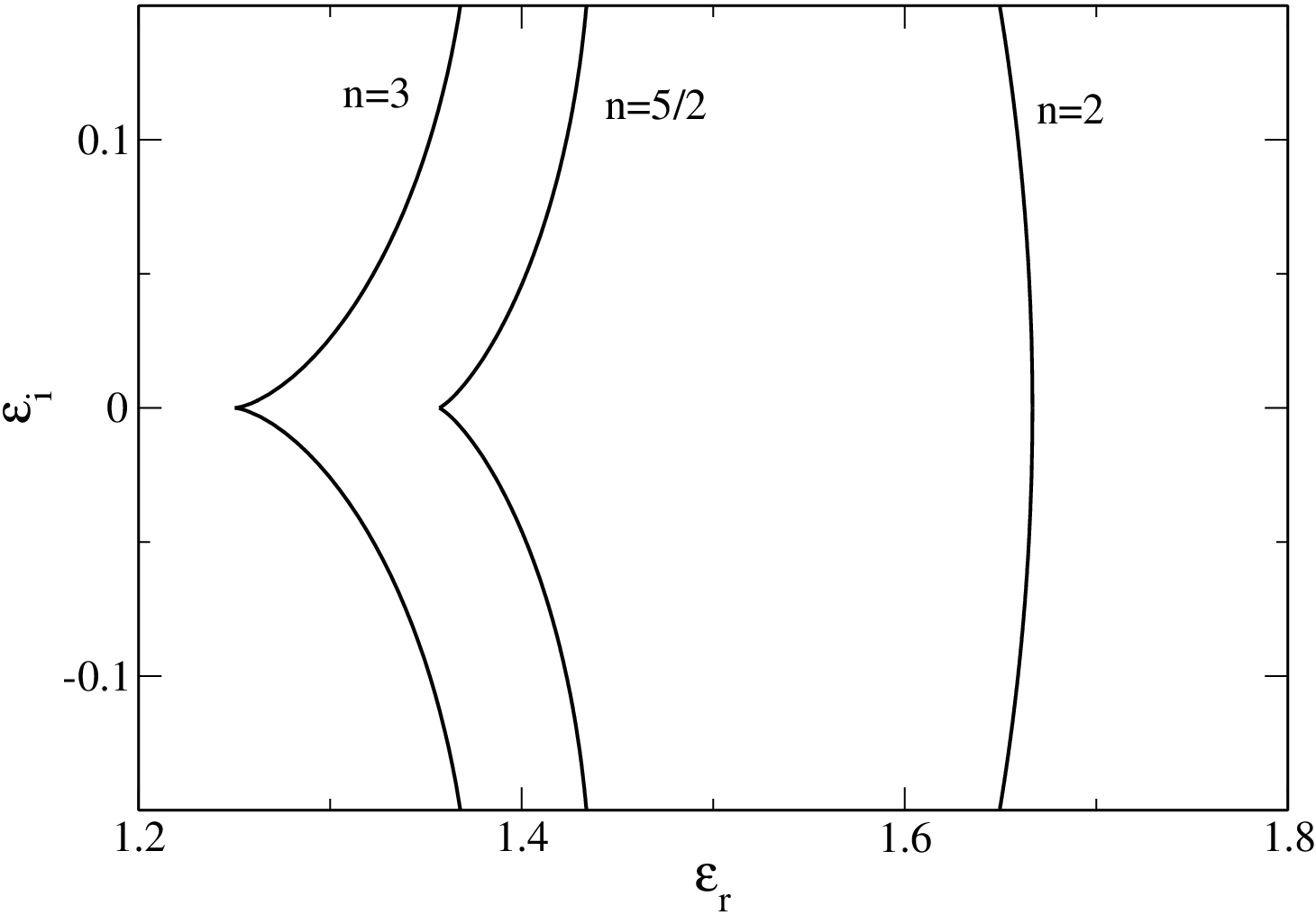}
\caption{Behavior of the Nyquist curve close to the point $(\epsilon_{r}(V_{m}),0)$. For $n>5/2$ (specifically $n=3$) the tangent is horizontal, for $n=5/2$ the tangent  has a finite slope and for $3/2<n<5/2$ (specifically $n=2$) the tangent is vertical.  }
\label{tot}
\end{center}
\end{figure}

$\bullet$ For $n>3/2$, the Tsallis distributions have a single maximum
at $v=0$ and they vanish at $v=\pm v_{m}$ with a horizontal tangent:
$f'(\pm v_{m})=0$ (see Fig. \ref{tsallisH}). Therefore, when
$\Omega_r\rightarrow \pm V_{m}$, the Nyquist curve
$(\epsilon_{r}(\Omega_{r}),\epsilon_{i}(\Omega_{r}))$ tends to the
point $(\epsilon_{r}(V_{m})=1+(\eta /\eta_c)\zeta_n,0)$ where
$\zeta_n\equiv -W_{r}^{(n)}(\sqrt{2(n+1)})\ge 0$ \footnote{We note
that $\epsilon_{r}(V_{m})\ge 1$ (it is equal to unity for
$n\rightarrow \infty$) so that $\Omega_r=\pm V_{m}$ can never
correspond to a marginal mode (it satisfies $\epsilon_i=0$ but it can
never satisfy $\epsilon_r=0$).}.  This parameter is plotted as a
function of $n$ in Fig. \ref{zetan} of Appendix \ref{sec_wrn}. We note that
$\epsilon_{i}(\Omega_{r})>0$ for $\Omega_{r}\rightarrow -V_{m}^{+}$
and $\epsilon_{i}(\Omega_{r})<0$ for $\Omega_{r}\rightarrow
V_{m}^{-}$. We need now to distinguish three sub-cases. For $n>5/2$,
$f''(v_{m})=0$: this implies $\epsilon_{i}'(V_{m})=0$ so that the
Nyquist curve has a horizontal tangent at
$(\epsilon_{r}(V_{m}),0)$. For $3/2<n<5/2$, $f''(v_{m})=+\infty$: this
implies $\epsilon_{i}'(V_{m})=\infty$, so that the Nyquist curve has a
vertical tangent at $(\epsilon_{r}(V_{m}),0)$. For $n=5/2$,
$f''(v_{m})=15\rho/(14\sqrt{7}T^{3/2})$ and
$\epsilon_{i}'(V_{m})=15\pi\eta^{3/2}/(14\sqrt{7})$
 are finite so that the
Nyquist curve has a finite slope at $(\epsilon_{r}(V_{m}),0)$. These
three behaviors are represented in Figs. \ref{tsallis5},
\ref{tsallis2} and \ref{tot}. On the other
hand, for $\Omega_r<-V_{m}$, the imaginary
part of the dielectric function $\epsilon_i(\Omega_r)=0$ and the real
part $\epsilon_r(\Omega_r)$ varies from $1$ (for $\Omega_r\rightarrow -\infty$) to $\epsilon_{r}(V_{m})=1+(\eta/\eta_{c})\zeta_n $ (for $\Omega_r\rightarrow - V_m^{-}$); see Appendix \ref{sec_wrn}.
For $\Omega_r>V_{m}$, the imaginary
part of the dielectric function $\epsilon_i(\Omega_r)=0$ and the real
part $\epsilon_r(\Omega_r)$ varies from $\epsilon_{r}(V_{m})=1+(\eta/\eta_{c})\zeta_n $ (for $\Omega_r\rightarrow +V_m^{+}$) to $1$ (for $\Omega_r\rightarrow +\infty$). This corresponds to a segment
on the $x$-axis between $(1,0)$ and $(1+(\eta/\eta_{c})\zeta_n,0)$ in Figs. \ref{tsallis5} and
\ref{tsallis2} (for clarity, we have slightly shifted the segment from the $x$-axis in Fig. \ref{tsallis2}
to show the complete path). Finally, considering the value
$\Omega_r=0$, we see that the Nyquist curve crosses the $x$-axis at
$(\epsilon_r(0)=1-\eta/\eta_c,0)$. For $\eta < \eta_c$ (i.e. $T >
T_c$), the Nyquist curve does not encircle the origin so that the
Tsallis distribution is stable. For $\eta > \eta_c$ (i.e. $T < T_c$)
the Nyquist curve encircles the origin so that the Tsallis
distribution is unstable.  For $\eta = \eta_c$ (i.e. $T = T_c$) the
Nyquist curve passes through the origin so that the Tsallis
distribution is marginally stable. The Nyquist curve of polytropic
distributions with index $n>5/2$ is very similar to the Nyquist curve of Maxwellian distributions (recovered for $n=+\infty$). They
essentially differ in the presence of a segment between $(1,0)$ and
$(1+(\eta/\eta_{c})\zeta_n,0)$ and in the value of the critical
temperature $\eta_{c}$. For $n\le 5/2$, the Nyquist curves of
polytropic and isothermal distributions become sensibly different.

\begin{figure}
\begin{center}
\includegraphics[clip,scale=0.3]{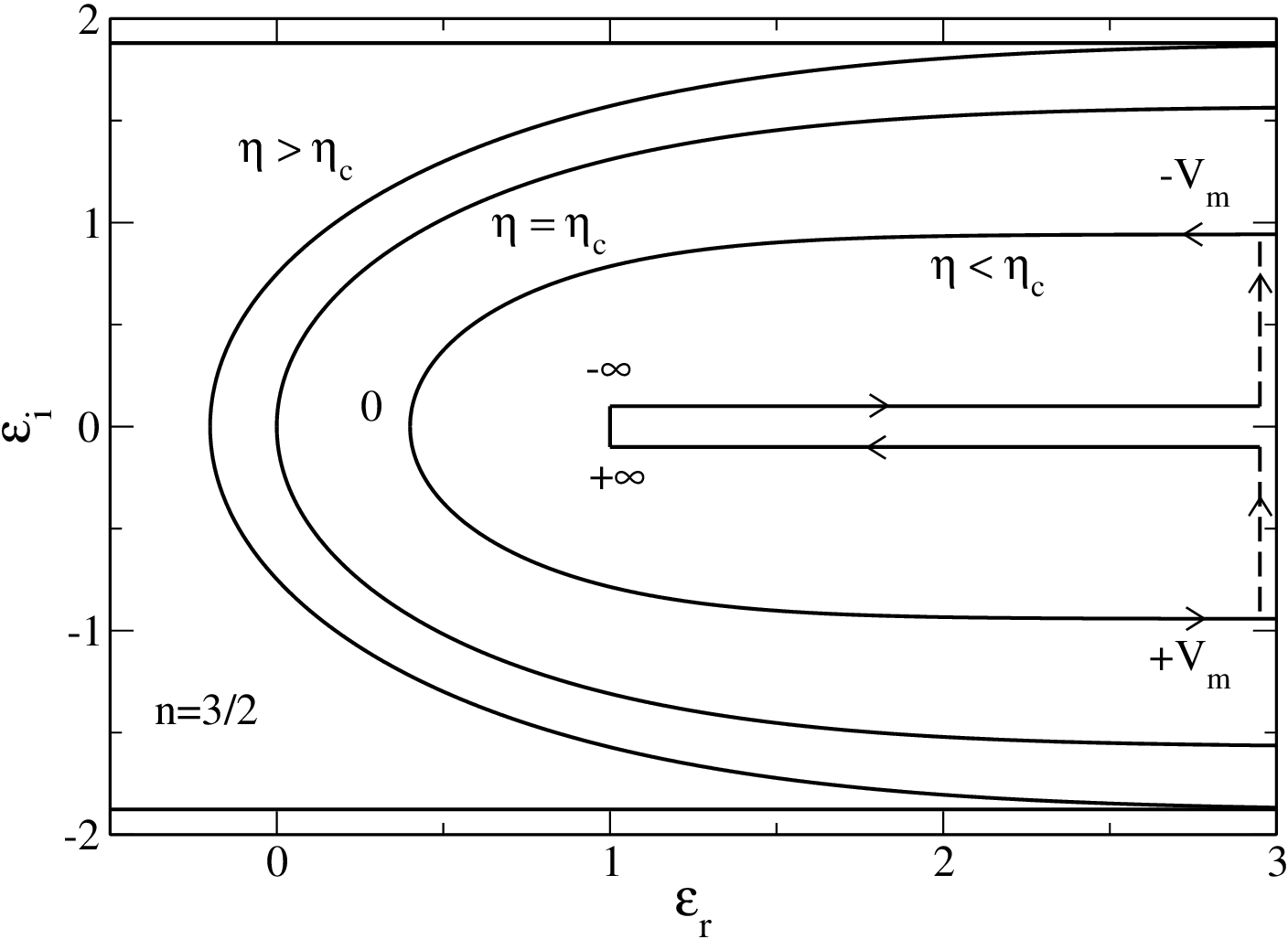}
\caption{Nyquist curve for the Tsallis distribution with $n=3/2$ yielding $\eta_{c}=5/3$. The DF is stable for $\eta<\eta_c$, marginally stable for $\eta=\eta_c$ and unstable for $\eta>\eta_c$. We have taken $\eta= 2, 5/3, 0.5$ from the outer to the inner curve.}
\label{tsallis32}
\end{center}
\end{figure}

\begin{figure}
\begin{center}
\includegraphics[clip,scale=0.3]{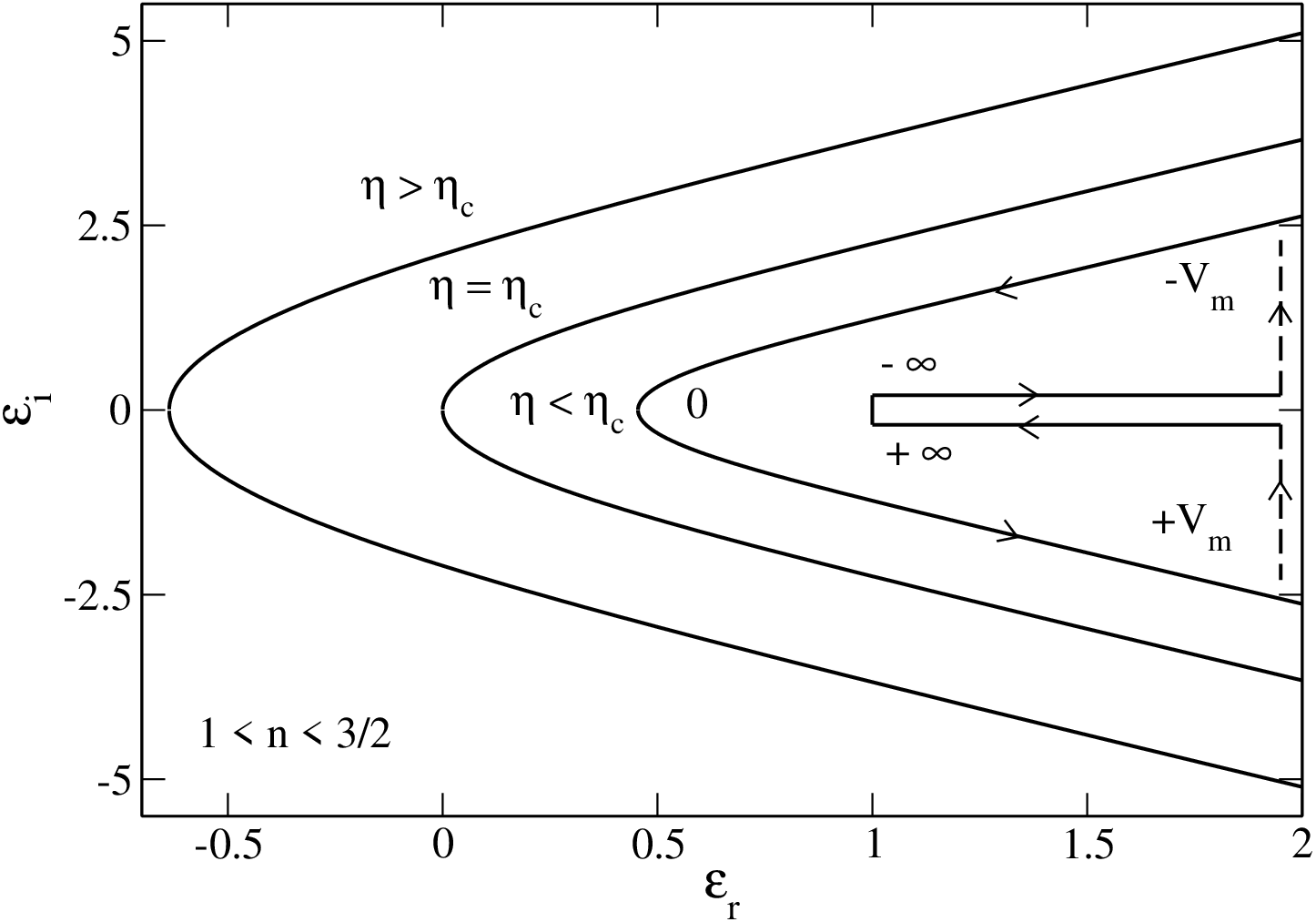}
\caption{Nyquist curve for the Tsallis distribution with $1<n<3/2$ (specifically $n=1.2$). The DF is stable for $\eta<\eta_c$, marginally stable for $\eta=\eta_c$ and unstable for $\eta>\eta_c$.}
\label{tsallis1p2}
\end{center}
\end{figure}

\begin{figure}
\begin{center}
\includegraphics[clip,scale=0.3]{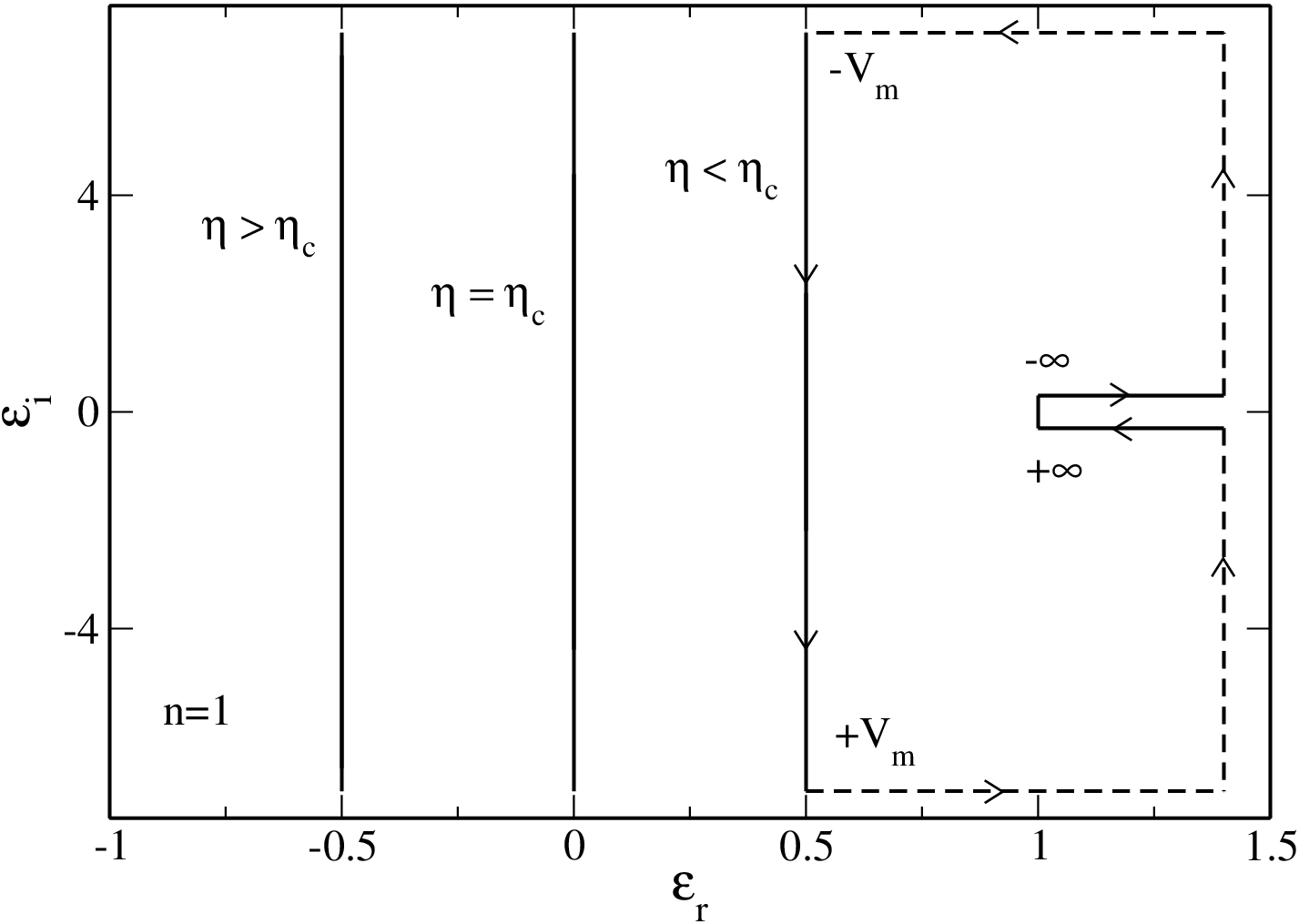}
\caption{Nyquist curve for the Tsallis distribution with $n=1$ yielding $\eta_{c}=2$. The DF is stable for $\eta<\eta_c$, marginally stable for $\eta=\eta_c$ and unstable for $\eta>\eta_c$. We have taken $\eta= 3, 2, 0.5$ from the outer to the inner curve.}
\label{tsallis1}
\end{center}
\end{figure}

\begin{figure}
\begin{center}
\includegraphics[clip,scale=0.3]{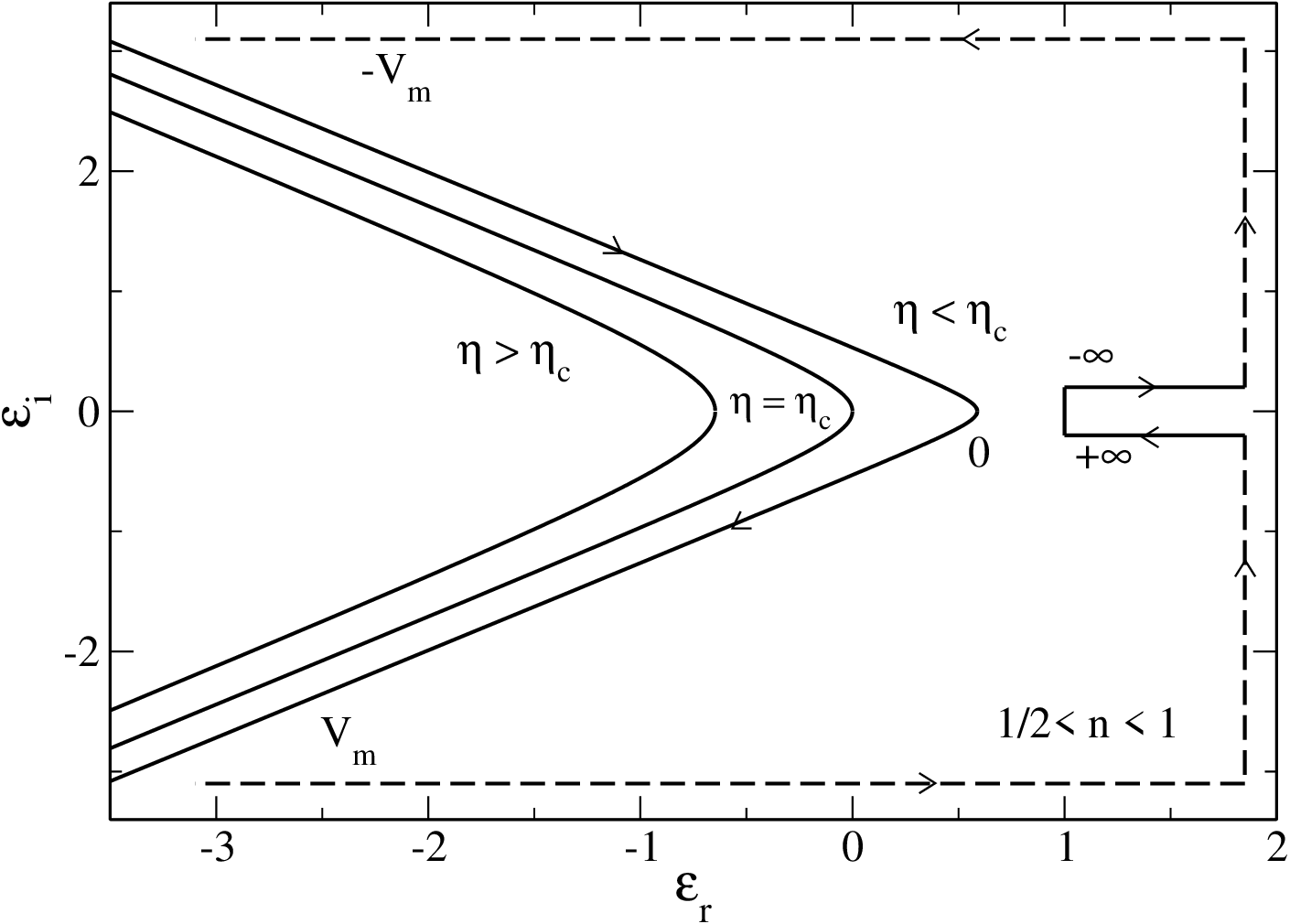}
\caption{Nyquist curve for the Tsallis distribution with $1/2<n<1$ (specifically $n=0.7$). The DF is stable for $\eta<\eta_c$, marginally stable for $\eta=\eta_c$ and unstable for $\eta>\eta_c$.}
\label{tsallisp7}
\end{center}
\end{figure}

\begin{figure}
\begin{center}
\includegraphics[clip,scale=0.3]{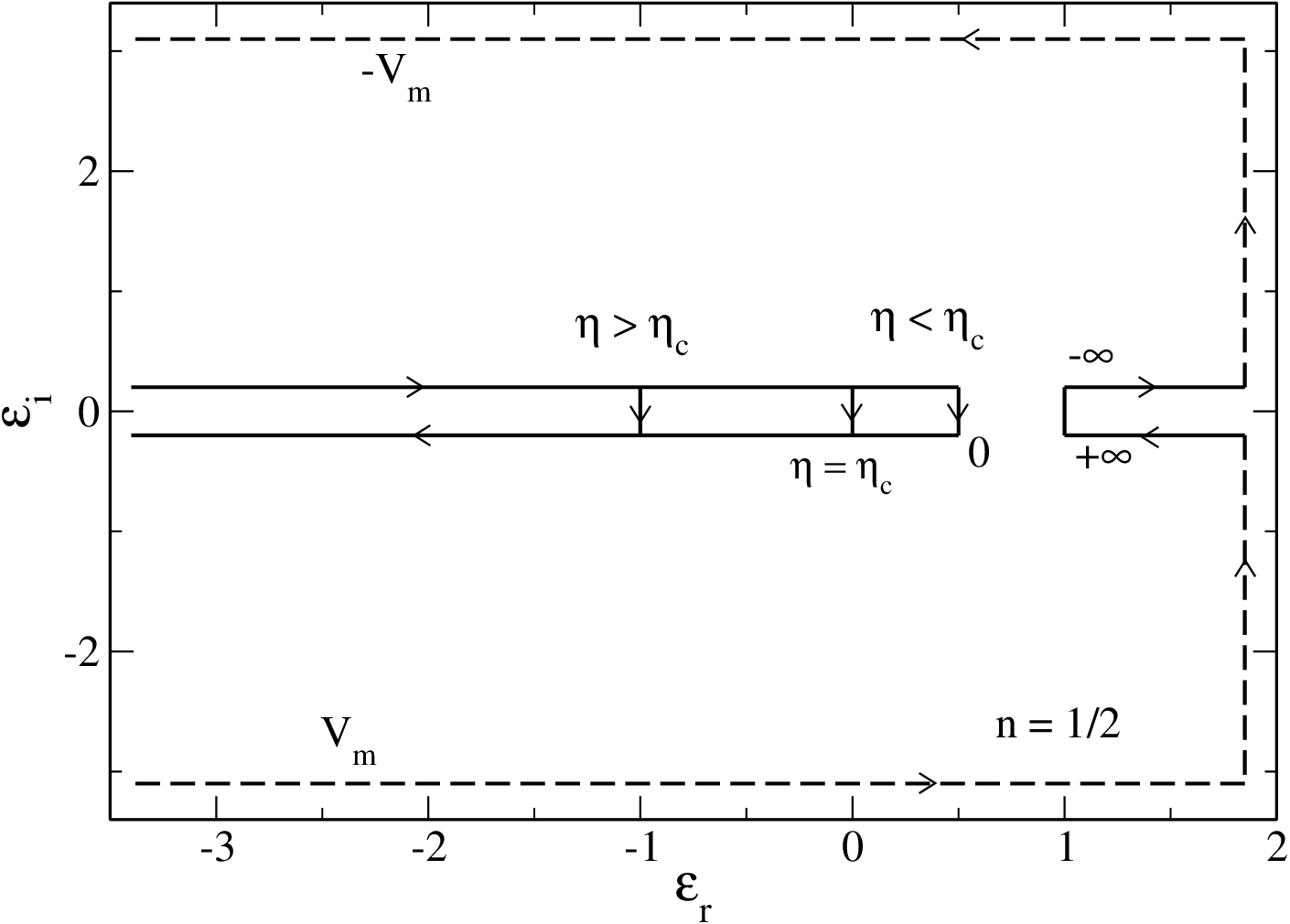}
\caption{Nyquist curve for the Tsallis distribution with $n=1/2$ yielding $\eta_c=3$. The DF is stable for $\eta<\eta_c$, marginally stable for $\eta=\eta_c$ and unstable for $\eta>\eta_c$.}
\label{waterbag}
\end{center}
\end{figure}

$\bullet$ For $1/2 <n \le 3/2 $, the Tsallis distributions have a
single maximum at $v=0$ and they vanish at $v=\mp v_{m}$ with a
vertical tangent: $f'(\mp v_{m})=\pm\infty$ (see Fig. \ref{tsallisV};
for $n=3/2$, the slope $f'(\mp v_{m})=\pm 3\rho/(10T)$ is
finite).  Therefore, the imaginary part of the dielectric function
$\epsilon_i(\Omega_r)\rightarrow \pm\infty$ when $\Omega_r\rightarrow
\mp V_{m}$ (for $n=3/2$, the imaginary part of the
dielectric function $\epsilon_i(\Omega_r)\rightarrow \pm 3\pi\eta/10$
 when $\Omega_r\rightarrow
\mp V_{m}$). We need now to distinguish three subcases. For $1<n\le
3/2$, the real part of the dielectric function
$\epsilon_r(\Omega_r)\rightarrow +\infty$ when $\Omega_r\rightarrow
\pm V_{m}$. For $n=1$, the dielectric function $\epsilon_r(\Omega_r)$
is independent on $\Omega_r\in [-V_m,V_m]$. For $1/2<n<1$, the
dielectric function $\epsilon_r(\Omega_r)\rightarrow -\infty$ when
$\Omega_r\rightarrow \pm V_{m}$; see Appendix \ref{sec_wrn}.  This leads to the Nyquist curves
represented in Figs. \ref{tsallis32},
\ref{tsallis1p2}, \ref{tsallis1} and \ref{tsallisp7}. On the other
hand, for $\Omega_r<-V_{m}$, the imaginary part of the dielectric
function $\epsilon_i(\Omega_r)=0$ and the real part
$\epsilon_r(\Omega_r)$ varies from $1$ (for $\Omega_r\rightarrow
-\infty$) to $+\infty$ (for $\Omega_r\rightarrow - V_m^{-}$).
For $\Omega_r>V_{m}$, the imaginary part of the dielectric
function $\epsilon_i(\Omega_r)=0$ and the real part
$\epsilon_r(\Omega_r)$ varies from $+\infty$ (for $\Omega_r\rightarrow
+V_m^{+}$) to $1$ (for $\Omega_r\rightarrow +\infty$). This
corresponds to a segment on the $x$-axis between $(1,0)$ and
$(+\infty,0)$ in Figs. \ref{tsallis32},
\ref{tsallis1p2}, \ref{tsallis1} and \ref{tsallisp7}  (for clarity, we have slightly shifted the segment from the $x$-axis to show the complete path). Finally, considering the value $\Omega_r=0$, we see that the Nyquist
curve crosses the $x$-axis at $(\epsilon_r(0)=1-\eta/\eta_c,0)$. For
$\eta < \eta_c$ (i.e. $T > T_c$), the Nyquist curve does not encircle
the origin so that the Tsallis distribution is stable. For $\eta >
\eta_c$ (i.e. $T < T_c$) the Nyquist curve encircles the origin so
that the Tsallis distribution is unstable.  For $\eta = \eta_c$
(i.e. $T = T_c$) the Nyquist curve passes through the origin so that
the Tsallis distribution is marginally stable. Note that it is
important to determine the complete path of evolution, from
$\Omega_r=-\infty$ to $\Omega_r=+\infty$, in order to apply the
Nyquist criterion.

$\bullet$ The index $n=1/2$ deserves a particular attention. In that
case, $f=\eta_{0}$ for $|v|\le v_{m}$ and $f=0$ otherwise. It
corresponds to the Fermi distribution or to the water-bag model.  The
constant $\eta_0$ is determined by the normalization condition
$\rho=\int f \, dv$ yielding $\rho=2\eta_{0}v_{m}$. The kinetic
temperature $T=\langle v^2\rangle$ is given by $T=v_{m}^{2}/3$. The
derivative of the distribution function is $f'(v)=\eta_{0}\left\lbrack
\delta(v+v_0)-\delta(v-v_0)\right\rbrack$. According to Eq. (\ref{m6}), the
critical temperature is $T_c={kM}/{12\pi}$ i.e. $\eta_{c}=3$ in
complete agreement with Eqs. (\ref{t4}) and (\ref{t12}). More
precisely, for $n=1/2$, we can obtain an analytical expression of the
dielectric function in the form \cite{cvb}:
\begin{equation}
\label{t13}
\epsilon(\Omega)=1-\frac{1}{3}\eta W^{(1/2)}(\sqrt{\eta}\Omega),
\end{equation}
with
\begin{equation}
\label{t14}
W^{(1/2)}(z)=\frac{1}{1-\frac{1}{3}z^2}.
\end{equation}
The condition $\epsilon(\omega)=0$ determines the complex
pulsation. For $T>T_{c}$, the system is stable and
\begin{equation}
\label{tnew1}
\omega=\pm\sqrt{3}(T-T_c)^{1/2}.
\end{equation}
For $T<T_{c}$, the system is unstable and
\begin{equation}
\label{tnew2}
\omega=\pm i\sqrt{3}(T_c-T)^{1/2}.
\end{equation}
On the other hand, for
$\Omega_{i}=0$, we get
\begin{equation}
\label{t15}
\epsilon_{r}(\Omega_{r})=1-\frac{1}{3}\eta \frac{1}{1-\frac{1}{3}\eta\Omega_{r}^{2}}, \qquad \epsilon_{i}(\Omega_{r})=0.
\end{equation}
Therefore, the Nyquist curve is made of two  segments $]-\infty, 1-\eta/3]$ and $[1,+\infty[$ as represented in Fig. \ref{waterbag}.

\section{The symmetric double-humped distribution}
\label{sec_sdh}

\subsection{Determination of the extrema}
\label{sec_ext}

\begin{figure}
\begin{center}
\includegraphics[clip,scale=0.3]{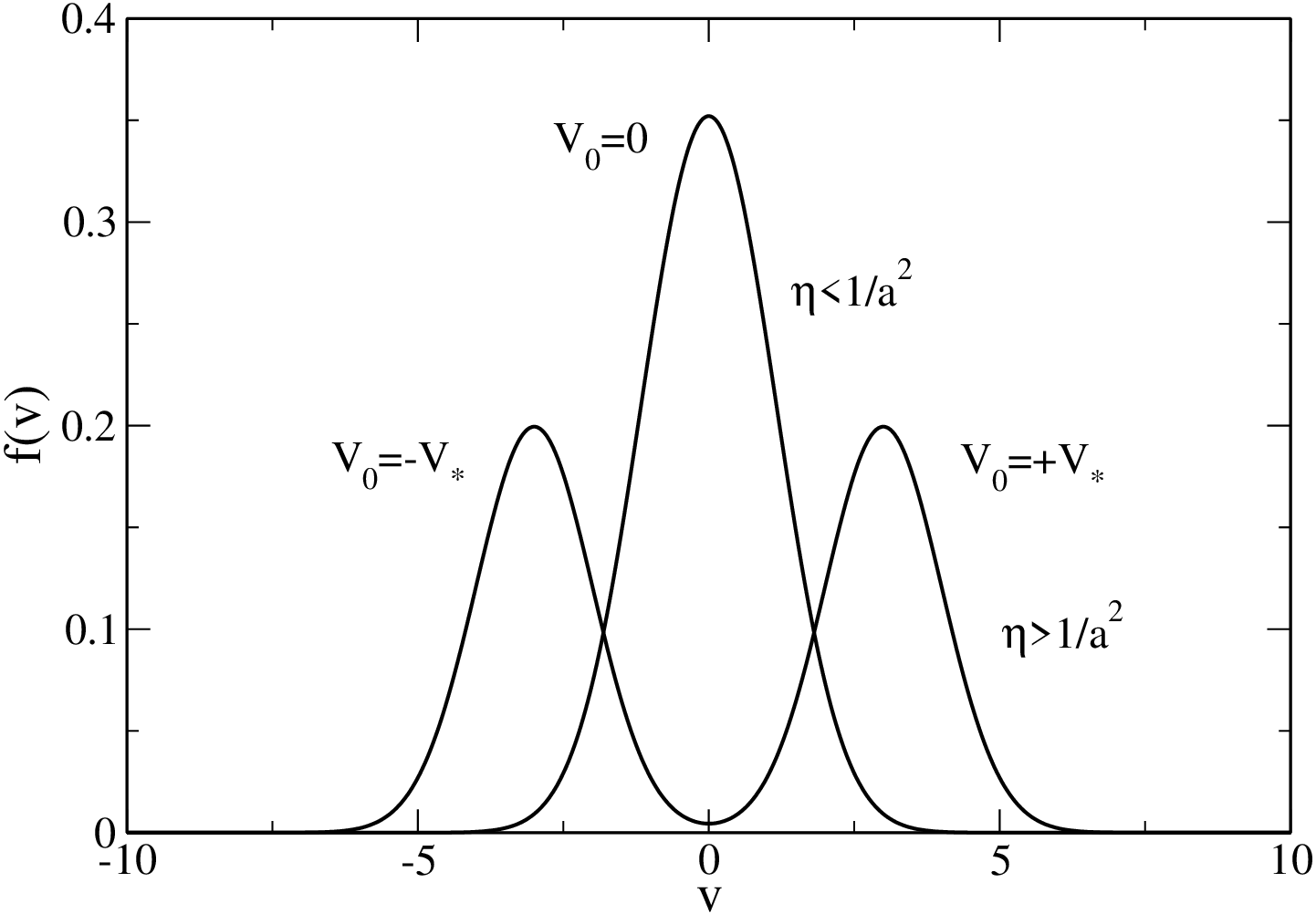}
\caption{Symmetric double-humped distribution made of two Maxwellian with separation $a$. If $\eta>1/a^2$, the distribution has two maxima at $\pm V_{*}$ and one minimum at $V_{0}=0$ while for  $\eta<1/a^2$, it has only one maximum at $V_{0}=0$.}
\label{dbumpdist}
\end{center}
\end{figure}

\begin{figure}
\begin{center}
\includegraphics[clip,scale=0.3]{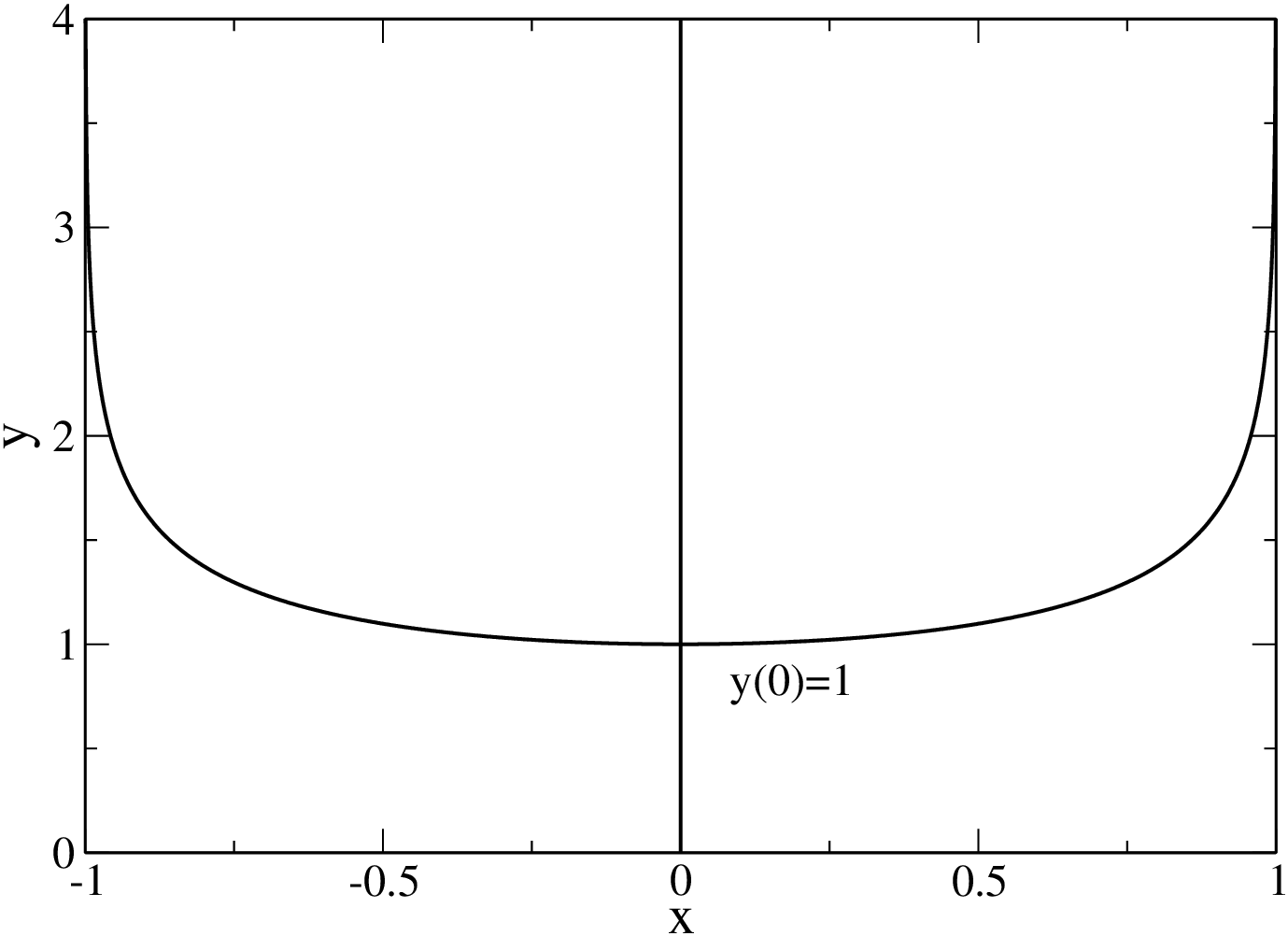}
\caption{The function $y(x)$ for the symmetric double-humped distribution.}
\label{ysym}
\end{center}
\end{figure}

We consider a symmetric double-humped distribution of the form
\begin{equation}
\label{e1}
f(v)=\sqrt{\frac{\beta}{2\pi}} \frac{\rho}{2} \left [  e^{-\frac{\beta}{2}(v-v_a)^2} + e^{-\frac{\beta}{2}(v+v_a)^2}  \right ].
\end{equation}
It corresponds to the superposition of two Maxwellian distributions
with temperature $T=1/\beta$ centered in $v_a$ and $-v_a$ respectively
(see Fig. \ref{dbumpdist}). This distribution models two streams of
particles in opposite direction. The average velocity is $\langle
v\rangle=0$ and the kinetic temperature $T_{kin}\equiv \langle
v^2\rangle=T+v_{a}^{2}$. The velocities $v_0$ at which the
distribution function $f(v)$ is extremum satisfy $f'(v_0)=0$. They are
determined by the equation
\begin{equation}
\label{e2}
e^{-2\beta v_{a} v_0}=\frac{v_a-v_0}{v_a+v_0}.
\end{equation}
Introducing the dimensionless temperature (\ref{mm4}), the dimensionless velocity (\ref{mm5}) and the dimensionless separation
\begin{equation}
\label{e3}
a=v_a \left (\frac{4\pi}{kM}\right )^{1/2},
\end{equation}
Eq. (\ref{e2}) can be rewritten
\begin{equation}
\label{e4}
\eta=\frac{1}{2aV_0}\ln\left (\frac{a+V_0}{a-V_0}\right ).
\end{equation}
For a given value of the inverse temperature $\eta$ and separation $a$, this equation determines the velocities $V_0$ where $f(V)$ is extremum. We note that $V_0\in \lbrack -a, +a\rbrack$. It is convenient to introduce the variables
\begin{equation}
\label{e5}
x=V_0/a, \qquad  y=\eta a^2.
\end{equation}
For fixed $a$, the parameter $y$ plays the role of the inverse
temperature and the parameter $x$ plays the role of the
velocity. Then, we have to study the function
\begin{equation}
\label{e6}
y(x)=\frac{1}{2x}\ln\left (\frac{1+x}{1-x}\right ),
\end{equation}
for $x \in ] -1, +1[$. This function is plotted in Fig. \ref{ysym}. It has the
following properties:
\begin{equation}
\label{e7}
y(-x)=y(x),
\end{equation}
\begin{equation}
\label{e8}
y(x)\sim -\frac{1}{2}\ln (1-x), \qquad (x\rightarrow 1^{-}),
\end{equation}
\begin{equation}
\label{e9}
y(0)=1.
\end{equation}
The extrema of the distribution function (\ref{e1}) can be determined
from the study of this function. First, considering Eq. (\ref{e4}), we
note that $f(V)$ always has an extremum at $V_0=0$, for any value of
$\eta$ and $a$. This is a ``degenerate'' solution of Eq. (\ref{e6})
corresponding to the vertical line $x=0$ in Fig. \ref{ysym}. On the
other hand, if $y>1$, i.e $\eta>1/a^2$, there exists two other extrema
$V_{0}=\pm V_{*}$ where $V_{*}=a x_*$ with $x_*=y^{-1}(\eta a^2)$.

In conclusion, for a given separation $a$:

$\bullet$ if $\eta>1/a^2$, the distribution function $f(V)$ has two maxima at $V_{0}=\pm V_{*}$  and one minimum at $V_{0}=0$.

$\bullet$ if $\eta\le 1/a^2$, the distribution function $f(V)$ has only one
maximum at $V_{0}=0$ (the limit case $\eta=1/a^2$ corresponds to $f''(0)=0$).

{\it Remark:} We can note the formal analogy with the mean field theory (ferromagnetic transition) of the one dimensional Ising model without magnetic field (compare Eq. (\ref{e4}) with Eq. (14) of \cite{bellac}).

\subsection{The condition of marginal stability}
\label{sec_sms}

The dielectric function associated to the symmetric double-humped
distribution (\ref{e1}) is
\begin{equation}
\label{sms1}
\epsilon(\Omega)= 1-\frac{\eta}{2} \left [ W(\sqrt{\eta}(\Omega - a))+W(\sqrt{\eta}(\Omega + a))   \right ],
\end{equation}
where $W(z)$ is defined in Eq. (\ref{mm7}). When $\Omega_{i}=0$, the real and imaginary parts of the dielectric function  $\epsilon(\Omega_{r})=\epsilon_{r}(\Omega_{r})+i\epsilon_{i}(\Omega_{r})$ can be written
\begin{eqnarray}
\label{sms2}
\epsilon_{r}(\Omega_{r})=1-\frac{\eta}{2} \left [ W_r(\sqrt{\eta}(\Omega_{r} - a))+W_r(\sqrt{\eta}(\Omega_{r} + a))   \right ],\nonumber\\
\end{eqnarray}
\begin{equation}
\epsilon_{i}(\Omega_{r})=-\frac{\eta}{2} \left [ W_i(\sqrt{\eta}(\Omega_{r} - a))+W_i(\sqrt{\eta}(\Omega_{r} + a))   \right ],
\end{equation}
where $W_r(z)$ and $W_i(z)$ are defined in
Eqs. (\ref{mm11})-(\ref{mm12}) where $z$ is here a real number. The
condition of marginal stability corresponds to
$\epsilon_{r}(\Omega_{r})=\epsilon_{i}(\Omega_{r})=0$. The condition
$\epsilon_{i}(\Omega_{r})=0$ is equivalent to
\begin{equation}
\label{sms3}
f'(\Omega_{r})=0.
\end{equation}
The condition $\epsilon_{r}(\Omega_{r})=0$ leads to
\begin{equation}
\label{sms4}
1-\frac{\eta}{2} \left [ W_{r}(\sqrt{\eta}(\Omega_{r} - a))+W_{r}(\sqrt{\eta}(\Omega_{r} + a))   \right ]=0.
\end{equation}
Therefore, according to Eq. (\ref{sms3}), the real pulsation
$\Omega_{r}$ is equal to a velocity $V_{0}$ at which the distribution
(\ref{e1}) is extremum. The second equation (\ref{sms4}) determines
the value(s) $\eta_{c}(a)$ of the temperature at which the
distribution is marginally stable.

\subsubsection{The case $\omega_{r}=0$}

\begin{figure}
\begin{center}
\includegraphics[clip,scale=0.3]{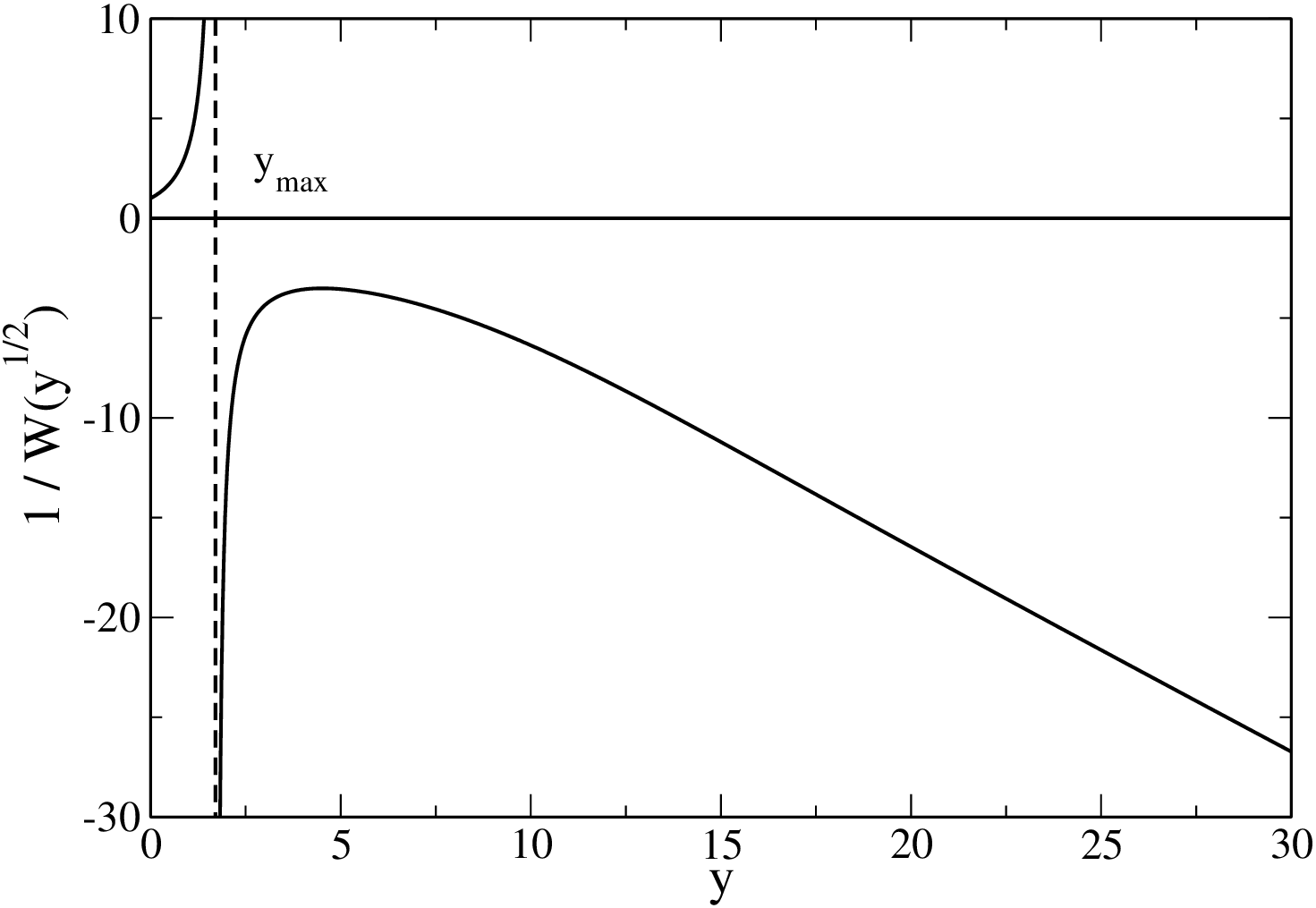}
\caption{$\eta=1/W_r(\sqrt{y})$  as a function of $y= \eta a^2$.}
\label{w3}
\end{center}
\end{figure}

Let us first consider the value $\Omega_{r}=0$ that is solution of
Eq. (\ref{sms3}) for any $a$ and $\eta$. In that case, Eq. (\ref{sms4}) becomes
\begin{equation}
\label{sms5}
\eta W_{r}(\sqrt{\eta}a)=1,
\end{equation}
where we have used Eq. (\ref{wr2}). For fixed $a$,
this equation determines the temperature(s) $\eta_{c}^{(0)}(a)$
corresponding to a mode of marginal stability with
$\Omega_{r}=0$. Introducing $y=\eta a^2\ge 0$, the foregoing equation
can be rewritten in the parametric form
\begin{equation}
\label{sms6}
a^2=y W_{r}(\sqrt{y}),
\end{equation}
\begin{equation}
\label{sms7}
\eta=\frac{1}{W_{r}(\sqrt{y})}.
\end{equation}
The function defined by Eq. (\ref{sms6}) is plotted in
Fig. \ref{w2}. {It vanishes at $y=y_{max}=z_{c}^{2}$ where
$z_{c}=1.307$ is the zero of $W_{r}(z)$ (see Appendix
\ref{sec_wr}). For $y>y_{max}=1.708$, the function (\ref{sms6}) is
negative so we restrict ourselves to $y\in[0,y_{max}]$.  In this
interval, it has a maximum at $y=y_{M}=0.724$ and its value is
$a_{M}=0.556$. Therefore, for $a>a_{M}$, Eq. (\ref{sms6}) has no
solution. For $a=a_{M}$, Eq. (\ref{sms6}) has one solution $y=y_{M}$
determining through Eq. (\ref{sms7}) a temperature
$\eta_{c}^{(0)(M)}=2.338$}. For $a<a_{M}$, Eq. (\ref{sms6}) has two
solutions $y_{1,2}$. Then, Eq. (\ref{sms7}) determines two
temperatures $\eta_{c}^{(0)(1,2)}(a)$ as illustrated in Fig. \ref{w3}.
These are the solutions of Eq. (\ref{sms5}) for $a\le a_{M}$.

\begin{figure}
\begin{center}
\includegraphics[clip,scale=0.3]{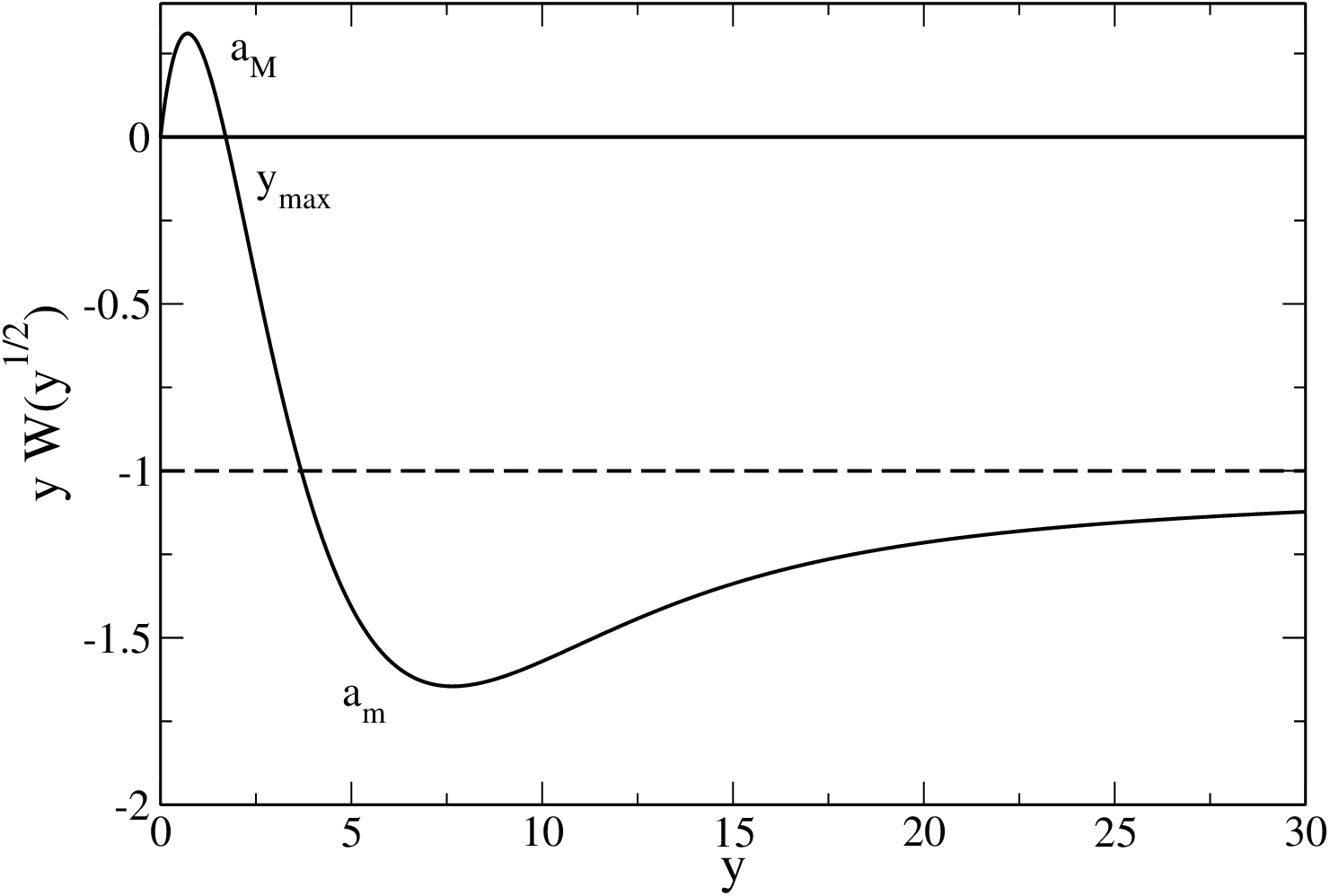}
\caption{$a^2=y W_{r}(\sqrt{y})$ as a function of $y= \eta a^2$.}
\label{w2}
\end{center}
\end{figure}

Let us consider the limit $a\rightarrow 0$. In that case, we see in
Fig. \ref{w2} that $y_{1}\rightarrow 0$ and $y_{2}\rightarrow
y_{max}$. In the first case $y_{1}\rightarrow 0$, using Eq. (\ref{wr9}), we
find from Eq. (\ref{sms7}) that
\begin{equation}
\label{sms8}
\eta_{c}^{(0)(1)}(a)\rightarrow 1, \qquad (a\rightarrow 0).
\end{equation}
This result is to be expected since, for $a=0$, the distribution
(\ref{e1}) reduces to the Maxwellian. We thus recover the critical
temperature (\ref{mm13}).  In the second case $y_{2}\rightarrow
y_{max}$, we find from Eq. (\ref{sms6}) that $a^2\sim
y_{max}W_r(\sqrt{y_{2}})$. Substituting this equivalent in
Eq. (\ref{sms7}), and recalling that $y_{max}=z_{c}^{2}$, we get
\begin{equation}
\label{sms9}
\eta_{c}^{(0)(2)}(a)\sim \frac{z_{c}^{2}}{a^2}, \qquad (a\rightarrow 0).
\end{equation}
For $a\rightarrow 0$, this solution $\eta_{c}^{(0)(2)}(a)\rightarrow
+\infty$ so that it does not appear in the study of the Maxwellian
distribution corresponding to $a=0$.

In conclusion:

$\bullet$ if $a<a_M$, there exists two critical temperatures
$\eta_{c}^{(0)(1)}(a)$ and $\eta_{c}^{(0)(2)}(a)$ determined by
Eqs. (\ref{sms6})-(\ref{sms7}) corresponding to a marginal mode
$(\Omega_{r}=0,\Omega_{i}=0)$. These two branches merge at the point
$(a_{M},
\eta_{c}^{(0)(M)})$.

$\bullet$ if $a>a_M$, there is no marginal mode
$(\Omega_{r}=0,\Omega_{i}=0)$.

The curves $\eta_{c}^{(0)(1)}(a)$ and $\eta_{c}^{(0)(2)}(a)$ corresponding to the
marginal mode with zero pulsation $\Omega_r=0$ are plotted in Fig. \ref{ba}.

\subsubsection{The case $\omega_{r}\neq 0$}

We now consider the pulsations $\Omega_{r}=\pm V_{*}$ that are
solutions of Eq. (\ref{sms3}) for $\eta>1/a^2$. To determine the
temperature(s) at which the distribution (\ref{e1}) is marginally stable, we
have to solve
\begin{equation}
\label{sms10}
1-\frac{\eta}{2} \left [ W_{r}(\sqrt{\eta}(V_{*} - a))+W_{r}(\sqrt{\eta}(V_{*}+ a))   \right ]=0,
\end{equation}
where $V_{*}$ is given by
\begin{equation}
\label{sms11}
\eta=\frac{1}{2aV_{*}}\ln\left (\frac{a+V_{*}}{a-V_{*}}\right ).
\end{equation}
Eliminating $V_*$ between these two expressions, we obtain the critical temperature(s) $\eta_{c}^{(\pm)}(a)$ as a function of $a$. However, it is easier to proceed differently. Setting $x=V_*/a$ and $y=\eta a^2$, we obtain the equations
\begin{equation}
\label{sms12}
y=\frac{1}{2x}\ln\left (\frac{1+x}{1-x}\right ),
\end{equation}
\begin{equation}
\label{sms13}
\eta=\frac{2}{\left [ W_{r}(\sqrt{y}(x - 1))+W_{r}(\sqrt{y}(x+ 1))   \right ]},
\end{equation}
\begin{equation}
\label{sms14}
a^2=\frac{y}{\eta}.
\end{equation}
For given $x$, we can obtain $y$ from Eq. (\ref{sms12}) [see also
Fig. \ref{ysym}], $\eta$ from Eq. (\ref{sms13}) and $a$ from
Eq. (\ref{sms14}). Varying $x$ in the interval $]-1,1[$ yields the
full curve $\eta_{c}^{(\pm)}(a)$. By symmetry, we can restrict
ourselves to the interval $x\in [0,1[$.

The limit $a\rightarrow +\infty$ (i.e. $y\rightarrow +\infty$)
corresponds to $V_{*}\rightarrow a^{-}$ (i.e. $x\rightarrow
1^{-}$). For $x\rightarrow 1^{-}$, Eqs. (\ref{sms12}) and
(\ref{sms13}) reduce to
\begin{equation}
\label{sms15}
y\sim -\frac{1}{2}\ln\left (1-x\right )\rightarrow +\infty,
\end{equation}
and
\begin{equation}
\label{sms16}
\eta\simeq \frac{2}{\left [ W_{r}(\sqrt{y}(x - 1))+W_{r}(2\sqrt{y})   \right ]}.
\end{equation}
Using Eqs. (\ref{wr8}) and (\ref{wr9}), we obtain at leading order
\begin{equation}
\label{sms17}
\eta\simeq \frac{2}{1-\frac{1}{4y}}\simeq 2 \left (1+\frac{1}{4y}\right )\simeq 2 \left (1+\frac{1}{4\eta a^2}\right ),
\end{equation}
so that
\begin{equation}
\label{sms18}
\eta_{c}^{(\pm)}(a)=2+\frac{1}{4a^2}+... \qquad (a\rightarrow +\infty).
\end{equation}
The result $\eta_{c}^{(\pm)}(a)\rightarrow 2$ can be understood
simply. For $a\rightarrow +\infty$, the two humps are far away from
each other so that we expect that the critical temperature
$\eta_{c}^{(\pm)}$ will coincide with the critical temperature of a
single Maxwellian since they do not ``see'' each other. Noting that
the mass of a single hump is $M/2$, the corresponding critical
temperature is $T_{c}=\frac{k(M/2)}{4\pi}=\frac{kM}{8\pi}$ leading to
$\eta_{c}^{(\pm)}=2$.

On the other hand, the branch $\eta_{c}^{(\pm)}(a)$ starts at the
point $(a_{*},\eta_{*})$, corresponding to $\Omega_{r}=\pm V_{*}=0$
(i.e. $x=0$), connecting the branch $\eta_{c}^{(0)}(a)$ along which
$\Omega_{r}=0$. {Now, for $x=0$, we have $y=1$ leading to
$\eta_*=1/W_r(1)=3.633$ and $a_*=\sqrt{W_r(1)}=0.524$.}  Therefore, the
branch $\eta_{c}^{(\pm)}(a)$ corresponding to $\Omega_{r}=\pm V_{*}$
starts at the intersection between the branch $\eta_{c}^{(0)}(a)$
corresponding to $\Omega_{r}=0$ and the hyperbole $\eta=1/a^2$
separating the regions where the DF has one or two maxima.

In conclusion:

$\bullet$ if $a>a_*$, there exists a single critical temperature
$\eta_{c}^{(\pm)}(a)$ determined by Eqs. (\ref{sms12})-(\ref{sms14})
corresponding to a marginal mode $(\Omega_r=\pm V_*,\Omega_i=0)$. Note
that the modes $\Omega_r=+V_*$ and $\Omega_r=-V_{*}$ are
degenerate. This degeneracy can be raised by a small asymmetry
(symmetry breaking) in the distribution (see Sec. \ref{sec_adh}).

$\bullet$ if $a<a_*$, there is no marginal mode $(\Omega_r=\pm V_*,\Omega_i=0)$.

The curve $\eta_{c}^{(\pm)}(a)$  corresponding to the
marginal mode with pulsation $\Omega_r=\pm V_{*}$ is plotted in Fig. \ref{ba}.

\subsection{The stability diagram}
\label{sec_sdsy}

\begin{figure}
\begin{center}
\includegraphics[clip,scale=0.3]{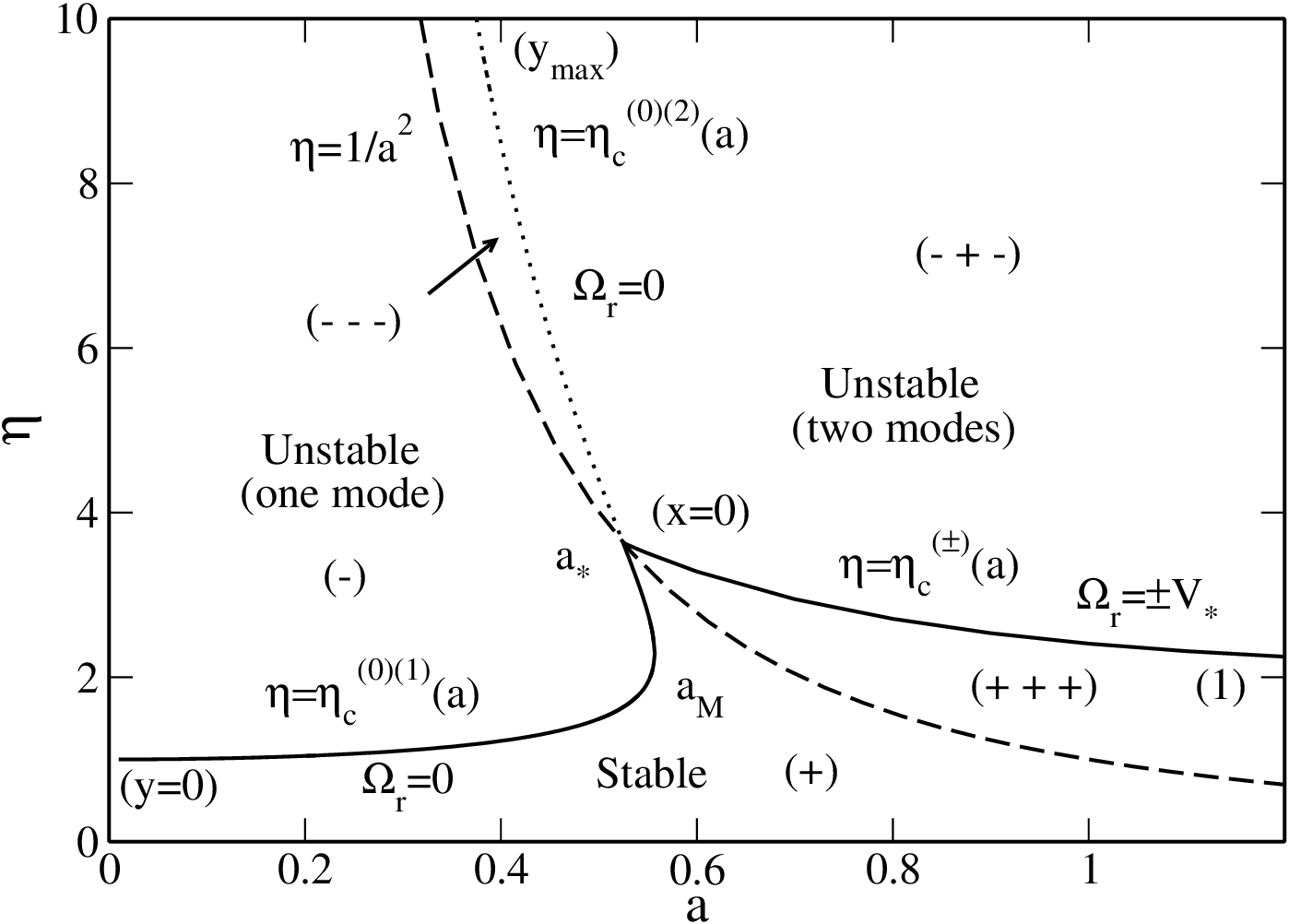}
\caption{Stability diagram of the symmetric double-humped
distribution (\ref{e1}). The solid line corresponds to the critical
line: below this line the DF is stable and above this line the DF is
unstable. On the left panel (delimited by the solid line and the
dotted line), there is one mode of instability and on the right panel
there are two modes of instability. For $a_*<a<a_M$, there exists a
``re-entrant" phase as described in the text. The dashed line
corresponds to the hyperbole $\eta=1/a^2$: below this line the DF has
one maximum and above this line the DF has two maxima and one
minimum. The symbols in parenthesis give the values of $x$ or $y$ that
parametrize the marginal curves. The notations like $(-+-)$ give the
positions of the $\epsilon_{r}(v_{ext})$'s in the Nyquist curves as
defined in Sec. \ref{sec_doubh}.}
\label{ba}
\end{center}
\end{figure}

The critical temperatures $\eta_c(a)$ corresponding to marginal
stability determined previously are represented as a function of the
separation $a$ in Fig. \ref{ba}. We have also plotted the hyperbole
$\eta=1/a^2$. Below this curve, the DF has a single maximum at $V_0=0$
and above this curve, the DF has two maxima at $V_0=\pm V_*$ and a
minimum at $V_0=0$. In order to investigate the stability of the
solutions in the different regions, we can use the Nyquist method.

\begin{figure}
\begin{center}
\includegraphics[clip,scale=0.3]{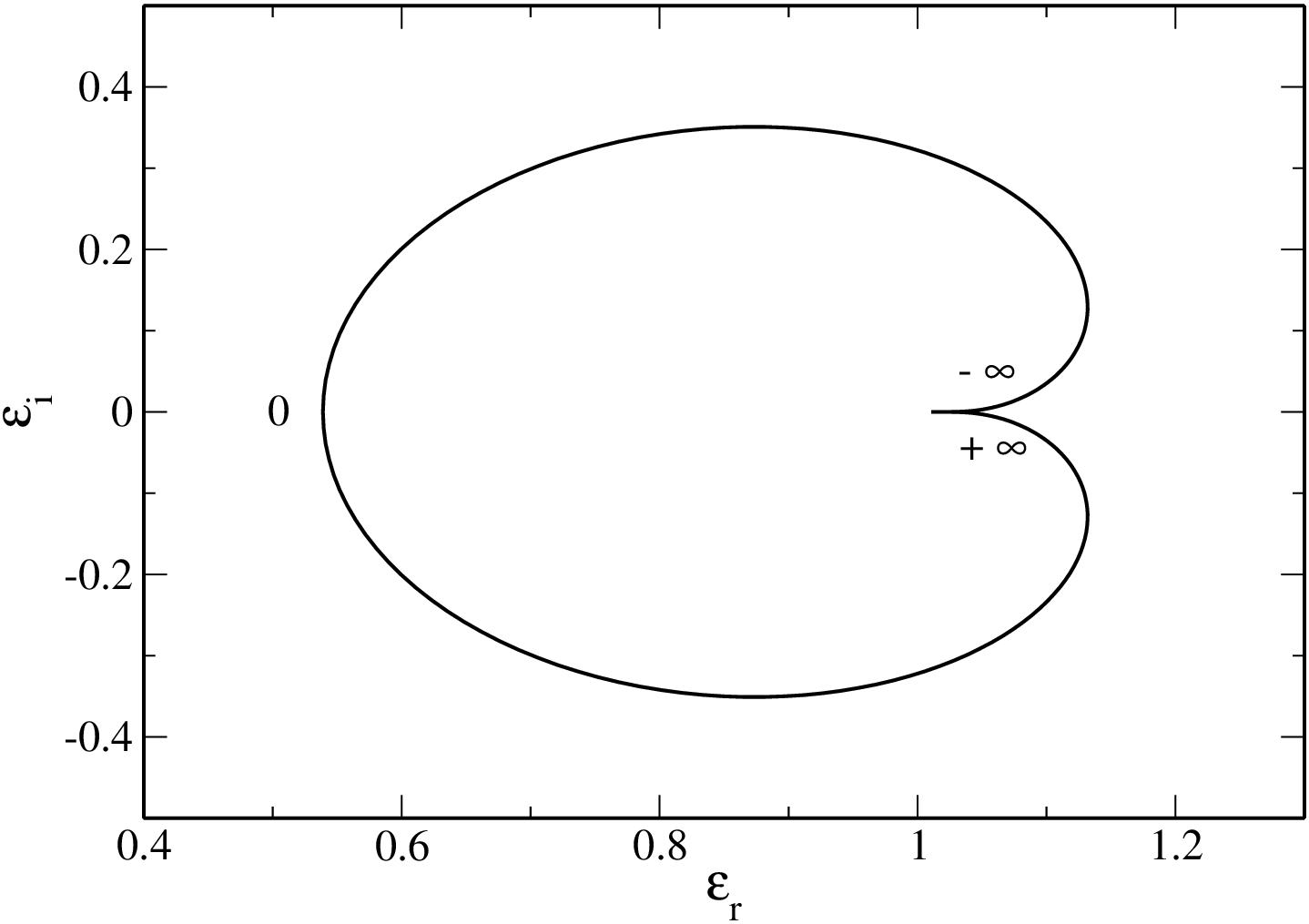}
\caption{Nyquist curve for $a<a_{*}$ and $\eta<\eta_{c}^{(0)(1)}<1/a^2$ (specifically $a=0.4$ and $\eta=0.5$). The DF has only one maximum at $V_{0}=0$. It is stable because the Nyquist curve does not encircle the origin. Case $(+)$.}
\label{Ny3}
\end{center}
\end{figure}

\begin{figure}
\begin{center}
\includegraphics[clip,scale=0.3]{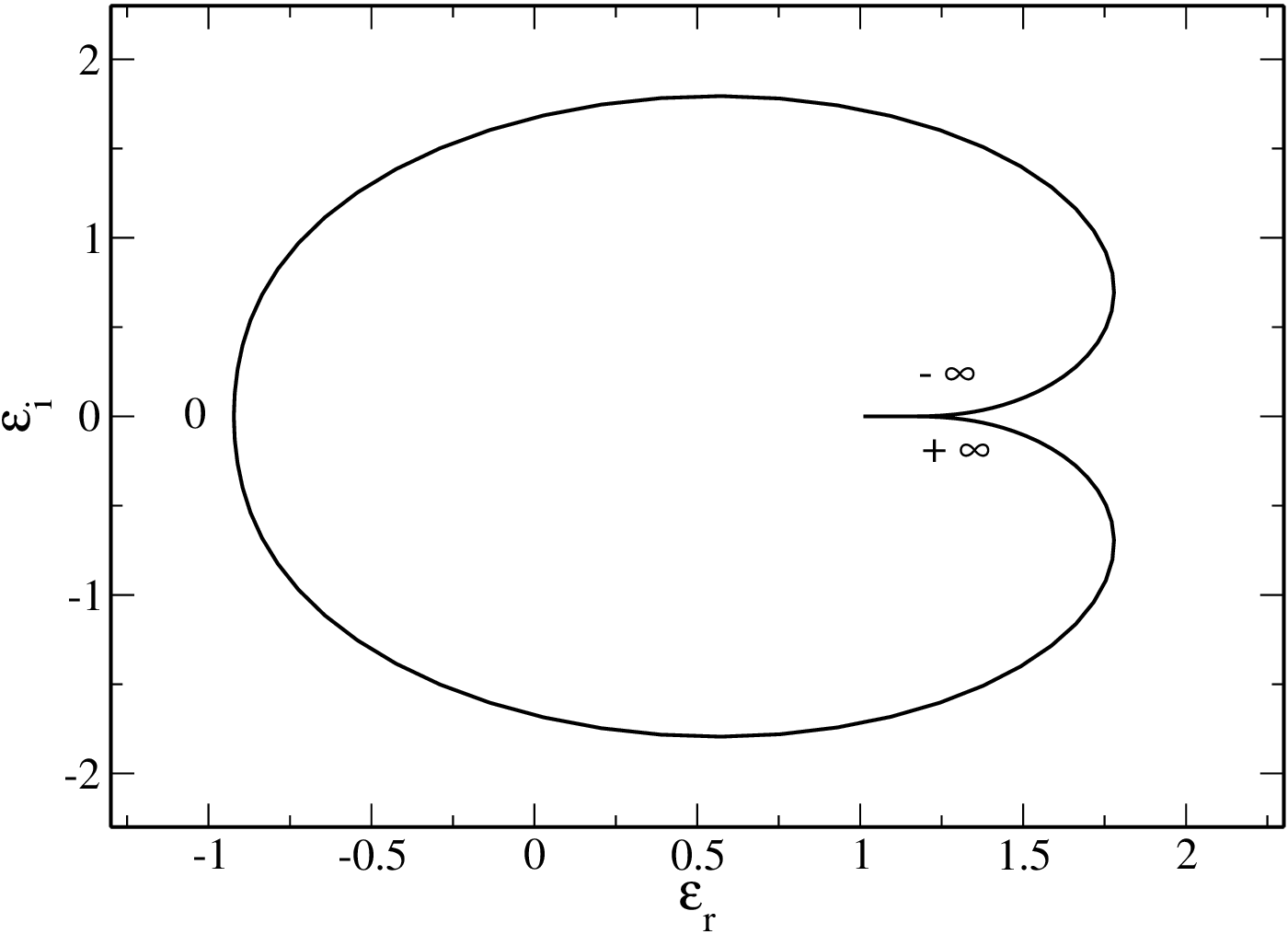}
\caption{Nyquist curve for $a<a_{*}$ and $\eta_{c}^{(1)(0)}<\eta<1/a^2<\eta_{c}^{(0)(2)}$ (specifically $a=0.4$ and $\eta=4$). The DF has only one maximum at $V_{0}=0$. The DF is unstable because the Nyquist curve encircles the origin. Case $(-)$.}
\label{Ny4}
\end{center}
\end{figure}

\begin{figure}
\begin{center}
\includegraphics[clip,scale=0.3]{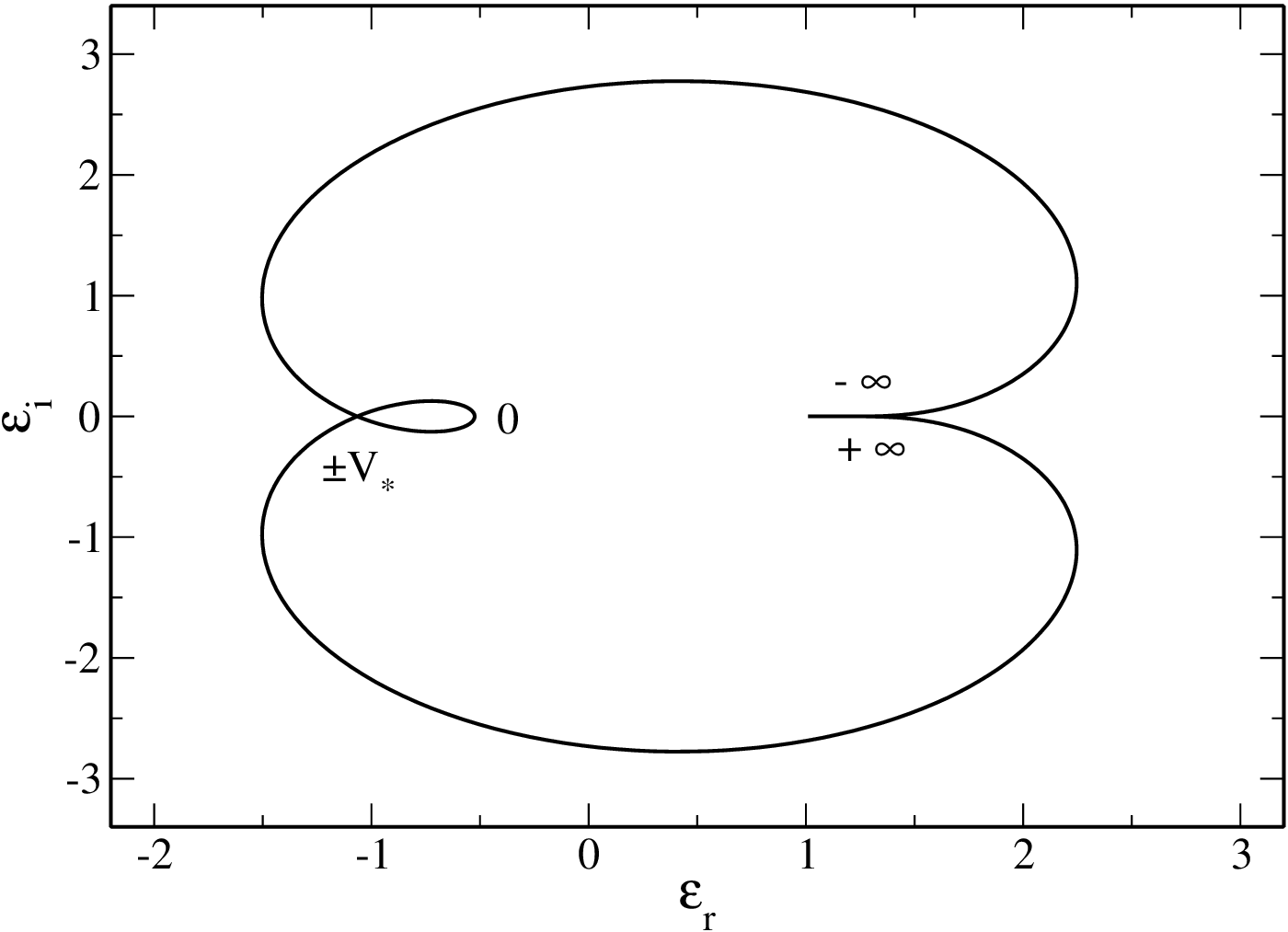}
\caption{Nyquist curve for $a<a_{*}$ and $1/a^2<\eta<\eta_{c}^{(0)(2)}$ (specifically $a=0.4$ and $\eta=7$). The DF has two maxima at $V_{0}=\pm V_{*}$ and one minimum at $V_{0}=0$. The DF is unstable because the Nyquist curve encircles the origin. Since it rotates only once around the origin, this implies that there is $N=1$ unstable mode $(\omega_r,\omega_i)$ with $\omega_{i}>0$. Case $(---)$.}
\label{Ny5}
\end{center}
\end{figure}

\begin{figure}
\begin{center}
\includegraphics[clip,scale=0.3]{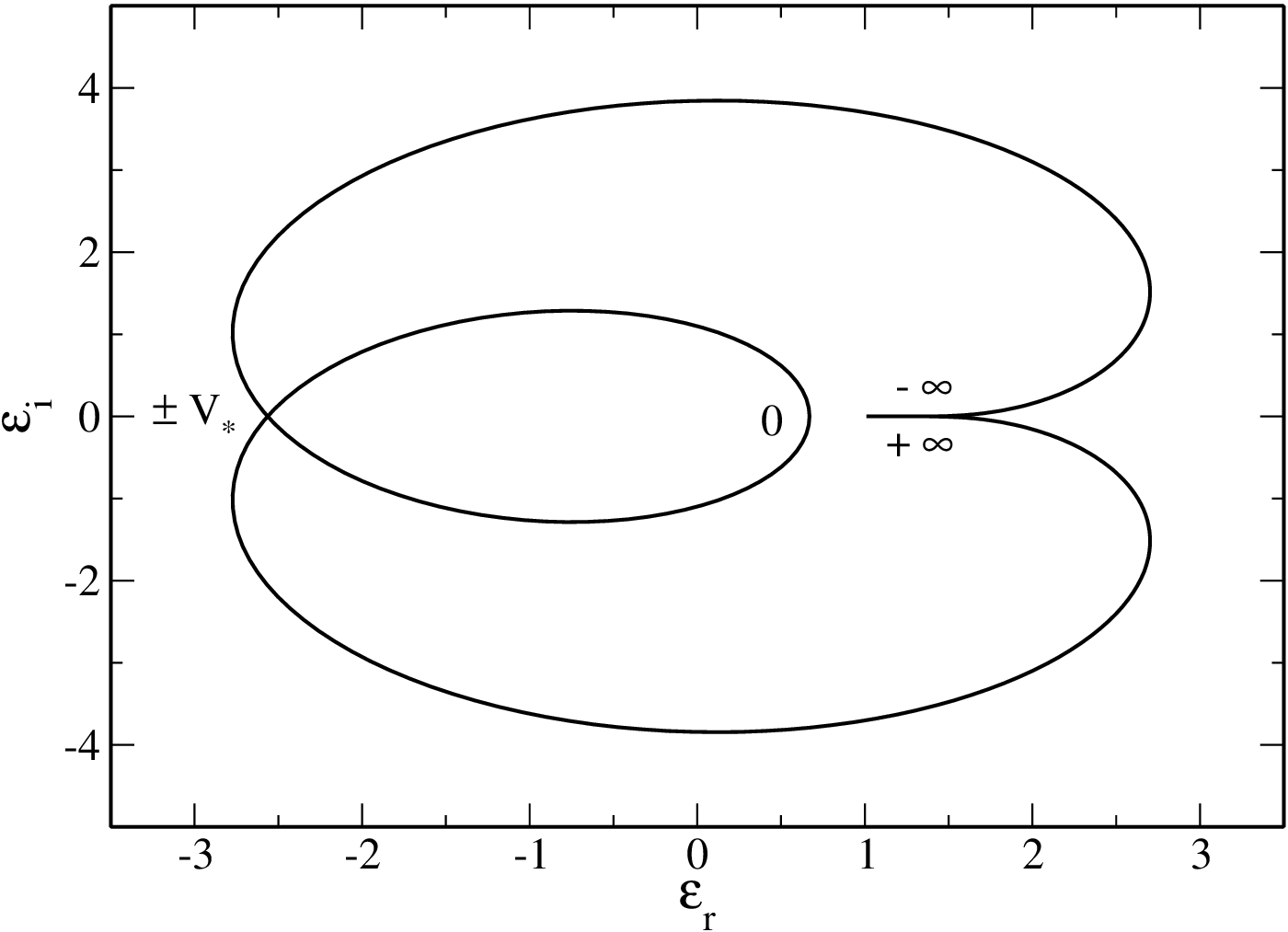}
\caption{Nyquist curve for $a<a_{*}$ and $\eta>\eta_{c}^{(0)(2)}$ (specifically $a=0.4$ and $\eta=10$).  The DF has two maxima at $V_{0}=\pm V_{*}$ and one minimum at $V_{0}=0$. The DF is unstable because the Nyquist curve encircles the origin. Since it rotates twice around the origin, this implies that there are  $N=2$ unstable modes $(\omega_r,\omega_i)$ with $\omega_{i}>0$. Case $(-+-)$.}
\label{Ny6}
\end{center}
\end{figure}

For $a<a_*$, there exists two temperatures $\eta_{c}^{(0)(1)}$ and
$\eta_{c}^{(0)(2)}$ at which the DF is marginally stable.  For
$\eta=\eta_{c}^{(0)(1)}$, the DF has a single maximum at $V_0=0$ and
for $\eta=\eta_{c}^{(0)(2)}$ the DF has a minimum at $V_0=0$ and two
maxima at $\pm V_*$. In both cases, the marginal perturbation has zero
pulsation $\Omega_r=0$. By considering the Nyquist curves in this
region (see Figs. \ref{Ny3}-\ref{Ny6}), we find that the DF is stable
for $\eta<\eta_{c}^{(0)(1)}$ and unstable for
$\eta<\eta_{c}^{(0)(1)}$.

\begin{figure}
\begin{center}
\includegraphics[clip,scale=0.3]{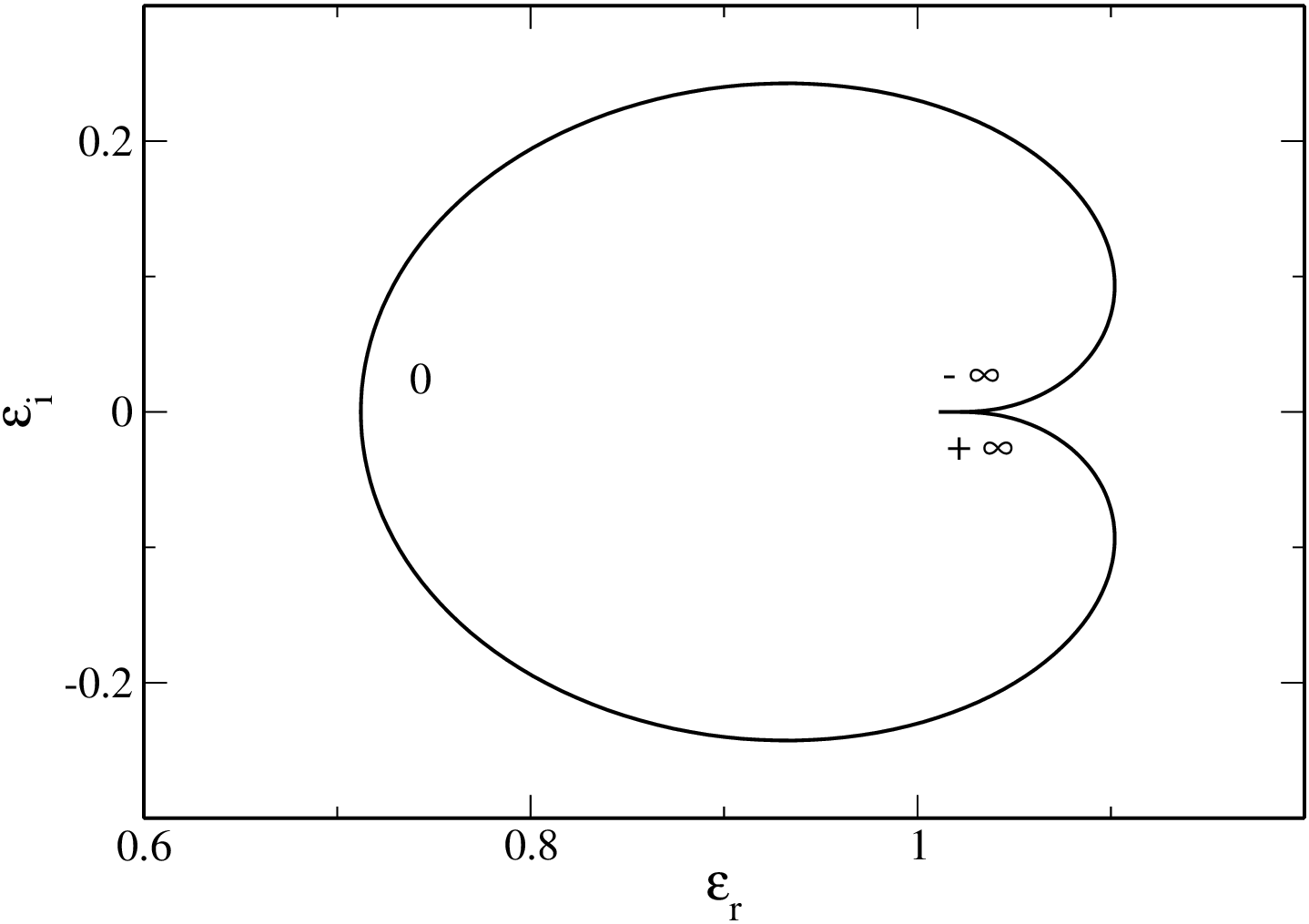}
\caption{Nyquist curve for $a>a_{M}$ and $\eta<1/a^2<\eta_{c}^{(\pm)}$ (specifically $a=1$ and $\eta=0.5$).  The DF has only one maximum at  $V_{0}=0$. The DF is stable because the Nyquist curve does not encircle the origin. Case $(+)$.}
\label{Ny1}
\end{center}
\end{figure}

\begin{figure}
\begin{center}
\includegraphics[clip,scale=0.3]{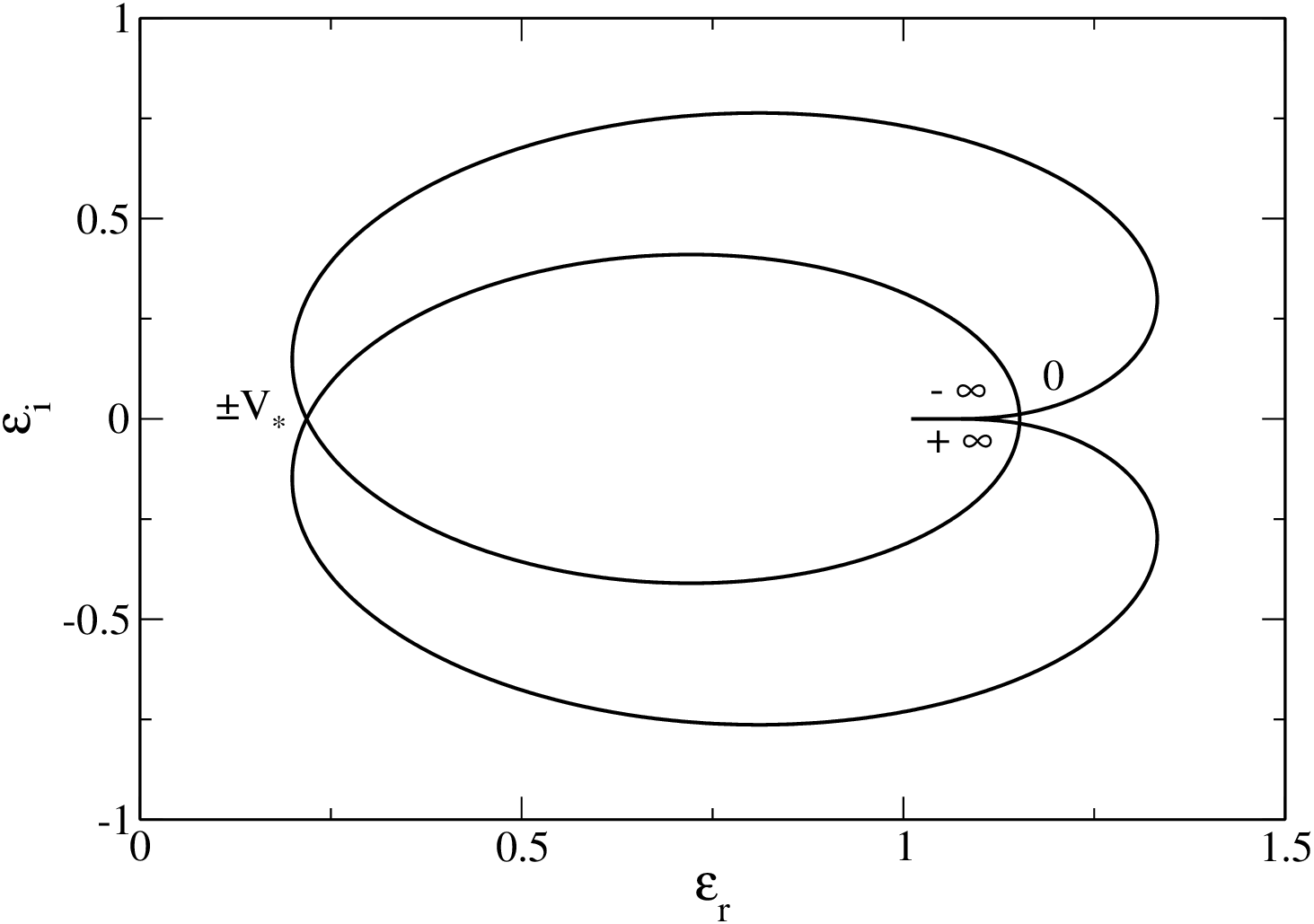}
\caption{Nyquist curve for $a>a_{M}$ and $1/a^2<\eta<\eta_{c}^{(\pm)}$ (specifically $a=1$ and $\eta=2$).  The DF has two maxima at $V_{0}=\pm V_{*}$ and one minimum at $V_{0}=0$. The DF is stable because the Nyquist curve does not encircle the origin. Case $(+++)$.}
\label{Ny1b}
\end{center}
\end{figure}

\begin{figure}
\begin{center}
\includegraphics[clip,scale=0.3]{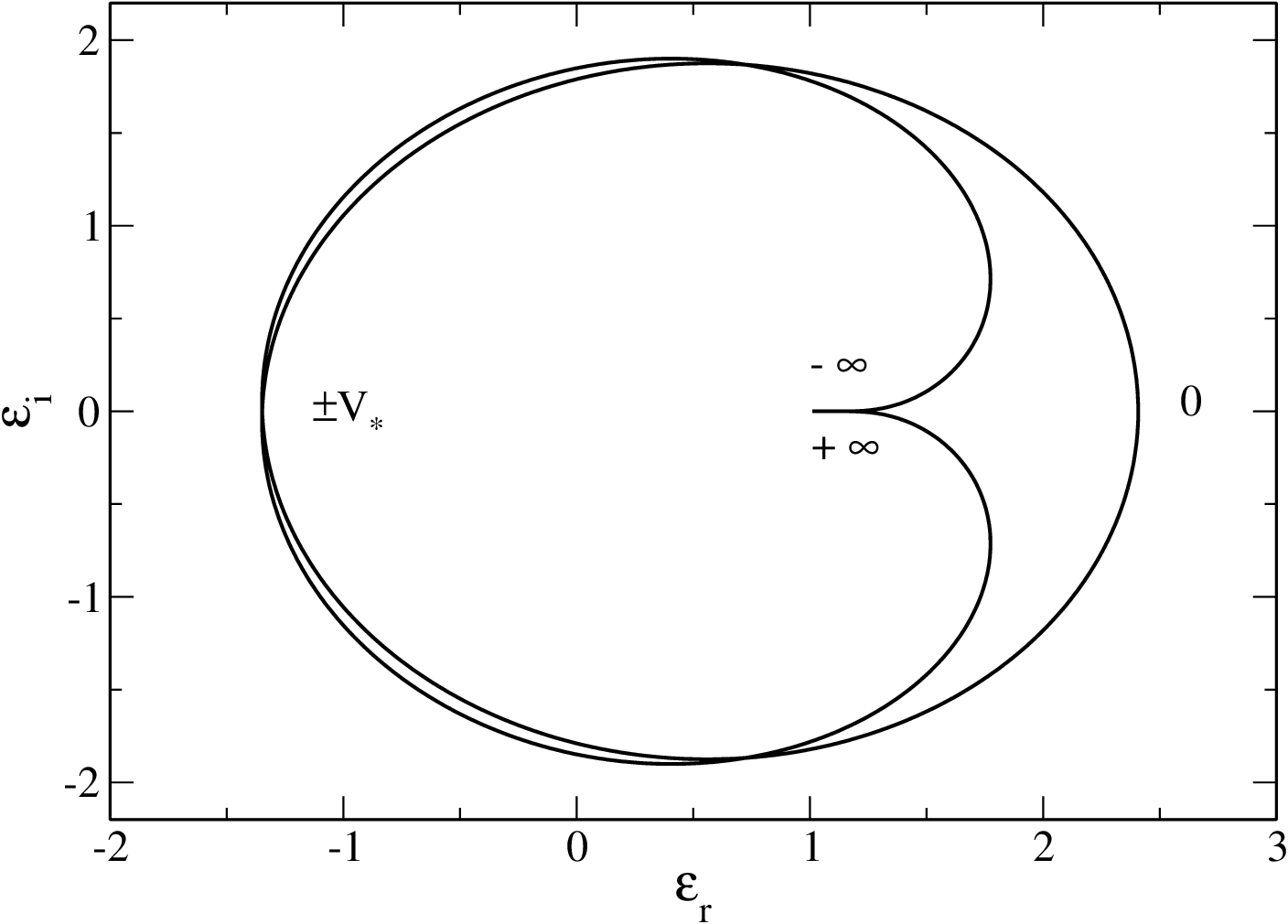}
\caption{Nyquist curve for $a>a_{M}$ and $\eta>\eta_{c}^{(\pm)}>1/a^2$ (specifically $a=1$ and $\eta=5$).  The DF has two maxima at $V_{0}=\pm V_{*}$ and one minimum at $V_{0}=0$. The DF is unstable because the Nyquist curve  encircles the origin. Since it rotates twice around the origin, this implies that there are  $N=2$ unstable modes $(\omega_r,\omega_i)$ with $\omega_{i}>0$.  Case $(-+-)$.}
\label{Ny2}
\end{center}
\end{figure}

For $a>a_M$, there exists one temperature $\eta_{c}^{(\pm)}$ at which
the DF is marginally stable. For $\eta=\eta_{c}^{(\pm)}$, the DF has
two maxima at $V_0=\pm V_*$ and one minimum at $V_0=0$. The marginal
perturbation evolves with a pulsation $\Omega_r=\pm V_*$. By
considering the Nyquist curves in this region (see
Figs. \ref{Ny1}-\ref{Ny2}), we find that the DF is stable for
$\eta<\eta_{c}^{(\pm)}$ and unstable for $\eta>\eta_{c}^{(\pm)}$.

\begin{figure}
\begin{center}
\includegraphics[clip,scale=0.3]{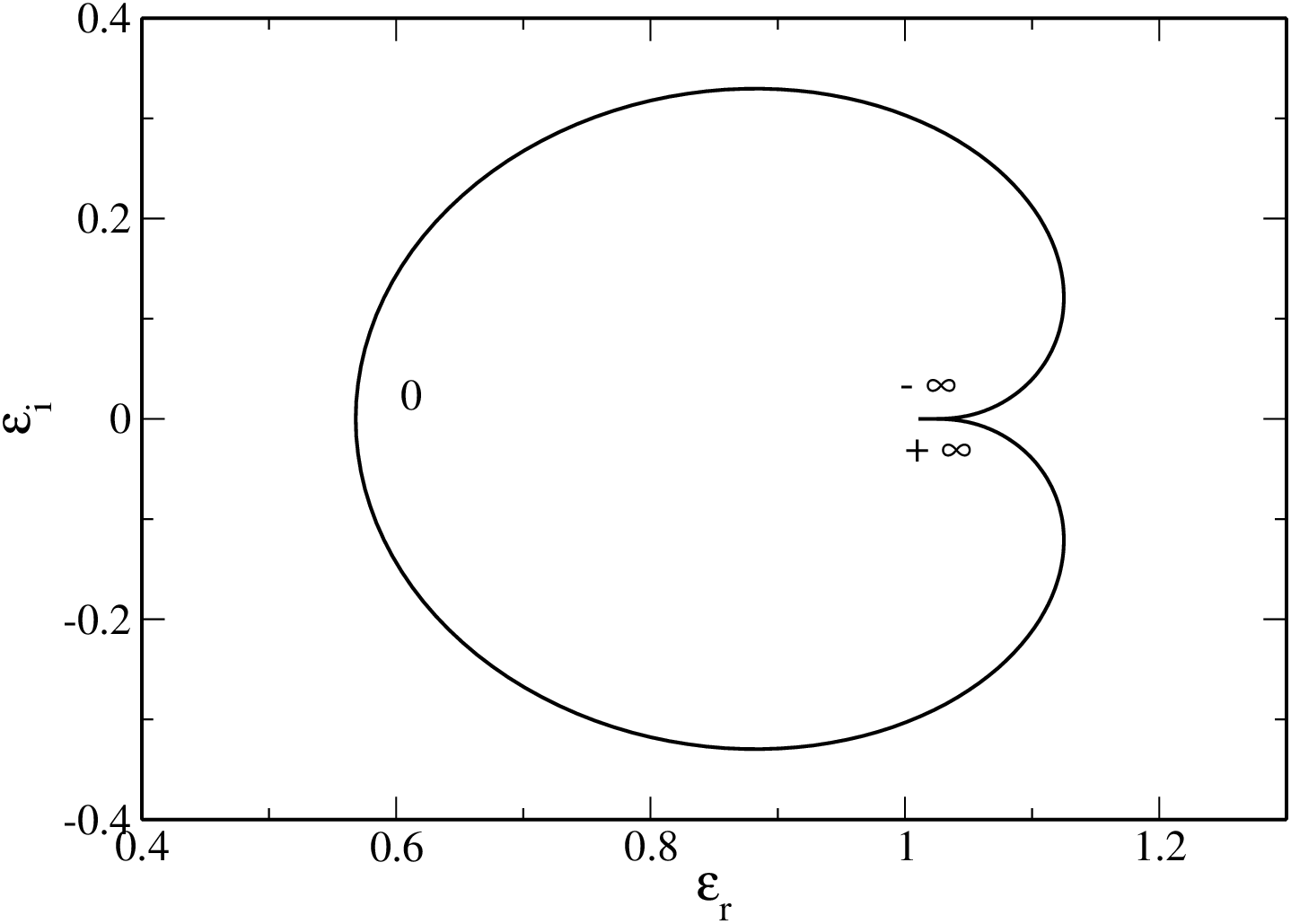}
\caption{Nyquist curve for $a_{*}<a<a_{M}$ and $\eta<\eta_{c}^{(0)(1)}<1/a^2$ (specifically $a=0.535$ and $\eta=0.5$).  The DF has only one maximum at $V_{0}=0$. The DF is stable because the Nyquist curve  does not encircle the origin. Case $(+)$.}
\label{Ny7}
\end{center}
\end{figure}

\begin{figure}
\begin{center}
\includegraphics[clip,scale=0.3]{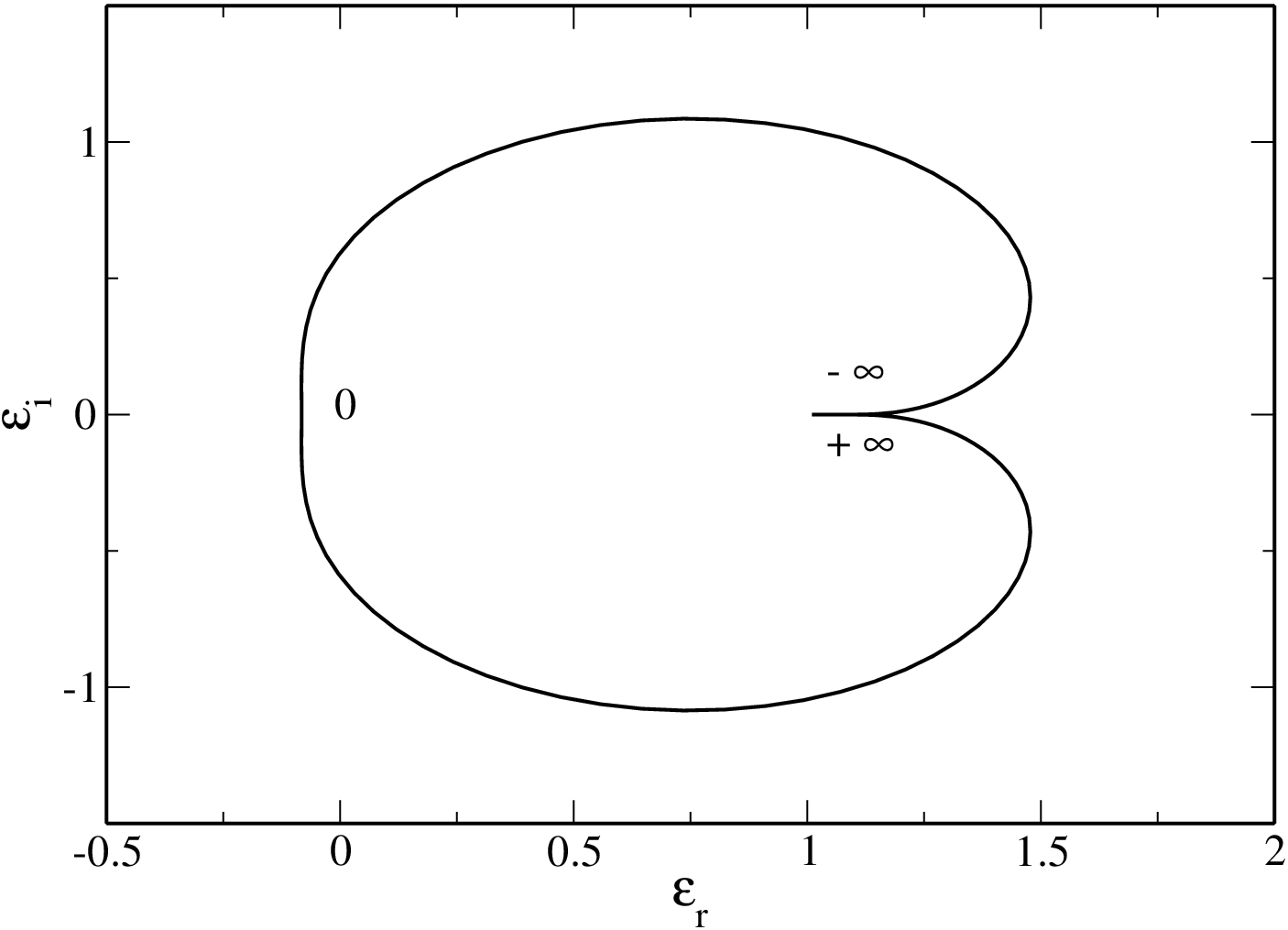}
\caption{Nyquist curve for $a_{*}<a<a_{M}$ and $\eta_{c}^{(0)(1)}<\eta<\eta_{c}^{(0)(2)}<1/a^2$ (specifically $a=0.535$ and $\eta=2.5$).  The DF has only one maximum at $V_{0}=0$. The DF is unstable because the Nyquist curve encircles the origin once. There is $N=1$ mode of instability. Case $(-)$.}
\label{Ny8}
\end{center}
\end{figure}

\begin{figure}
\begin{center}
\includegraphics[clip,scale=0.3]{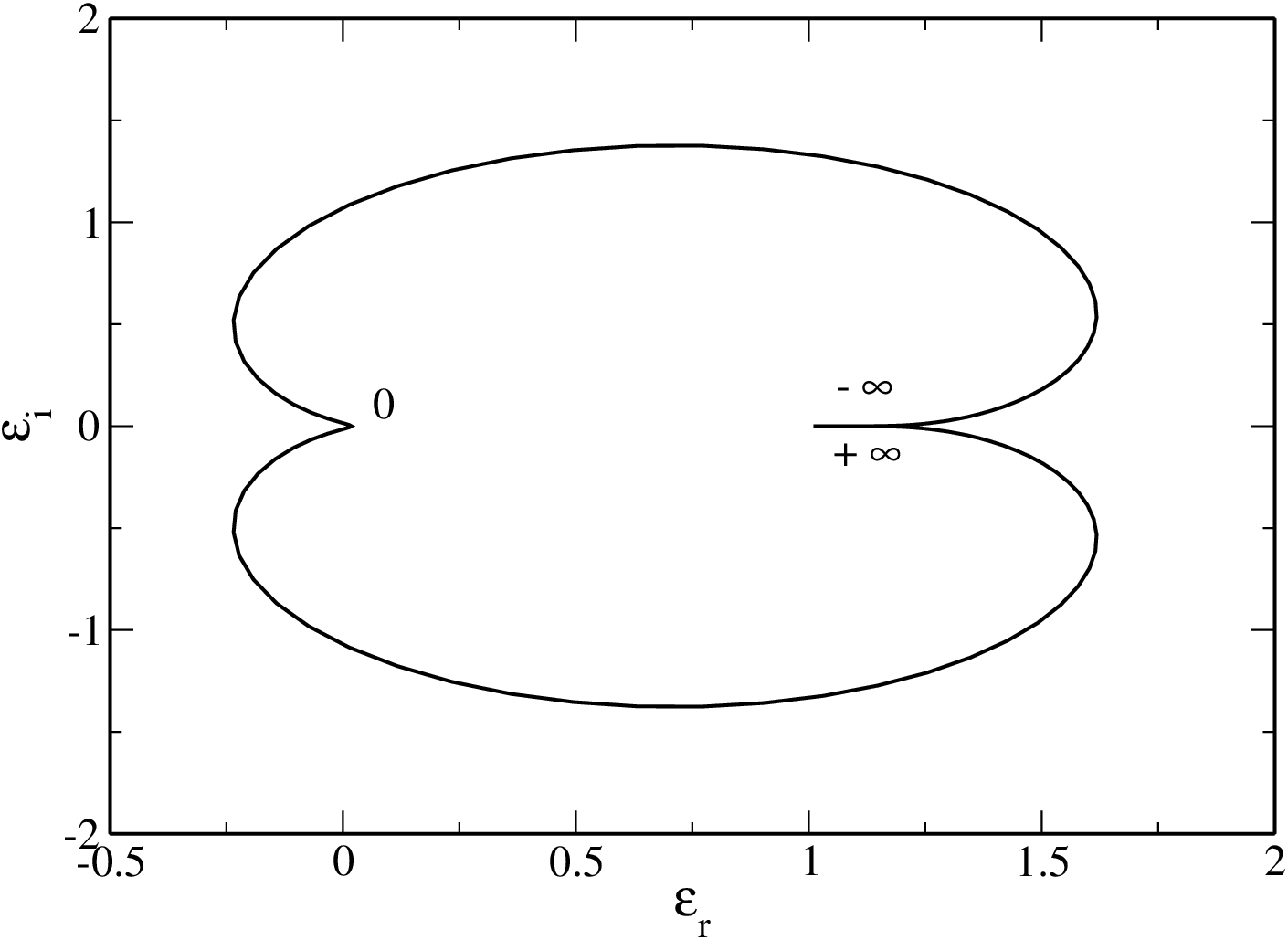}
\caption{Nyquist curve for $a_{*}<a<a_{M}$ and $\eta_{c}^{(0)(2)}<\eta<1/a^2<\eta_{c}^{(\pm)}$ (specifically $a=0.535$ and $\eta=3.4$).  The DF has only one maximum at $V_{0}=0$. The DF is stable because the Nyquist curve does not encircle the origin.  Case $(+)$. Note that when $\eta\rightarrow 1/a^2$, the tangent at $(\epsilon_{r}(\omega_r=0),0)$ is no more vertical since $f''(0)=0$ (see remark at the end of Sec. \ref{sec_gp}). 
This corresponds precisely the transition between one maximum and two
maxima and a minimum.  }
\label{Ny9}
\end{center}
\end{figure}

\begin{figure}
\begin{center}
\includegraphics[clip,scale=0.3]{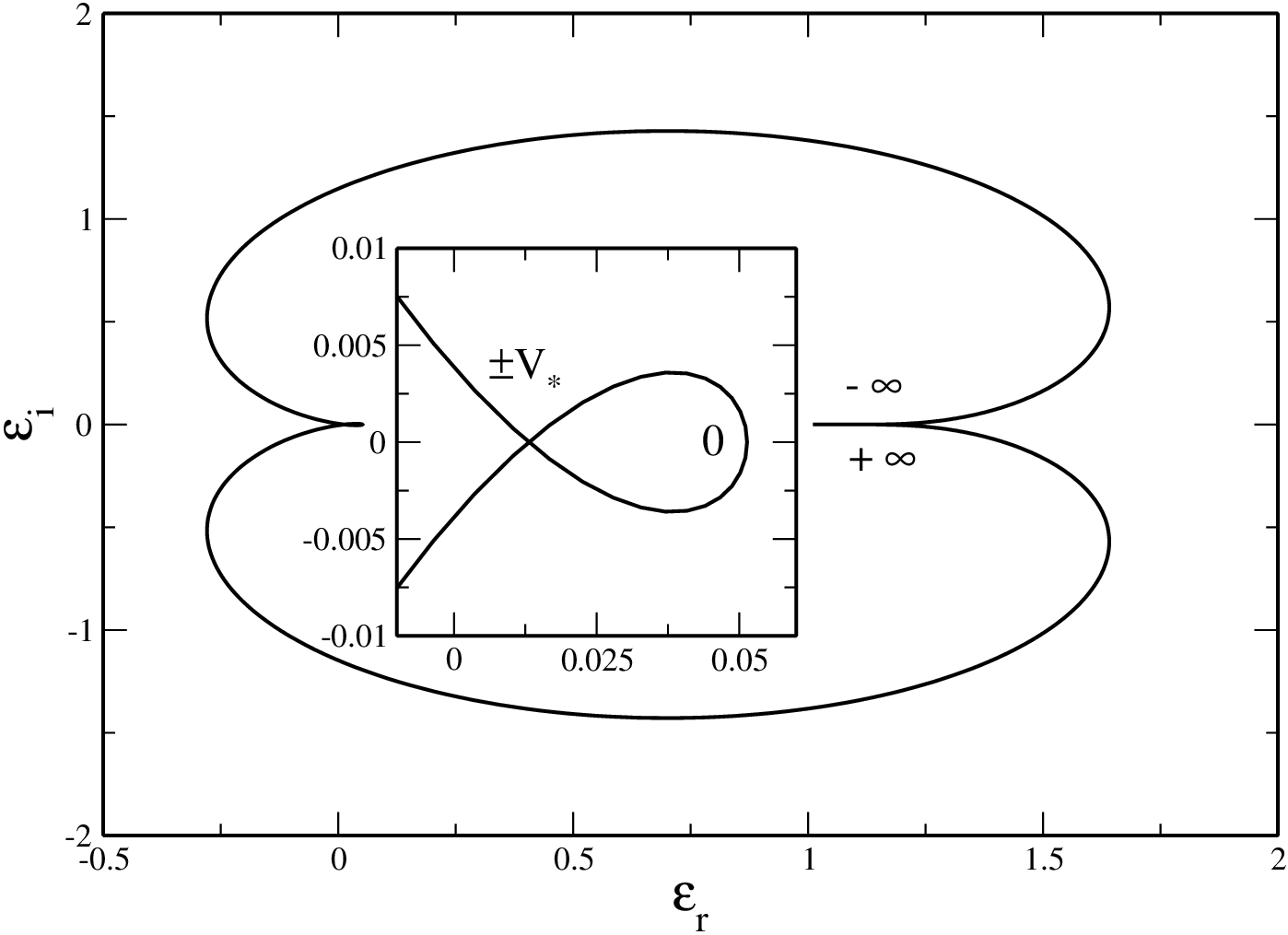}
\caption{Nyquist curve for $a_{*}<a<a_{M}$ and $\eta_{c}^{(0)(2)}<1/a^2<\eta<\eta_{c}^{(\pm)}$ (specifically $a=0.535$ and $\eta=3.55$).  The DF has two maxima at $V_{0}=\pm V_{*}$ and one minimum at $V_{0}=0$. The DF is stable because the Nyquist curve does not encircle the origin.  Case $(+++)$.}
\label{Ny10}
\end{center}
\end{figure}

\begin{figure}
\begin{center}
\includegraphics[clip,scale=0.3]{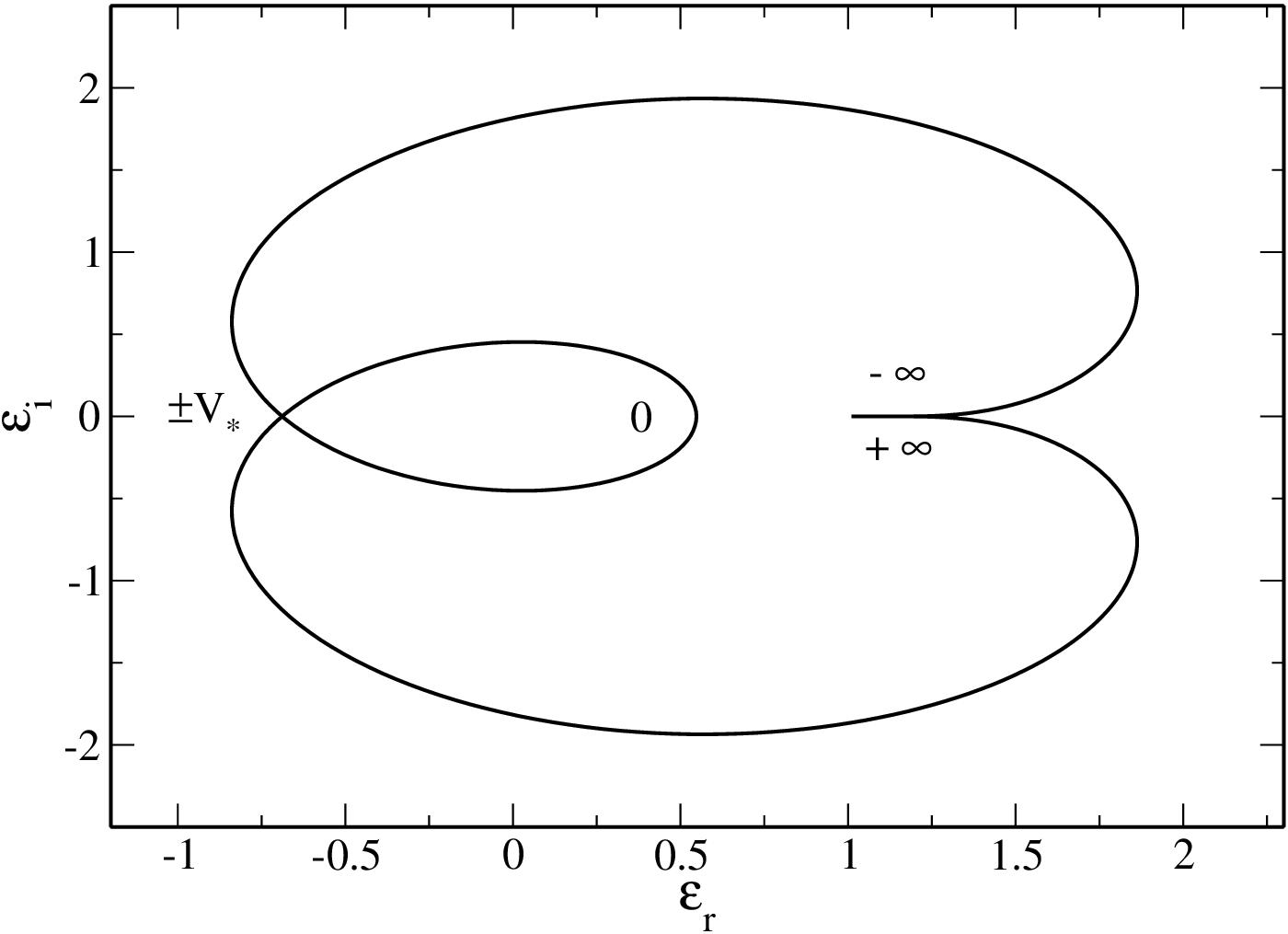}
\caption{Nyquist curve for $a_{*}<a<a_{M}$ and $\eta>\eta_{c}^{(\pm)}>1/a^2$ (specifically $a=0.535$ and $\eta=5$).  The DF has two maxima at $V_{0}=\pm V_{*}$ and one minimum at $V_{0}=0$. The DF is unstable because the Nyquist curve encircles the origin twice. There are $N=2$ modes of instability. Case $(-+-)$.}
\label{Ny11}
\end{center}
\end{figure}

For $a_*<a<a_M$, there exists three temperatures $\eta_{c}^{(0)(1)}$,
$\eta_{c}^{(0)(2)}$ and $\eta_{c}^{(\pm)}$ at which the DF is
marginally stable. For $\eta=\eta_{c}^{(0)(1)}$ and
$\eta=\eta_{c}^{(0)(2)}$, the DF has a single maximum at $V_0=0$. The
marginal perturbation has zero pulsation $\Omega_r=0$. For
$\eta=\eta_{c}^{(\pm)}$, the DF has two maxima at $V_0=\pm V_*$ and
one minimum at $V_0=0$. The marginal perturbation has a pulsation
$\Omega_r=\pm V_*$.  By considering the Nyquist curves in this
region (see Figs. \ref{Ny7}-\ref{Ny11}), we find that the DF is stable
for $\eta<\eta_{c}^{(0)(1)}$ unstable for
$\eta_{c}^{(0)(1)}<\eta<\eta_{c}^{(0)(2)}$, stable again for
$\eta_{c}^{(0)(2)}<\eta<\eta_{c}^{(\pm)}$ and unstable again for
$\eta>\eta_{c}^{(\pm)}$. This corresponds to a re-entrant phase
\footnote{Interestingly, a re-entrant phase
has also been observed in the stability diagram of the Lynden-Bell (or
Fermi-Dirac) distributions  for the HMF model \cite{epjb,reentrant}.}.

In conclusion, considering the stability diagram of Fig. \ref{ba}, the
symmetric double-humped distribution is stable below the solid line
and unstable above it. When the solution is unstable, we expect that
the system will evolve through violent relaxation towards a QSS that
corresponds, in case of efficient mixing, to a Lynden-Bell statistical
equilibrium state.

\section{The asymmetric double-humped distribution}
\label{sec_adh}

\subsection{Determination of the extrema}
\label{sec_ae}

\begin{figure}
\begin{center}
\includegraphics[clip,scale=0.3]{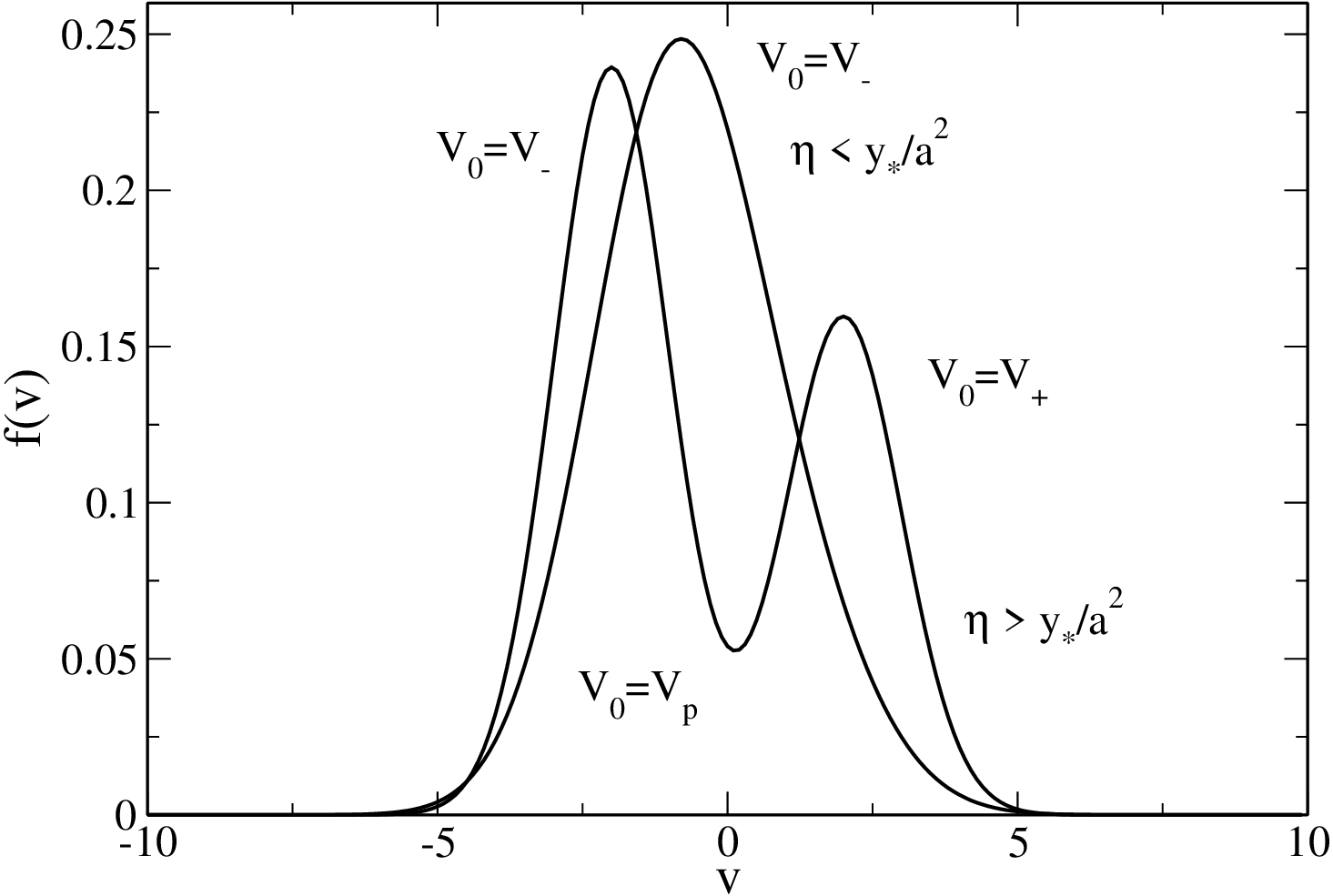}
\caption{Asymmetric double-humped distribution made of two Mawellians with separation $a$ and asymmetry $\Delta>1$. If $\eta>y_{*}/a^2$, the DF has one global maximum at $V_{0}=V_{-}<0$, a minimum at $V_{0}=V_{p}>0$ and a local maximum at $V_{0}=V_{+}>0$. If $\eta<y_{*}/a^2$, the DF has only one maximum at $V_{0}=V_{-}<0$.}
\label{asymdistr}
\end{center}
\end{figure}

\begin{figure}
\begin{center}
\includegraphics[clip,scale=0.3]{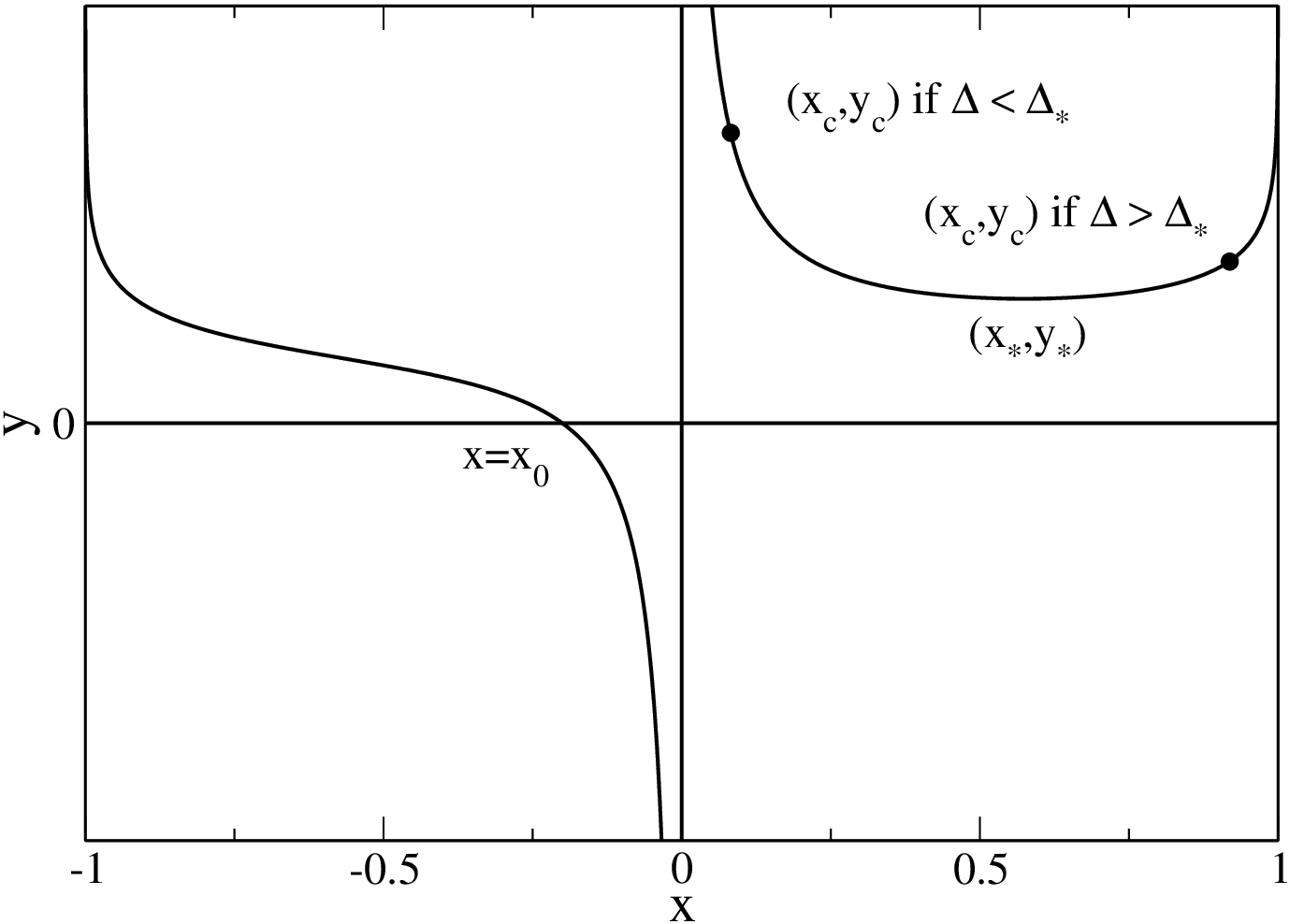}
\caption{The function $y(x)$ for the asymmetric double-humped distribution.}
\label{yasym}
\end{center}
\end{figure}

\begin{figure}
\begin{center}
\includegraphics[clip,scale=0.3]{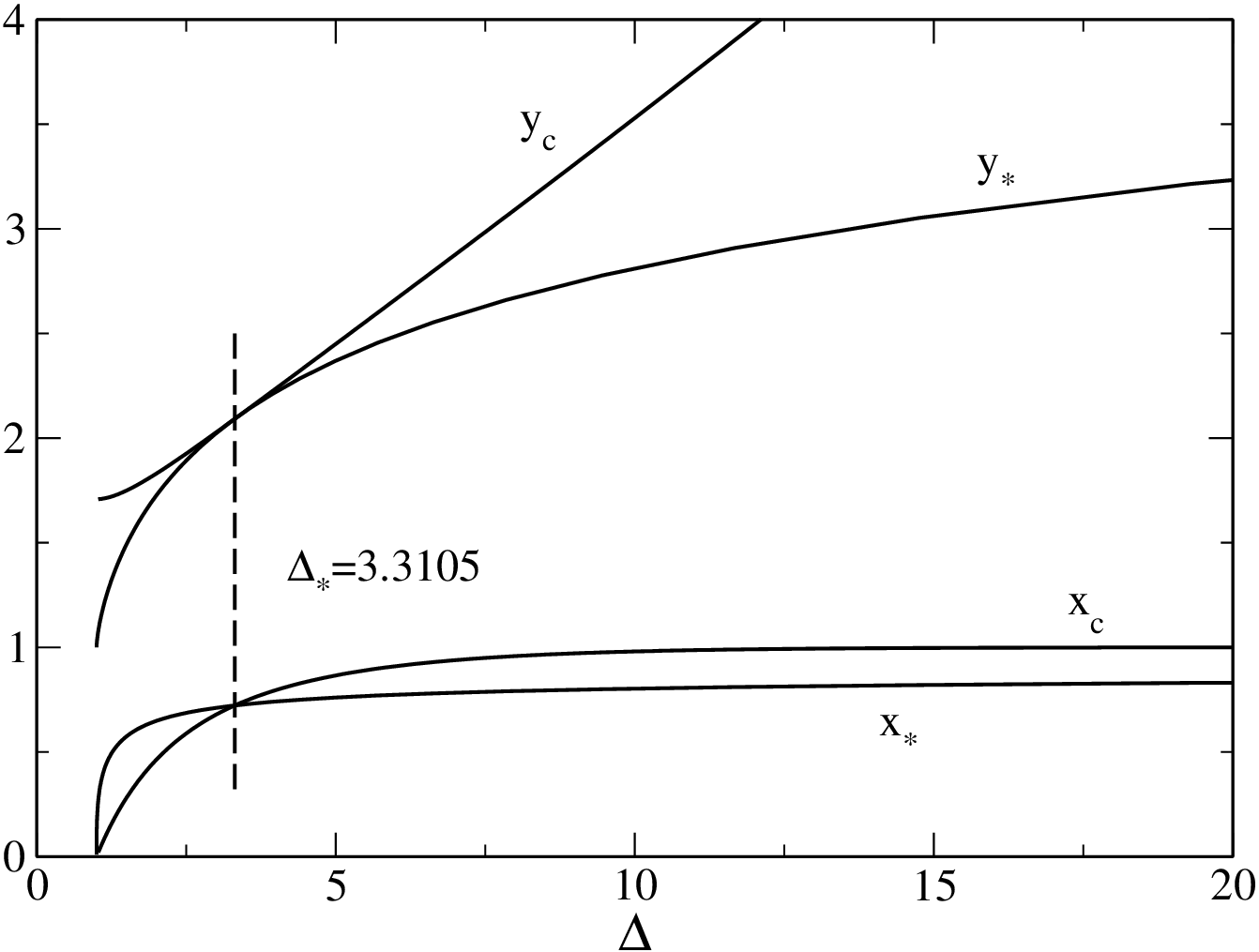}
\caption{Evolution of $x_*$, $y_*$, $x_c$ and $y_c$ as a function of $\Delta$. The curves intersect each other at $\Delta_{*}=3.3105$.}
\label{xynew}
\end{center}
\end{figure}

We consider an asymmetric
double-humped distribution of the form
\begin{equation}
\label{ae1}
f(v)=\sqrt{\frac{\beta}{2\pi}} \frac{\rho}{1+\Delta} \left [  e^{-\frac{\beta}{2}(v-v_a)^2} + \Delta e^{-\frac{\beta}{2}(v+v_a)^2}  \right ],
\end{equation}
where $T=1/\beta$ is the temperature of the Maxwellians and $\Delta$
is the asymmetry parameter (we assume here that $\Delta>1$). This distribution  is plotted in Fig. \ref{asymdistr}. The symmetric case is recovered for
$\Delta=1$. The average velocity is $\langle
v\rangle=-[(\Delta-1)/(\Delta+1)]v_a$ and the kinetic temperature
$T_{kin}\equiv \langle (v-\langle v\rangle)^2\rangle=T+[4\Delta/(\Delta+1)^2]v_{a}^{2}$.  The velocities $v_0$ at
which the distribution function $f(v)$ is extremum satisfy
$f'(v_0)=0$. They are determined by the equation
\begin{equation}
\label{ae2}
e^{-2\beta v_{a} v_0}=\frac{1}{\Delta}\frac{v_a-v_0}{v_a+v_0}.
\end{equation}
Introducing the dimensionless temperature (\ref{mm4}), the
dimensionless velocity (\ref{mm5}) and the dimensionless separation
(\ref{e3}), Eq. (\ref{ae2}) can be rewritten
\begin{equation}
\label{ae3}
\eta=\frac{1}{2aV_0}\ln\left (\frac{a+V_0}{a-V_0}\right )+\frac{\ln(\Delta)}{2 a V_0}.
\end{equation}
For a given value of inverse temperature $\eta$, separation $a$ and
asymmetry  $\Delta$, this equation determines the velocities
$V_0$ where $f(V)$ is extremum. We note that $V_0\in \rbrack -a,
+a\lbrack$. It is convenient to introduce the variables
\begin{equation}
\label{ae4}
x=V_0/a, \qquad  y=\eta a^2.
\end{equation}
For fixed $a$, the parameter $y$ plays the role of the inverse
temperature and the parameter $x$ plays the role of the
velocity. Then, we have to study the function
\begin{equation}
\label{ae5}
y(x)=\frac{1}{2x}\ln\left (\frac{1+x}{1-x}\right )+\frac{\ln(\Delta)}{2x},
\end{equation}
for $x \in ] -1, +1[$. This function is plotted in
Fig. \ref{yasym}. It has the following properties
\begin{equation}
\label{ae6}
y(x)\sim -\frac{1}{2}\ln (1-x), \qquad (x\rightarrow 1^{-}),
\end{equation}
\begin{equation}
\label{ae7}
y(x)\sim -\frac{1}{2}\ln (1+x), \qquad (x\rightarrow -1^{+}),
\end{equation}
\begin{equation}
\label{ae8}
y(x)\sim \frac{\ln(\Delta)}{2x}, \qquad (x\rightarrow 0).
\end{equation}
Considering the negative velocities $x<0$, we note that $y\ge 0$ iff $x\le x_0$ with
\begin{equation}
\label{ae9}
x_{0}=\frac{\Delta-1}{\Delta+1}.
\end{equation}
Considering the positive velocities $x>0$, we note that the curve $y(x)$ is
minimum at $x=x_*$ where $x_{*}$ is solution of
\begin{equation}
\label{ae10}
\frac{2x_{*}}{1-x_{*}^{2}}-\ln\left (\frac{1+x_{*}}{1-x_{*}}\right )=\ln(\Delta).
\end{equation}
This function is represented in Fig. \ref{xynew}. We note that
Eq. (\ref{ae10}) has a unique solution $x_*$ for each value of
$\Delta>1$. Therefore, the function $y(x)$ has a single minimum at
$x=x_{*}$. The value of this minimum is
\begin{equation}
\label{ae11}
y_{*}=\frac{1}{1-x_{*}^{2}}>1.
\end{equation}
The extrema of the distribution $f(V)$ can be determined from the
study of the function (\ref{ae5}). If $y<y_{*}$, i.e. $\eta<y_{*}/a^2$, the
distribution $f(V)$ has a single maximum at $V_{0}=V_{-}<0$.  If
$y>y_{*}$, i.e. $\eta>y_{*}/a^2$, the distribution $f(V)$ has two
maxima at $V_{0}=V_{-}<0$ and $V_{0}=V_{+}>0$ and one minimum at
$V_{0}=V_{p}>0$. These different values are given by $V_{0}=a y^{-1}(\eta
a^2)$.

In conclusion, for a given asymmetry $\Delta$ and separation $a$:

$\bullet$ if  $\eta>y_{*}/a^2$, the distribution function $f(V)$ has a global maximum at $V_{0}=V_{-}<0$, a local maximum  at $V_{0}=V_{+}>0$  and one minimum at  $V_{0}=V_{p}>0$.

$\bullet$ if  $\eta<y_{*}/a^2$, the distribution function $f(V)$ has only one
maximum at $V_{0}=V_{-}<0$.

{\it Remark:} We can note the formal analogy with the mean field
theory (ferromagnetic transition) of the one dimensional Ising model
with magnetic field where the asymmetry $\Delta$ plays the role of the
magnetic field (compare Eq. (\ref{ae3}) with Eq. (14) of
\cite{bellac}).

\subsection{The condition of marginal stability}
\label{sec_ams}

\begin{figure}
\begin{center}
\includegraphics[clip,scale=0.3]{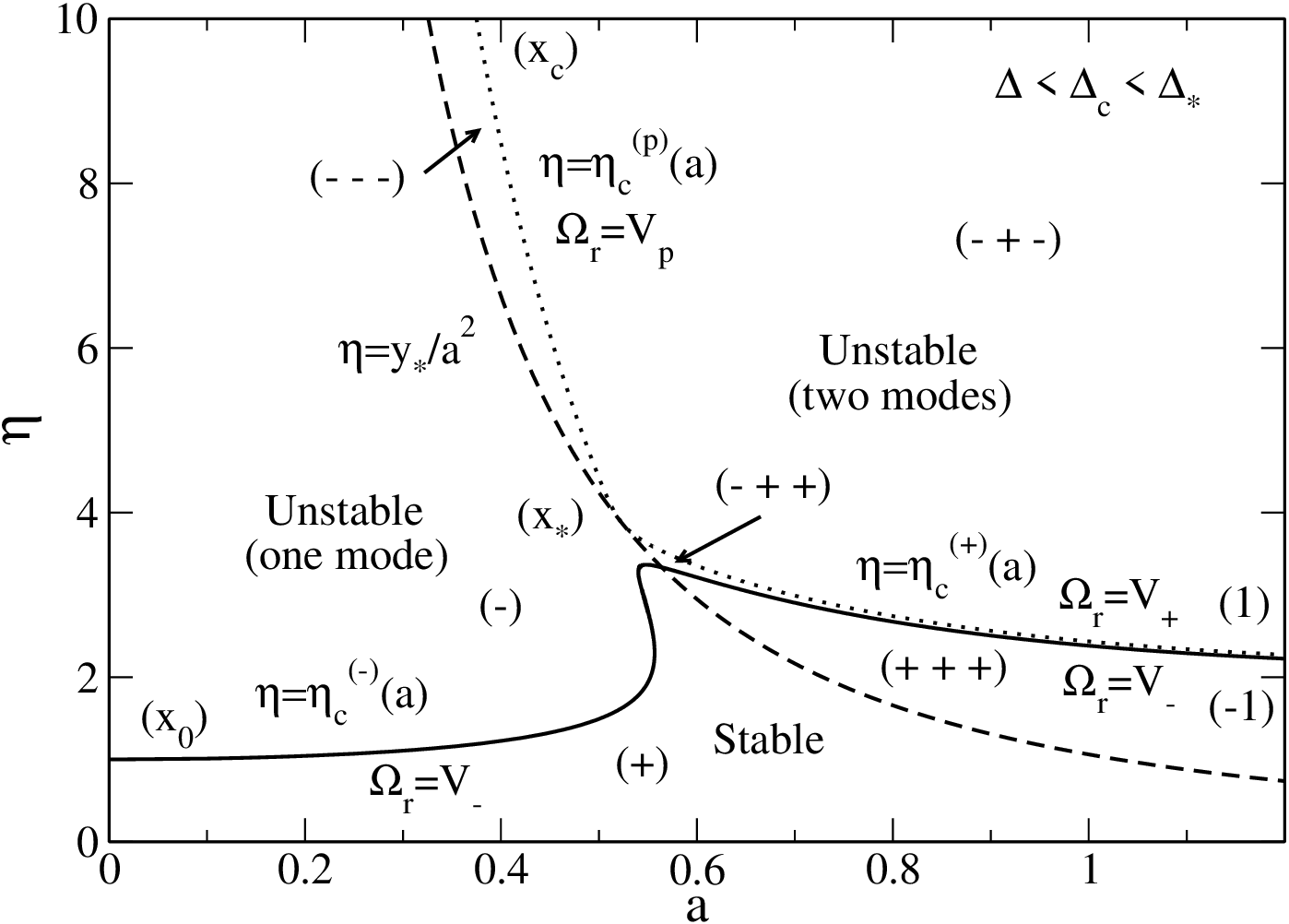}
\caption{Stability diagram of the asymmetric double-humped distribution for $\Delta<\Delta_c<\Delta_{*}$ (specifically $\Delta=1.02$) showing a re-entrant phase in continuity with the symmetric case. Since $\Delta<\Delta_{*}$, there exists three marginal branches $\eta_{c}^{(-)}(a)$, $\eta_{c}^{(p)}(a)$ and $\eta_{c}^{(+)}(a)$. }
\label{diagasym1}
\end{center}
\end{figure}

\begin{figure}
\begin{center}
\includegraphics[clip,scale=0.3]{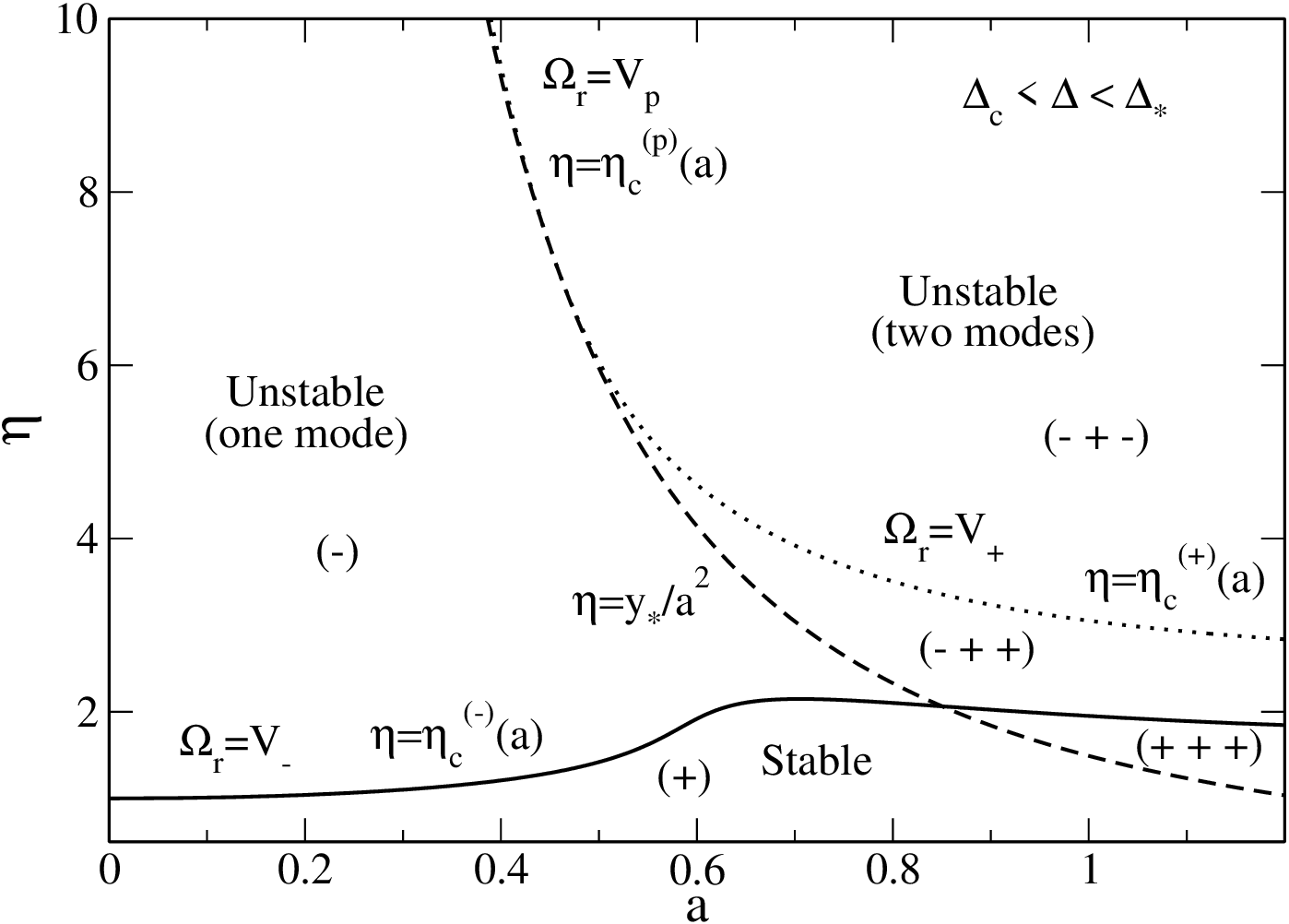}
\caption{Stability diagram of the asymmetric double-humped distribution for $\Delta_c<\Delta<\Delta_{*}$ (specifically $\Delta=1.5$) showing the disappearance of the re-entrant phase.  }
\label{diagasym2}
\end{center}
\end{figure}

\begin{figure}
\begin{center}
\includegraphics[clip,scale=0.3]{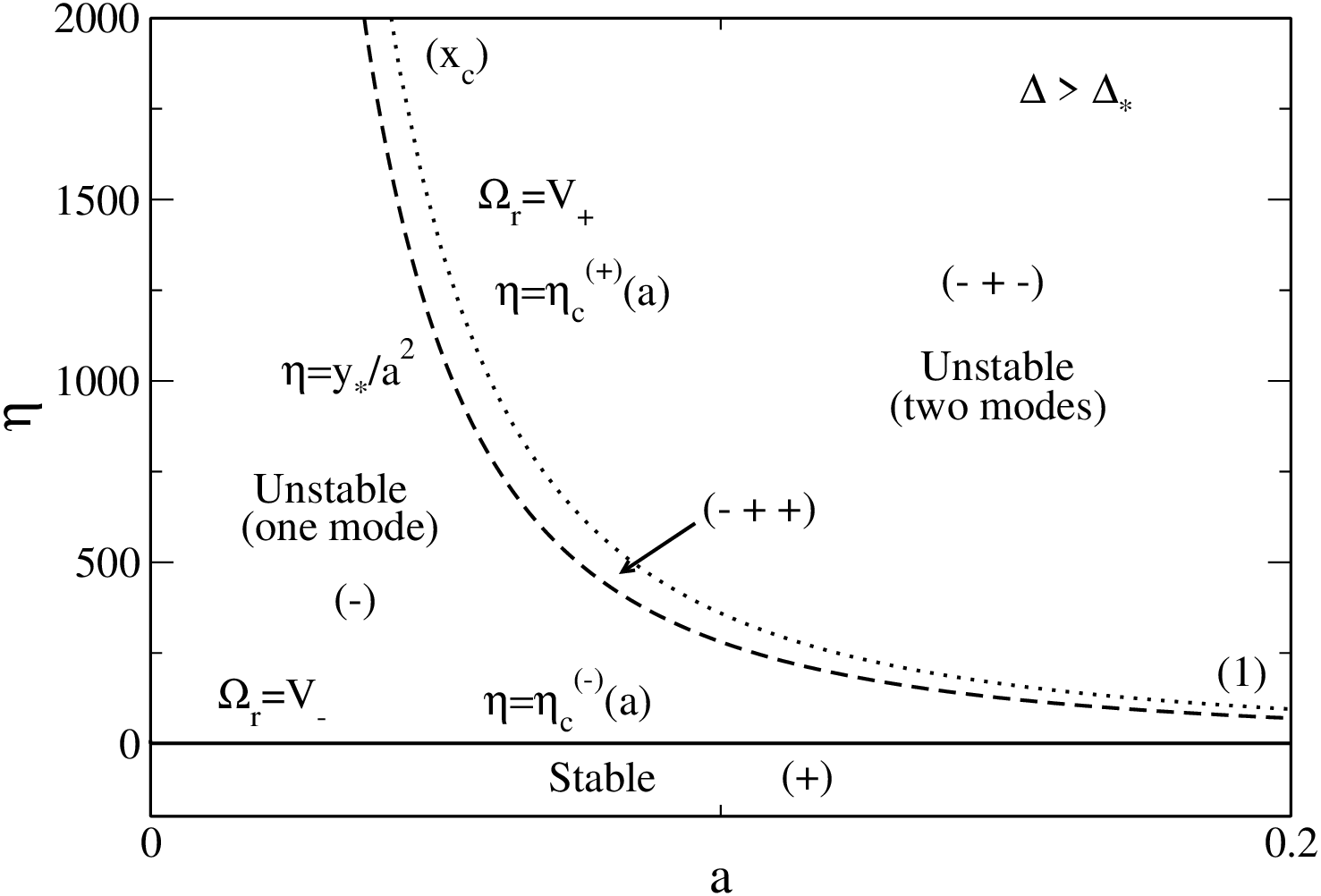}
\caption{Stability diagram of the asymmetric double-humped distribution for $\Delta>\Delta_{*}$ (specifically $\Delta=10$). Since $\Delta>\Delta_{*}$, there exists only two marginal branches   $\eta_{c}^{(-)}(a)$ and $\eta_{c}^{(+)}(a)$.  }
\label{diag10}
\end{center}
\end{figure}

The dielectric function associated to the asymmetric double-humped
distribution is
\begin{eqnarray}
\label{ams1}
\epsilon(\Omega)= 1-\frac{\eta}{1+\Delta} \left [ W(\sqrt{\eta}(\Omega - a))+\Delta W(\sqrt{\eta}(\Omega + a))   \right ],\nonumber\\
\end{eqnarray}
where $W(z)$ is defined in Eq.  (\ref{mm7}). When $\Omega_{i}=0$, the real and imaginary parts of the dielectric function  $\epsilon(\Omega_{r})=\epsilon_{r}(\Omega_{r})+i\epsilon_{i}(\Omega_{r})$ can be written
\begin{eqnarray}
\label{ams2}
\epsilon_{r}(\Omega_{r})=1-\frac{\eta}{1+\Delta}  [ W_r(\sqrt{\eta}(\Omega_{r} - a))\nonumber\\
+\Delta W_r(\sqrt{\eta}(\Omega_{r} + a))    ],
\end{eqnarray}
\begin{eqnarray}
\label{ams3}
\epsilon_{i}(\Omega_{r})=-\frac{\eta}{1+\Delta}  [ W_i(\sqrt{\eta}(\Omega_{r} - a))\nonumber\\
+\Delta W_i(\sqrt{\eta}(\Omega_{r} + a))   ],
\end{eqnarray}
where $W_r(z)$ and $W_i(z)$ are defined in Eqs. (\ref{mm11})-(\ref{mm12})
where $z$ is here a real number. The condition of marginal stability corresponds to $\epsilon_{r}(\Omega_{r})=\epsilon_{i}(\Omega_{r})=0$. The condition $\epsilon_{i}(\Omega_{r})=0$ is equivalent  to
\begin{equation}
\label{ams4}
f'(\Omega_{r})=0.
\end{equation}
The condition $\epsilon_{r}(\Omega_{r})=0$ leads to
\begin{eqnarray}
\label{ams5}
1-\frac{\eta}{1+\Delta} \left [ W_{r}(\sqrt{\eta}(\Omega_{r} - a))+\Delta W_{r}(\sqrt{\eta}(\Omega_{r} + a))   \right ]=0.\nonumber\\
\end{eqnarray}
According to Eq. (\ref{ams4}), the real pulsation $\Omega_{r}$ is
equal to a velocity $V_{0}$ at which the distribution (\ref{ae1}) is
extremum. The second equation (\ref{ams5}) determines the value(s)
$\eta_{c}(a)$ of the temperature at which the distribution is marginally
stable. Therefore, we have to solve
\begin{eqnarray}
\label{ams6}
1-\frac{\eta}{1+\Delta} \left [ W_{r}(\sqrt{\eta}(V_{0} - a))+\Delta W_{r}(\sqrt{\eta}(V_{0}+ a))   \right ]=0,\nonumber\\
\end{eqnarray}
where $V_{0}$ is given by
\begin{equation}
\label{ams7}
\eta=\frac{1}{2aV_{0}}\ln\left (\frac{a+V_{0}}{a-V_{0}}\right )+\frac{\ln(\Delta)}{2 a V_{0}}.
\end{equation}
Eliminating $V_0$ between these two expressions yields the
critical temperature(s) $\eta_{c}(a)$ as a function of
$a$. However, it is easier to proceed differently. Setting $x=V_0/a$
and $y=\eta a^2$, we obtain the equations
\begin{equation}
\label{ams8}
y=\frac{1}{2x}\ln\left (\frac{1+x}{1-x}\right
)+\frac{\ln(\Delta)}{2x},
\end{equation}
\begin{equation}
\label{ams9}
\eta=\frac{1+\Delta}{\left [ W_{r}(\sqrt{y}(x - 1))+\Delta W_{r}(\sqrt{y}(x+ 1))   \right ]},
\end{equation}
\begin{equation}
\label{ams10}
a^2=\frac{y}{\eta}.
\end{equation}
For given $x$, we can obtain $y$ from Eq. (\ref{ams8}) [see also
Fig. \ref{yasym}], $\eta$ from Eq. (\ref{ams9}) and $a$ from
Eq. (\ref{ams10}). Varying $x$ in the interval $]-1,1[$ yields the
full curve $\eta_c(a)$. We have three types of solutions. For $x\in
]-1,x_0]$, we obtain a branch $\eta_c^{(-)}(a)$ where the pulsation of
the marginal mode is negative: $\Omega_{r}=V_{-}<0$ corresponding to
the global maximum of $f(v)$. For $x\in ]0,x_*]$, we obtain a branch
$\eta_c^{(p)}(a)$ where the pulsation of the marginal mode is
positive: $\Omega_{r}=V_{p}>0$ corresponding to the minimum of
$f(v)$. For $x\in [x_*,1[$, we obtain a branch $\eta_c^{(+)}(a)$ where
the pulsation of the marginal mode is positive: $\Omega_{r}=V_{+}>0$
corresponding to the local maximum of $f(v)$. This leads to the curves
reported in Figs. \ref{diagasym1}, \ref{diagasym2} and \ref{diag10}
for three values of the asymmetry factor $\Delta$.

The limit $a\rightarrow +\infty$ (i.e. $y\rightarrow +\infty$)
corresponds to $V_{0}\rightarrow a^{-}$ (i.e. $x\rightarrow 1^{-}$) or
$V_{0}\rightarrow -a^{+}$ (i.e. $x\rightarrow -1^{+}$). For
$x\rightarrow 1^{-}$, Eqs. (\ref{ams8}) and (\ref{ams9}) reduce to
\begin{equation}
\label{ams11}
y\sim -\frac{1}{2}\ln (1-x) \rightarrow +\infty,
\end{equation}
and
\begin{equation}
\label{ams12}
\eta\simeq \frac{1+\Delta}{\left [ W_{r}(\sqrt{y}(x - 1))+\Delta W_{r}(2\sqrt{y})   \right ]}.
\end{equation}
Using Eqs. (\ref{wr8}) and (\ref{wr9}), we obtain at leading order
\begin{eqnarray}
\label{ams13}
\eta\simeq \frac{1+\Delta}{1-\frac{\Delta}{4y}}\simeq (1+\Delta) \left (1+\frac{\Delta}{4y}\right )\simeq (1+\Delta) \left (1+\frac{\Delta}{4\eta a^2}\right ),\nonumber\\
\end{eqnarray}
so that
\begin{equation}
\label{ams14}
\eta_{c}^{(+)}(a)=1+\Delta+\frac{\Delta}{4a^2}+... \qquad (a\rightarrow +\infty).
\end{equation}
For $x\rightarrow -1^{+}$, Eqs.  (\ref{ams8}) and (\ref{ams9}) reduce to
\begin{equation}
\label{ams15}
y\sim -\frac{1}{2}\ln (1+x) \rightarrow +\infty,
\end{equation}
and
\begin{equation}
\label{ams16}
\eta\simeq \frac{1+\Delta}{\left [ W_{r}(2\sqrt{y})+\Delta W_{r}(\sqrt{y}(1+x))   \right ]}.
\end{equation}
Using Eqs. (\ref{wr8}) and (\ref{wr9}), we obtain at leading order
\begin{eqnarray}
\label{ams17}
\eta\simeq \frac{1+\Delta}{\Delta-\frac{1}{4y}}\simeq \frac{1+\Delta}{\Delta} \left (1+\frac{1}{4\Delta y}\right )\simeq \frac{1+\Delta}{\Delta} \left (1+\frac{1}{4\Delta \eta a^2}\right ),\nonumber\\
\end{eqnarray}
so that
\begin{equation}
\label{ams18}
\eta_{c}^{(-)}(a)=\frac{1+\Delta}{\Delta}+\frac{1}{4\Delta a^2}+... \qquad (a\rightarrow +\infty).
\end{equation}

The limit $a\rightarrow 0$ with $\eta$ finite corresponds to $y\rightarrow 0$, i.e. $x\rightarrow x_0$. We find from Eq.  (\ref{ams9}) that $\eta\rightarrow 1/W_{r}(0)=1$ so that
\begin{equation}
\label{ams19}
\eta_{c}^{(-)}(a)\rightarrow 1 \qquad (a\rightarrow 0).
\end{equation}
This returns the critical temperature (\ref{mm13}) associated with the
Maxwellian distribution (\ref{mm1}) corresponding to $a=0$. The limit
$a\rightarrow 0$ with $\eta\rightarrow +\infty$ corresponds to
\begin{equation}
\label{ams20}
\eta_{c}^{(P)}(a)\sim\frac{y_{c}}{a^2}, \qquad (a\rightarrow 0),
\end{equation}
with $y_{c}$ finite ($P=p$ if $\Delta<\Delta_{*}$ and $P=+$ if
$\Delta>\Delta_{*}$ as will become clear below). Since $\eta\rightarrow
+\infty$, this constant $y_c$ is determined by
\begin{equation}
\label{ams21}
y_{c}=\frac{1}{2x_{c}}\ln\left (\frac{1+x_{c}}{1-x_{c}}\right
)+\frac{\ln(\Delta)}{2x_{c}},
\end{equation}
\begin{equation}
\label{ams22}
W_{r}(\sqrt{y_{c}}(x_{c} - 1))+\Delta W_{r}(\sqrt{y_{c}}(x_{c}+ 1))=0.
\end{equation}
Note that there is no physical solution to
Eqs. (\ref{ams8})-(\ref{ams10}) when $0<x<x_{c}$ ($\eta$ would be
negative) so that the branch $\eta_{c}^{(P)}(a)$ starts at
$(0,+\infty)$ corresponding to $x=x_{c}$.  The evolution of $x_c$ and
$y_c$ with $\Delta$ is studied in Fig. \ref{xynew} and is compared
with the evolution of $x_*$ and $y_*$. The curves intersect each other
at $\Delta_*=3.3105$.  For $\Delta<\Delta_*$, $x_c<x_*$ so that the
stability diagram displays three marginal branches
$\eta_{c}^{(-)}(a)$, $\eta_{c}^{(p)}(a)$ and $\eta_{c}^{(+)}(a)$ (see
Fig. \ref{diagasym1}). The branches $\eta_{c}^{(p)}(a)$ and
$\eta_{c}^{(+)}(a)$ connect each other at $(a_{*},\eta_{*})$
corresponding to $x=x_{*}$. At that point they touch the hyperbole
$\eta=y_{*}/a^2$ separating distributions with one or two
maxima. Since $y_c> y_*$, the curve $\eta_{c}^{(p)}(a)$ for
$a\rightarrow 0$ is above the hyperbole $\eta=y_{*}/a^2$. For
$\Delta=\Delta_*$, $x_c=x_*$ so that the branch $\eta_{c}^{(p)}(a)$ is
rejected at infinity and only the branches $\eta_{c}^{(-)}(a)$ and
$\eta_{c}^{(+)}(a)$ remain. Since $y_c=y_*$, the branch
$\eta_{c}^{(+)}(a)$ coincides with the hyperbole $\eta=y_{*}/a^2$ for
$a\rightarrow 0$. For $\Delta>\Delta_*$, $x_c>x_*$ so that the phase
diagram displays only two marginal branches $\eta_{c}^{(-)}(a)$ and
$\eta_{c}^{(+)}(a)$ (see Fig. \ref{diag10}). The curve
$\eta_{c}^{(+)}(a)$ is stictly above the hyperbole $\eta=y_{*}/a^2$
without any intersection.

\subsection{The stability diagram}

\begin{figure}
\begin{center}
\includegraphics[clip,scale=0.3]{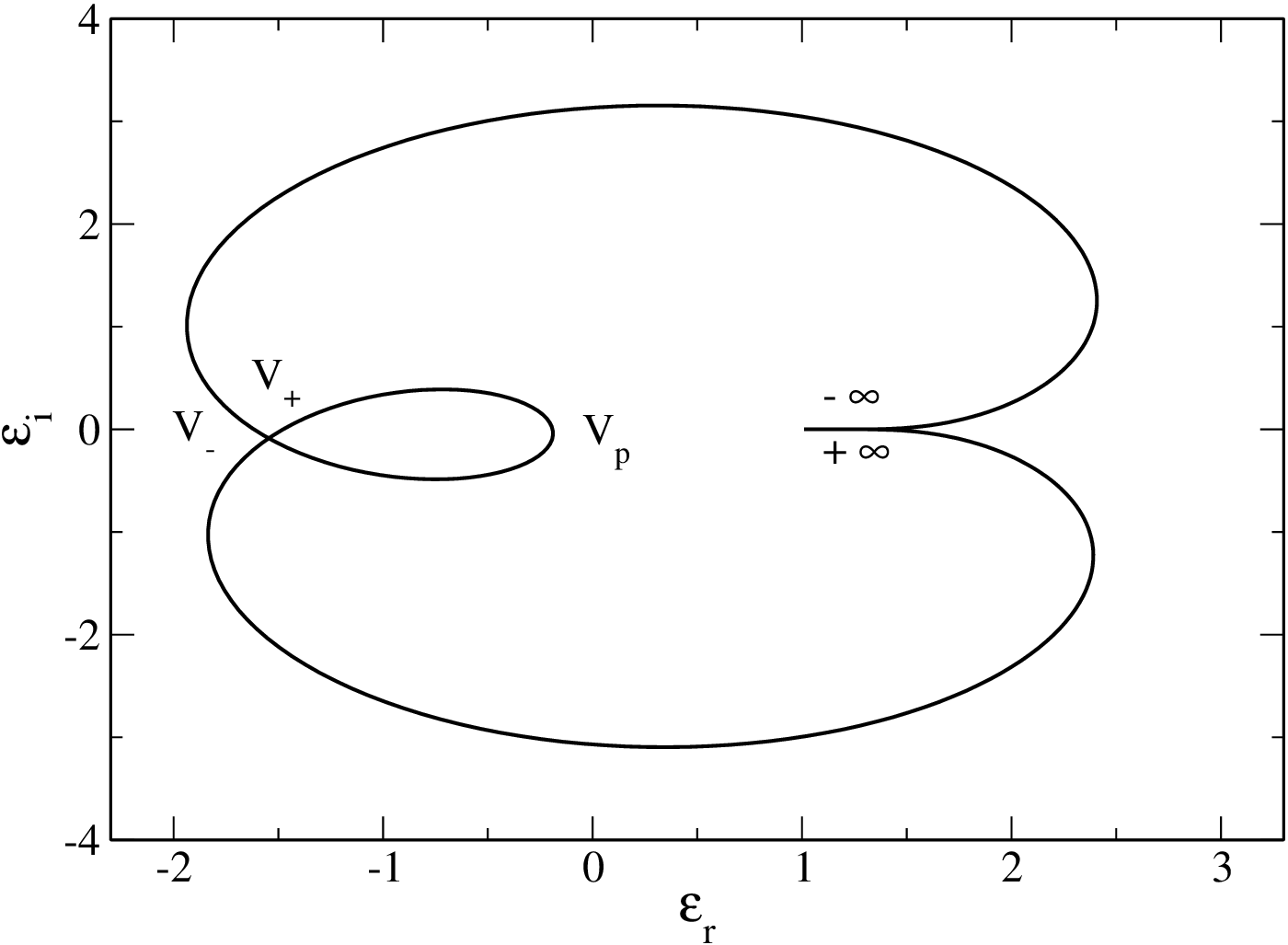}
\caption{For $\Delta<\Delta_{c}<\Delta_{*}$: Nyquist curve for $a<a_{*}$ and $y_{*}/a^2<\eta<\eta_{c}^{(p)}$ (specifically $\Delta=1.02$, $a=0.4$ and $\eta=8$).  The DF has two  maxima at $V_{-}$ and $V_{+}$ and one minimum at $V_{P}$. The DF is unstable because the Nyquist curve encircles the origin once. There is $N=1$ unstable mode. Case $(---)$. }
\label{aa1}
\end{center}
\end{figure}

\begin{figure}
\begin{center}
\includegraphics[clip,scale=0.3]{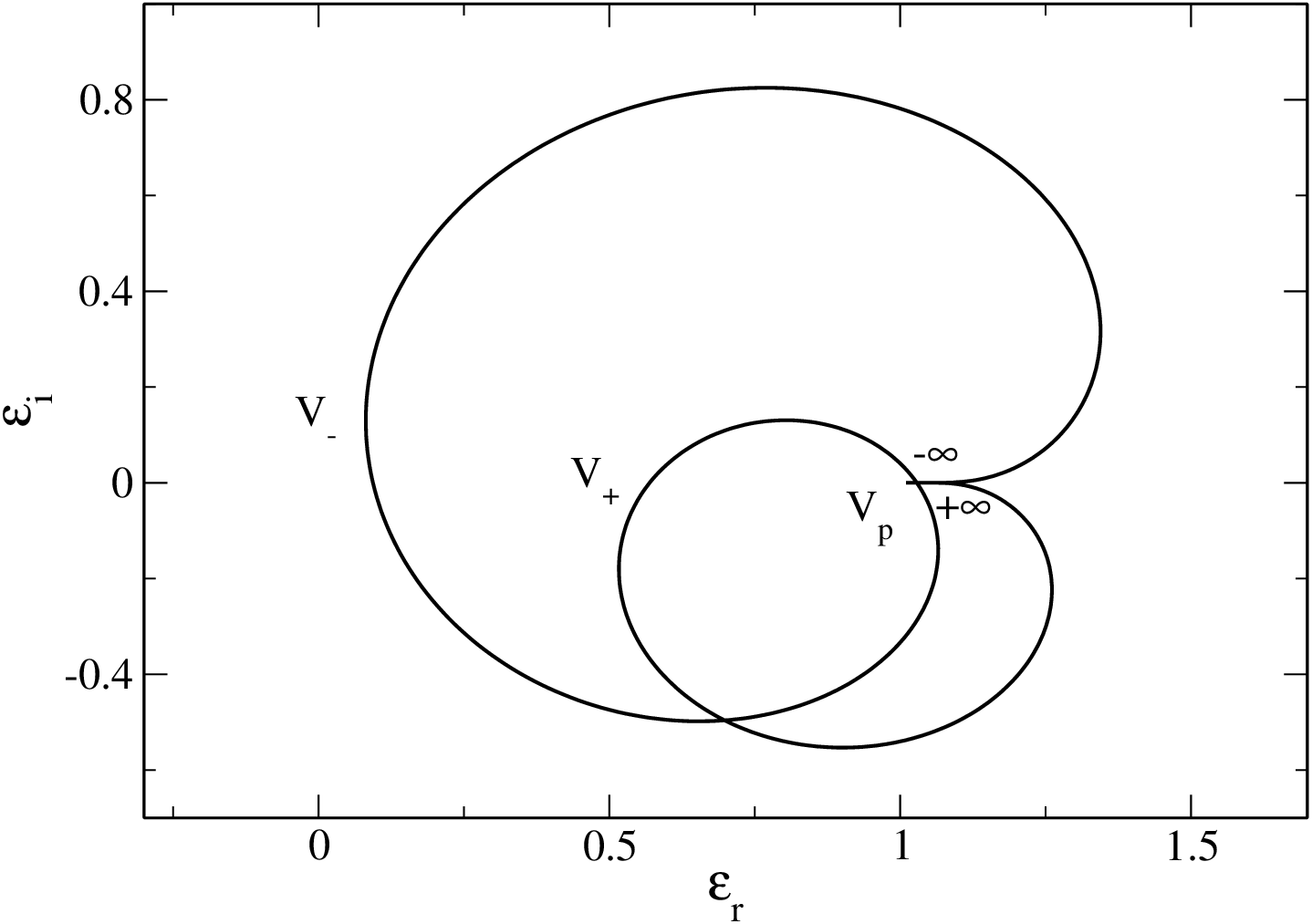}
\caption{For $\Delta_c<\Delta<\Delta_{*}$: Nyquist curve for  $y_{*}/a^2<\eta<\eta_{c}^{(-)}$ (specifically  $\Delta=1.5$, $a=1$ and $\eta=1.8$).  The DF has two  maxima at $V_{-}$ and $V_{+}$ and one minimum at $V_{P}$. The DF is stable because the Nyquist curve does not encircle the origin. Case $(+++)$.}
\label{aa2}
\end{center}
\end{figure}

\begin{figure}
\begin{center}
\includegraphics[clip,scale=0.3]{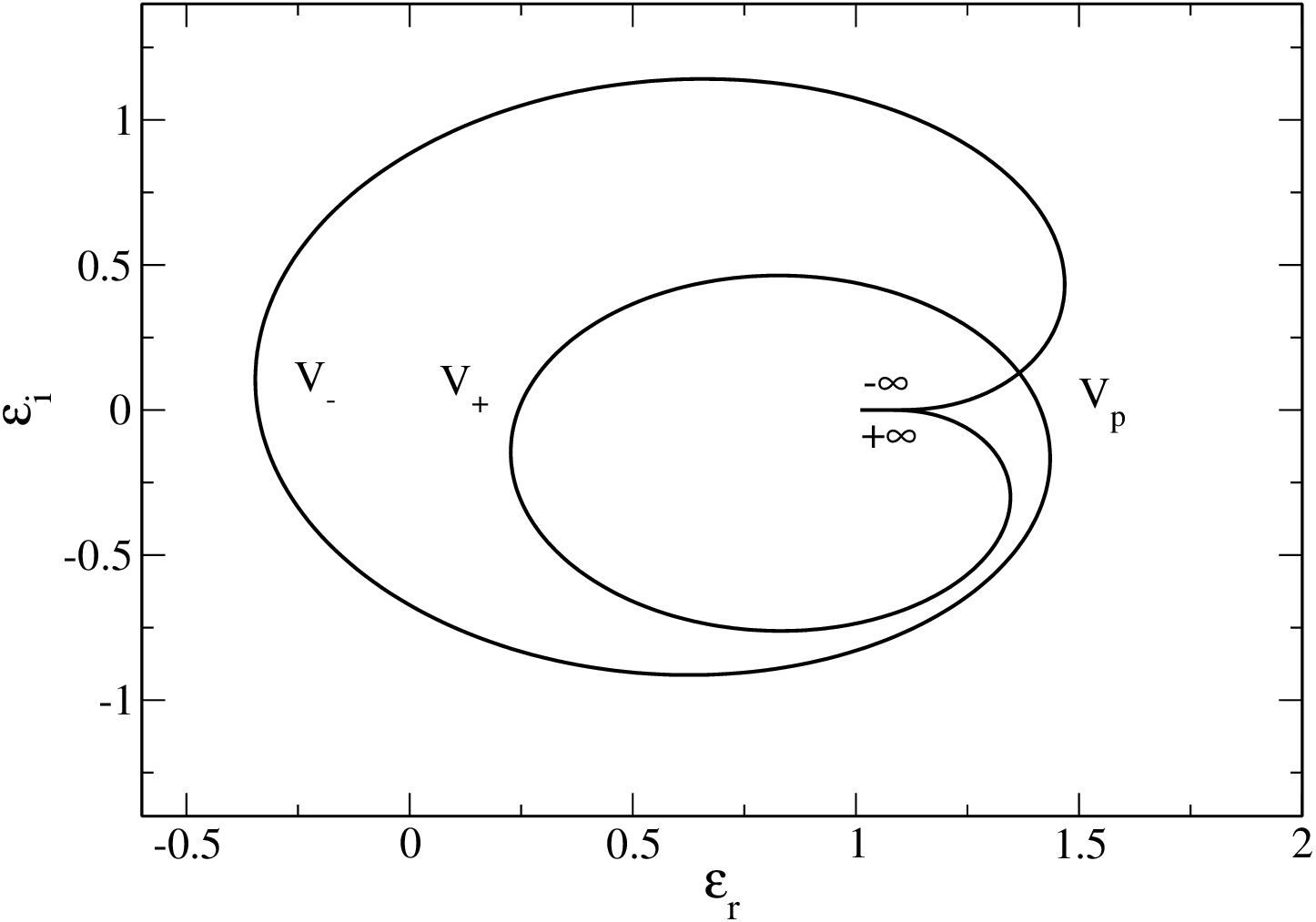}
\caption{For $\Delta_c<\Delta<\Delta_{*}$: Nyquist curve for $y_{*}/a^2<\eta_{c}^{(-)}<\eta<\eta_{c}^{(+)}$ (specifically $\Delta=1.5$, $a=1$ and $\eta=2.5$).  The DF has two  maxima at $V_{-}$ and $V_{+}$ and one minimum at $V_{P}$. The DF is unstable because the Nyquist curve encircles the origin once. There is $N=1$ unstable mode. Case $(-++)$.}
\label{aa3}
\end{center}
\end{figure}

The critical temperatures $\eta_c(a)$ corresponding to marginal
stability determined previously are represented as a function of the
separation $a$ in Figs. \ref{diagasym1}, \ref{diagasym2} and
\ref{diag10} for three values of the asymmetry factor $\Delta$. We
recall that $\eta_{c}^{(-)}(a)$ corresponds to the temperature
associated with a marginal mode with pulsation $\Omega_{r}=V_{-}<0$,
$\eta_{c}^{(p)}(a)$ corresponds to the temperature associated with a
marginal mode with pulsation $\Omega_{r}=V_{p}>0$ and
$\eta_{c}^{(+)}(a)$ corresponds to the temperature associated with a
marginal mode with pulsation $\Omega_{r}=V_{+}>0$. We have also
plotted the hyperbole $\eta=y_*/a^2$. Below this curve, the DF has a
single maximum at $V_{-}<0$ and above this curve, the DF has two
maxima at $V_{-}<0$ and $V_{+}>0$ and a minimum at $V_{p}>0$. In order
to investigate the stability of the solutions in the different regions
of the parameter space, we have used the Nyquist criterion. {There
exists a critical value $\Delta_{c}=1.09$ of the asymmetry parameter.}
For $\Delta<\Delta_{c}$, the stability diagram displays a re-entrant
phase (in continuity with the case $\Delta=1$) but for
$\Delta>\Delta_{c}$ the re-entrant phase disappears.  The description
of the stability diagram is similar to the one given in
Sec. \ref{sec_sdsy} and the different possible cases can be understood
directly from the reading of Figs. \ref{diagasym1}, \ref{diagasym2}
and \ref{diag10}. The best way is to fix $a$ and progressively
increase the value of $\eta$. Below the dashed line, the distribution
has only one maximum at $V_{0}=V_{-}$ so the Nyquist curve has one
intersection with the $x$-axis (in addition to the limit point
$(1,0)$). For $\eta\rightarrow 0$, we find that $\epsilon_{r}(V_{-})>0$
so the system is stable. As we increase $\eta$ and pass above the
dashed line, the distribution has two maxima at $V_{0}=V_{-}$ and
$V_{0}=V_{+}$ and one minimum at $V_{0}=V_{p}$. At each intersection
with a marginal line, one of the values $\epsilon_{r}(V_{-})$,
$\epsilon_{r}(V_{p})$ or $\epsilon_{r}(V_{+})$ changes sign. We have indicated by symbols like $(-++)$ the repsective signs of  $\epsilon_{r}(V_{-})$,
$\epsilon_{r}(V_{p})$ and $\epsilon_{r}(V_{+})$. We
can then easily draw by hands the corresponding Nyquist curve.
Therefore, it is not necessary to show all the possibilities and we
have only indicated a few representative cases in
Figs. \ref{aa1}-\ref{aa3} for illustration.

\begin{figure}
\begin{center}
\includegraphics[clip,scale=0.3]{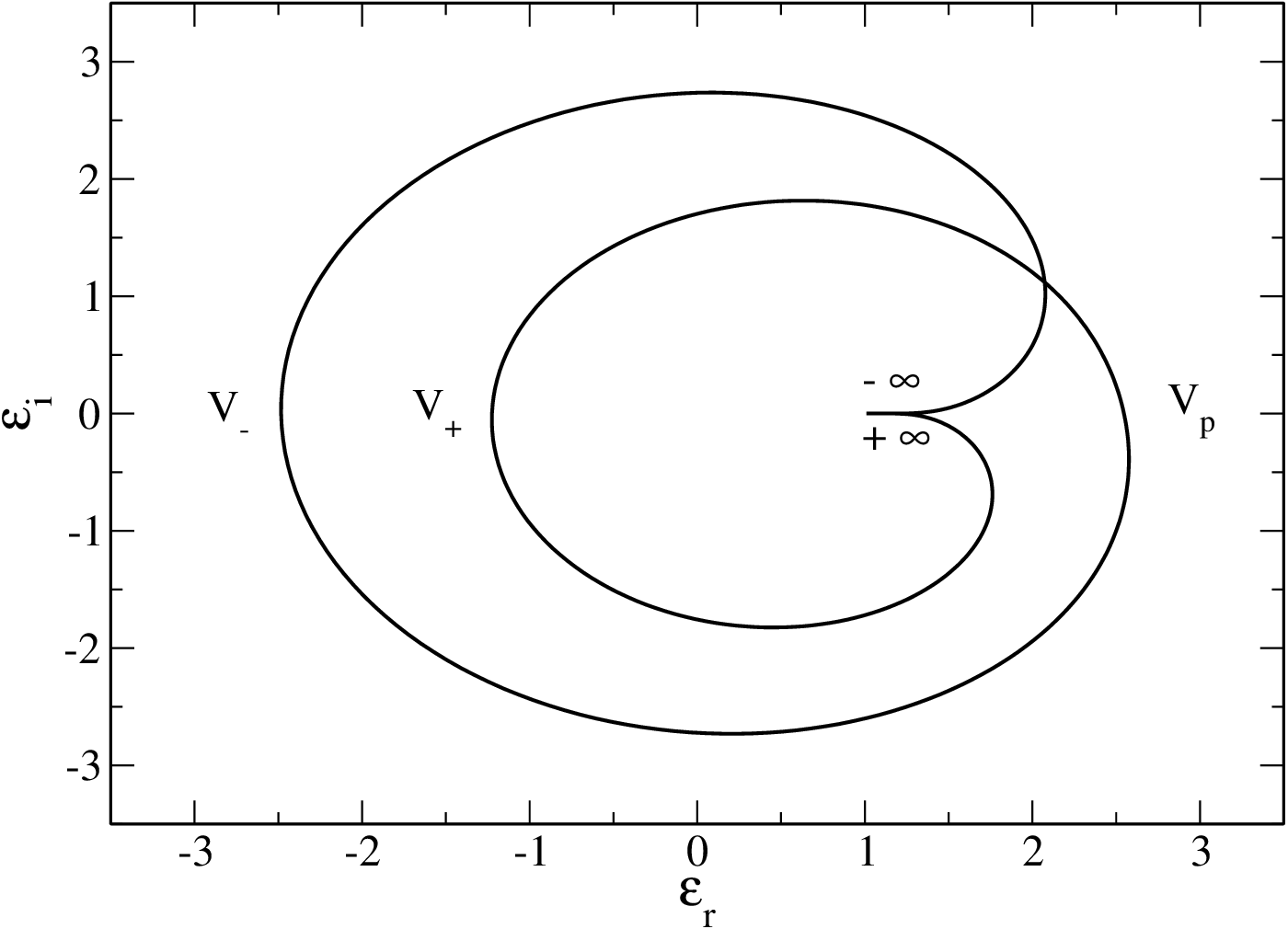}
\caption{For $\Delta_c<\Delta<\Delta_{*}$: Nyquist curve for $\eta>\eta_{c}^{(+)}>y_{*}/a^2$ (specifically $\Delta=1.5$, $a=1$ and $\eta=6$).  The DF has two  maxima at $V_{-}$ and $V_{+}$ and one minimum at $V_{P}$. The DF is unstable because the Nyquist curve encircles the origin twice. There are $N=2$ unstable modes. Case $(-+-)$.}
\label{aa4}
\end{center}
\end{figure}

\section{The repulsive HMF model}
\label{sec_rep}

\subsection{General results}
\label{sec_rgr}

We shall now consider the HMF model with a repulsive interaction
between particles. It is described by the Hamiltonian
\begin{eqnarray}
H=\sum_{i=1}^{N}\frac{v_{i}^{2}}{2}+\frac{k}{2\pi}\sum_{i<j}\cos(\theta_{i}-\theta_{j}),
\label{rgr1}
\end{eqnarray}
with $k>0$. In this
paper, we shall investigate the linear dynamical stability
\footnote{The nonlinear regime of the repulsive HMF model leading to a
bicluster (for low energies) has been studied in
\cite{bicluster}.} of stationary solutions of the Vlasov equation of
the form $f=f(v)$. In the repulsive case, the dispersion relation
becomes
\begin{eqnarray}
\epsilon(\omega)\equiv 1-{k\over 2}\int_{C} {{f'(v)}\over v-{\omega}}dv=0.
\label{rgr2}
\end{eqnarray}

When $\omega_i=0$, the real and imaginary parts of the dielectric
function
$\epsilon(\omega_{r})=\epsilon_{r}(\omega_{r})+i\epsilon_{i}(\omega_{r})$
are
\begin{eqnarray}
\epsilon_{r}(\omega_{r})=1-{k\over 2}{P}\int_{-\infty}^{+\infty} {{f'(v)}\over v-\omega_{r}}dv,
\label{rgr3}
\end{eqnarray}
\begin{eqnarray}
\epsilon_{i}(\omega_{r})=-\pi {k\over 2} f'(\omega_{r}).
\label{rgr4}
\end{eqnarray}
To apply the Nyquist method, we have to plot the curve
$(\epsilon_{r}(\omega_{r}),\epsilon_{i}(\omega_{r}))$ parameterized by
$\omega_{r}$ going from $-\infty$ to $+\infty$. Let us consider the
asymptotic behavior for $\omega_{r}\rightarrow \pm\infty$. Since $f(v)$
tends to zero for $v\rightarrow \pm\infty$, we conclude that
$\epsilon_{i}(\omega_{r})\rightarrow 0$ for $\omega_{r}\rightarrow
\pm\infty$ and that $\epsilon_{i}(\omega_{r})<0$ for
$\omega_{r}\rightarrow -\infty$ while $\epsilon_{i}(\omega_{r})>0$ for
$\omega_{r}\rightarrow +\infty$.  On the other hand, for
$\omega_{r}\rightarrow \pm\infty$, we obtain at leading order
\begin{eqnarray}
\epsilon_{r}(\omega_{r})\simeq 1-{k\over 2}\frac{\rho}{\omega_{r}^{2}}, \qquad (\omega_{r}\rightarrow \pm\infty).
\label{rgr5}
\end{eqnarray}
From these results, we conclude that the behavior of the curve close to the point $(1,0)$ is the one represented in Fig. \ref{mt}.
Let $v_{ext}$ be the velocity corresponding to an extremum of the distribution $(f'(v_{ext})=0)$. Then, we have
\begin{eqnarray}
\epsilon_{r}(v_{ext})=1+{k\over 2}\int_{-\infty}^{+\infty} \frac{f(v_{ext})-f(v)}{(v-v_{ext})^{2}}\, dv >1.
\label{rgr6}
\end{eqnarray}
If $v_{Max}$ denotes the velocity corresponding to the global maximum of the distribution, we clearly have
\begin{eqnarray}
\epsilon_{r}(v_{Max})> 1.
\label{rgr6b}
\end{eqnarray}
Finally, for a repulsive interaction, we note that if the zero of
$f'(v)$ corresponds to a maximum (resp. minimum) of $f$ then the
hodograph crosses the real axis upward (resp. downward).

\subsubsection{Single-humped distributions}
\label{sec_shr}

\begin{figure}
\begin{center}
\includegraphics[clip,scale=0.3]{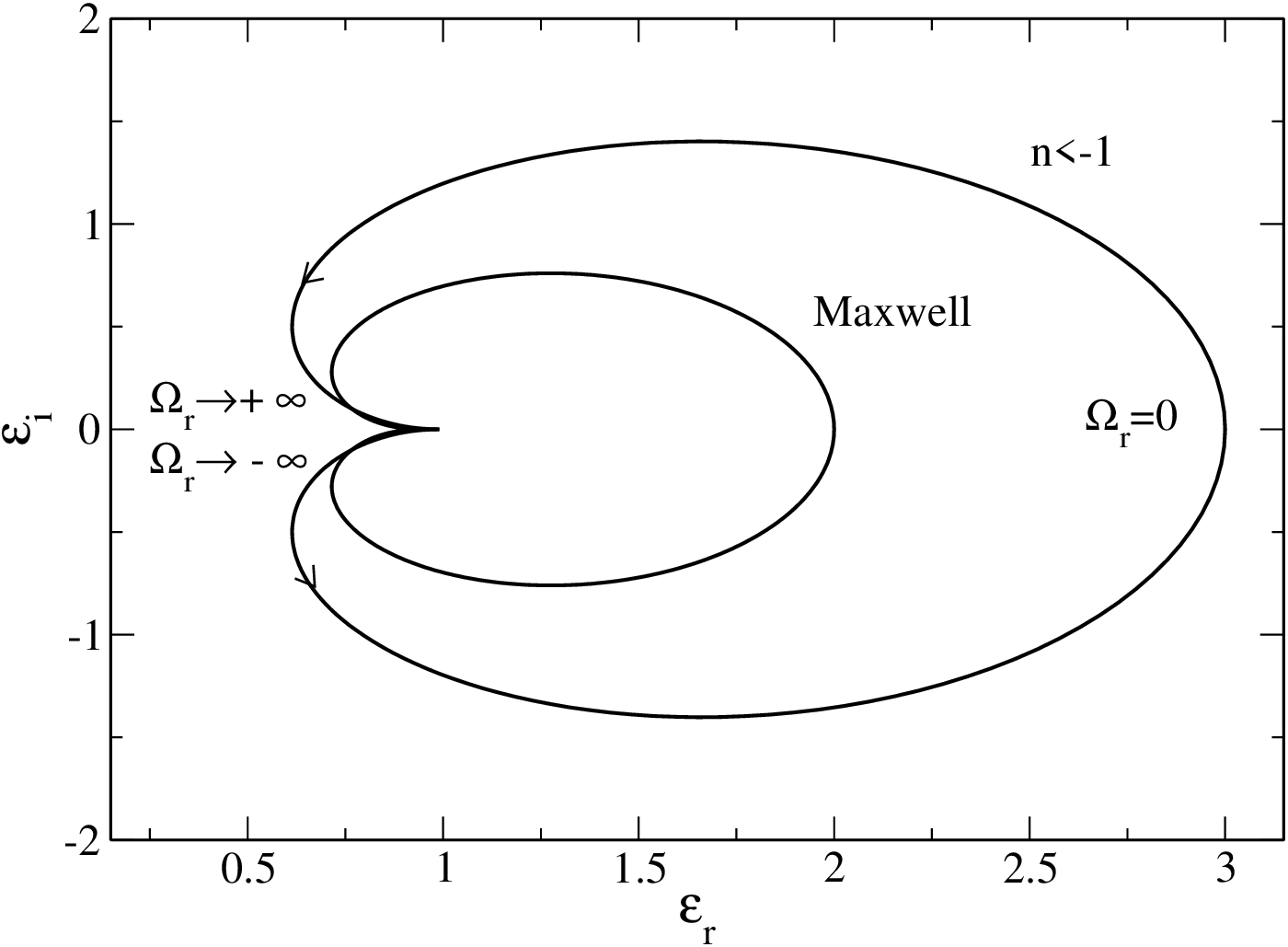}
\caption{Nyquist curve for the Maxwell and the Tsallis distribution with $n<-1$ (specifically $n=-2$). }
\label{mt}
\end{center}
\end{figure}

\begin{figure}
\begin{center}
\includegraphics[clip,scale=0.3]{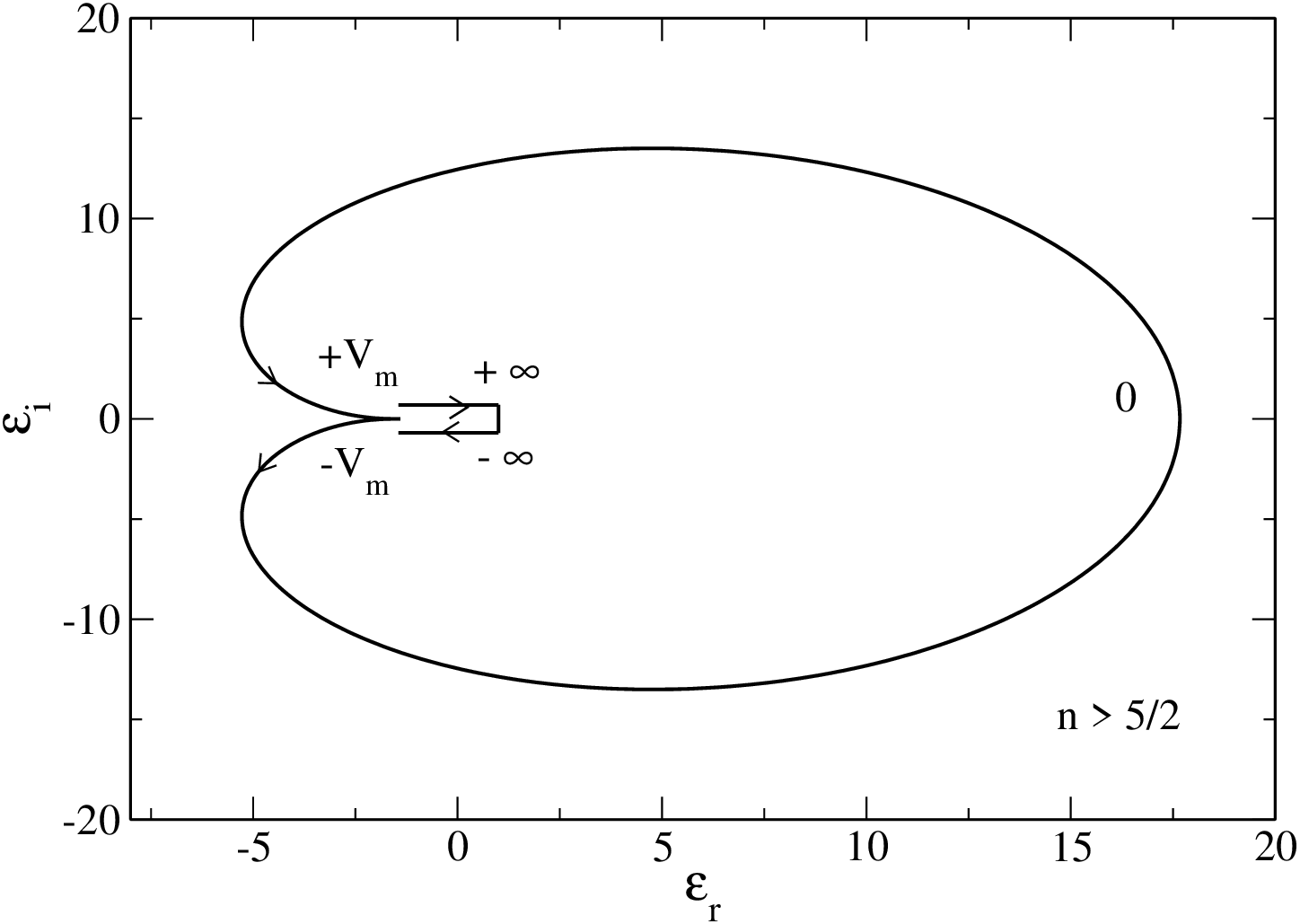}
\caption{Nyquist curve for the Tsallis distribution with $n>5/2$ (specifically $n=5$).}
\label{r5}
\end{center}
\end{figure}

\begin{figure}
\begin{center}
\includegraphics[clip,scale=0.3]{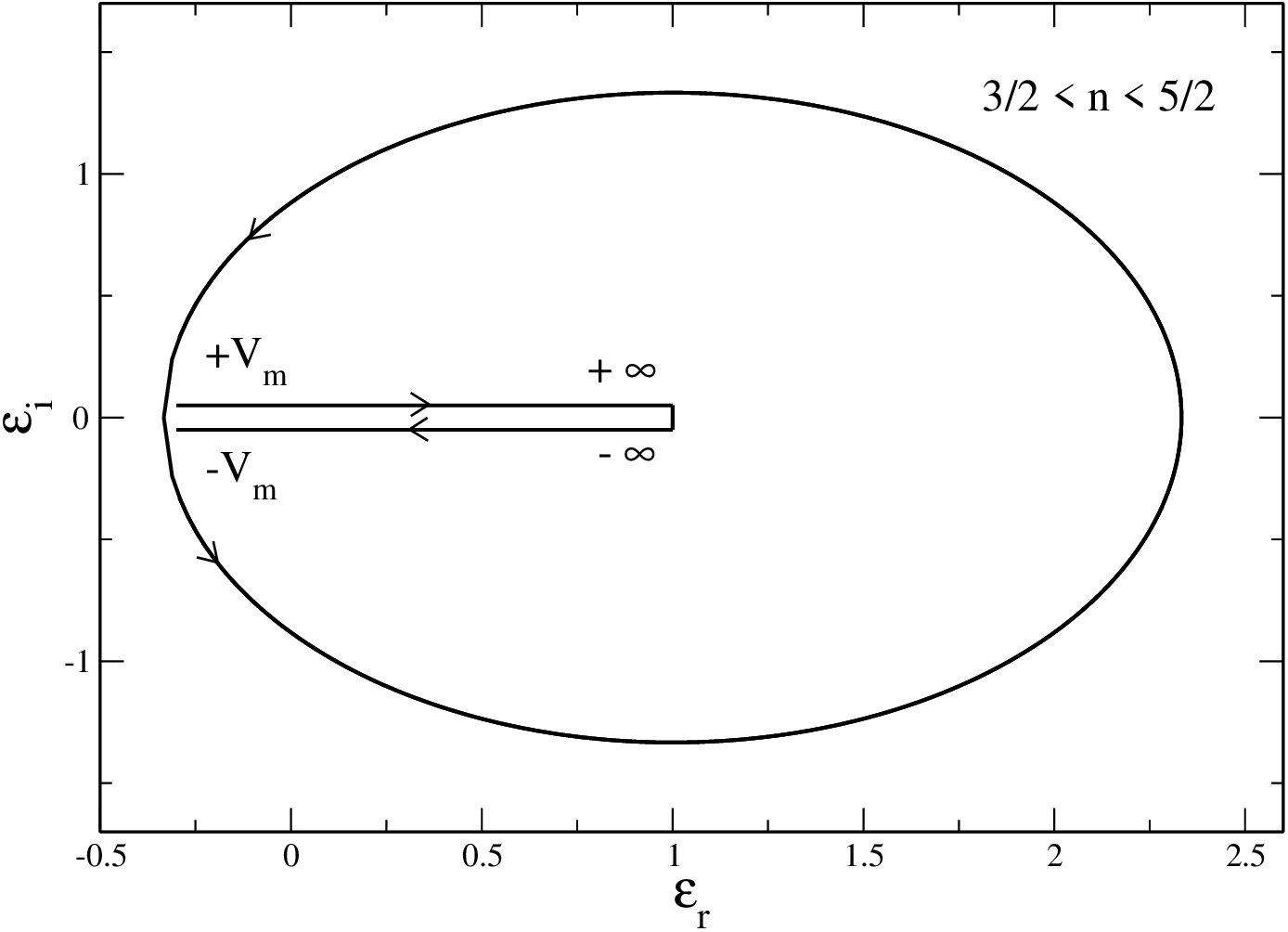}
\caption{Nyquist curve for the Tsallis distribution with $3/2<n<5/2$ (specifically $n=2$).}
\label{r2}
\end{center}
\end{figure}

\begin{figure}
\begin{center}
\includegraphics[clip,scale=0.3]{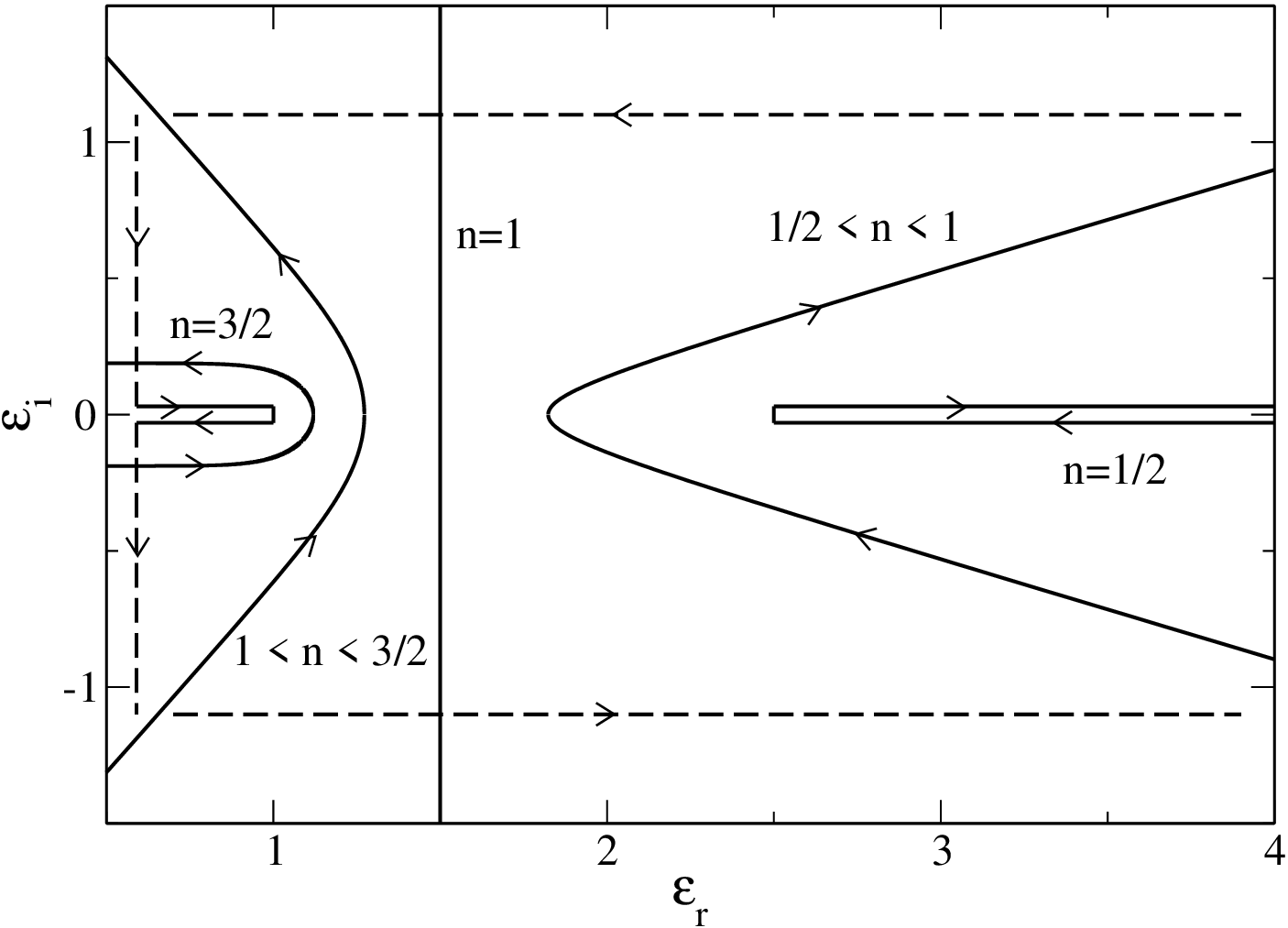}
\caption{Nyquist curve for the Tsallis distribution with $1/2<n<3/2$ showing the different cases (specifically $n=3/2$, $n=1.2$, $n=1$, $n=0.7$ and $n=1/2$).}
\label{rp3}
\end{center}
\end{figure}

Let us assume that the distribution $f(v)$ has a single maximum at
$v=v_{0}$ (so that $f'(v_0)=0$) and tends to zero for $v\rightarrow
\pm \infty$. Then, the Nyquist curve cuts the $x$-axis ($\epsilon_{i}(\omega_{r})$ vanishes) at the limit point
$(1,0)$ when $\omega_{r}\rightarrow \pm \infty$ and at the point
$(\epsilon_{r}(v_0),0)$ when $\omega_{r}=v_{0}$. Due to its behavior
close to the limit point $(1,0)$, the fact that it rotates in the
counterclockwise sense, and the property that $\epsilon_r(v_0)>1$ according to Eq. (\ref{rgr6b}),  the Nyquist curve must necessarily behave like in
Fig. \ref{mt}.  Therefore, the Nyquist curve starts on the
real axis at $\epsilon_r(\omega_r) =1$ for $\omega_r \rightarrow
-\infty$, then going in counterclockwise sense it crosses the real
axis at the point $\epsilon_r(v_0)>1$ and returns on the real axis at
$\epsilon_r(\omega_r) =1$ for $\omega_r \rightarrow +
\infty$. Therefore, it cannot encircle the origin. According to the Nyquist criterion exposed in
Sec. \ref{sec_nyquist}, we conclude that a single-humped distribution
is always linearly stable. For illustration, the Nyquist curves for
the Maxwell and the Tsallis distributions are represented in
Figs. \ref{mt}-\ref{rp3}.

For the water-bag model ($n=1/2$), we can obtain an analytical
expression of the dielectric function in the form
\begin{equation}
\label{rgr7}
\epsilon(\Omega)=1+\frac{1}{3}\eta W^{(1/2)}(\sqrt{\eta}\Omega),
\end{equation}
where $W^{(1/2)}(z)$ is given by Eq. (\ref{t14}).  The condition
$\epsilon(\omega)=0$ determines the complex pulsation. We have
\begin{equation}
\label{rgr8}
\omega^2=\frac{kM}{4\pi}+3T.
\end{equation}
The pulsation is real so the mode is purely oscillating without being
damped. For $T=0$, we find that $\omega=\omega_p$ where $\omega_p$ is
the proper pulsation defined in Eq. (\ref{landau9}). Note, in passing,
that the dimensionless pulsation can be written
$\Omega=\omega/\omega_p$. On the other hand, for $\Omega_{i}=0$, we
get
\begin{equation}
\label{rgr9}
\epsilon_{r}(\Omega_{r})=1+\frac{1}{3}\eta \frac{1}{1-\frac{1}{3}\eta\Omega_{r}^{2}}, \qquad \epsilon_{i}(\Omega_{r})=0.
\end{equation}
Therefore, the Nyquist curve is made of two segments $]-\infty, 1]$
and $[1+\eta/3,+\infty[$ as represented in Fig. \ref{rp3}.

\subsubsection{Double-humped distributions}
\label{sec_dhhr}

Let us consider a double-humped distribution with a global maximum at
$v_{Max}$, a minimum at $v_{min}$ and a local maximum at $v_{max}$. We
assume $v_{Max}<v_{min}<v_{max}$. The Nyquist curves starts at
$(1,0)$, progresses in the counterclockwise sense and crosses the
$x$-axis at $\epsilon_{r}(v_{Max})>1$, then at $\epsilon_{r}(v_{min})$
and $\epsilon_{r}(v_{max})$. We can convince ourselves by making drawings of the following results. If 

$(+++)$: $\epsilon_{r}(v_{Max})>0$, $\epsilon_{r}(v_{min})>0$, $\epsilon_{r}(v_{max})>0$,

$(+--)$: $\epsilon_{r}(v_{Max})>0$, $\epsilon_{r}(v_{min})<0$, $\epsilon_{r}(v_{max})<0$,

$(++-)$: $\epsilon_{r}(v_{Max})>0$, $\epsilon_{r}(v_{min})>0$, $\epsilon_{r}(v_{max})<0$,

\noindent the Nyquist curve does not encircle the origin so the
system is stable. If

$(+-+)$: $\epsilon_{r}(v_{Max})>0$, $\epsilon_{r}(v_{min})<0$, $\epsilon_{r}(v_{max})>0$,

\noindent the Nyquist curve rotates one time
around the origin so that there is one mode of instability. Since
$\epsilon_{r}(v_{Max})>0$ there is no mode of marginal stability with
$\omega_{r}=v_{Max}$. Cases $(+++)$, $(+-+)$ and $(+--)$ are
observed in Sec. \ref{sec_rass} for an asymmetric double-humped
distribution made of two Maxwellians.

If the double-humped distribution is symmetric with respect to the
origin with two maxima at $\pm v_{*}$ and a minimum at $v=0$, we get
the same results as above with the additional properties
$\epsilon_{r}(v_{Max})=\epsilon_{r}(v_{max})=\epsilon_{r}(v_{*})>1$
and $\epsilon_{r}(v_{min})=\epsilon_{r}(0)$. We have only two cases
$(+++)$ and $(+-+)$. They are observed in Sec. \ref{sec_rsh} for a
symmetric double-humped distribution made of two Maxwellians. Since
$\epsilon_{r}(v_{*})>0$, there is no mode of marginal stability with
$\omega_{r}=\pm v_{*}$.

\subsection{The symmetric double-humped distribution}
\label{sec_rsh}

Let us now consider the symmetric double-humped distribution (\ref{e1}). In the repulsive case, the dielectric function is
\begin{equation}
\label{rsh1}
\epsilon(\Omega)= 1+\frac{\eta}{2} \left [ W(\sqrt{\eta}(\Omega - a))+W(\sqrt{\eta}(\Omega + a))   \right ].
\end{equation}
The condition of marginal stability corresponds to $\epsilon(\Omega)=0$ and $\omega_i=0$. The condition $\epsilon_i(\omega_r)=0$ is equivalent to $f'(\omega_r)=0$ so that the real pulsations correspond to the velocities where the distribution is extremum: $\omega_r=v_0$. Then, the temperatures at which the distribution  is marginally stable are obtained by solving $\epsilon_r(\omega_r=v_0)=0$. Proceeding as in Sec. \ref{sec_sdh} and introducing the parameters $x=V_0/a$ and $y=\eta a^2$, we find that the equations determining the critical temperatures
$\eta_c(a)$ are given by
\begin{equation}
\label{rsh2}
y=\frac{1}{2x}\ln\left (\frac{1+x}{1-x}\right ),
\end{equation}
\begin{equation}
\label{rsh3}
\eta=\frac{-2}{\left [ W_{r}(\sqrt{y}(x - 1))+W_{r}(\sqrt{y}(x+ 1))   \right ]},
\end{equation}
\begin{equation}
\label{rsh3b}
a^2=\frac{y}{\eta}.
\end{equation}
We note that only the sign in Eq. (\ref{rsh3}) changes with respect to the study of the attractive case, so we can readily adapt the results of Sec. \ref{sec_sdh} to the present situation by simply reverting the sign. For  $\Omega_{r}=\pm V_*$, corresponding to $x\neq 0$, there is no physical solution to Eqs. (\ref{rsh2})-(\ref{rsh3b}) with positive temperature $\eta>0$. Therefore, in the repulsive case, there is no marginal mode with non zero pulsation for the symmetric double-humped distribution (in agreement with the general discussion of Sec. \ref{sec_dhhr}). We now consider the marginal mode with $\Omega_{r}=0$. This corresponds to the ``degenerate" solution  $x=0$ (for any $y$) for which Eqs. (\ref{rsh2})-(\ref{rsh3b}) reduce to
\begin{equation}
\label{rsh4}
\eta=\frac{-1}{W_{r}(\sqrt{y})},
\end{equation}
\begin{equation}
\label{rsh5}
a^2=\frac{y}{\eta}=-y W_{r}(\sqrt{y}).
\end{equation}
According to Fig. \ref{w2}, physical solutions exist only for $y\ge
y_{max}=z_{c}^{2}$. We note that the range of parameters that was
forbidden in the attractive case is now allowed in the repulsive case
and vice versa. We thus consider the range $y\in \lbrack
y_{max},+\infty\lbrack$. {The function (\ref{rsh5}) has a maximum $a_m=1.282$ at
$y=y_{m}=7.642$ and for $y\rightarrow +\infty$ it tends to
$a=1$.} Therefore, for $a>a_m$, Eqs. (\ref{rsh4})-(\ref{rsh5}) have no solution. {For $a=a_m$,
Eq. (\ref{rsh5}) has one solution $y=y_m$ determining, through Eq. (\ref{rsh4}), a
temperature $\eta_c^{(0)(m)}=4.647$}. For $1<a<a_m$, Eq. (\ref{rsh5}) has two
solutions $y=y_{1,2}$. Then, Eq. (\ref{rsh4}) determines two temperatures
$\eta_c^{(0)(1)}$ and $\eta_c^{(0)(2)}$ as illustrated in
Fig. \ref{w3}. Finally, for $0<a<1$, Eq. (\ref{rsh5}) has one solution $y=y_1$
determining, through Eq. (\ref{rsh4}), a temperature $\eta_c^{(0)(1)}$.

The limit $y\rightarrow y_{max}$ corresponds to $a\rightarrow 0$. Since $a^2\sim -y_{max} W_{r}(\sqrt{y})$
and $\eta\sim -1/W_{r}(\sqrt{y})$, we get
\begin{equation}
\label{rsh6}
\eta_{c}^{(0)(1)}(a)\sim \frac{z_c^2}{a^2}, \qquad (a\rightarrow 0).
\end{equation}
This result is to be expected since, for $a=0$, the distribution (\ref{e1}) reduces to the Maxwellian that is stable for a repulsive interaction (hence $\eta_c=+\infty$).

Using Eq. (\ref{wr8}), the limit $y\rightarrow +\infty$ corresponds to $a\rightarrow 1^{+}$. Then, we get 
\begin{equation}
\label{rsh7}
\eta_{c}^{(0)(2)}(a)\sim \frac{3}{2(a-1)}, \qquad (a\rightarrow 1^{+}).
\end{equation}

\begin{figure}
\begin{center}
\includegraphics[clip,scale=0.3]{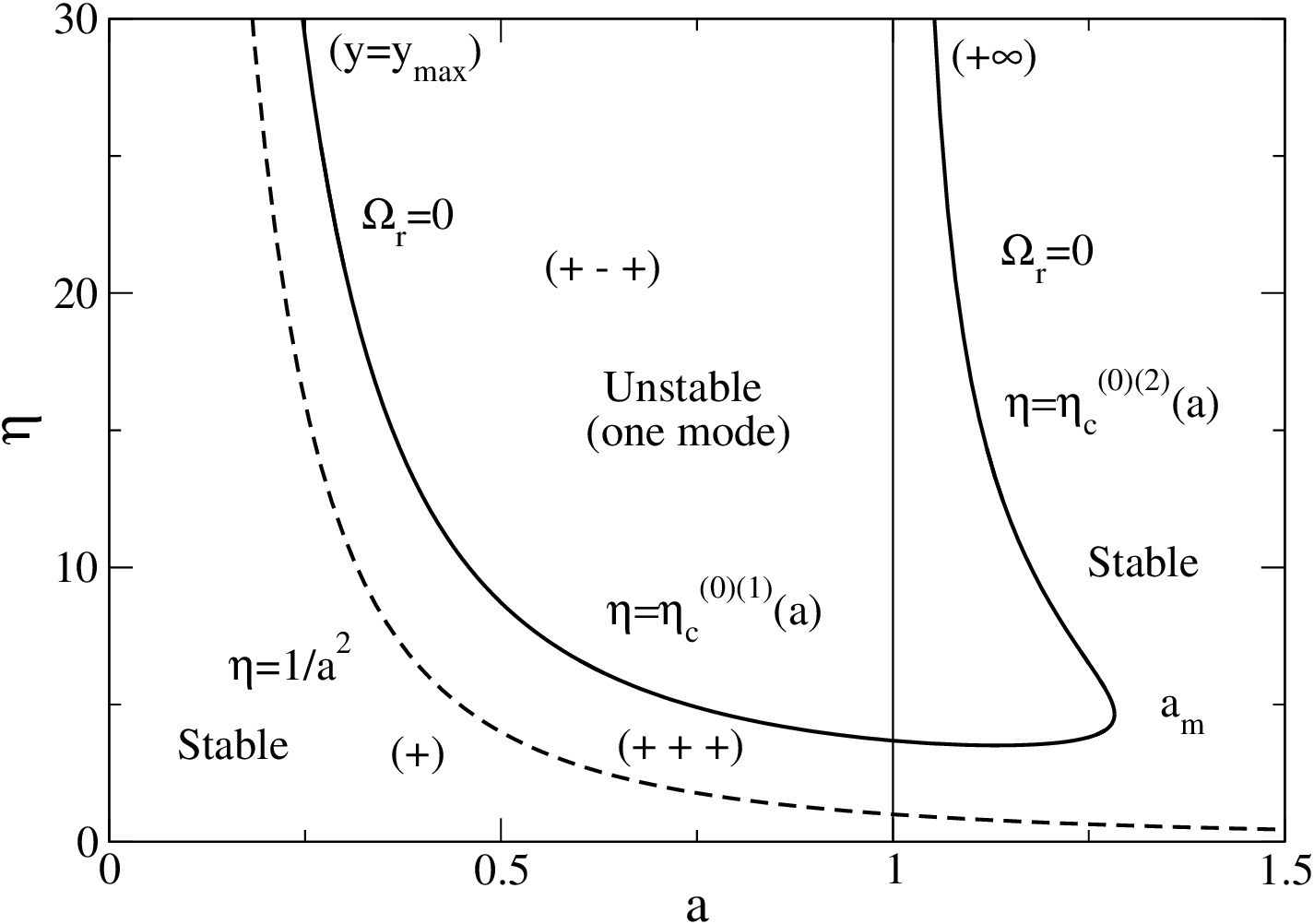}
\caption{Stability diagram of the symmetric double-humped distribution for the repulsive HMF model. There is a re-entrant phase for $1<a<a_m$. }
\label{rphs}
\end{center}
\end{figure}

The critical temperature(s) $\eta_c(a)$ corresponding to marginal stability determined previously are represented as a function of the separation $a$ in Fig. \ref{rphs}. We have also plotted  the hyperbole $\eta=1/a^2$. Below this curve, the distribution has a single maximum at $V_0=0$ and above this curve, the distribution has two maxima at $V_0=\pm V_*$ and a minimum at $V_0=0$. In order to investigate the stability of the solutions in the different regions, we have  used the Nyquist criterion.

\begin{figure}
\begin{center}
\includegraphics[clip,scale=0.3]{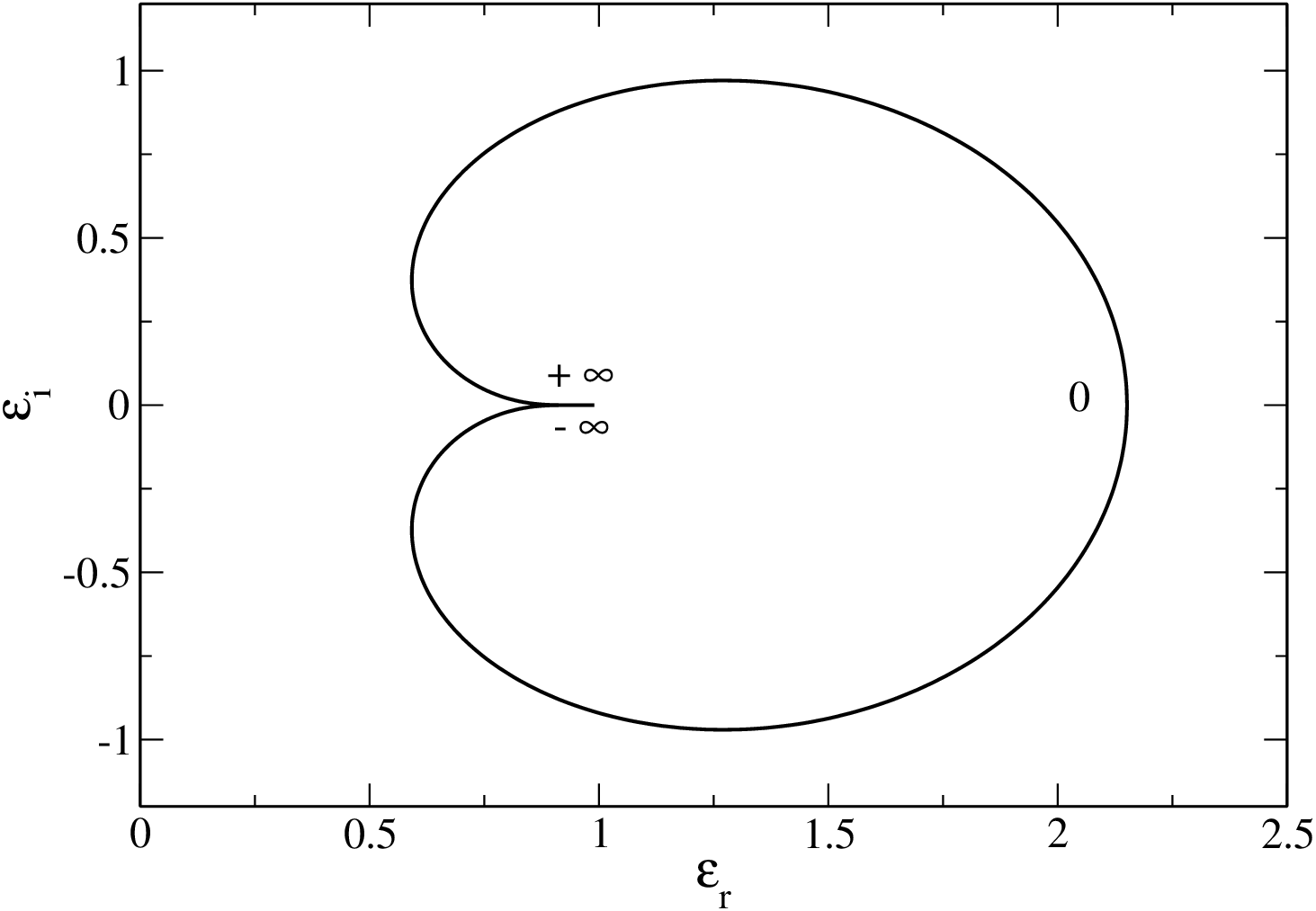}
\caption{Nyquist curve for $a<1$ and $\eta<1/a^2<\eta_c^{(0)(1)}(a)$ (specifically $a=0.5$, $\eta=2$). The system is stable. Case $(+)$.}
\label{rsp52}
\end{center}
\end{figure}

\begin{figure}
\begin{center}
\includegraphics[clip,scale=0.3]{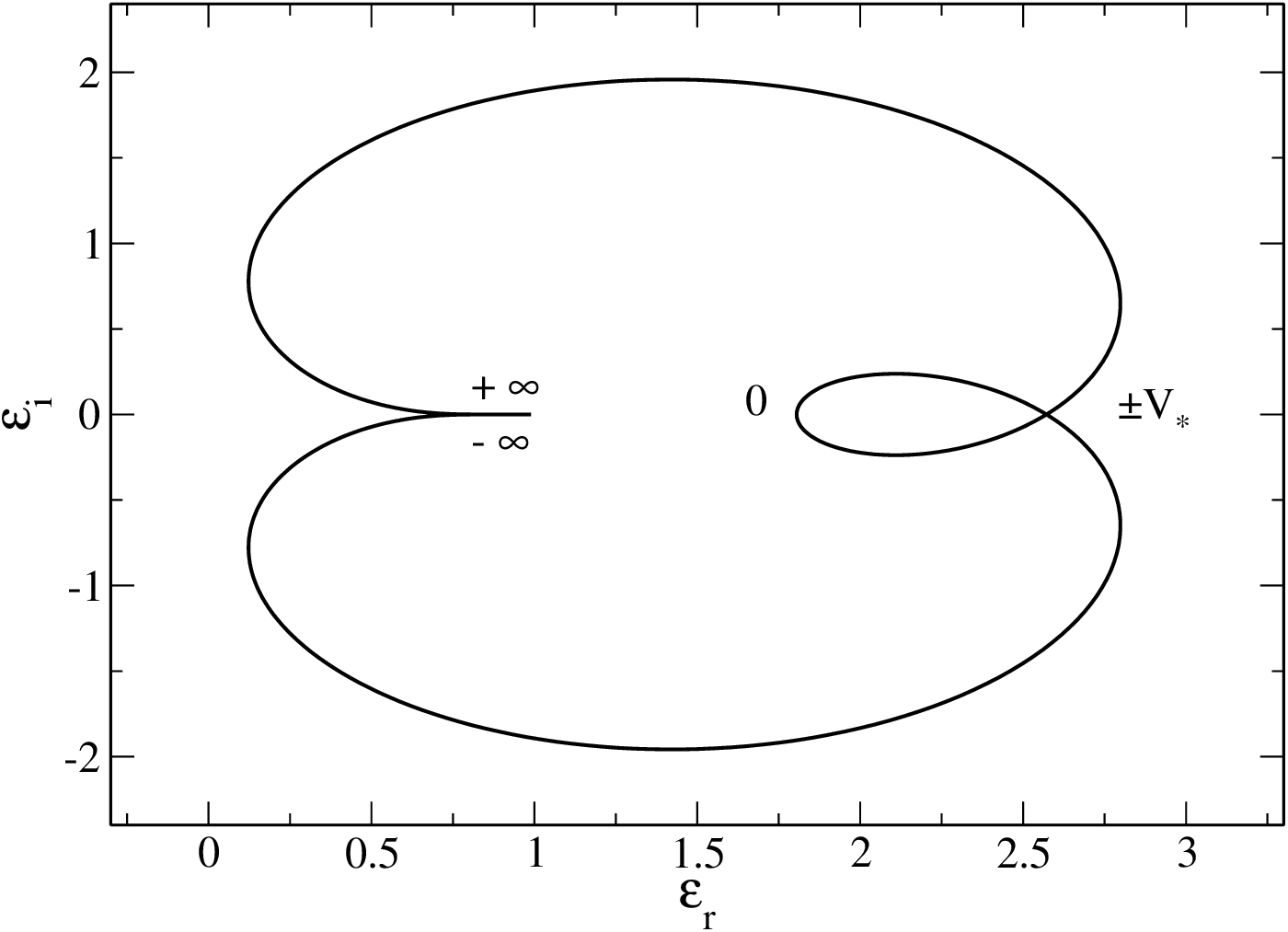}
\caption{Nyquist curve for $a<1$ and $1/a^2<\eta<\eta_c^{(0)(1)}(a)$ (specifically $a=0.5$, $\eta=5$). The system is stable. Case $(+++)$.}
\label{rsp55}
\end{center}
\end{figure}

\begin{figure}
\begin{center}
\includegraphics[clip,scale=0.3]{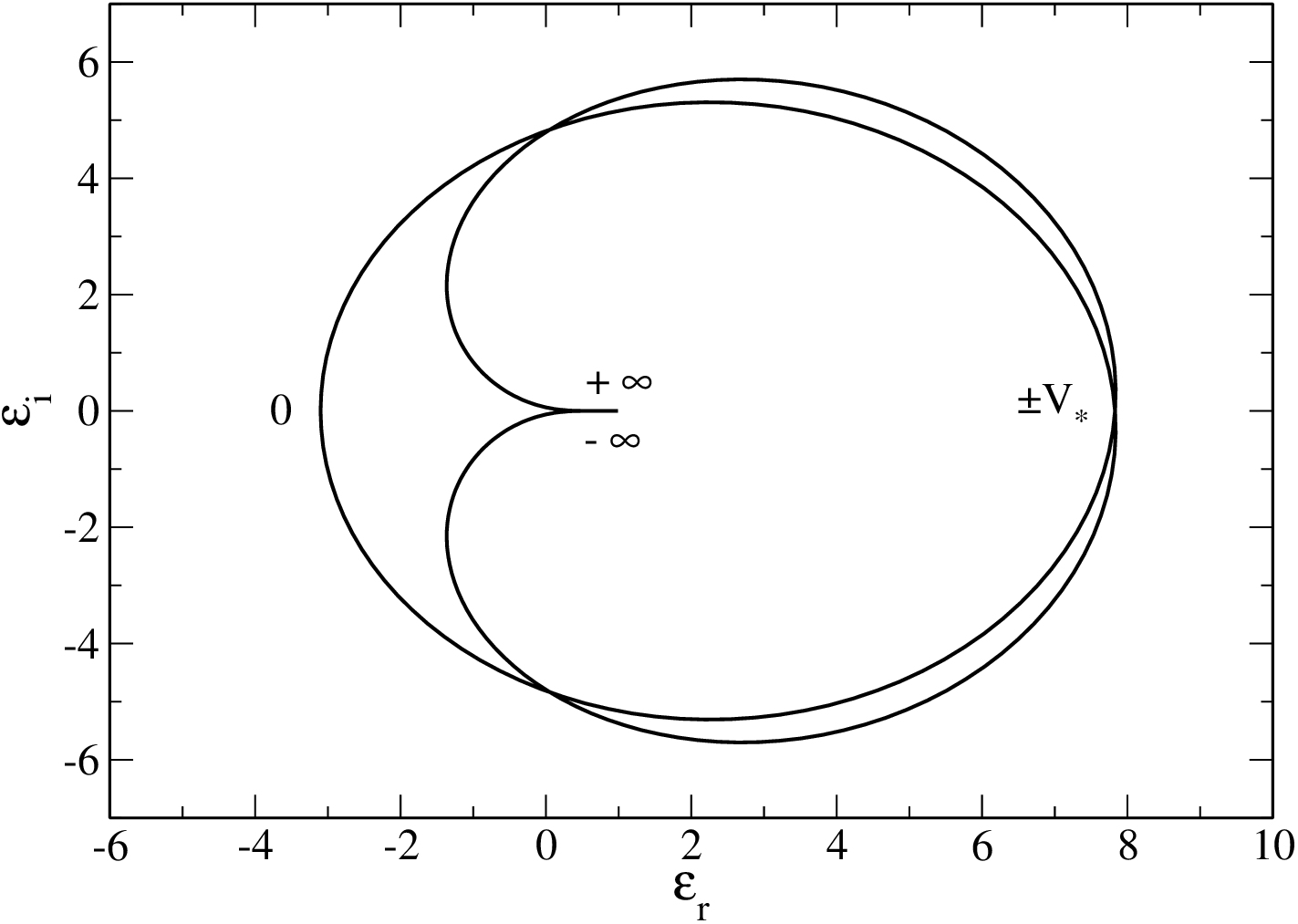}
\caption{Nyquist curve for $a<1$ and $\eta>\eta_c^{(0)(1)}(a)>1/a^2$ (specifically $a=0.5$, $\eta=15$). The system is unstable. Case $(+-+)$. }
\label{rsp515}
\end{center}
\end{figure}

For $a<1$, there exists one temperature $\eta_{c}^{(0)(1)}$ at which
the distribution is marginally stable.  For $\eta=\eta_{c}^{(0)(1)}$, the distribution has
a minimum at $V_0=0$ and two maxima at $\pm V_*$. The marginal
perturbation has zero pulsation $\Omega_r=0$. By considering the
Nyquist curves in this region (see Figs. \ref{rsp52}-\ref{rsp515}), we
find that the DF is stable for $\eta<\eta_{c}^{(0)(1)}$ and unstable
for $\eta>\eta_{c}^{(0)(1)}$.

\begin{figure}
\begin{center}
\includegraphics[clip,scale=0.3]{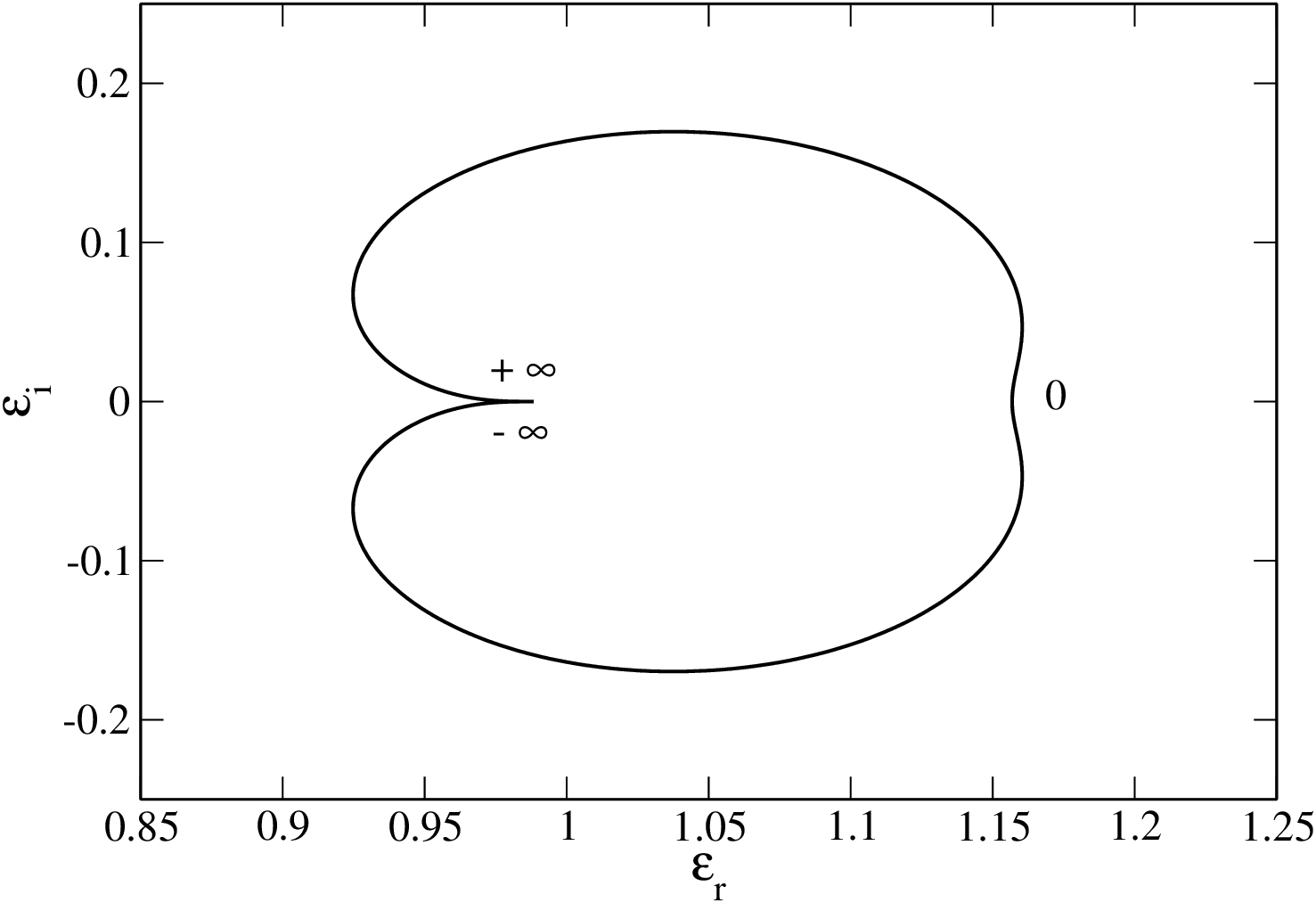}
\caption{Nyquist curve for $a>a_m$ and $\eta<1/a^2$ (specifically $a=1.4$, $\eta=0.4$). The system is stable.  Case $(+)$.}
\label{rspm1}
\end{center}
\end{figure}

\begin{figure}
\begin{center}
\includegraphics[clip,scale=0.3]{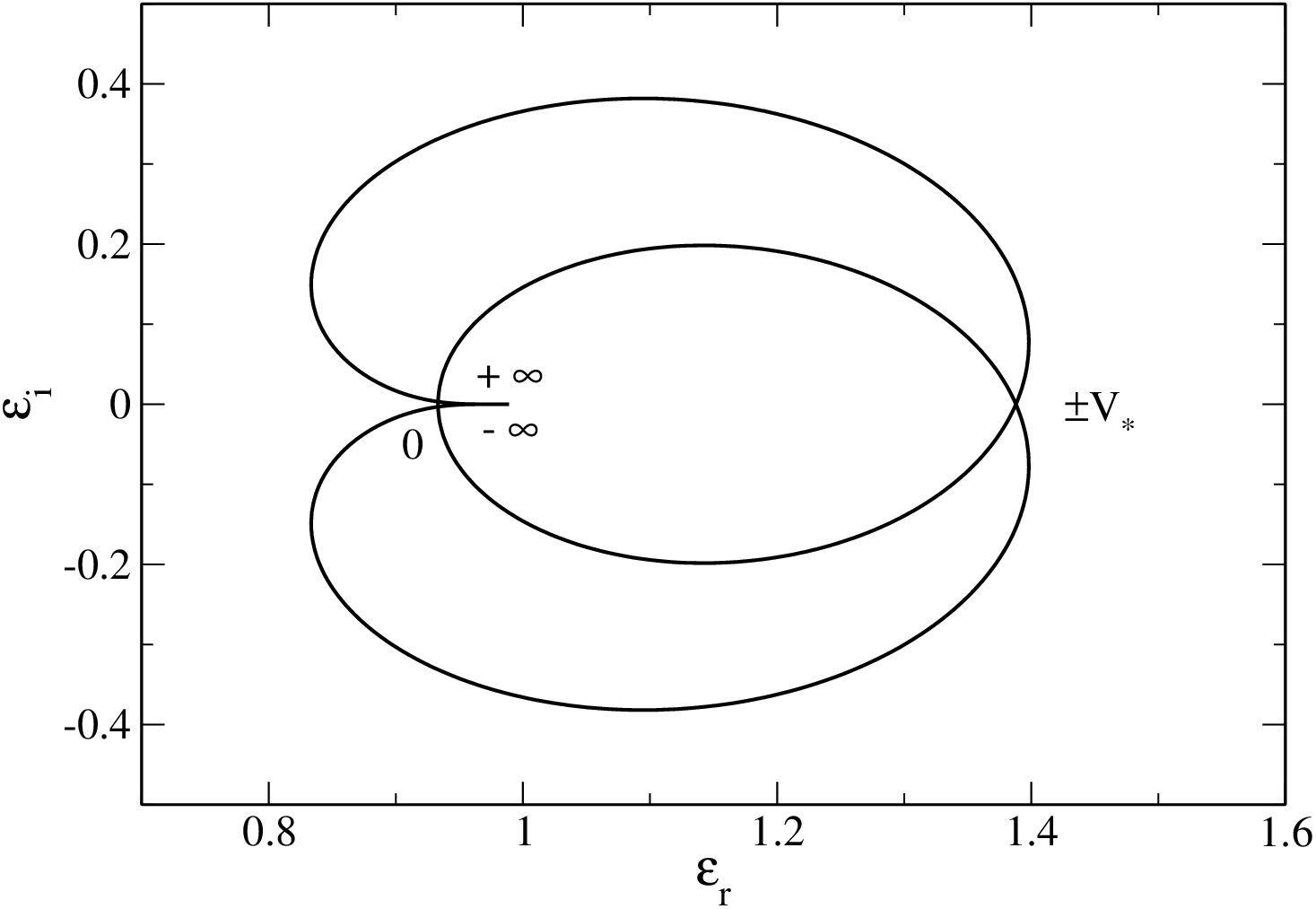}
\caption{Nyquist curve for $a>a_m$ and $\eta>1/a^2$ (specifically $a=1.4$, $\eta=1$). The system is stable. Case $(+++)$.}
\label{rspm2}
\end{center}
\end{figure}

For $a>a_m$, there is no marginal mode and the distribution is always
stable (see Figs. \ref{rspm1}-\ref{rspm2}). This result is to be expected because, for $a\rightarrow
+\infty$, the two humps do not ``see" each other and behave as two
independent single-humps distributions that are stable in the repulsive
case.

\begin{figure}
\begin{center}
\includegraphics[clip,scale=0.3]{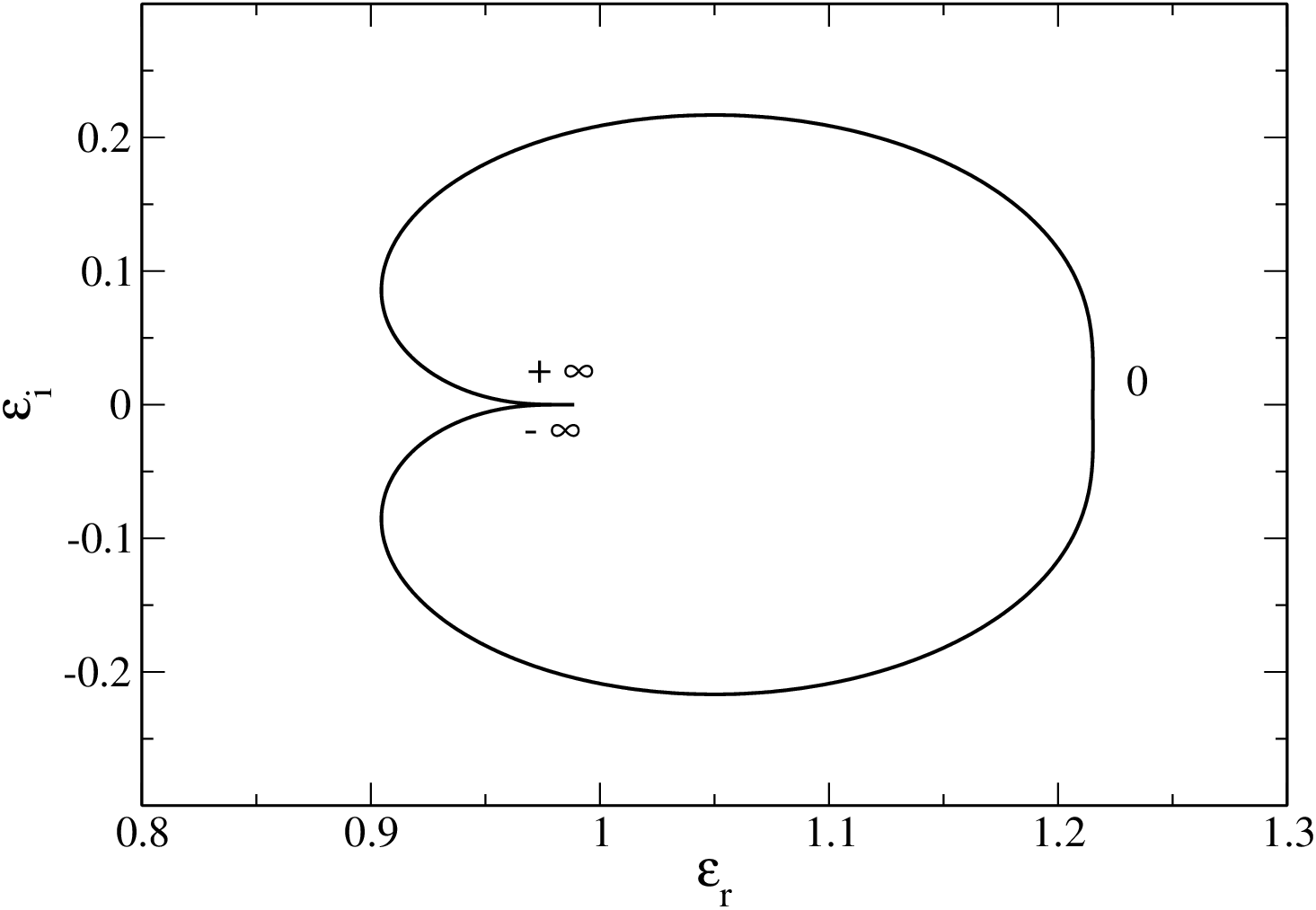}
\caption{Nyquist curve for $1<a<a_m$ and $\eta<1/a^2$ (specifically $a=1.2$, $\eta=0.5$). The system is stable. Case $(+)$.}
\label{rr1}
\end{center}
\end{figure}

\begin{figure}
\begin{center}
\includegraphics[clip,scale=0.3]{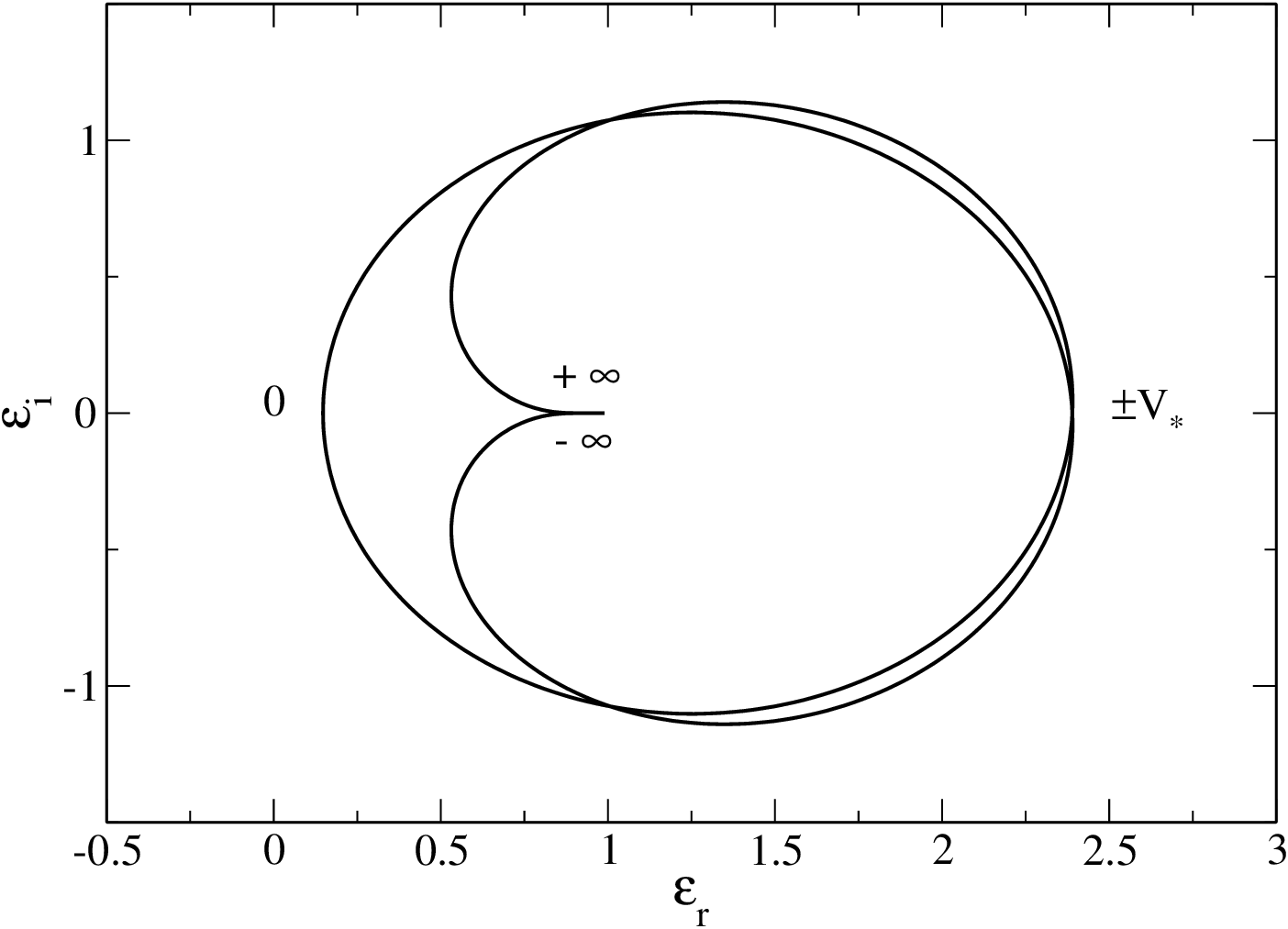}
\caption{Nyquist curve for $1<a<a_m$ and $1/a^2<\eta<\eta_c^{(0)(1)}(a)$ (specifically $a=1.2$, $\eta=3$). The system is stable. Case $(+++)$.}
\label{rr2}
\end{center}
\end{figure}

\begin{figure}
\begin{center}
\includegraphics[clip,scale=0.3]{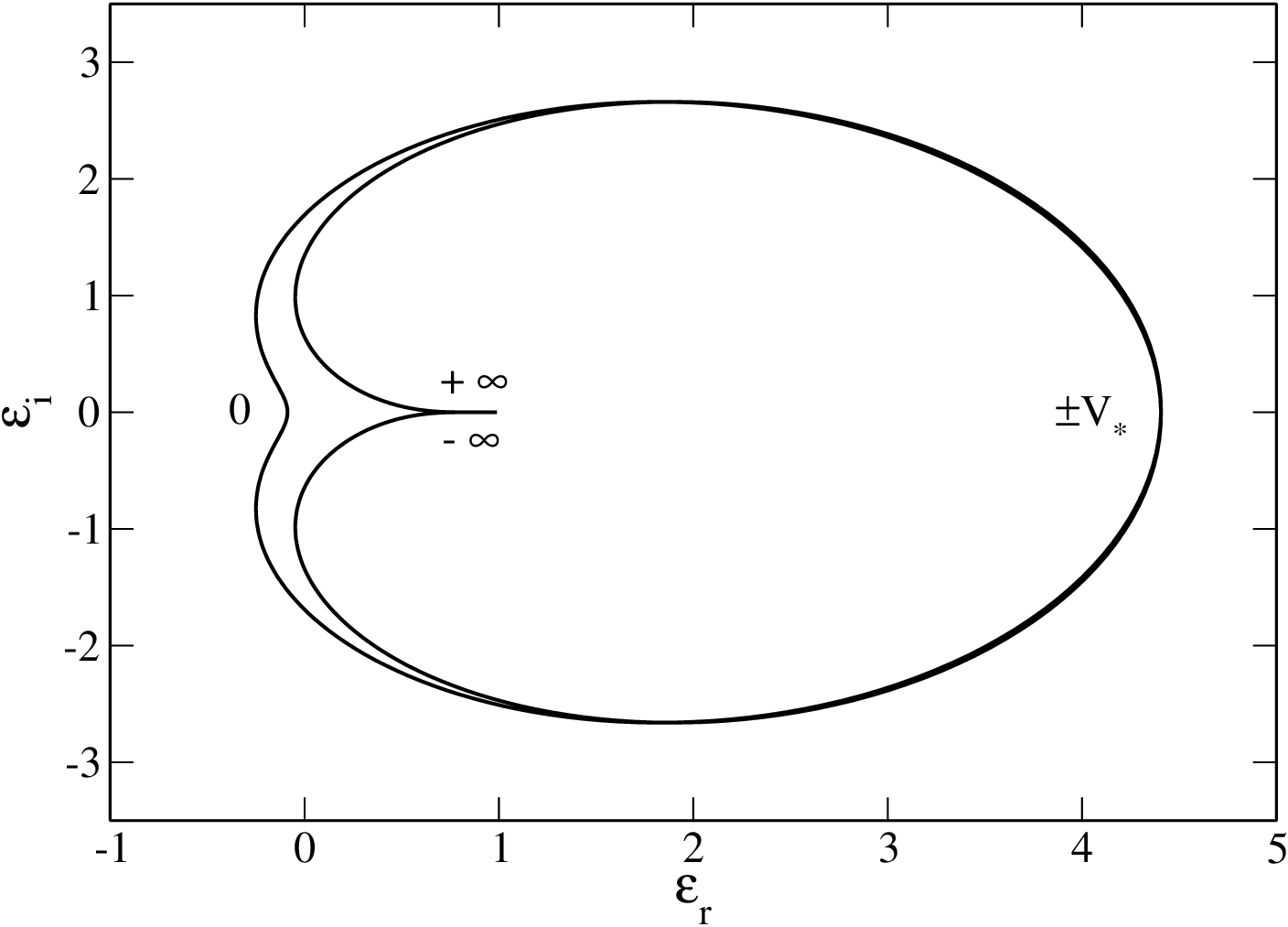}
\caption{Nyquist curve for $1<a<a_m$ and $1/a^2<\eta_c^{(0)(1)}(a)<\eta<\eta_c^{(0)(2)}(a)$ (specifically $a=1.2$, $\eta=7$). The system is unstable. Case $(+-+)$.}
\label{rr3}
\end{center}
\end{figure}

\begin{figure}
\begin{center}
\includegraphics[clip,scale=0.3]{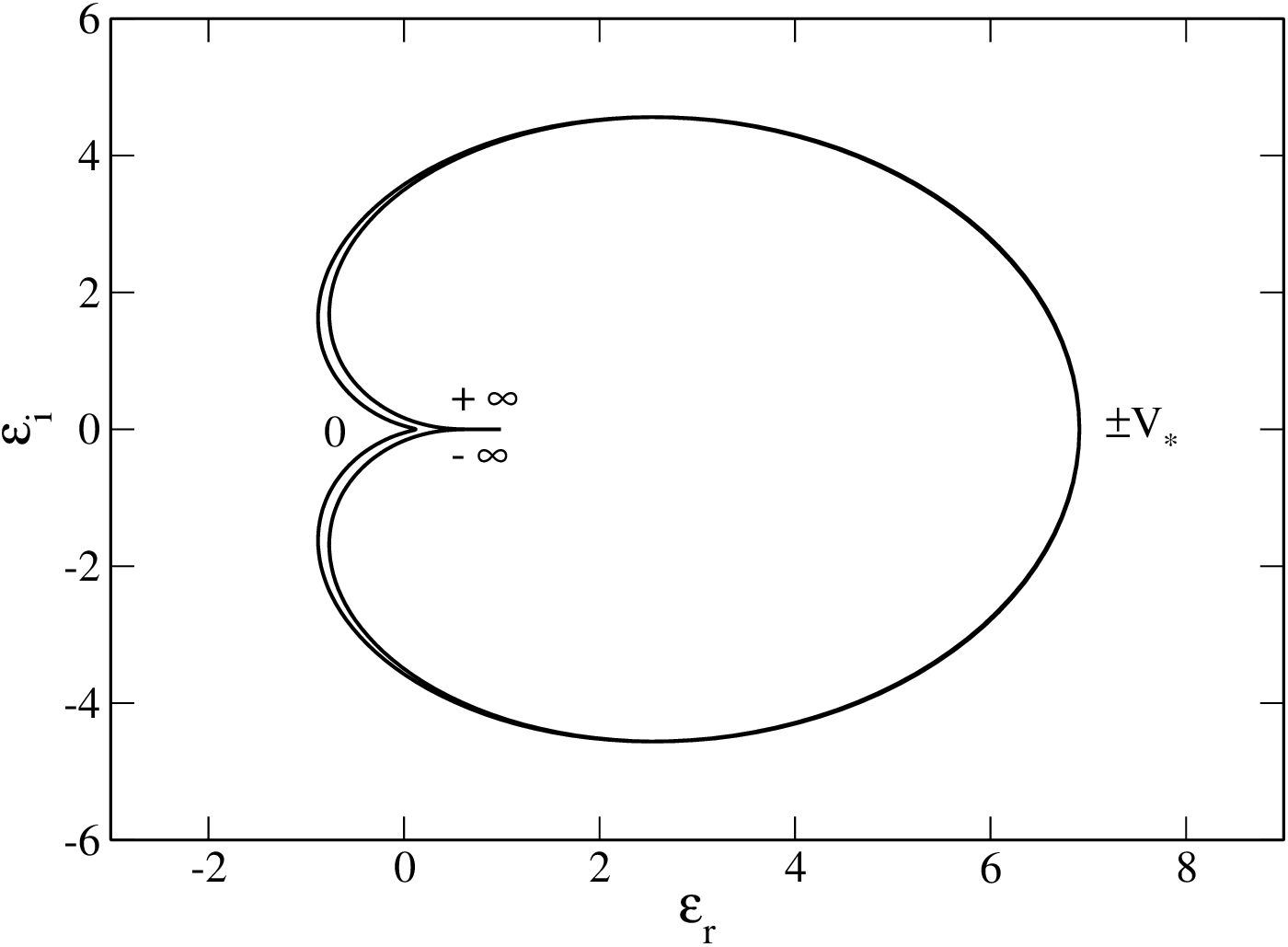}
\caption{Nyquist curve for $1<a<a_m$ and $\eta>\eta_c^{(0)(2)}(a)>1/a^2$ (specifically $a=1.2$, $\eta=12$). The system is stable. Case $(+++)$.}
\label{rr4}
\end{center}
\end{figure}

For $1<a<a_m$, there exists two temperatures $\eta_{c}^{(0)(1)}$, and
$\eta_{c}^{(0)(2)}$ at which the distribution is marginally stable. These two
branches merge at the point $(a_{m}, \eta_{c}^{(0)(m)})$. For
$\eta=\eta_{c}^{(0)(1)}$ and $\eta=\eta_{c}^{(0)(2)}$, the DF has a
minimum at $V_0=0$ and two maxima at $\pm V_*$.  The marginal
perturbation has zero pulsation $\Omega_r=0$.  By considering the
Nyquist curves in this region (see Figs. \ref{rr1}-\ref{rr4}), we
find that the distribution is stable for $\eta<\eta_{c}^{(0)(1)}$, unstable for
$\eta_{c}^{(0)(1)}<\eta<\eta_{c}^{(0)(2)}$ and  stable again for
$\eta>\eta_{c}^{(0)(2)}$. This corresponds to a re-entrant phase.

\subsection{The asymmetric double-humped distribution}
\label{sec_rass}

\begin{figure}
\begin{center}
\includegraphics[clip,scale=0.3]{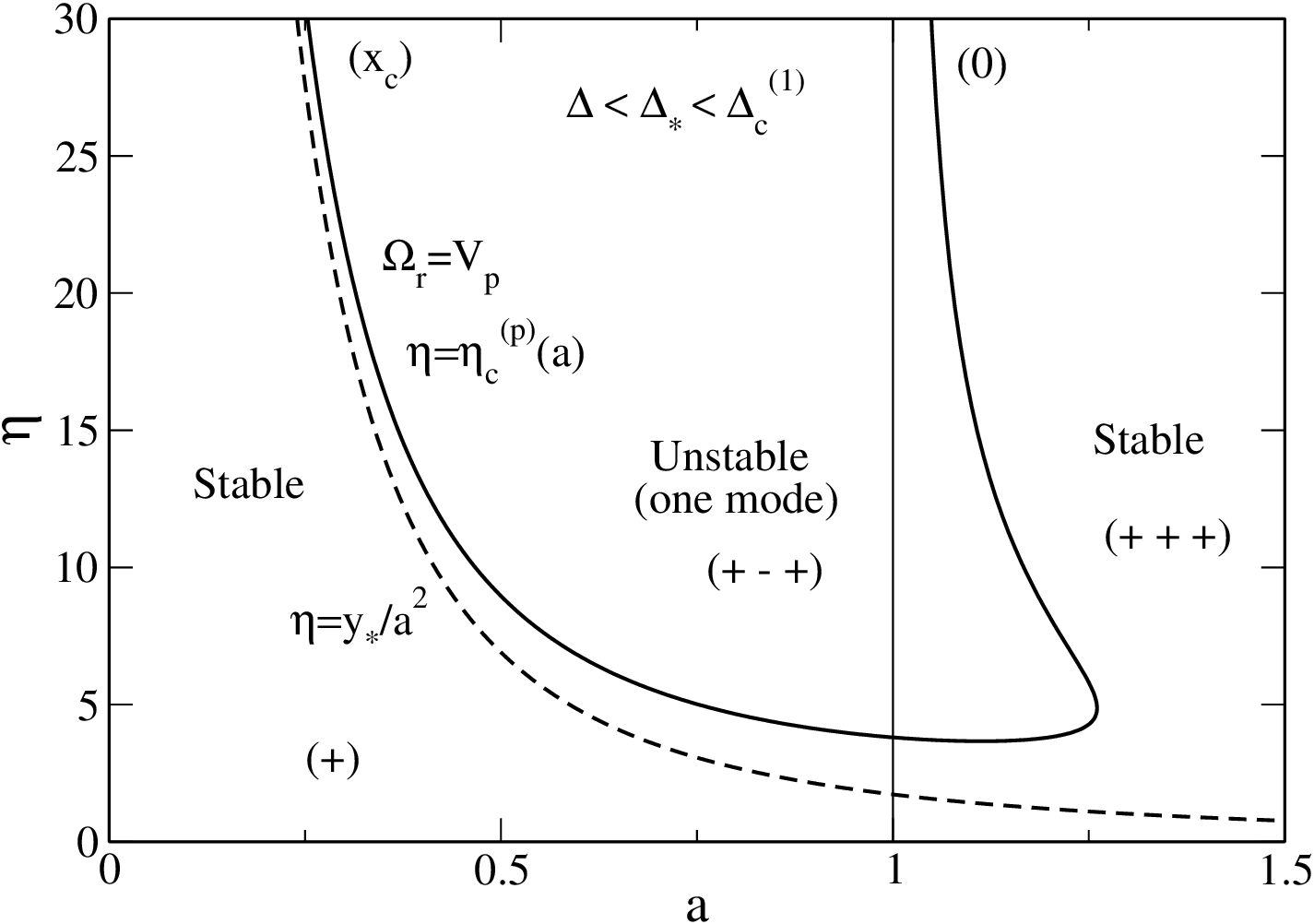}
\caption{Stability diagram for the asymmetric double-humped distribution with $\Delta<\Delta_*<\Delta_{c}^{(1)}$ (specifically $\Delta=2$). We have also plotted  the hyperbole $\eta=y_*/a^2$. Below this curve, the distribution has a single maximum at $V_p>0$ and above this curve, the distribution has a global  maximum at $V_0=V_-<0$, a local maximum at  $V_0=V_+>0$ and a minimum at $V_p>0$. Since $\Delta<\Delta_*$, there exists only one marginal branch $\eta_{c}^{(p)}$ which is always stricly above the hyperbole $\eta=y_*/a^2$. }
\label{dred2}
\end{center}
\end{figure}

\begin{figure}
\begin{center}
\includegraphics[clip,scale=0.3]{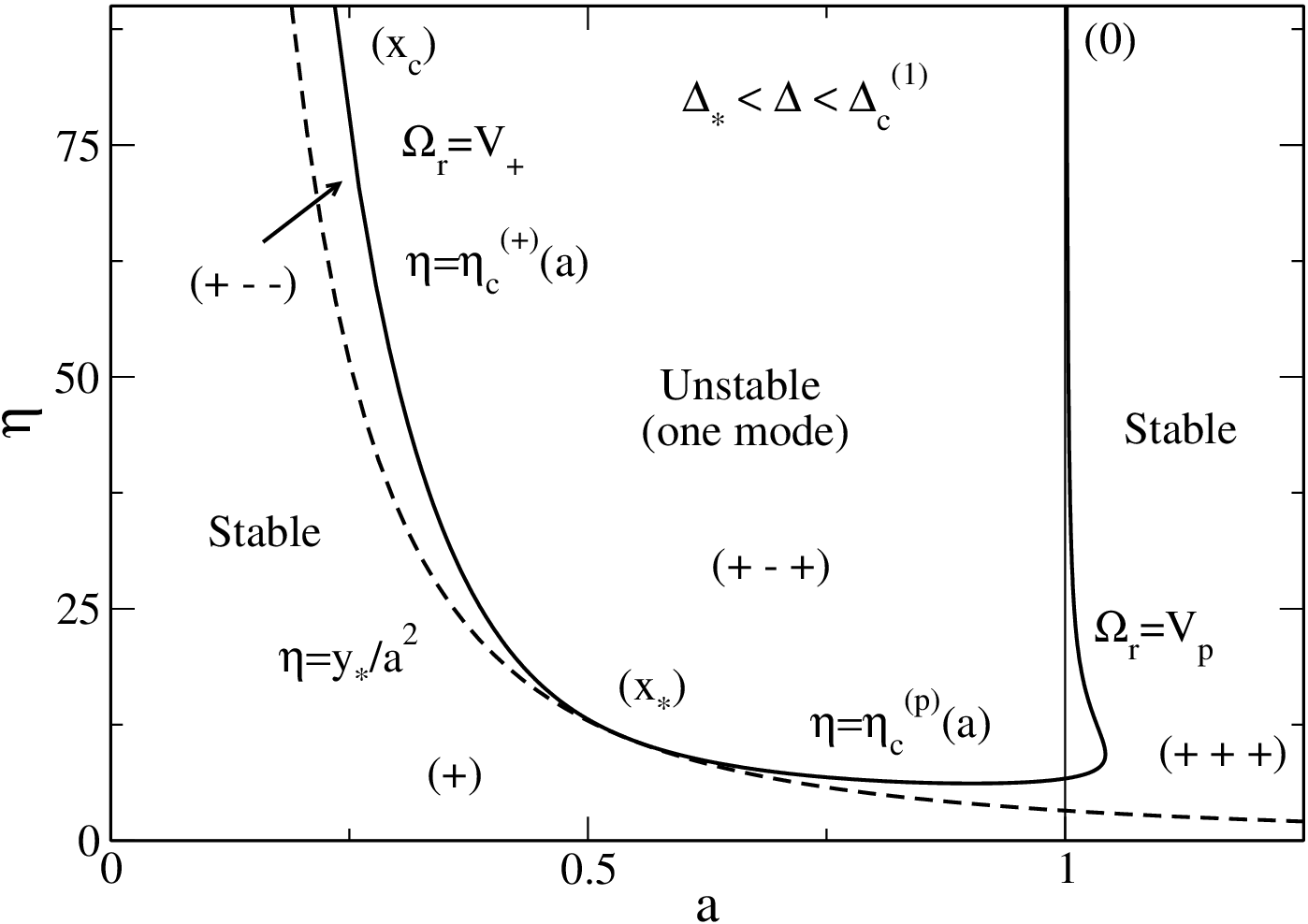}
\caption{Stability diagram for the asymmetric double-humped distribution with $\Delta_{*}<\Delta<\Delta_{c}^{(1)}$ (specifically $\Delta=20$).  Since $\Delta>\Delta_*$, there exists two marginal branch $\eta_{c}^{(p)}$ and $\eta_{c}^{(+)}$ that connect each other at the point of contact with the hyperbole $\eta=y_*/a^2$. }
\label{dred20}
\end{center}
\end{figure}

The dielectric function associated to the asymmetric double-humped
distribution (\ref{ae1}) in the repulsive case is
\begin{eqnarray}
\label{rass1}
\epsilon(\Omega)= 1+\frac{\eta}{1+\Delta} \left [ W(\sqrt{\eta}(\Omega - a))+\Delta W(\sqrt{\eta}(\Omega + a))   \right ].\nonumber\\
\end{eqnarray}
Proceeding as in Sec. \ref{sec_adh} and introducing the parameters
$x=V_0/a$ and $y=\eta a^2$, we find that the equations determining the
critical temperatures $\eta_c(a)$ are given by
\begin{equation}
\label{rass2}
y=\frac{1}{2x}\ln\left (\frac{1+x}{1-x}\right
)+\frac{\ln(\Delta)}{2x},
\end{equation}
\begin{equation}
\label{rass3}
\eta=-\frac{1+\Delta}{\left [ W_{r}(\sqrt{y}(x - 1))+\Delta W_{r}(\sqrt{y}(x+ 1))   \right ]}.
\end{equation}
\begin{equation}
\label{rass3b}
a^2=\frac{y}{\eta}.
\end{equation}
Equations (\ref{rass2}) and (\ref{rass3b}) determine the extrema of
the distribution and Eq. (\ref{rass3}) determines the temperatures
corresponding to the modes of marginal stability. As in
Sec. \ref{sec_adh}, the curve $\eta_c(a)$ can be obtained by varying
$x$ between $-1$ and $+1$. In the repulsive case, there exists
physical solutions with positive temperature only for $0<x\le x_c$.
For $\Delta<\Delta_{*}$, we get only one marginal branch
$\eta_{c}^{(p)}(a)$ corresponding to the mode $\Omega_r=V_{p}>0$ (see
Fig. \ref{dred2}). For $\Delta>\Delta_{*}$, we get two marginal
branches $\eta_{c}^{(p)}(a)$ and $\eta_{c}^{(+)}(a)$ corresponding to
the modes $\Omega_r=V_{p}>0$ and $\Omega_r=V_{+}>0$ (see
Fig. \ref{dred20}).  They connect each other at $(a_{*},\eta_{*})$
corresponding to $x=x_{*}$. At that point they touch the hyperbole
$\eta=y_{*}/a^2$ separating distributions with one or two maxima.

For $x\rightarrow 0^+$, using Eq. (\ref{wr8}), we find that
$a\rightarrow 1$. Then, proceeding carefully, we get 
\begin{equation}
\label{rass4}
\eta_c^{(p)}(a)\sim \left (\frac{3}{2}+\frac{1-\Delta}{1+\Delta}\frac{\ln\Delta}{2}\right )\frac{1}{a-1}, \qquad (a\rightarrow 1).
\end{equation}
For $x\rightarrow x_c$, we find that $a\rightarrow 0$ and
\begin{equation}
\label{rass5}
\eta_c^{(P)}(a)\sim \frac{y_c}{a^2}, \qquad (a\rightarrow 0).
\end{equation}

\begin{figure}
\begin{center}
\includegraphics[clip,scale=0.3]{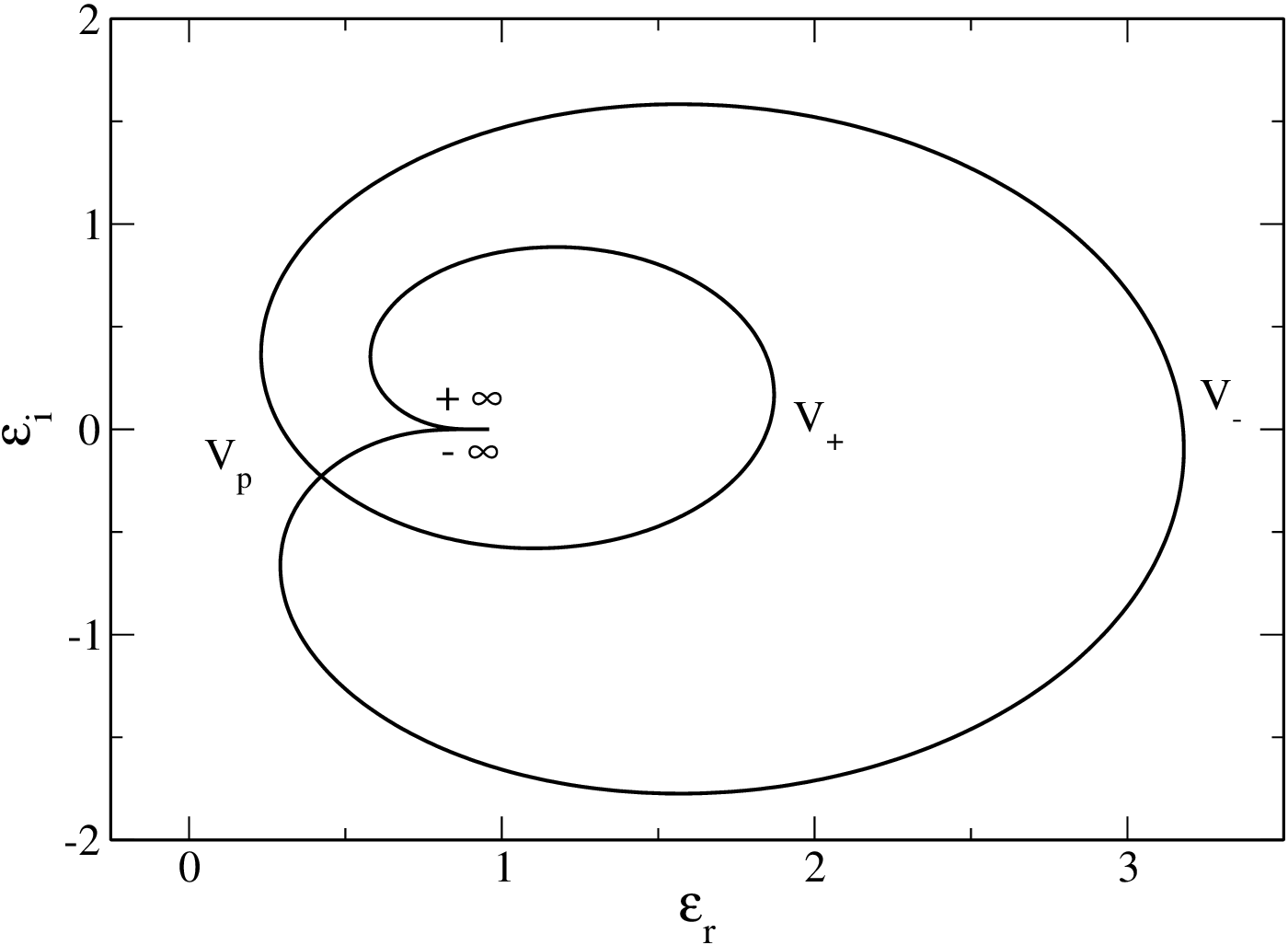}
\caption{For $\Delta<\Delta_{*}<\Delta_{c}^{(1)}$: Nyquist curve for  $y_{*}/a^2<\eta<\eta_{c}^{(p)}(a)$ (specifically $\Delta=2$, $a=0.9$ and $\eta=3.5$).  The system is stable. Case $(+++)$.   }
\label{ga1}
\end{center}
\end{figure}

\begin{figure}
\begin{center}
\includegraphics[clip,scale=0.3]{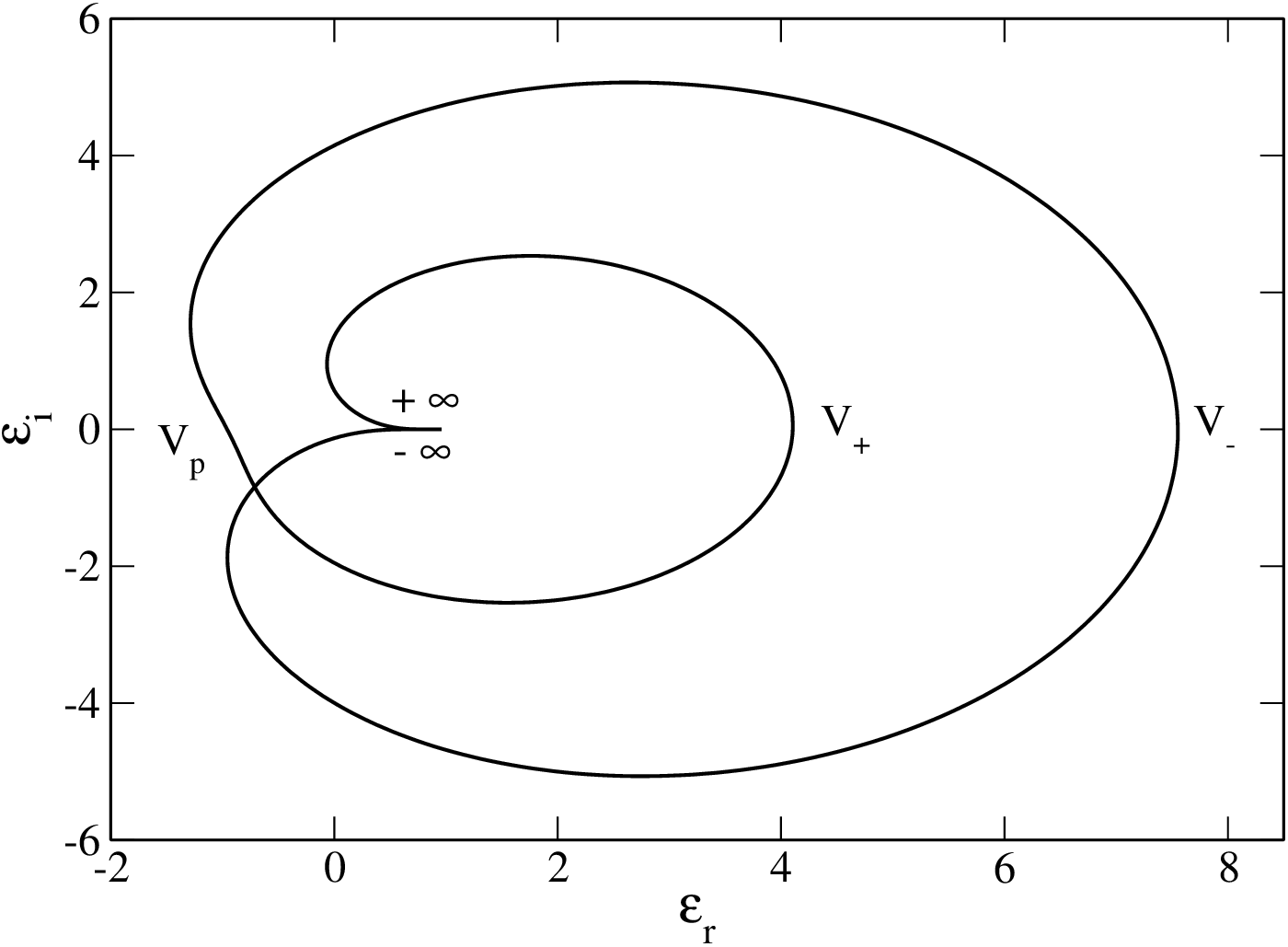}
\caption{For $\Delta<\Delta_{*}<\Delta_{c}^{(1)}$: Nyquist curve for  $y_{*}/a^2<\eta_{c}^{(p)}(a)<\eta$ (specifically  $\Delta=2$, $a=0.9$ and $\eta=10$).  The system is unstable. Case $(+-+)$. }
\label{ga2}
\end{center}
\end{figure}

\begin{figure}
\begin{center}
\includegraphics[clip,scale=0.3]{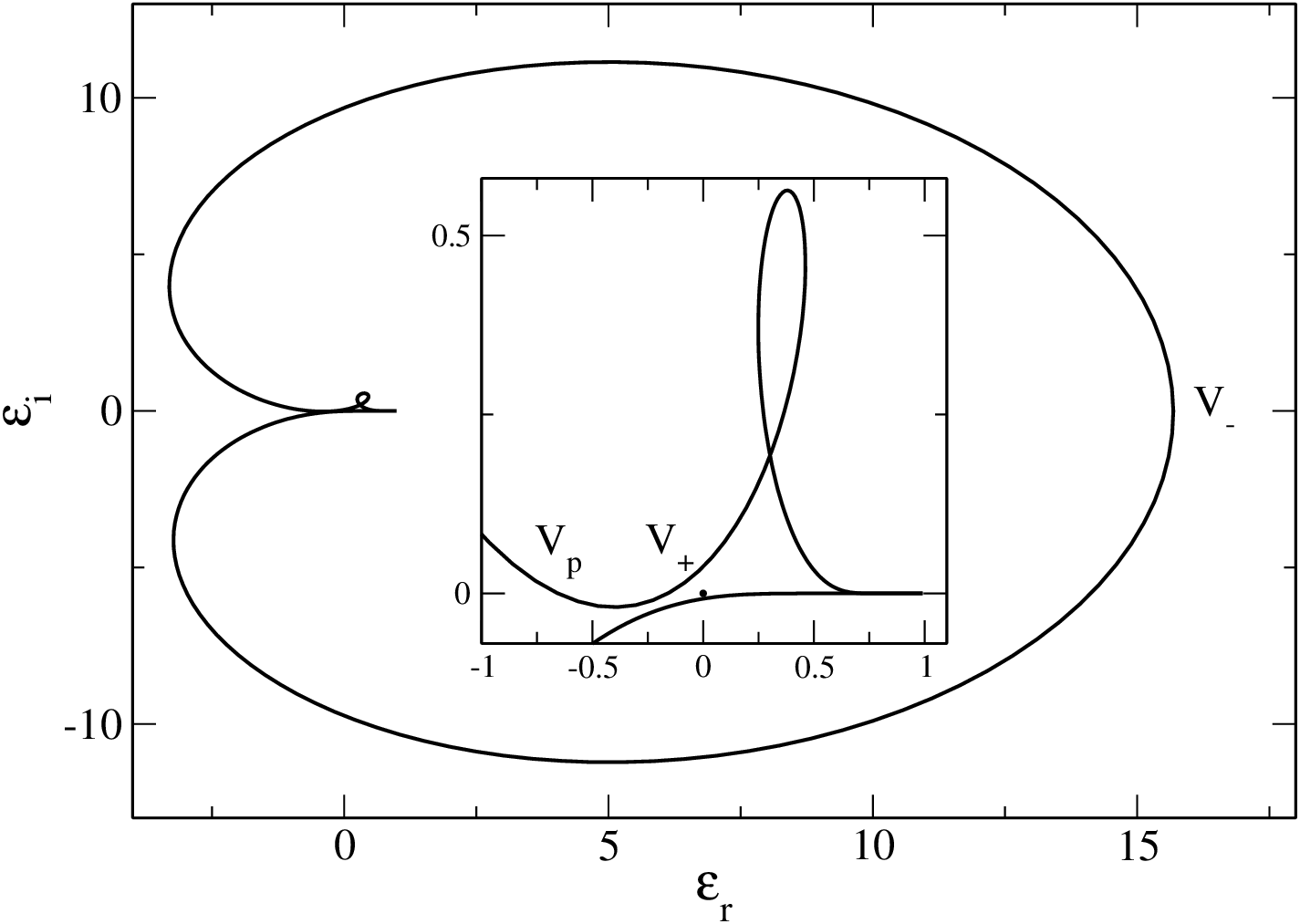}
\caption{For $\Delta_{*}<\Delta<\Delta_{c}^{(1)}$: Nyquist curve for  $y_{*}/a^2<\eta<\eta_{c}^{(+)}(a)$ (specifically  $\Delta=20$, $a=0.46$ and $\eta=15.5$).  The system is stable. Case $(+--)$.}
\label{ga3}
\end{center}
\end{figure}

\begin{figure}
\begin{center}
\includegraphics[clip,scale=0.3]{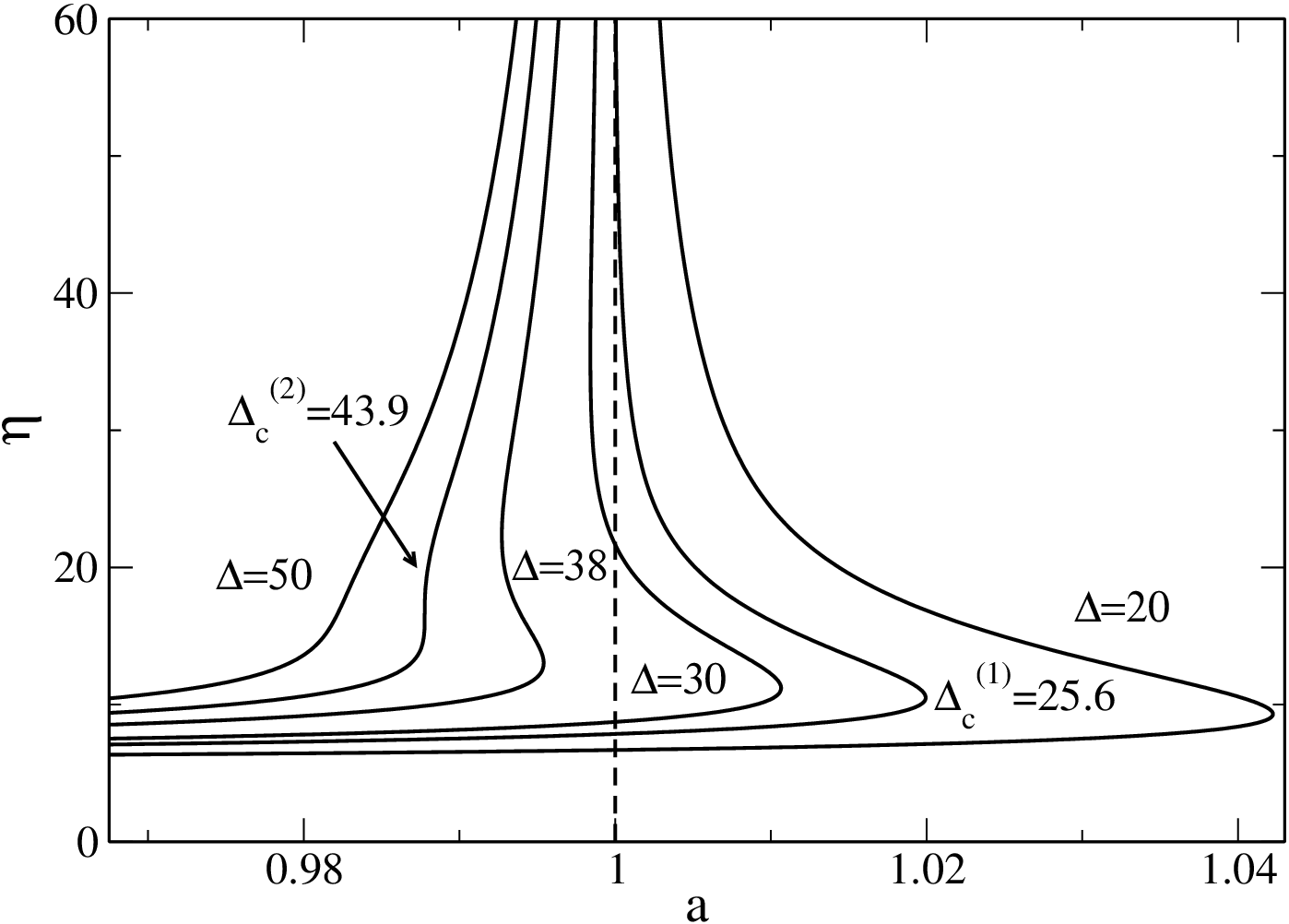}
\caption{Stability diagram for different values of the asymmetry $\Delta$ showing the appearance of a double re-entrant phase at $\Delta_c^{(1)}=25.6$  and their  disappearance  above $\Delta_c^{(2)}=43.9$. }
\label{allq}
\end{center}
\end{figure}

The stability diagrams corresponding to the asymmetric double-humped
distribution with $\Delta=2$ and $\Delta=20$ are represented in
Figs. \ref{dred2} and \ref{dred20}. In continuity with the symmetric
case $\Delta=1$, they display a re-entrant phase. Some representative
Nyquist curves are represented in Figs. \ref{ga1}-\ref{ga3}. In
Fig. \ref{allq}, we determine how the stability diagram changes with
the asymmetry $\Delta$. We find that a double re-entrant phase appears
for $\Delta_c^{(1)}=25.6$ and that they both disappear above
$\Delta_c^{(2)}=43.9$. In fact, the critical value $\Delta_{c}^{(1)}$
corresponds to the case where the asymptote passes from $a=1^{+}$ to
$a=1^{-}$. Therefore, it corresponds to the vanishing of the term in
parenthesis in Eq. (\ref{rass4}). A numerical solution of $3+\ln\Delta
(1-\Delta)/(1+\Delta)=0$ then gives $\Delta_c^{(1)}=25.6268...$.

\section{General results for an arbitrary potential of interaction}
\label{sec_genr}

In this section, we show that the previous results obtained for the
attractive or repulsive HMF models can have applications for more
general potentials of interaction. We also make the connexion between
linear and formal nonlinear stability, and between dynamical stability
with respect to the the Euler and Vlasov equations. 

\subsection{Linear stability for the barotropic Euler equation}
\label{sec_bel}

We consider the dynamics of a gas governed by the Euler equations
\begin{eqnarray}
{\partial \rho\over\partial t}+\nabla\cdot (\rho {\bf u})=0,
\label{bel1}
\end{eqnarray}
\begin{eqnarray}
\rho\biggl \lbrack {\partial {\bf u}\over\partial t}+({\bf u}\cdot \nabla){\bf u}\biggr \rbrack=-\nabla p-\rho\nabla\Phi,
\label{bel2}
\end{eqnarray}
\begin{eqnarray}
\Phi({\bf r},t)=\int u(|{\bf r}-{\bf r}'|)\rho({\bf r}',t)d{\bf r}',
\label{bel3}
\end{eqnarray}
where $u(|{\bf r}-{\bf r}'|)$ is an arbitrary binary potential of interaction. We assume that the gas is described by a  barotropic equation of state
$p=p(\rho)$.  Clearly, a spatially homogeneous distribution $\rho=\Phi={\rm Cst.}$ with ${\bf u}={\bf 0}$ is a stationary solution of Eq. (\ref{bel1})-(\ref{bel3})
provided that $\Phi=\rho U$ with $U=\int u({\bf x})d{\bf x}$ (for the gravitational potential, this last condition is not realized and we have to make the Jeans swindle \cite{bt}). We want to investigate the linear dynamical stability of this spatially homogeneous distribution. Linearizing the Euler  equations (\ref{bel1})-(\ref{bel3}) around this steady state and decomposing the perturbations in normal modes $\delta\rho_{{\bf k}\omega}\sim \delta \Phi_{{\bf k}\omega}\sim\delta {\bf u}_{{\bf k}\omega}\sim e^{i({\bf k}\cdot {\bf r}-\omega t)}$  we obtain the dispersion relation \cite{cvb}:
\begin{eqnarray}
\omega^{2}=c_{s}^{2}k^{2}+(2\pi)^{d}\hat{u}({k}) k^{2} \rho,
\label{bel4}
\end{eqnarray}
where we have introduced the velocity of sound $c_{s}^{2}=p'(\rho)$. We note that the pulsation $\omega$ is either real ($\omega^2>0$) or purely imaginary ($\omega^2<0$). As a result, the system is linearly stable with respect to  a perturbation with wavenumber $k$ if
\begin{eqnarray}
c_{s}^{2}+(2\pi)^{d}\hat{u}({k})\rho>0,
\label{bel5}
\end{eqnarray}
and linearly unstable otherwise (when 
$\omega^2<0$, the mode $\omega=+i\sqrt{-\omega^2}$ grows exponentially rapidly).

For repulsive potentials satisfying $\hat{u}({k})>0$, the homogeneous distribution is always stable.
For Coulombian plasmas in $d=3$, using Eq. (\ref{ana1}), the dispersion relation can
be written \cite{nicholson}:
\begin{eqnarray}
\omega^{2}=\omega_p^2+c_{s}^{2}k^{2},
\label{bel4b}
\end{eqnarray}
where $\omega_p$ is the plasma pulsation Eq. (\ref{ana2}). For the repulsive HMF model, using $\hat{u}_n=\frac{k}{4\pi}\delta_{n,\pm 1}$,  the dispersion relation is \cite{cvb}:
\begin{eqnarray}
\omega^{2}=c_{s}^{2}n^2+\frac{kM}{4\pi}n^2\delta_{n,\pm 1}.
\label{tard1}
\end{eqnarray}
The modes $n\neq \pm 1$ oscillate with a pulsation $\omega=\pm c_s n$. The pulsation of the modes $n=\pm 1$ is $\omega^{2}=\omega_p^2+c_{s}^{2}$ where $\omega_p$ is the proper pulsation (\ref{landau9}).

For attractive potentials satisfying $\hat{u}({k})<0$, the homogeneous distribution is linearly stable if
\begin{eqnarray}
c_{s}^{2}>(c_{s}^{2})_{crit}\equiv (2\pi)^{d}\rho |\hat{u}({k})|_{max},
\label{bel6}
\end{eqnarray}
and linearly unstable if $c_{s}^{2}<(c_{s}^{2})_{crit}$.  In this last
case, the unstable wavelengths are determined by the converse of inequality (\ref{bel5}).
For the gravitational
interaction in $d=3$, using Eq. (\ref{ana7}),  the
dispersion relation can be written \cite{bt}:
\begin{eqnarray}
\omega^{2}=c_{s}^{2}k^{2}-4\pi G\rho.
\label{bel7}
\end{eqnarray}
The system is always unstable ($(c_{s}^{2})_{crit}=\infty$) for sufficiently small wavenumbers
\begin{eqnarray}
k<k_{J}\equiv \biggl ({4\pi G\rho \over c_{s}^{2}}\biggr )^{1/2},
\label{bel8}
\end{eqnarray}
where $k_{J}$ is the Jeans wavenumber for a barotropic gas. The growth rate is maximum for $k\rightarrow 0$. For the attractive HMF model, using $\hat{u}_n=-\frac{k}{4\pi}\delta_{n,\pm 1}$,  the dispersion relation is \cite{cvb}:
\begin{eqnarray}
\omega^{2}=c_{s}^{2}n^2-\frac{kM}{4\pi}n^2\delta_{n,\pm 1}.
\label{tard2}
\end{eqnarray}
The modes $n\neq \pm 1$ oscillate with a pulsation $\omega=\pm c_s n$. The complex pulsation of the modes $n=\pm 1$ is $\omega^{2}=c_{s}^{2}-\frac{kM}{4\pi}$. 
The system is stable for $c_{s}^{2}>\frac{kM}{4\pi}$ and unstable for $c_{s}^{2}<\frac{kM}{4\pi}$.

\subsection{Formal stability for the barotropic Euler equation}
\label{sec_formeuler}

The barotropic Euler equation conserve the mass $M=\int \rho\, d{\bf r}$ and the energy functional \cite{bt}:
\begin{eqnarray}
{\cal W}[\rho,{\bf u}]=\int \rho\int ^{\rho}\frac{p(\rho')}{\rho^{'2}}\, d\rho'd{\bf r}+\frac{1}{2}\int \rho\Phi\, d{\bf r}+\int \rho \frac{{\bf u}^{2}}{2}\, d{\bf r}.\nonumber\\
\label{fe1}
\end{eqnarray}
Therefore, a minimum of the energy functional at fixed mass determines
a steady state of the Euler equation that is nonlinearly dynamically
stable \footnote{In astrophysics, this is called the Chandrasekhar energy principle \cite{bt}.}. The
qualitative idea is the following. If the system is an energy minimum,
and since the energy is conserved by the Euler equation, then it
cannot evolve away from that point. Therefore, it is dynamically
stable. Alternatively, if the flow is a saddle point of energy, then
it can evolve away from that point by following an iso-energy line. In
that case, it is not dynamically stable.  Here, we have stability in
the sense of Lyapunov. This means that the size of the perturbation is
bounded by the size of the initial perturbation for all times.  We are
led therefore to considering the minimization problem
\begin{eqnarray}
\min_{\rho,{\bf u}}\ \lbrace {\cal W}[\rho,{\bf u}]\quad |\quad M[\rho]=M \rbrace.
\label{fe2}
\end{eqnarray}
In this paper we shall consider {\it small} perturbations and determine conditions under which the system is a local minimum of energy ($\delta {\cal W}=0$ and $\delta^{2}{\cal W}$  positive definite). Thus, we restrict ourselves to formal nonlinear stability \cite{holm}. Since $\delta^{2}{\cal W}$ provides a norm preserved by the linearized equations, formal stability implies linear stability (which in turn implies spectral stability). However, for infinite dimensional systems, formal nonlinear stability need not imply nonlinear stability.

The critical points of energy at fixed mass are given by $\delta{\cal W}-\mu\delta M=0$ where $\mu$ is a Lagrange multiplier. This leads  to ${\bf u}={\bf 0}$ and to the condition of hydrostatic equilibrium
\begin{eqnarray}
\nabla p+\rho\nabla\Phi={\bf 0}.
\label{fe3}
\end{eqnarray}
Therefore, the critical points of energy at fixed mass are steady
states of the Euler equation (\ref{bel1})-(\ref{bel3}). For a barotropic gas $p=p(\rho)$, the
condition of hydrostatic equilibrium (\ref{fe3}) leads to $\rho=\rho(\Phi)$ and
$p=p(\Phi)$ with $p'(\Phi)=-\rho(\Phi)$. This leads to the identity
\begin{eqnarray}
p'(\rho)=-\frac{\rho}{\rho'(\Phi)}.
\label{fe4}
\end{eqnarray}
Since $p'(\rho)=c_s^2>0$, this relation implies that $\rho'(\Phi)<0$. 

A critical point of energy at fixed mass is a (local) minimum iff
\begin{eqnarray}
\delta^{2}{\cal W}=\int \frac{p'(\rho)}{2\rho}(\delta \rho)^{2}\, d{\bf r}+\frac{1}{2}\int \delta\rho\delta\Phi\, d{\bf r}>0
\label{fe5}
\end{eqnarray}
for all perturbations $\delta\rho$ that conserve mass. Using Eq. (\ref{fe4}), this  can be rewritten
\begin{eqnarray}
\delta^{2}{\cal W}=-\frac{1}{2}\int \frac{(\delta \rho)^{2}}{\rho'(\Phi)}\, d{\bf r}+\frac{1}{2}\int \delta\rho\delta\Phi\, d{\bf r}>0
\label{fe6}
\end{eqnarray}
for all perturbations $\delta\rho$ that conserve mass. We are led therefore to considering the eigenvalue equation
\begin{eqnarray}
p'(\rho)\delta\rho+\rho\delta\Phi=\lambda\delta\rho,
\label{fe7}
\end{eqnarray}
with $\delta\Phi=u* \delta\rho$, for perturbations such that $\int
\delta\rho\, d{\bf r}=0$. For these eigenmodes, we have
\begin{eqnarray}
\delta^{2}{\cal W}=\frac{1}{2}\lambda\int \frac{(\delta \rho)^{2}}{\rho}\, d{\bf r}.
\label{fe8}
\end{eqnarray}
Therefore, a critical point of energy at fixed mass  is a (local) minimum iff all the eigenvalues of the equation (\ref{fe7}) are positive.

Let us specifically consider spatially homogeneous
systems. Solving Eq. (\ref{fe7}) in Fourier space and using
$\delta\hat{\Phi}=(2\pi)^{d}\hat{u}(k)\delta\hat{\rho}$ since the
integral in Eq. (\ref{bel3}) is a convolution, we find that the eigenvalues are
given by
\begin{eqnarray}
\lambda(k)=c_{s}^{2}+(2\pi)^{d}\hat{u}(k)\rho.
\label{fe9}
\end{eqnarray}
Therefore, a spatially homogeneous system is a (local)
 minimum of energy at fixed mass iff
\begin{eqnarray}
c_{s}^{2}+(2\pi)^{d}\hat{u}(k)\rho>0, \quad \forall k.
\label{fe10}
\end{eqnarray}
When condition (\ref{fe10}) is fulfilled, we conclude that the system is
formally nonlinearly stable with respect to the barotropic Euler
equations (\ref{bel1})-(\ref{bel3}). We see that, for spatially homogeneous systems, the condition of
formal stability (\ref{fe10}) coincides with the condition of linear stability
(\ref{bel5}).  On the other hand, when condition (\ref{fe10}) is not fulfilled the
system is linearly unstable according to Sec. \ref{sec_bel}. In conclusion, {\it for spatially homogeneous systems described by the barotropic
Euler equation, formal nonlinear stability coincides with linear stability} \footnote{This result may remain true for inhomogeneous systems. In particular, it is shown in \cite{aaantonov} that the neutral modes of linear and formal stability for  barotropic stars are determined by the same condition.}.

\subsection{Linear stability for the Vlasov equation }
\label{sec_vl}

We consider the dynamics of a kinetic system described by the Vlasov equation
\begin{eqnarray}
{\partial f\over\partial t}+{\bf v}\cdot {\partial f\over\partial {\bf r}}-\nabla\Phi\cdot {\partial f\over\partial {\bf v}}=0,
\label{vl1}
\end{eqnarray}
\begin{eqnarray}
\Phi({\bf r},t)=\int u(|{\bf r}-{\bf r}'|)\rho({\bf r}',t)d{\bf r}',
\label{vel2}
\end{eqnarray}
where $u(|{\bf r}-{\bf r}'|)$ is an arbitrary binary potential of
interaction. Clearly, a distribution function $f=f({\bf v})$ that
depends only on the velocity is a stationary solution of the Vlasov
equation with $\Phi=\rho U$. We want to investigate the linear dynamical stability of
this spatially homogeneous distribution. Linearizing the Vlasov
equation around this steady state and using Laplace-Fourier transform (writing the perturbations in the
form $\delta f_{{\bf k}\omega}\sim \delta\Phi_{{\bf k}\omega}\sim e^{i({\bf k}\cdot {\bf r}-\omega
t)}$), we obtain the dispersion relation \cite{cvb}:
\begin{eqnarray}
\epsilon({\bf k},\omega)\equiv 1-(2\pi)^{d}\hat{u}({\bf k})\int {{\bf k}\cdot{\partial f\over\partial {\bf v}}\over {\bf k}\cdot {\bf v}-\omega}d{\bf v}=0.
\label{vel3}
\end{eqnarray}
If we take the wavevector ${\bf k}$ along the  $z$-axis \footnote{If the distribution function is isotropic, there is no restriction in making this choice.}, the dispersion relation becomes
\begin{eqnarray}
\epsilon({k},\omega)\equiv 1-(2\pi)^{d}\hat{u}({k})\int_{C} {{k}{\partial f\over\partial {v_z}}\over {k} {v_z}-\omega}d{v}_x d{v}_y d{v}_z=0,\nonumber\\
\label{vel4}
\end{eqnarray}
where the integral must be performed along the Landau contour $C$. We define $g(v_z)=\int f dv_x dv_y$. In the following, we shall note $v$ instead of $v_z$ and $f$ instead of $g$. With these conventions, the dispersion relation (\ref{vel4}) can be rewritten
\begin{eqnarray}
\epsilon({k},\omega)\equiv 1-(2\pi)^{d}\hat{u}({k})\int_{C} {f'(v)\over {v}-\frac{\omega}{k}}d{v}=0.
\label{vel5}
\end{eqnarray}
We are led therefore to a one-dimensional problem like the one investigated in the first part of this paper for the HMF model.

For $\omega_{i}=0$, the real and imaginary parts of the dielectric function $\epsilon(k,\omega_{r})=\epsilon_r(k,\omega_{r})+i\epsilon_i(k,\omega_{r})$ are given by
\begin{eqnarray}
\epsilon_r({k},\omega_r)= 1-(2\pi)^{d}\hat{u}({k})P\int_{-\infty}^{+\infty} {f'(v)\over {v}-\frac{\omega_r}{k}}d{v},
\label{vel6}
\end{eqnarray}
\begin{eqnarray}
\epsilon_i({k},\omega_r)= -\pi (2\pi)^{d}\hat{u}({k})f'\left (\frac{\omega_r}{k}\right ).
\label{vel7}
\end{eqnarray}
The condition of marginal stability corresponds to $\epsilon_r({k},\omega_r)=\epsilon_i({k},\omega_r)=0$. The condition $\epsilon_i({k},\omega_r)=0$ is equivalent to $f'\left (\frac{\omega_r}{k}\right )=0$ so that $\omega_r/k=v_0$ is equal to a velocity where $f(v)$ is extremum ($f'(v_0)=0$). The condition $\epsilon_r({k},\omega_r=v_0)=0$ then gives
\begin{eqnarray}
1-(2\pi)^{d}\hat{u}({k})\int_{-\infty}^{+\infty} {f'(v)\over {v}-v_0}d{v}=0.
\label{vel8}
\end{eqnarray}
This determines the wavenumber(s) $k_{c}$ corresponding to marginal stability.

On the other hand, as a direct consequence of the Nyquist criterion, if $f(v)$ has a unique maximum at $v=v_0$, then the DF is linearly stable with respect to a perturbation with wavenumber $k$ iff
$\epsilon_r({k},\omega_r=v_0)>0$ i.e. iff
\begin{eqnarray}
1-(2\pi)^{d}\hat{u}({k})\int_{-\infty}^{+\infty} {f'(v)\over {v}-v_0}d{v}>0.
\label{vel9}
\end{eqnarray}
This stability criterion generalizes Eq. (\ref{sh1}) to the case of an arbitrary
potential of interaction. If $\hat{u}({k})>0$ (repulsive potential), a
single-humped distribution is always stable. This is the case in particular for a Coulombian plasma \cite{nicholson}.
If $\hat{u}({k})<0$ (attractive
potential), the distribution is stable to all wavenumbers if
\begin{eqnarray}
\frac{\rho}{\int_{-\infty}^{+\infty} {f'(v)\over {v}-v_0}d{v}}>(2\pi)^{d}\rho|\hat{u}({k})|_{max},
\label{vel9b}
\end{eqnarray}
and unstable otherwise. In that last case, the range of unstable wavenumbers is determined by the converse of  inequality (\ref{vel9}).   For the gravitational interaction in
$d=3$, the stability criterion (\ref{vel9}) becomes
\begin{eqnarray}
1+{4\pi G\over k^{2}}\int_{-\infty}^{+\infty} {f'(v)\over v-v_0}dv>0.
\label{vel11}
\end{eqnarray}
The system is always unstable for sufficiently small wavenumbers:
\begin{eqnarray}
k<k_{J}=\left (-4\pi G\int_{-\infty}^{+\infty} \frac{f'(v)}{v-v_{0}}\, dv\right )^{1/2},
\label{vel12}
\end{eqnarray}
where $k_{J}$ is the Jeans wavenumber for a stellar system. We shall
make the connexion between the stability of a kinetic system and the stability of the corresponding barotropic gas  in the next section. In particular, using identity (\ref{fo21}), we will show that Eqs. (\ref{vel9}) and (\ref{vel9b}) are equivalent to Eqs. (\ref{bel5}) and (\ref{bel6}).

\subsection{Formal stability for the Vlasov equation }
\label{sec_formal}

The Vlasov equation conserves the mass $M=\int f\, d{\bf r}d{\bf v}$,
the energy $E=\int f \frac{v^2}{2}\, d{\bf r}d{\bf v}+\frac{1}{2}\int
\rho\Phi\, d{\bf r}$ and the Casimirs $I_{h}=\int h(f)\, d{\bf r}d{\bf
v}$ where $h$ is an arbitrary continuous function. Let us introduce the
functionals $S=-\int C(f)\, d{\bf r}d{\bf v}$ where $C$ is an
arbitrary convex function ($C''>0$). They will be called ``pseudo
entropies''. We also introduce the ``pseudo free energies'' $J=S-\beta
E$ or $F=E-TS$  and the ``pseudo grand potential'' $G=S-\beta E-\alpha M$ where $\beta=1/T$ and $\alpha$ are some constants. Since $E$, $M$,
$I_{h}$, $S$, $J$, $F$ and $G$ are conserved by the Vlasov equation, the
following optimization principles \footnote{A review of these variational principles will be given in a forthcoming paper.}
\begin{eqnarray}
\min_{f}\ \lbrace E[f]\quad |\quad {\rm all\  the \ Casimirs\ } I_{h}\rbrace,
\label{fo1}
\end{eqnarray}
\begin{eqnarray}
\min_{f}\ \lbrace E[f]\quad |\quad M[f]=M,\quad S[f]=S \rbrace,
\label{fo2}
\end{eqnarray}
\begin{eqnarray}
\max_{f}\ \lbrace S[f]\quad |\quad M[f]=M,\quad E[f]=E \rbrace,
\label{fo3}
\end{eqnarray}
\begin{eqnarray}
\max_{f}\ \lbrace J[f]\quad |\quad M[f]=M \rbrace,
\label{fo4}
\end{eqnarray}
\begin{eqnarray}
\min_{f}\ \lbrace F[f]\quad |\quad M[f]=M \rbrace,
\label{fo5}
\end{eqnarray}
\begin{eqnarray}
\max_{f}\ \lbrace G[f]\rbrace,
\label{fo5b}
\end{eqnarray}determine steady states of the Vlasov equation that are nonlinearly
dynamically stable. The first criterion is the most refined stability
criterion that has been introduced in the literature \footnote{In
astrophysics, this is called the Antonov energy principle
\cite{bt}. Recently, developing these arguments, Lemou {\it et al.} 
\cite{lemougrav} have proven for the first time 
that all DF $f=f(\epsilon)$ with $f'(\epsilon)<0$ are 
{\it nonlinearly} stable with respect to the
Vlasov-Poisson system.}. The other criteria provide only {\it
sufficient} conditions of dynamical stability. Since the solution of
an optimization problem is always the solution of a more constrained
optimization problem, we have the relations
\begin{eqnarray}
(\ref{fo5b}) \Rightarrow  (\ref{fo5}) \Leftrightarrow (\ref{fo4}) \Rightarrow (\ref{fo3}) \Leftrightarrow (\ref{fo2}) \Rightarrow (\ref{fo1}).\nonumber\\
\label{fo6}
\end{eqnarray}
We note that (\ref{fo3}) is similar to a condition of microcanonical
(two constraints) stability, (\ref{fo4}) or (\ref{fo5}) is similar to
a condition of canonical (one constraint) stability, and (\ref{fo5b})
is similar to a condition of grand canonical (no constraint) stability
\footnote{We could also introduce a grand microcanonical ensemble
corresponding to the maximization of $K=S-\alpha M$ at fixed energy
\cite{aa3}.} in thermodynamics \cite{aa3,aaantonov}. Therefore, the
inclusion $(\ref{fo5b}) \Rightarrow (\ref{fo5}) \Leftrightarrow
(\ref{fo4}) \Rightarrow (\ref{fo3})$ is similar to the fact that
``grand canonical stability implies canonical stability which implies
microcanonical stability'' in thermodynamics. However, the reciprocal
is wrong in case of ``ensemble inequivalence" that is generic for
systems with long-range interactions \cite{ellis,bb,ijmpb}. In the
following, we shall mainly be interested by formal dynamical
stability \cite{holm}. We shall consider small perturbations and
determine under which conditions the first variations of the
functional vanish (critical point) and the second variations are
positive definite (for problems (\ref{fo1}), (\ref{fo2}) and
(\ref{fo5})) or negative definite (for problems (\ref{fo3}),
(\ref{fo4}) and (\ref{fo5b})).

{\it Critical points:} considering problem (\ref{fo1}), we find that a DF is a
critical point of energy for symplectic perturbations (i.e
perturbations that conserve all the Casimirs) iff $f({\bf r},{\bf v})$
is a steady state of the Vlasov equation \cite{bartholomew,kandrup}. On the other hand, the
critical points of problems (\ref{fo2})-(\ref{fo5b}) lead to DF of the form $f=f(\epsilon)$
with $f'(\epsilon)<0$, i.e. the DF depends only on the individual energy
$\epsilon=v^2/2+\Phi({\bf r})$ and is monotonically decreasing \cite{aaantonov}.

{\it Formal stability:} considering problem (\ref{fo1}), a critical point of energy for symplectic perturbations is a (local) minimum iff
\begin{eqnarray}
-\int \frac{(\delta f)^{2}}{f'(\epsilon)}\, d{\bf r}d{\bf v}+\int \delta f\delta\Phi\, d{\bf r}d{\bf v}>0,
\label{fo7}
\end{eqnarray}
for all perturbations
$\delta f$ that conserve the energy and all the Casimirs at first
order $\delta E=\delta I_{h}=0$. Considering problems (\ref{fo2}) and (\ref{fo3}), a critical point of energy (resp. pseudo entropy) at fixed mass and pseudo entropy (resp. energy) is a (local) minimum (resp. maximum) iff inequality (\ref{fo7}) is satisfied all perturbations $\delta f$ that conserve mass and energy at first order $\delta M=\delta E=0$. Considering problems (\ref{fo4}) and (\ref{fo5}), a critical point of pseudo free energy $J$  (resp. pseudo free energy $F$) at fixed mass is a (local) maximum (resp. minimum) if inequality (\ref{fo7}) is satisfied for all perturbations $\delta f$ that conserve mass  $\delta
M=0$. Considering problem (\ref{fo5b}), a critical point of pseudo grand potential $G$ is a (local) maximum if inequality (\ref{fo7}) is satisfied for all perturbations. Problem (\ref{fo1}) is the most refined stability criterion because it tells that, in order to settle the dynamical stability of the system, we just need considering symplectic (i.e. dynamically accessible) perturbations. Of course, if inequality (\ref{fo7}) is satisfied by a larger class of perturbations, as implied by problems (\ref{fo2})-(\ref{fo5b}), the system will be stable a fortiori. Therefore, we have the implications (\ref{fo6}). Problems  (\ref{fo2})-(\ref{fo5b}) provide sufficient (but not necessary) conditions of dynamical stability. A steady state can be stable according to  (\ref{fo1}) while it does not satisfy (\ref{fo2})-(\ref{fo3}), or it can be stable according to (\ref{fo2})-(\ref{fo3}) while it does not satisfy (\ref{fo4})-(\ref{fo5}), or it can be stable according to (\ref{fo4})-(\ref{fo5}) while it does not satisfy (\ref{fo5b}). Therefore, the stability criterion (\ref{fo1}) is more refined than (\ref{fo2})-(\ref{fo3}) which is itself more refined than (\ref{fo4})-(\ref{fo5}) which is itself more refined than (\ref{fo5b}).

Let us consider the minimization problem
\begin{eqnarray}
\min_{f}\ \lbrace F[f]\quad |\quad M[f]=M \rbrace,
\label{fo8}
\end{eqnarray}
in more detail. As we have seen, it provides a sufficient condition of dynamical stability for the Vlasov equation. The critical points of pseudo free energy $F$ at fixed
mass $M$ are given by
\begin{eqnarray}
\delta F+\alpha T\delta M=0,
\label{fo9}
\end{eqnarray}
where $\alpha$ is a Lagrange multiplier. This yields
\begin{eqnarray}
C'(f)=-\beta\epsilon-\alpha.
\label{fo10}
\end{eqnarray}
Since $C$ is convex this relation can be reversed to give
\begin{eqnarray}
f=F(\beta\epsilon+\alpha),
\label{fo11}
\end{eqnarray}
where $F(x)=(C')^{-1}(-x)$. Differentiating Eq. (\ref{fo10}), we obtain
\begin{eqnarray}
C''(f)=-\frac{\beta}{f'(\epsilon)},
\label{fo12}
\end{eqnarray}
so that $f'(\epsilon)$ is monotonic. We shall assume that $f'(\epsilon)<0$ so that the pseudo temperature $T$ is positive.  A critical point of pseudo free energy at fixed mass is a (local) minimum iff
\begin{eqnarray}
\delta^{2}F=\frac{T}{2}\int C''(f)(\delta f)^{2}\, d{\bf r}d{\bf v}+\frac{1}{2}\int \delta\rho\delta\Phi\, d{\bf r}>0\quad
\label{fo13}
\end{eqnarray}
for all perturbations that conserve mass. Using Eq. (\ref{fo12}), this can be rewritten
\begin{eqnarray}
\delta^{2}F=-\frac{1}{2}\int \frac{(\delta f)^{2}}{f'(\epsilon)}\, d{\bf r}d{\bf v}+\frac{1}{2}\int \delta\rho\delta\Phi\, d{\bf r}>0
\label{fo14}
\end{eqnarray}
for all perturbations that conserve mass.

For any distribution $f=f({\bf r},{\bf v})$, the density and the
pressure are given by $\rho=\int f\, d{\bf v}$ and $p=\frac{1}{d}\int
f v^2\, d{\bf v}$. If $f=f(\epsilon)$, we have $\rho=\rho(\Phi)$ and
$p=p(\Phi)$.  Eliminating $\Phi({\bf r})$ between these expressions we
find that the pressure is a function of the density:
$p=p(\rho)$. Therefore, to any kinetic system described by a distribution function
$f=f(\epsilon)$, we can associate a barotropic gas with an equation of
state $p(\rho)$ that is entirely specified by the function
$C(f)$. We now introduce the functional
\begin{eqnarray}
F[\rho]=\int \rho\int ^{\rho}\frac{p(\rho')}{\rho^{'2}}\, d\rho'd{\bf r}+\frac{1}{2}\int \rho\Phi\, d{\bf r},
\label{fo15}
\end{eqnarray}
and consider the minimization problem
\begin{eqnarray}
\min_{\rho}\ \lbrace F[\rho]\quad |\quad M[\rho]=M \rbrace.
\label{fo16}
\end{eqnarray}
It can be shown \cite{aaantonov,assiseph} that
\begin{eqnarray}
(\ref{fo8})  \Leftrightarrow (\ref{fo16}).
\label{fo17}
\end{eqnarray}
Therefore, $f$ is a minimum of $F[f]$ at fixed mass iff $\rho$ is a
minimum of $F[\rho]$ at fixed mass. As a result, (\ref{fo16}) provides a
sufficient condition of nonlinear dynamical stability for the Vlasov
equation. However, it is not a necessary condition of nonlinear
dynamical stability. For example, a DF can be nonlinearly stable
according to (\ref{fo1}) or (\ref{fo3}) although it does not satisfy (\ref{fo8}) or (\ref{fo16}). This
is similar to ``ensemble inequivalence'' in thermodynamics \cite{ellis,bb,ijmpb}.

It is also obvious that
\begin{eqnarray}
(\ref{fe2}) \Leftrightarrow (\ref{fo16}).
\label{fo18}
\end{eqnarray}
Therefore, if $\rho$ is nonlinearly stable with respect to the Euler
equation then it solves (\ref{fe2}) and (\ref{fo16}). According to (\ref{fo17}), the
corresponding distribution function $f$ solves (\ref{fo8}) so it is
nonlinearly stable with respect to the Vlasov equation. Therefore, {\it a
distribution function $f=f(\epsilon)$ with $f'(\epsilon)<0$ is
nonlinearly stable with respect to the Vlasov equation if the
corresponding density profile $\rho$ is nonlinearly stable with
respect to the Euler equation}. In astrophysics, this corresponds to
the (nonlinear) Antonov first law: ``a stellar system
$f=f(\epsilon)$ with $f'(\epsilon)<0$ is (nonlinearly) stable with
respect to the Vlasov equation if the corresponding barotropic star
with an equation of state $p=p(\rho)$ is (nonlinearly) stable with
respect to the Euler equations''.  However, the reciprocal is wrong in
general.  The distribution function $f$ can be nonlinearly stable
according to (\ref{fo1}) or (\ref{fo3}) although it does not satisfy (\ref{fo8}) so that the
density profile $\rho$ does not satisfy (\ref{fo16}) or (\ref{fe2}). We see therefore
that the nonlinear Antonov first law is similar to the notion of
``ensemble inequivalence'' in thermodynamics \cite{aaantonov}.

According to the previous discussion, $f$ is a minimum of $F[f]$ at
fixed mass iff $\rho$ is a minimum of $F[\rho]$ or ${\cal W}[\rho,{\bf u}]$ at fixed mass. We can therefore
use  the results of Sec. \ref{sec_formeuler}. In particular, for a spatially
homogeneous system, the distribution $f({\bf v})$ is a
(local) minimum of $F[f]$ at fixed mass iff
\begin{eqnarray}
c_{s}^{2}+(2\pi)^{d}\hat{u}(k)\rho>0, \quad \forall k,
\label{fo19}
\end{eqnarray}
where $c_{s}^{2}=p'(\rho)$ is the velocity of sound in the corresponding barotropic gas. Now, using Eq. (\ref{fe4}) and the fact that $\rho=\int f(\epsilon)\, d{\bf v}$, we have for any distribution  $f=f(\epsilon)$:
\begin{eqnarray}
c_{s}^2({\bf r})=-\frac{\rho({\bf r})}{\int f'(\epsilon)\, d{\bf v}}=-\frac{\rho({\bf r})}{\int \frac{\frac{\partial f}{\partial v_z}}{v_z} d{\bf v}}=-\frac{\rho({\bf r})}{\int_{-\infty}^{+\infty} \frac{\frac{\partial g}{\partial v_z}}{v_z} d{v}_{z}},\nonumber\\
\label{fo20}
\end{eqnarray}
where $g(v_z)=\int f({\bf v}) \, dv_xdv_y$.  For a spatially homogeneous system, we obtain
\begin{eqnarray}
c_{s}^2=-\frac{\rho}{\int_{-\infty}^{+\infty} \frac{f'(v)}{v}\, d{v}},
\label{fo21}
\end{eqnarray}
where we have noted $v$ for $v_z$ and $f$ for $g$ like in
Sec. \ref{sec_vl}. Substituting this relation in Eq. (\ref{fo19}), we find that a homogeneous distribution is a (local) minimum of $F$ at fixed mass iff
\begin{eqnarray}
1-(2\pi)^{d}\hat{u}({k})\int_{-\infty}^{+\infty} {f'(v)\over {v}}d{v}>0, \quad \forall k.
\label{fo22}
\end{eqnarray}
When condition (\ref{fo22}) is fulfilled, we conclude that the system
is formally nonlinearly stable with respect to the Vlasov equation
[according to criterion (\ref{fo5})]. We see that, for spatially
homogeneous systems, the condition of formal stability (\ref{fo22})
coincides with the condition of linear stability (\ref{vel9})
\footnote{Since $f=f(\epsilon)$ with $f'(\epsilon)<0$ where
$\epsilon=v^2/2+\Phi({\bf r})$, we deduce that, for a spatially
homogeneous system ($\Phi={\rm cst.}$), $f(v)$ is a symmetric function
with a single maximum at $v_{0}=0$. Therefore, Eq. (\ref{vel9}) reduces to
Eq. (\ref{fo22}).}. When condition (\ref{fo22}) is not fulfilled the
system is linearly unstable according to Sec. \ref{sec_vl}. Therefore,
for homogeneous systems, criterion (\ref{fo5}) is a sufficient and
necessary condition of dynamical stability (in that case, we have
ensemble equivalence). In conclusion, {\it for spatially homogeneous
systems described by the Vlasov equations, formal nonlinear stability
coincides with linear stability.} On the other hand, Eq. (\ref{fo21})
shows that the integral appearing in Eqs. (\ref{fo22}) and
(\ref{vel9}) is directly related to the sound velocity in the
corresponding barotropic gas \footnote{This is true for any
distribution function, not only for the Maxwell and Tsallis
distributions. However, the relation (\ref{fo21}) can be checked
explicitly for these simple distributions \cite{cvb}.}. This implies
that the condition of (linear and formal) dynamical stability for the
Vlasov equation coincide with the condition of (linear and formal)
dynamical stability for the Euler equation. In conclusion, {\it a
spatially homogeneous system is dynamically stable with respect to the
Vlasov equation iff the corresponding barotropic gas is dynamically
stable with respect to the Euler equations}. This is the proper
formulation of the Antonov first law for homogeneous systems. In that
case, we have ``ensemble equivalence''.  This is not true in general
for spatially inhomogeneous systems.

{\it Remark:} we have obtained the condition of formal stability
(\ref{fo22}) by a method different from the one developed by Yamaguchi
{\it et al.} \cite{yamaguchi}. Our result is more general because it
is valid for an arbitrary potential of interaction. Our method
exploits the link between the variational problems acting on the
distribution function $f$ and those acting on the spatial density
$\rho$. We have used the fact that a distribution function $f$ is
stable with respect to the Vlasov equation if the corresponding
density profile $\rho$ is stable with respect to the Euler
equation. This is similar to the Antonov first law in astrophysics and
this is related to a notion of ensemble inequivalence in
thermodynamics (for spatially homogenous systems, we have ensemble
equivalence).  As a by-product, our approach gives a physical
interpretation of the stability criterion (\ref{fo22}) in terms on a
condition on the velocity of sound (\ref{fo19}) in the corresponding
barotropic gas. Furthermore, our method can be extended to spatially
inhomogeneous systems (see Appendix
\ref{sec_inho}). Indeed, if all the eigenvalues $\lambda$ of the
eigenvalue equation (\ref{fe7}) are positive, then $\rho$ is formally
stable with respect to the Euler equation, implying by (\ref{fo18})
and (\ref{fo17}) that the corresponding distribution $f$ is formally
stable with respect to the Vlasov equation.

\section{Conclusion}

In this paper, we have carried out an exhaustive study of the
dynamical stability of systems with long-range interactions described
by the Vlasov equation. For illustration, we have considered the HMF
model and we have treated the case of single-humped (Maxwell,
Tsallis,...) as well as double-humped (symmetric and asymmetric)
distributions. The stability of the system has been investigated by
using the Nyquist method.  Interestingly, the stability diagrams of
these systems are very rich, exhibiting re-entrant phases in the case
of double-humped distributions. For the HMF model, the potential of
interaction is truncated to one Fourier mode so that only the modes
$n=\pm 1$ can propagate. As a result, the stability of the system is
expressed as a condition on the temperature $T$ while for other
systems (plasmas, gravitational systems,...) it is expressed as a
condition on the wavenumber $k$ (for a given value of the temperature
$T$). On the other hand, for the HMF model, we can consider attractive
as well as repulsive interactions and there exists spatially
homogeneous distributions in each case allowing the application of the
Nyquist method. Therefore, by changing the sign of the interaction,
this simple toy model exhibits features similar to Coulombian or
gravitational plasmas. Noticably, this is the first time that the
Nyquist method is applied to a system with {\it attractive}
interactions without making any (Jeans) swindle. Finally, we have shown
that the results and methods developed for the HMF model are in fact
applicable to more general potentials of interaction giving further
interest to our study. This opens the way to many numerical and/or
analytical investigations that will be considered in future works.

\appendix

\section{The function $W_{r}(z)$}
\label{sec_wr}

\begin{figure}
\begin{center}
\includegraphics[clip,scale=0.3]{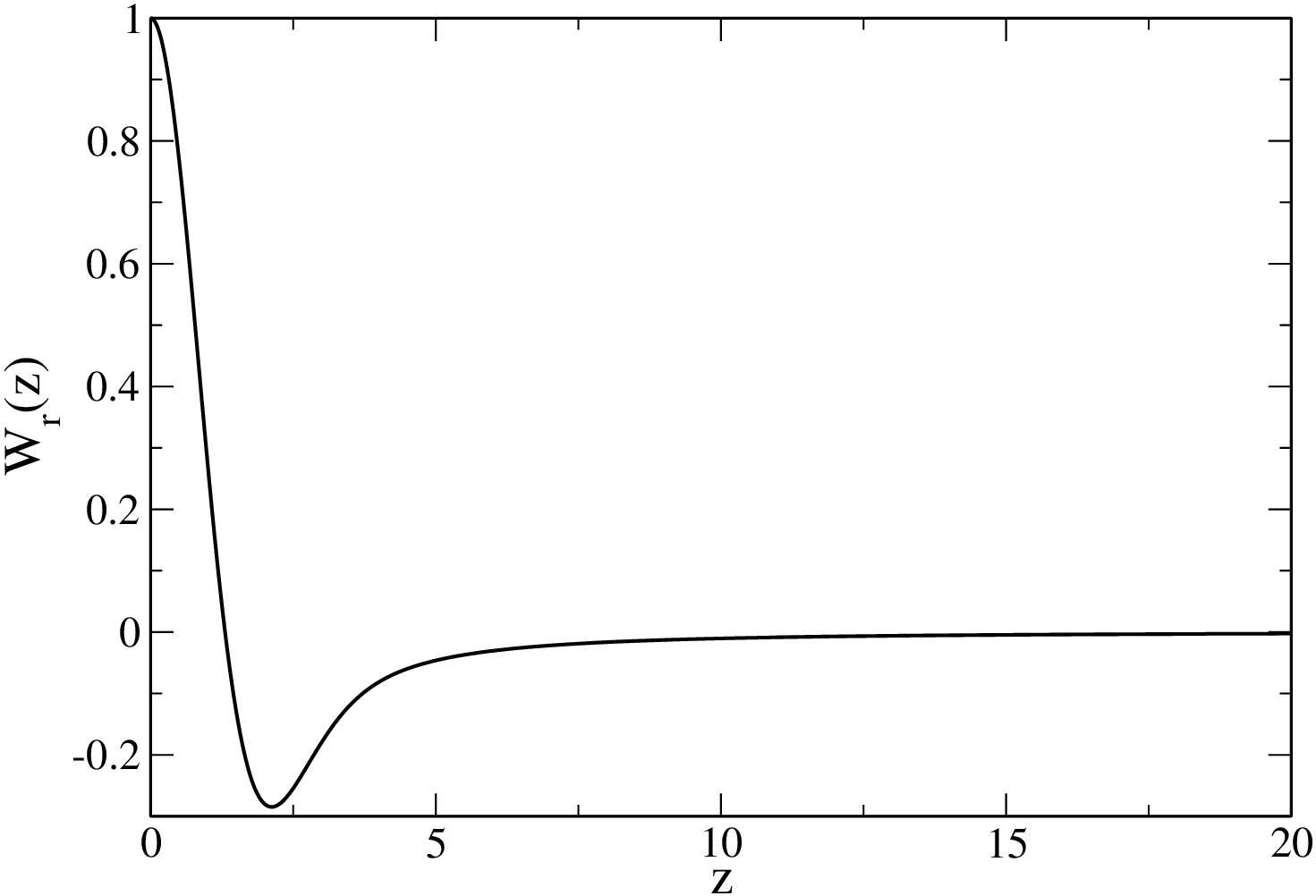}
\caption{The function $W_{r}(z)$ for $z$ real.}
\label{wer}
\end{center}
\end{figure}

Let us study the function
\begin{equation}
\label{wr1}
W_{r}(z)=1-z e^{-\frac{z^2}{2}}\int_{0}^{z}e^{\frac{x^2}{2}}\, dx,
\end{equation}
for $z$ real. We clearly have
\begin{equation}
\label{wr2}
W_{r}(-z)=W_{r}(z),
\end{equation}
so we can restrict ourselves to $z\ge 0$. This function is plotted in Fig. \ref{wer}. {It starts from $1$ for $z=0$, crosses the $x$-axis for $z=z_c=1.307$, achieves a minimum value $-0.2847$ for $z=2.124$ and tends to $0$ for $z\rightarrow +\infty$}. In order to determine its asymptotic behavior, let us consider the integral
\begin{equation}
\label{wr3}
I=\int_{0}^{z}e^{\frac{x^2}{2}}\, dx.
\end{equation}
It can be rewritten
\begin{equation}
\label{wr4}
I=e^{\frac{z^2}{2}} \int_{0}^{z}e^{-\frac{z^2-x^2}{2}}\, dx.
\end{equation}
With the change of variables $y=\sqrt{z^{2}-x^{2}}$, we get
\begin{equation}
\label{wr5}
I=\frac{1}{z}e^{\frac{z^2}{2}} \int_{0}^{z}e^{-\frac{y^2}{2}} \frac{y}{\sqrt{1-\frac{y^{2}}{z^{2}}}}\, dy.
\end{equation}
For $z\rightarrow +\infty$, we obtain
\begin{equation}
\label{wr6}
I=\frac{1}{z}e^{\frac{z^2}{2}} \int_{0}^{z}dy \, e^{-\frac{y^2}{2}} y \left (1+\frac{y^{2}}{2z^{2}}+\frac{3y^{4}}{8z^{4}}+...\right ),
\end{equation}
so that finally
\begin{equation}
\label{wr7}
I=\frac{1}{z}e^{\frac{z^2}{2}}\left (1+\frac{1}{z^{2}}+\frac{3}{z^{4}}+... \right ).
\end{equation}
Therefore, we get
\begin{equation}
\label{wr8}
W_{r}(z)= -\frac{1}{z^{2}}-\frac{3}{z^{4}}-..., \qquad (z\rightarrow +\infty).
\end{equation}
On the other hand, for $z\rightarrow 0$, we have
\begin{equation}
\label{wr9}
W_{r}(z)=1-z^{2}-..., \qquad (z\rightarrow 0).
\end{equation}

\section{The Tsallis and Cauchy distributions}
\label{sec_tc}

The Tsallis distributions are usually written in the form
\cite{cstsallis}:
\begin{eqnarray}
\label{tc1}
f=C_q\left \lbrack 1-(q-1)\frac{v^2}{2\theta}\right\rbrack_{+}^{\frac{1}{q-1}},
\end{eqnarray}
where $C_q$ is a normalization constant (see below). The parameter
$\theta>0$ has the dimension of a temperature but it is {\it not} the
kinetic temperature $T=\langle v^2\rangle$. To arrive at the form (\ref{t1}),
we can either follow the steps described in \cite{cvb,cstsallis} or
directly proceed as follows. First, we introduce notations similar to
those used in astrophysics \cite{bt} and define the polytropic index $n$ through
the relation $1/(q-1)=n-1/2$. Equation (\ref{tc1}) can then be rewritten
\begin{eqnarray}
\label{tc2}
f=A_n \frac{\rho}{\sqrt{2\pi\theta}}\left \lbrack 1-\frac{1}{n-\frac{1}{2}}\frac{v^2}{2\theta}\right \rbrack_{+}^{n-1/2}.
\end{eqnarray}
This distribution is normalizable provided that $n>1/2$ or $n<0$. Using $\rho=\int f\, dv$, we get
\begin{eqnarray}
\label{tc3}
A_n&=&\frac{\Gamma(n+1)}{\Gamma(n+1/2)(n-1/2)^{1/2}}, \qquad n>{1\over 2},\\
A_n&=&\frac{\Gamma(1/2-n)}{\Gamma(-n)[-(n-1/2)]^{1/2}}, \qquad n<0.
\end{eqnarray}
According to criterion (\ref{sh2}), the Tsallis distributions  (\ref{tc2}) are linearly stable with respect to the Vlasov equation iff
\begin{eqnarray}
\label{tc4}
\theta\ge\theta_{c}\equiv \frac{n}{n-1/2}\frac{k M}{4\pi}.
\end{eqnarray}
On the other hand, the kinetic temperature (variance) $T=\langle v^2\rangle$ exists if $n>1/2$ or $n<-1$. In that case, we have
\begin{eqnarray}
\label{tc5}
T=\frac{n-1/2}{n+1}\theta.
\end{eqnarray}
Substituting this relation in Eq. (\ref{tc2}) yields expression (\ref{t1}) of the distribution function where $T$ has a clear physical meaning as
being the kinetic temperature (or the velocity
dispersion).

\begin{figure}
\begin{center}
\includegraphics[clip,scale=0.3]{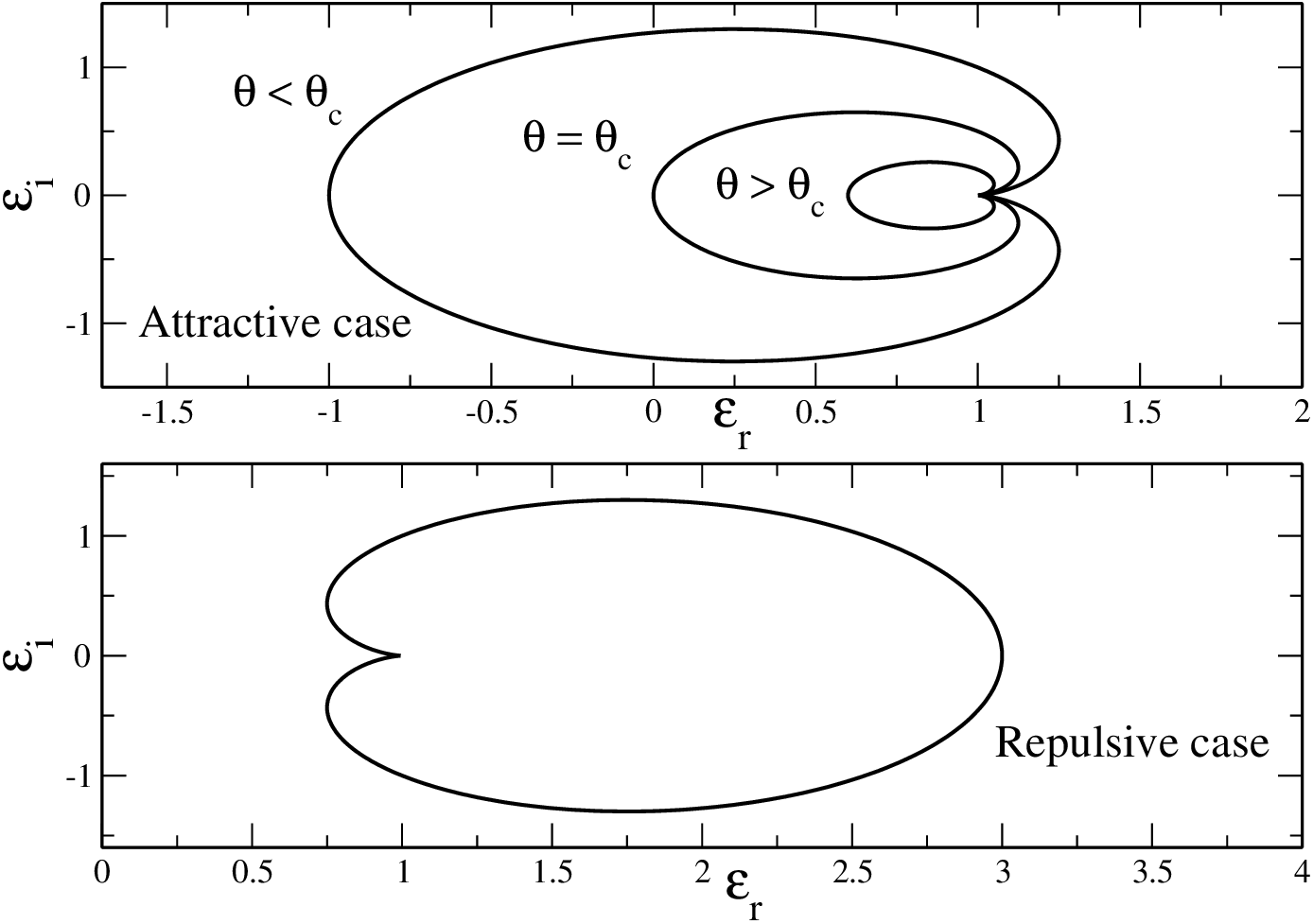}
\caption{Nyquist curve (analytical) for the Cauchy distribution  for the attractive and repulsive HMF models.}
\label{cauchy}
\end{center}
\end{figure}

The Tsallis distributions with index $-1<n<0$ have not been considered
in Sec. \ref{sec_tsallis} because they have infinite kinetic energy
$K=N\langle v^2\rangle/2$, so they are not physical. Yet, they can be
studied at a mathematical level. For these distributions, the
stability criterion is given by Eq. (\ref{tc4}) and the Nyquist curves
are similar to those reported in Sec. \ref{sec_tsallis} for $n<-1$. Of
particular interest is the Cauchy distribution corresponding to
$n=-1/2$. The Cauchy distribution is
\begin{eqnarray}
\label{tc6}
f=\frac{1}{\pi}\frac{\rho}{\sqrt{2\theta}}\frac{1}{1+\frac{v^2}{2\theta}},
\end{eqnarray}
and the corresponding  dielectric function (\ref{dr2}) can be written
\begin{eqnarray}
\label{tc7}
\epsilon(\omega)=1-\frac{kM}{4\pi^2\theta}\int_{C}\frac{x}{(1+x^2)^2(x-\frac{\omega}{\sqrt{2\theta}})}\, dx.
\end{eqnarray}
The integral can be computed analytically by completing the domain  of integration  by a large semi-circle in the lower-half plane of the complex plane and using the Cauchy residue theorem for a function with a pole of order $2$ at $x=-i$. This yields
\begin{eqnarray}
\label{tc8}
\epsilon(\omega)=1+\frac{kM}{8\pi\theta}\frac{1}{\left (i+\frac{\omega}{\sqrt{2\theta}}\right )^{2}}.
\end{eqnarray}
The condition $\epsilon(\omega)=0$ gives the complex pulsations
\begin{eqnarray}
\label{tc9}
\omega_{\pm}=-i\sqrt{2\theta}\left (1\pm \sqrt{\frac{kM}{8\pi\theta}}\right ).
\end{eqnarray}
For the attractive HMF model (see Sec. \ref{sec_dyn}), the pulsations
are purely imaginary. The mode $\omega_{+}$ is always damped
(stable). The mode $\omega_{-}$ is damped (stable) if
$\theta>\theta_{c}$ while it grows exponentially with time (unstable)
if $\theta<\theta_{c}$ with
\begin{eqnarray}
\label{tc10}
\theta_{c}=\frac{kM}{8\pi},
\end{eqnarray}
in agreement with Eq. (\ref{tc4}). On the other hand, if we take $\omega_i=0$, we find that
\begin{eqnarray}
\label{tc11}
\epsilon_{r}(\omega_{r})=1-\frac{kM}{8\pi\theta}\frac{1-\frac{\omega_{r}^{2}}{2\theta}}{\left (1+\frac{\omega_{r}^{2}}{2\theta}\right )^{2}},
\end{eqnarray}
\begin{eqnarray}
\label{tc12}
\epsilon_{i}(\omega_{r})=-\frac{kM}{4\pi\theta}\frac{\frac{\omega_{r}}{\sqrt{2\theta}}}{\left (1+\frac{\omega_{r}^{2}}{2\theta}\right )^{2}}.
\end{eqnarray}
The corresponding Nyquist curve is represented in Fig. \ref{cauchy}. It is
very similar to the one obtained for other
values of $n<0$ but, for the Cauchy distribution, it is obtained analytically.

The case of the repulsive HMF model (see Sec. \ref{sec_rep}) is obtained by
changing $+k$ in $-k$ in the preceding relations. The condition
$\epsilon(\omega)=0$ gives the complex pulsations
\begin{eqnarray}
\label{tc13}
\omega_{\pm}=\pm \sqrt{\frac{kM}{4\pi}}-i\sqrt{2\theta}.
\end{eqnarray}
These modes are always stable. They evolve with the (proper) pulsation
$\omega_{p}=\pm \sqrt{kM/(4\pi)}$ (see Appendix \ref{sec_landau})
independent on the temperature and they are damped at a rate
$\gamma=\sqrt{2\theta}$. For $\omega_{i}=0$, the real and imaginary
parts of the dielectric function are given by Eqs. (\ref{tc11}) and
(\ref{tc12}) with $k$ replaced by $-k$. The corresponding Nyquist
curve is represented in Fig. \ref{cauchy}.

\section{The function  $W_{r}^{(n)}(z)$}
\label{sec_wrn}

\begin{figure}
\begin{center}
\includegraphics[clip,scale=0.3]{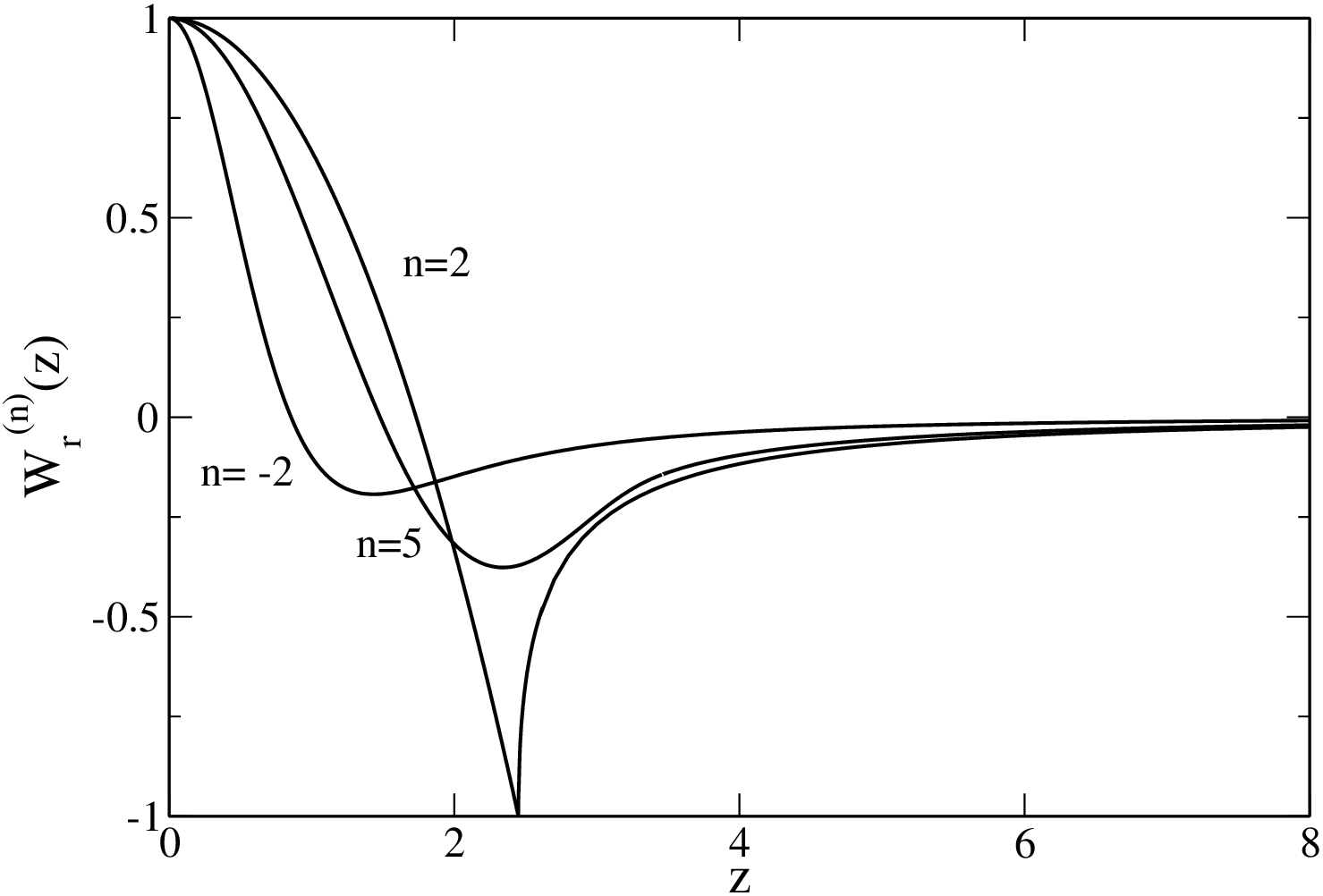}
\caption{The function $W_{r}^{(n)}(z)$ for $n<-1$ and $n>3/2$. }
\label{wts1}
\end{center}
\end{figure}

\begin{figure}
\begin{center}
\includegraphics[clip,scale=0.3]{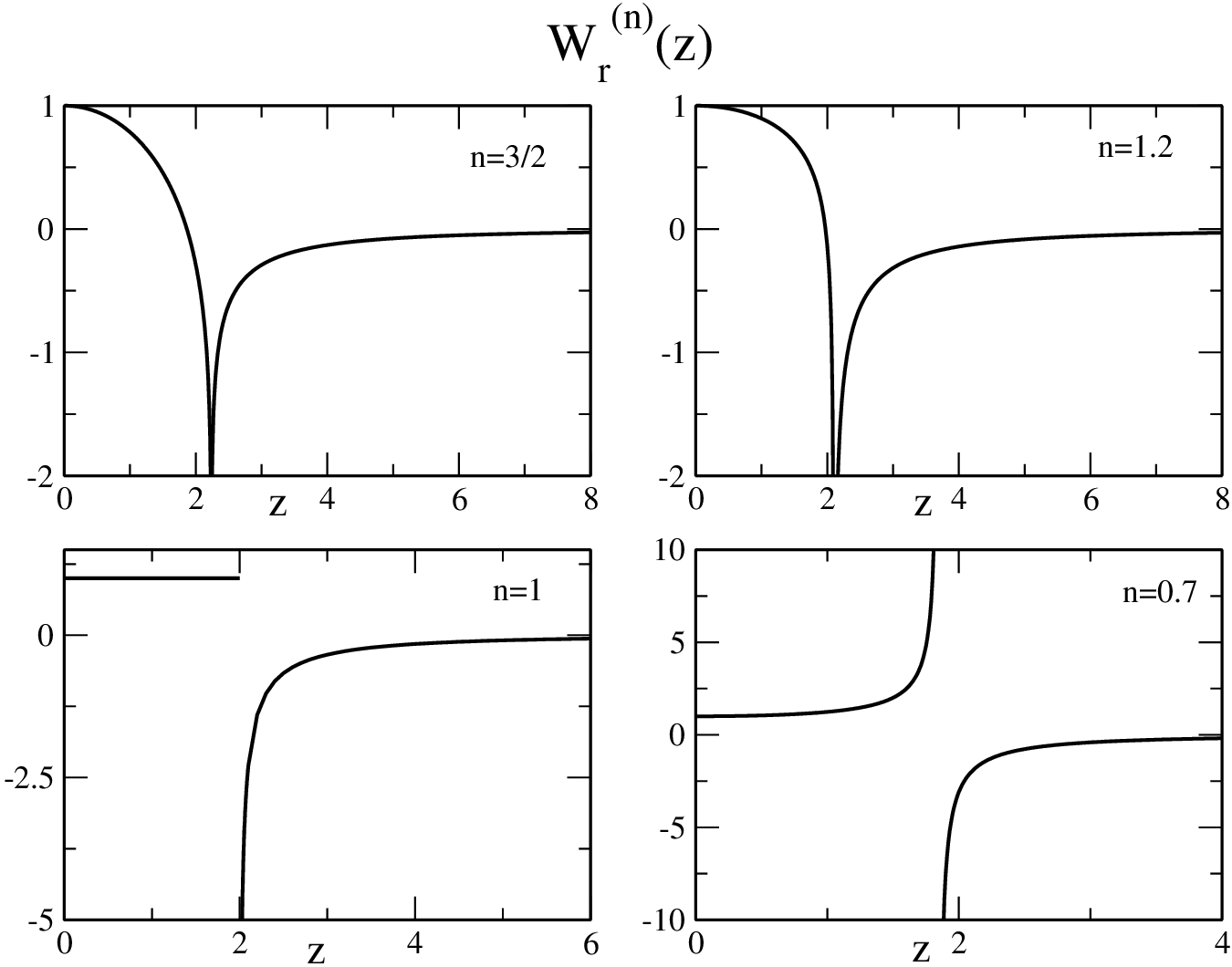}
\caption{The function $W_{r}^{(n)}(z)$ for $1/2<n<3/2$. }
\label{wts2}
\end{center}
\end{figure}

\begin{figure}
\begin{center}
\includegraphics[clip,scale=0.3]{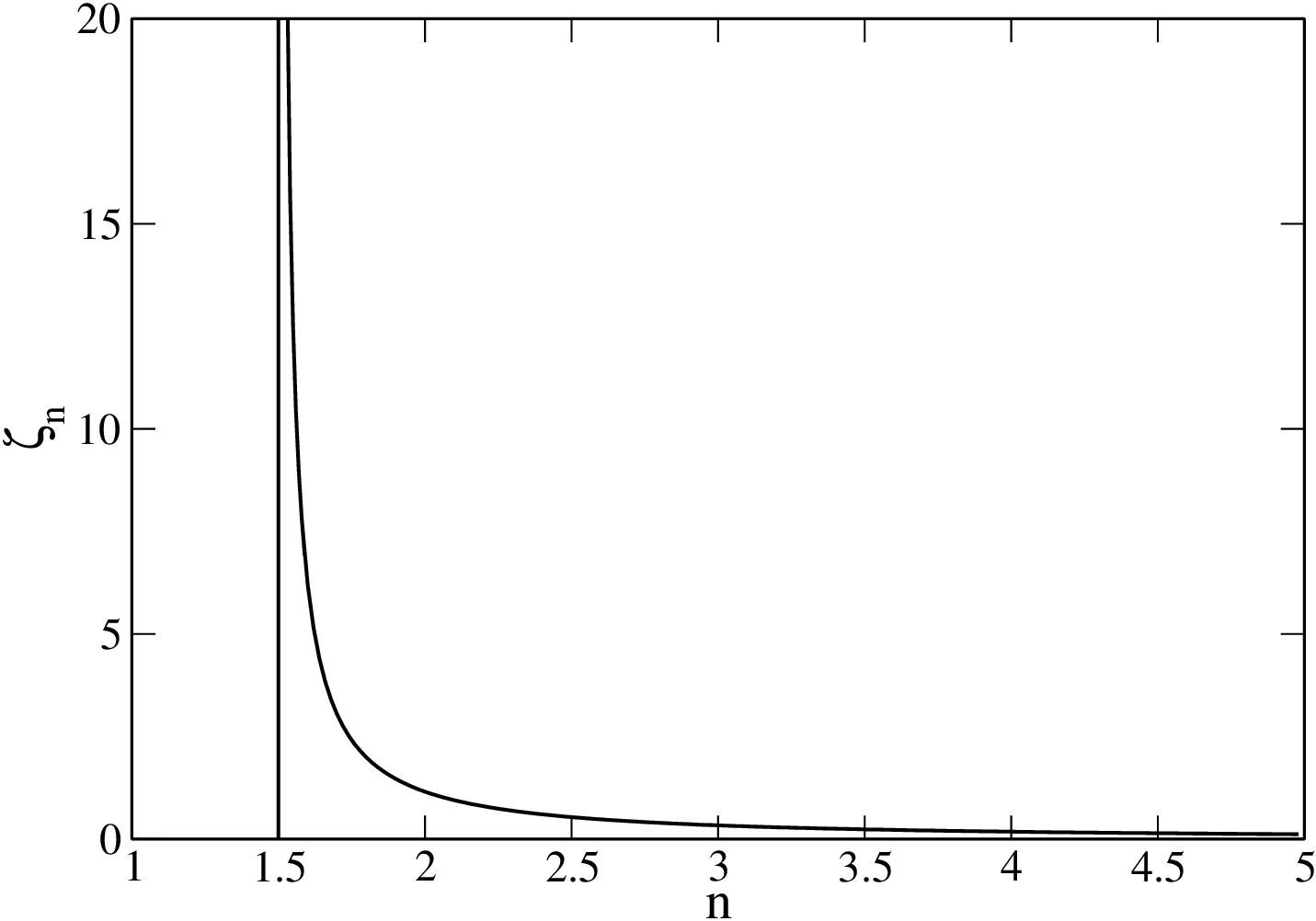}
\caption{The parameter $\zeta_{n}= -W_{r}^{(n)}(\sqrt{2(n+1)})$ as a function of the polytropic index $n> 3/2$.}
\label{zetan}
\end{center}
\end{figure}

Let us consider the function
\begin{eqnarray}
\label{n1}
W_{r}^{(n)}(z)=\frac{1}{\sqrt{2\pi}}\frac{B_n}{n}\left (n-\frac{1}{2}\right )P\int_{-\infty}^{+\infty}\frac{x[1-\frac{x^2}{2(n+1)}]_{+}^{n-3/2}}{x-z}dx,\nonumber\\
\end{eqnarray}
for $z$ real and $n<-1$ or $n>1/2$. We clearly have $W_{r}^{(n)}(-z)=W_{r}^{(n)}(z)$. For $n<-1$ and $n>5/2$, this function is  defined for any $z$ and its derivative is continuous. For $3/2<n\le 5/2$, this function is  defined for any $z$ but its derivative is discontinuous at $z= \sqrt{2(n+1)}$. For $1/2<n\le 3/2$, this function diverges at $z=\sqrt{2(n+1)}$. This leads to the behaviors represented in Figs.  \ref{wts1} and \ref{wts2}. For $n>3/2$, we define $\zeta_n\equiv -W_{r}^{(n)}(\sqrt{2(n+1)})$ or, explicitly,
\begin{eqnarray}
\label{n2}
\zeta_n=\frac{1}{\sqrt{2\pi}}\frac{B_n}{n}\left (n-\frac{1}{2}\right )\int_{-\sqrt{2(n+1)}}^{\sqrt{2(n+1)}}\frac{x[1-\frac{x^2}{2(n+1)}]^{n-3/2}}{\sqrt{2(n+1)}-x}dx.\nonumber\\
\end{eqnarray}
Note that the principal value is not needed in that case. Note also that $\zeta_n\ge 0$. This inequality can be obtained  from Eq. (\ref{gp6}) by taking $v_{ext}=\sqrt{2(n+1)}$ and recalling that  $f(v_{ext})=0$. Setting $y=x/\sqrt{2(n+1)}$, Eq. (\ref{n2}) can be rewritten
\begin{eqnarray}
\label{n3}
\zeta_n=\frac{1}{\sqrt{2\pi}}\frac{B_n}{n}\left (n-\frac{1}{2}\right )\sqrt{2(n+1)}I_{n},
\end{eqnarray}
with
\begin{eqnarray}
\label{n4}
I_{n}=\int_{-1}^{+1} y(1-y)^{n-5/2}(1+y)^{n-3/2}\, dy.
\end{eqnarray}
It turns out that this integral can be computed analytically yielding
\begin{eqnarray}
\label{n5}
I_{n}=\frac{1}{4}\Gamma(n-3/2)\left\lbrack\frac{4^{n}\Gamma(n+1/2)}
{\Gamma(2n-1)}-\frac{4\sqrt{\pi}}{\Gamma(n-1)}\right \rbrack.\quad
\end{eqnarray}
For $n\rightarrow 3/2^{+}$, we have the equivalent $\zeta_{n}\sim
1/[2(n-3/2)]$. For $n\rightarrow +\infty$, $\zeta_n\rightarrow 0$.
This parameter is plotted as a function of $n$ in Fig. \ref{zetan}.

Finally, for the index $n=3/2$, the functions (\ref{t10})-(\ref{t11}) can be calculated analytically. For any real $z\neq \pm \sqrt{5}$, we find
\begin{eqnarray}
\label{n6}
W_{r}^{(3/2)}(z)=1+\frac{z}{2\sqrt{5}}\ln\left |\frac{\sqrt{5}-z}{\sqrt{5}+z}\right |.
\end{eqnarray}
On the other hand, for $z\in [-\sqrt{5},\sqrt{5}]$, we have
\begin{eqnarray}
\label{n6b}
W_{i}^{(3/2)}(z)=\frac{\pi}{2\sqrt{5}}z,
\end{eqnarray}
while $W_{i}^{(3/2)}(z)=0$ otherwise. Using these results, the Nyquist curve in Fig. \ref{tsallis32} can be obtained analytically.

Finally, for the index $n=3/2$ and for $z=i\gamma$ where $\gamma$ is a
real number, the function (\ref{t7}) can be calculated analytically.
 For $\gamma>0$, using Eq. (\ref{dr3}), we get
\begin{eqnarray}
\label{n6c}
W^{(3/2)}(i\gamma)=1-\frac{\gamma}{\sqrt{5}}\arctan \left (\frac{\sqrt{5}}{\gamma}\right ). 
\end{eqnarray}
For  $\gamma<0$, using Eq. (\ref{dr5}), we get
\begin{eqnarray}
\label{n6d}
W^{(3/2)}(i\gamma)=1-\frac{\gamma}{\sqrt{5}}\left\lbrack \arctan \left (\frac{\sqrt{5}}{\gamma}\right )+\pi\right\rbrack. 
\end{eqnarray}
Finally, for $\gamma=0$, using Eq. (\ref{dr4}), we have $W^{(3/2)}(0)=0$. Therefore, looking for a solution of the equation $\epsilon(\Omega)\equiv 1-(\eta/\eta_{c}) W^{(3/2)}(\sqrt{\eta}\Omega)=0$ of the form $\Omega=i\Omega_{i}$, we find that for $\eta>\eta_{c}=5/3$ (unstable), the perturbation grows with a rate $\Omega_{i}>0$ given by
\begin{eqnarray}
\label{n6e}
1-\frac{\eta}{\eta_c}\left\lbrack 1-\frac{\sqrt{\eta}\Omega_{i}}{\sqrt{5}}\arctan \left (\frac{\sqrt{5}}{\sqrt{\eta}\Omega_i}\right )\right\rbrack=0, 
\end{eqnarray}
while for  $\eta<\eta_{c}=5/3$ (stable), the perturbation is damped with a rate $\Omega_{i}<0$ given by
\begin{eqnarray}
\label{n6f}
1-\frac{\eta}{\eta_c}\left\lbrack 1-\frac{\sqrt{\eta}\Omega_{i}}{\sqrt{5}}\left\lbrace \arctan \left (\frac{\sqrt{5}}{\sqrt{\eta}\Omega_i}\right )+\pi\right\rbrace\right\rbrack=0. \qquad
\end{eqnarray}
Finally, for  $\eta=\eta_{c}=5/3$, we recover the case of marginal stability $\Omega_i=0$. For $\eta\rightarrow +\infty$, we recover Eq. (\ref{landau28}) and for $\eta\rightarrow 0$, we find $\Omega_{i}\sim -5\sqrt{5}/(3\pi\eta^{3/2})$.

\section{Analogies and differences between the HMF model, Coulombian  plasmas and self-gravitating systems}
\label{sec_ana}

For Coulombian plasmas, the potential of interaction is determined by the Poisson equation $\Delta u=-4\pi(e^2/m^2)\delta({\bf x})$ yielding
\begin{eqnarray}
\label{ana1}
(2\pi)^{3}\hat{u}(k)=\frac{4\pi e^{2}}{m^{2}k^{2}}>0.
\end{eqnarray}
We introduce the plasma frequency
\begin{eqnarray}
\label{ana2}
\omega_{p}^{2}\equiv \frac{4\pi \rho e^{2}}{m^{2}},
\end{eqnarray}
and the Debye wavenumber
\begin{eqnarray}
\label{ana3}
k_{D}^{2}\equiv \frac{4\pi e^2\beta\rho}{m}=\beta m \omega_{p}^{2}.
\end{eqnarray}
The dispersion relation (\ref{vel5}) can
be written
\begin{eqnarray}
\label{ana4}
\epsilon(k,\omega)=1-\frac{4\pi e^2}{m^2 k^{2}}\int_{C}\frac{f'(v)}{v-\frac{\omega}{k}}\, dv.
\end{eqnarray}
Comparing this expression with Eq. (\ref{rgr2}), we make the link between Coulombian plasmas and the repulsive HMF model by setting
\begin{eqnarray}
\label{ana5}
\frac{4\pi e^{2}}{m^{2}k^{2}}=\frac{\tilde k}{2}, \quad \beta m=\tilde\beta, \quad \frac{\omega}{k}=\tilde\omega, \quad \frac{\omega_{p}}{k}=\tilde\omega_{p},
\end{eqnarray}
where the tilde variables refer to the HMF model and the un-tilde variables to the plasmas. Then, we find that the dimensionless parameters for the plasmas are
\begin{eqnarray}
\label{ana6}
\eta=\frac{k_{D}^{2}}{k^{2}},\quad \Omega=\frac{\omega}{\omega_{p}}, \quad a=\frac{1}{\sqrt{\eta}}\sqrt{\beta m} \ v_{a}.
\end{eqnarray}
The results obtained in Sec. \ref{sec_rep} for the repulsive HMF model
can be applied to Coulombian plasmas provided that we interpret $\eta$
as a normalized squared wavelength and $\Omega$ as a normalized
pulsation according to Eq. (\ref{ana6}). In particular, single-humped
distributions are stable to all wavelengths \cite{nicholson} and the
Nyquist curves for Maxwell and Tsallis distributions are given in
Sec. \ref{sec_shr}. By contrast, the phase diagrams of the double-humped
distribution (in Sec. \ref{sec_dhhr}) are not directly applicable to plasmas
because, according to Eq. (\ref{ana6}), the dimensionless parameter
$a\propto 1/\sqrt{\eta}$ depends on $k$. Therefore, this is not the
correct dimensionless separation. For plasmas, the correct
dimensionless separation (which must be independent on the wavenumber
$k$) is $y=\eta a^2=\beta m v_{a}^{2}$ instead of $a$. Therefore, the
correct stability diagram corresponds to the curve(s) $\eta_{c}(y)$
obtained from Eqs. (\ref{rsh2})-(\ref{rsh3}). This will be discussed
in more detail in a specific paper.

For self-gravitating  systems, the potential of interaction is determined by $\Delta u=4\pi G\delta({\bf x})$ yielding
\begin{eqnarray}
\label{ana7}
(2\pi)^{3}\hat{u}(k)=-\frac{4\pi G}{k^{2}}<0.
\end{eqnarray}
We introduce the  Jeans wavenumber
\begin{eqnarray}
\label{ana9}
k_{J}^{2}=4\pi G\beta\rho m.
\end{eqnarray}
The dispersion relation
(\ref{vel5}) can be written
\begin{eqnarray}
\label{ana10}
\epsilon(k,\omega)=1+\frac{4\pi G}{k^{2}}\int_{C}\frac{f'(v)}{v-\frac{\omega}{k}}\, dv.
\end{eqnarray}
Comparing this expression with Eq. (\ref{dr2}), we make the link with
the attractive HMF model by setting
\begin{eqnarray}
\label{ana11}
\frac{4\pi G}{k^{2}}=\frac{\tilde k}{2}, \quad \beta m=\tilde\beta, \quad \frac{\omega}{k}=\tilde\omega,
\end{eqnarray}
where the tilde variables refer the HMF model and the un-tilde
variables to self-gravitating systems. Then, we find that the
dimensionless parameters for self-gravitating systems are
\begin{eqnarray}
\label{ana12}
\eta=\frac{k_{J}^{2}}{k^{2}},\quad \Omega=\frac{\omega}{\sqrt{4\pi G\rho}}, \quad a=\frac{1}{\sqrt{\eta}}\sqrt{\beta m} \ v_{a}.
\end{eqnarray}
The results obtained in Sec. \ref{sec_dyn} for the attractive HMF
model can be applied to (uniform) self-gravitating systems provided
that we interpret $\eta$ as a normalized squared wavelength and
$\Omega$ as a normalized pulsation according to Eq. (\ref{ana12}). In
particular, the Maxwell distribution is stable for $k>k_{J}$
(i.e. $\eta<1$) and unstable for $k<k_{J}$ (i.e. $\eta>1$) and the
Tsallis distributions are stable for $k>(\frac{n}{n+1})^{1/2}k_{J}$
(i.e. $\eta<\eta_c$) and unstable for $k<(\frac{n}{n+1})^{1/2}k_{J}$
(i.e. $\eta>\eta_c$).  The Nyquist curves for the Maxwell and Tsallis
distributions are given in Secs. \ref{sec_maxwell} and
\ref{sec_tsallis}. By contrast, the phase diagrams of the
double-humped distribution (in Secs. \ref{sec_sh} and \ref{sec_doubh})
are not directly applicable to self-gravitating systems because,
according to Eq. (\ref{ana12}), the dimensionless parameter $a\propto
1/\sqrt{\eta}$ depends on $k$. Therefore, this is not the correct
dimensionless separation. For self-gravitating systems, the correct
dimensionless separation (which must be independent on the wavenumber
$k$) is $y=\eta a^2=\beta m v_{a}^{2}$ instead of $a$. Therefore, the
correct stability diagram corresponds to the curve(s) $\eta_{c}(y)$
obtained from Eqs. (\ref{ams8})-(\ref{ams9}). This will be discussed
in more detail in a specific paper.

\section{Landau damping and Landau growth for the repulsive and attractive HMF models}
\label{sec_landau}

Like for Coulombian plasmas, the repulsive HMF model leads to Landau
damping at low temperatures. On the other hand, the attractive HMF
model can lead to ``Landau growth'' associated to the phase transition
(collapse) at low temperatures.

\subsection{The limit $T\rightarrow 0$}
\label{sec_landaulow}

Let us first consider the limit $T\rightarrow 0$ for the repulsive HMF
model. The case $T=0$ corresponds to $f=\rho\ \delta(v)$. In that
case, we obtain $\epsilon(\omega)=1-\frac{k\rho}{2\omega^{2}}$. The
condition $\epsilon(\omega)=0$ yields $\omega=\pm
(k\rho/2)^{1/2}=\pm\omega_p$ where $\omega_p$ is the proper pulsation
\begin{eqnarray}
\omega_{p}^{2}=\frac{k\rho}{2}=\frac{kM}{4\pi}.
\label{landau9}
\end{eqnarray}
This is the equivalent of the plasma pulsation in plasma
physics. Therefore, for $T=0$ the system is stable and the
perturbation oscillates with the pulsation $\omega_p$ without
attenuation. To determine the damping rate $\gamma=-\omega_i$ for
$T\ll 1$, we shall solve the equation (\ref{rgr2}) for the complex
pulsation $\omega=\omega_{r}+i\omega_{i}$ by assuming $\omega_{i}\ll
\omega_{r}$ (we adapt the method of plasma physics given in \cite{nicholson}). Expanding the dielectric function in Taylor series as
$\epsilon(\omega)=\epsilon(\omega_{r}+i\omega_{i})\simeq
\epsilon(\omega_r)+i\epsilon'(\omega_r)\omega_i$, the condition $\epsilon(\omega)=0$  gives to leading order
\begin{eqnarray}
\epsilon_{r}(\omega_{r})=0,
\label{landau1}
\end{eqnarray}
\begin{eqnarray}
\epsilon_{i}(\omega_{r})+\epsilon_{r}'(\omega_{r})\omega_{i}=0,
\label{landau2}
\end{eqnarray}
where $\epsilon_{r}(\omega_{r})$ and $\epsilon_{i}(\omega_{r})$ are
defined in Eqs. (\ref{rgr3}) and (\ref{rgr4}). Integrating by parts, Eq. (\ref{rgr3}) can be rewritten
\begin{eqnarray}
\epsilon_{r}(\omega_{r})=1-\frac{k}{2}P\int_{-\infty}^{+\infty}\frac{f(v)}{(v-\omega_{r})^{2}}\, dv.
\label{landau3}
\end{eqnarray}
For $v/\omega_{r}\ll 1$, we get
\begin{eqnarray}
\epsilon_{r}(\omega_{r})=1-\frac{k\rho}{2\omega_{r}^{2}}\left (1+\frac{2\langle v\rangle}{\omega_{r}}+\frac{3\langle v^2\rangle}{\omega_{r}^{2}}+...\right ).
\label{landau4}
\end{eqnarray}
If the distribution is symmetrical with respect to the origin and if the variance is finite, we obtain
\begin{eqnarray}
\epsilon_{r}(\omega_{r})=1-\frac{k\rho}{2\omega_{r}^{2}}\left (1+\frac{3\langle v^2\rangle}{\omega_{r}^{2}}+...\right ),
\label{landau5}
\end{eqnarray}
and
\begin{eqnarray}
\epsilon_{r}'(\omega_{r})=\frac{k\rho}{\omega_{r}^{3}}.
\label{landau6}
\end{eqnarray}
Substituting these relations in Eqs. (\ref{landau1})-(\ref{landau2}), we find that
\begin{eqnarray}
1-\frac{k\rho}{2\omega_{r}^{2}}\left (1+\frac{3\langle v^2\rangle}{\omega_{r}^{2}}\right )=0,
\label{landau7}
\end{eqnarray}
\begin{eqnarray}
-\pi \frac{k}{2}f'(\omega_r)+\frac{k\rho}{\omega_{r}^{3}}\omega_i=0.
\label{landau8}
\end{eqnarray}
This expansion is valid for $T=\langle v^2\rangle\rightarrow 0$. For
$T=0$, Eq. (\ref{landau7}) gives $\omega_{r}=\omega_{p}$.  Then, to next order, Eq. (\ref{landau7}) gives
\begin{eqnarray}
\omega_{r}^{2}=\omega_{p}^{2}+3T+...\qquad (T\rightarrow 0).
\label{landau10}
\end{eqnarray}
This asymptotic expansion for $T\rightarrow 0$ also gives the exact
pulsation (valid at any temperature) for the water-bag distribution
(see Sec. \ref{sec_shr}). Using the dimensionless variables introduced in
Sec. \ref{sec_maxwell}, Eq. (\ref{landau10}) can be rewritten \footnote{Let us recall that this solution only exists for the repulsive HMF model. For
the attractive HMF model, there is collapse when $T\rightarrow 0$ (see below).}
\begin{eqnarray}
\Omega_{r}^{2}=1+\frac{3}{\eta}+...\qquad (\eta\rightarrow +\infty).
\label{landau11}
\end{eqnarray}
On the other hand, to leading order when $T\rightarrow 0$ (or $\eta\rightarrow +\infty$), Eq. (\ref{landau8}) gives the damping rate \footnote{A more accurate expression is obtained by replacing $f'(\omega_{p})$ by $f'(\omega_{r})$.}
\begin{eqnarray}
\omega_i=\frac{\pi\omega_{p}^{3}}{2\rho}f'(\omega_{p}).
\label{landau12}
\end{eqnarray}
When $f'(\omega_{p})<0$, we obtain the expression of the Landau
damping for the repulsive HMF model. We stress that the relation
(\ref{landau10}) is the same for {\it any} distribution. By contrast,
the damping coefficient (\ref{landau12}) strongly depends on the form
of the distribution.  For the Maxwellian distribution (\ref{mm1}), we
find that
\begin{eqnarray}
\Omega_{i}=-\sqrt{\frac{\pi}{8}}\eta^{3/2}e^{-\eta/2}.
\label{landau13}
\end{eqnarray}
For the asymmetric double-humped distribution (\ref{ae1}), we get
\begin{eqnarray}
\Omega_{i}=-\frac{1}{1+\Delta}\sqrt{\frac{\pi}{8}}\eta^{3/2}\biggl\lbrack (1-a)e^{-\frac{\eta}{2}(1-a)^{2}}\nonumber\\
+\Delta (1+a)e^{-\frac{\eta}{2}(1+a)^{2}}\biggr\rbrack.
\label{landau14}
\end{eqnarray}
For the Tsallis distributions (\ref{t1}), we obtain
\begin{eqnarray}
\Omega_{i}=-\sqrt{\frac{\pi}{8}}B_{n}\eta^{3/2}\left (n-\frac{1}{2}\right )\frac{1}{n+1}\left\lbrack 1-\frac{\eta}{2(n+1)}\right \rbrack^{n-3/2}_{+}.\nonumber\\
\label{landau15}
\end{eqnarray}
For $n\rightarrow \pm\infty$, we recover Eq. (\ref{landau13}). For
$n>1/2$, the damping is zero if $\omega_{p}>v_{m}$,
i.e. $\eta>2(n+1)$.  For fixed $n$ in the range $]-\infty,-1[$ and
$\eta\rightarrow +\infty$, Eq. (\ref{landau15}) becomes
\begin{eqnarray}
\Omega_{i}=\sqrt{\frac{\pi}{8}}B_{n}\frac{1}{2^{n-\frac{3}{2}}}\left (n-\frac{1}{2}\right )\frac{1}{\lbrack -(n+1)\rbrack^{n-\frac{1}{2}}}\eta^{n},
\label{landau16}
\end{eqnarray}
so the damping rate behaves like $\Omega_{i}\propto -1/\eta^{|n|}$ for
$\eta\rightarrow +\infty$. Let us consider for illustration the index
$n=-3/2$ corresponding to the distribution
\begin{eqnarray}
f=\frac{2\rho}{\pi\sqrt{T}}\frac{1}{\left (1+\frac{v^{2}}{T}\right )^{2}}.
\label{landau17}
\end{eqnarray}
In that case, we have
\begin{eqnarray}
\Omega_{i}=-\frac{4}{\eta^{3/2}}.
\label{landau18}
\end{eqnarray}

Using the correspondance between Coulombian plasmas and the repulsive
HMF model (see Appendix \ref{sec_ana}), we find that the equivalent of Eqs. (\ref{landau10}) and  (\ref{landau12}) are
\begin{eqnarray}
\omega_{r}^2=\omega_{p}^2+3Tk^2, \qquad \omega_{i}=\frac{\pi\omega_{p}^{3}}{2\rho k^2}f'\left (\frac{\omega_{p}}{k}\right ).
\label{landau20}
\end{eqnarray}
These expressions are valid for $k/k_{D}\rightarrow 0$. For the Maxwellian distribution, we get
\begin{eqnarray}
\omega_{i}=-\sqrt{\frac{\pi}{8}}\omega_p \left (\frac{k_{D}}{k}\right )^{3}e^{-\frac{\beta m\omega_{p}^{2}}{2k^2}}.
\label{landau22}
\end{eqnarray}
For the Tsallis distributions, we get
\begin{eqnarray}
\omega_{i}=-\sqrt{\frac{\pi}{8}}B_{n}\omega_p \left (\frac{k_{D}}{k}\right )^{3}\nonumber\\
\times\left (n-\frac{1}{2}\right )\frac{1}{n+1}\left\lbrack 1-\frac{\beta m\omega_{p}^{2}}{2(n+1)k^2}\right \rbrack^{n-3/2}_{+}.
\label{landau23}
\end{eqnarray}
For $n\rightarrow \pm\infty$, we recover Eq. (\ref{landau22}). For $n>1/2$, the damping is zero if $\omega_{p}/k>v_{m}$, i.e. if $k/k_{J}<1/\sqrt{2(n+1)}$. For fixed $n$ in the interval $]-\infty,-1[$ and  $k/k_{D}\rightarrow 0$, we obtain
\begin{eqnarray}
\omega_{i}=\sqrt{\frac{\pi}{8}}B_{n}\omega_p\frac{1}{2^{n-\frac{3}{2}}}\left (n-\frac{1}{2}\right )\frac{1}{\lbrack -(n+1)\rbrack^{n-\frac{1}{2}}}\left (\frac{k_{D}}{k}\right )^{2n},\nonumber\\
\label{landau24}
\end{eqnarray}
so the damping behaves like $\omega_{i}\propto -k^{2|n|}$ for
$k/k_D\rightarrow 0$. For the distribution (\ref{landau17})
corresponding to $n=-3/2$, we have
\begin{eqnarray}
\omega_{i}=-\frac{4\omega_p}{k_{D}^{3}}k^3.
\label{landau25}
\end{eqnarray}

Let us now consider the limit $T\rightarrow 0$ for the attractive HMF
model. The case $T=0$ corresponds to $f=\rho\ \delta(v)$. In that
case, we obtain $\epsilon(\omega)=1+\frac{k\rho}{2\omega^{2}}=0$
yielding $\omega=\pm i (k\rho/2)^{1/2}$. Therefore, for $T=0$ the
system is unstable and the perturbation grows with a growth rate
$\omega_{i}=(kM/4\pi)^{1/2}$. Let us now consider the case $T\ll 1$.
We anticipate a solution with $\omega_{r}=0$ and $\omega_i>0$.  In that case,
the perturbation grows ($\omega_{i}>0$)
without oscillating. The growth rate is given by
Eq. (\ref{mon2}). Integrating Eq. (\ref{mon2}) by parts, we obtain
\begin{eqnarray}
1-\frac{k}{2}\int_{-\infty}^{+\infty}\frac{f(v)(\omega_{i}^{2}-v^2)}{(v^2+\omega_{i}^2)^{2}}\, dv=0.
\label{landau26}
\end{eqnarray}
Expanding the integrand in powers of $v/\omega_i\ll 1$ for $T\rightarrow 0$, we find that
\begin{eqnarray}
\omega_{i}^2=\frac{kM}{4\pi}-3T-... \qquad (T\rightarrow 0),
\label{landau27}
\end{eqnarray}
with $T=\langle v^2\rangle$. This asymptotic expansion for
$T\rightarrow 0$ also gives the exact pulsation (valid at any
temperature $T<T_c=kM/12\pi$) for the water-bag distribution (see
Sec. \ref{sec_tsallis}). Introducing dimensionless variables, we get
\begin{eqnarray}
\Omega_{i}^2=1-\frac{3}{\eta}-... \qquad (\eta\rightarrow +\infty).
\label{landau28}
\end{eqnarray}
For the Maxwellian distribution, we recover Eq. (136) of \cite{cvb}.

\subsection{The limit $T\rightarrow +\infty$}
\label{sec_landauhigh}

Let us consider the limit $T\rightarrow +\infty$ for the repulsive HMF
model (we adapt the method of plasma physics given in
\cite{balescubook}). We look for a solution of the equation
$\epsilon(\omega)=0$ of the form $\omega=\omega_r+i\omega_i$ with
$\omega_i<0$ (damping) and $\omega_i/\omega_r\gg 1$. Using
Eq. (\ref{dr5}) for $\omega_{i}<0$ , Eq. (\ref{rgr2}) can be written
\begin{eqnarray}
1-\frac{k}{2}\int_{-\infty}^{+\infty} {{f'(v)}\over v-{\omega}}dv-ik\pi f'(\omega)=0.
\label{gh1}
\end{eqnarray}
If $f$ decreases exponentially rapidly for large $v$, then for $\omega_i/\omega_r\gg 1$, the foregoing equation reduces to
\begin{eqnarray}
1-ik\pi f'(\omega)=0.
\label{gh2}
\end{eqnarray}
Separating real and imaginary parts, we obtain two transcendant equations
\begin{eqnarray}
{\rm Re}\left\lbrack ik\pi f'(\omega_{r}+i\omega_{i})\right\rbrack=1,
\label{gh3}
\end{eqnarray}
\begin{eqnarray}
{\rm Im}\left\lbrack i f'(\omega_{r}+i\omega_{i})\right\rbrack=0,
\label{gh4}
\end{eqnarray}
which strongly depend on the form of the distribution. For the Maxwellian (\ref{mm1}), they can be rewritten to leading order in the limit $\omega_i/\omega_r\gg 1$ as
\begin{eqnarray}
k\pi\left (\frac{\beta}{2\pi}\right )^{1/2}\rho\beta\omega_i e^{\frac{\beta}{2}\omega_{i}^{2}}\cos(\beta\omega_r\omega_i)=1,
\label{gh5}
\end{eqnarray}
\begin{eqnarray}
\sin(\beta\omega_r\omega_i)=0.
\label{gh6}
\end{eqnarray}
Equation (\ref{gh6}) implies $\beta \omega_r\omega_i=m\pi$. Eq.  (\ref{gh5}) will have a solution provided that $m$ is even. Let us take $\beta \omega_r\omega_i=\pi$. Then, Eq.  (\ref{gh5}) gives
\begin{eqnarray}
-k\pi\left (\frac{\beta}{2\pi}\right )^{1/2}\rho\beta\omega_i e^{\frac{\beta}{2}\omega_{i}^{2}}=1,
\label{gh7}
\end{eqnarray}
which determines $\omega_i$. By a graphical construction, it is easy to see that $|\omega_i|$ is an increasing function of $T$. For $T\rightarrow +\infty$, we find the asymptotic behavior
\begin{eqnarray}
\omega_i=-\sqrt{2T\ln T}, \qquad \omega_r=-\pi\sqrt{\frac{T}{2\ln T}}.
\label{gh8}
\end{eqnarray}
Since $\omega_{i}/\omega_{r}\sim \ln T\rightarrow +\infty$, our basic assumption is satisfied. We also note that $\omega_{r}<0$ in the repulsive case.
Introducing dimensionless variables, we have for $\eta\rightarrow 0$:
\begin{eqnarray}
\Omega_i=-\sqrt{\frac{2|\ln\eta|}{\eta}}, \qquad \Omega_r=\frac{-\pi}{\sqrt{2\eta |\ln\eta|}}.
\label{gh9}
\end{eqnarray}
Using the correspondance (\ref{ana6}) between the repulsive HMF model and Coulombian plasmas, we recover the results of plasma physics \cite{balescubook} for $k/k_D\rightarrow +\infty$:
\begin{eqnarray}
\omega_i=-\frac{2\omega_{p}}{k_{D}}k\sqrt{\ln k},  \qquad \omega_r=-\frac{\pi\omega_p}{2k_{D}}\frac{k}{\sqrt{\ln k}}.
\label{gh10}
\end{eqnarray}

Let us consider the limit $T\rightarrow +\infty$ for the attractive HMF
model. We again look for a solution of the equation
$\epsilon(\omega)=0$ of the form $\omega=\omega_r+i\omega_i$ with
$\omega_i<0$ (damping) and $\omega_i/\omega_r\gg 1$. Repeating the procedure
followed previously, we obtain two transcendant equations
\begin{eqnarray}
{\rm Re}\left\lbrack ik\pi f'(\omega_{r}+i\omega_{i})\right\rbrack=-1,
\label{gh11}
\end{eqnarray}
\begin{eqnarray}
{\rm Im}\left\lbrack i f'(\omega_{r}+i\omega_{i})\right\rbrack=0.
\label{gh12}
\end{eqnarray}
For the Maxwellian (\ref{mm1}), they can be rewritten to leading order in the limit $\omega_i/\omega_r\gg 1$ as
\begin{eqnarray}
k\pi\left (\frac{\beta}{2\pi}\right )^{1/2}\rho\beta\omega_i e^{\frac{\beta}{2}\omega_{i}^{2}}\cos(\beta\omega_r\omega_i)=-1,
\label{gh13}
\end{eqnarray}
\begin{eqnarray}
\sin(\beta\omega_r\omega_i)=0.
\label{gh14}
\end{eqnarray}
Equation (\ref{gh14}) implies $\beta \omega_r\omega_i=m\pi$. Eq.  (\ref{gh13}) will have a solution provided that $m$ is odd. Let us take $\omega_r=0$. Then, Eq.  (\ref{gh13}) reduces to Eq. (\ref{gh7}). Therefore, for $T\rightarrow +\infty$, we find the asymptotic behavior
\begin{eqnarray}
\omega_i=-\sqrt{2T\ln T}, \qquad \omega_r=0.
\label{gh8b}
\end{eqnarray}
This returns Eq. (141) of \cite{cvb}.

\section{Marginal stability of spatially inhomogeneous states}
\label{sec_inho}

For the HMF model, the eigenvalue equation associated with the
minimization problem (\ref{fe2}) can be written (see Eq. (77) of \cite{cvb})
\begin{eqnarray}
\frac{d}{d\theta}\left (\frac{p'(\rho)}{\rho}\frac{dq}{d\theta}\right )+\frac{k}{2\pi}\int_0^{2\pi}q(\theta')\cos(\theta-\theta')\, d\theta'=-2\lambda q,\nonumber\\
\label{inho1}
\end{eqnarray}
with $q(0)=q(2\pi)=0$, where
$q(\theta)=\int_0^{\theta}\delta\rho(\theta')d\theta'$ is the
perturbed integrated density \footnote{Note that the eigenvalue
equation associated with the linear stability problem (see Eq. (85) of
\cite{cvb}) has a similar form as Eq. (\ref{inho1}) and that they both 
coincide at the neutral point $\lambda=0$.}. This equation is valid
for homogeneous and inhomogenous equilibrium density profiles
$\rho(\theta)$. We have $\delta^{2}{\cal W}=\frac{1}{2}\int {\cal
L}[q]q\, d\theta$ where ${\cal L}$ is the operator defined by the
r.h.s. of Eq. (\ref{inho1}). Therefore, a critical point of energy
${\cal W}[\rho,{\bf u}]$ at fixed mass is a (local) minimum iff all
the eigenvalues $\lambda$ are positive. In that case, the density
profile $\rho(\theta)$ is formally nonlinearly stable with respect to
the Euler equation. On the other hand, since
$(\ref{fe2})\Leftrightarrow (\ref{fo16})
\Leftrightarrow (\ref{fo8})$, the corresponding distribution function
$f(\theta,v)$ (leading to an equation of state $p(\rho)$) is a (local)
minimum of pseudo free energy $F[f]$ at fixed mass, so that it is
formally nonlinearly stable with respect to the Vlasov
equation. Therefore, the positiveness of the eigenvalues of
Eq. (\ref{inho1}) is a sufficient condition of formal nonlinear stability
for the Vlasov equation. Unfortunately, it does not appear feasible to
solve Eq. (\ref{inho1}) analytically in the general case.  However, it
is possible to solve it analytically for the marginal case
$\lambda=0$. This criterion will give the point at which the series of
equilibria of the inhomogeneous equilibrium states becomes
unstable. This is the first result concerning the stability of {\it
inhomogeneous} states of the HMF model. The eigenvalue equation
(\ref{inho1}) can be rewritten
\begin{eqnarray}
\frac{d}{d\theta}\left (\frac{p'(\rho)}{\rho}\frac{dq}{d\theta}\right )-b_x \cos\theta-b_y\sin\theta =\frac{\lambda^2}{\rho}q,
\label{inho2}
\end{eqnarray}
with
\begin{eqnarray}
b_x=-\frac{k}{2\pi}\int_{0}^{2\pi}q(\theta)\cos\theta\, d\theta=\frac{k}{2\pi}\int_{0}^{2\pi}q'(\theta)\sin\theta\, d\theta,\nonumber\\
\label{inho3}
\end{eqnarray}
\begin{eqnarray}
b_y=-\frac{k}{2\pi}\int_{0}^{2\pi}q(\theta)\sin\theta\, d\theta=-\frac{k}{2\pi}\int_{0}^{2\pi}q'(\theta)\cos\theta\, d\theta,\nonumber\\
\label{inho4}
\end{eqnarray}
where we have used an integration by parts to obtain the second
equalities. For $\lambda=0$, Eq. (\ref{inho2}) is readily integrated
to yield
\begin{eqnarray}
\frac{dq}{d\theta}=\frac{\rho}{p'(\rho)}(b_x \sin\theta- b_y \cos\theta+C), 
\label{inho5}
\end{eqnarray}
where $C$ is a constant of integration. Another integration with the boundary condition $q(0)=0$ yields
\begin{eqnarray}
q(\theta)=b_x\int_{0}^{\theta} \frac{\rho}{p'(\rho)}\sin\theta'\, d\theta' \nonumber\\
- b_y \int_{0}^{\theta} \frac{\rho}{p'(\rho)}\cos\theta'\, d\theta'+C \int_{0}^{\theta} \frac{\rho}{p'(\rho)}\, d\theta'.
\label{inho6}
\end{eqnarray}
Then, the condition $q(2\pi)=0$ determines the constant
\begin{eqnarray}
C=b_y \frac{\int_{0}^{2\pi} \frac{\rho}{p'(\rho)}\cos\theta\, d\theta}{\int_{0}^{2\pi} \frac{\rho}{p'(\rho)}\, d\theta},
\label{inho7}
\end{eqnarray}
where we have assumed that the density profile $\rho(\theta)$ is
symmetric with respect to $\theta=0$ to simplify some terms
\footnote{Since $\rho=\rho(\Phi)$ at equilibrium with
$\Phi=B_{x}\cos\theta+B_{y}\sin\theta$ (where $B_{x},B_{y}$ are the
two components of the magnetization), it is always possible to choose
the $x$-axis such that the distribution profile is even
($B_{y}=0$).}. Substituting Eq. (\ref{inho5}) into Eq. (\ref{inho3}),
we find either that $b_x=0$ or that
\begin{eqnarray}
1=\frac{k}{2\pi}\int_{0}^{2\pi} \frac{\rho}{p'(\rho)}\sin^2\theta\, d\theta.
\label{inho8}
\end{eqnarray}
Substituting Eq. (\ref{inho5}) into Eq. (\ref{inho4}) we find either that $b_y=0$ or that
\begin{eqnarray}
1=\frac{k}{2\pi}\int_{0}^{2\pi} \frac{\rho}{p'(\rho)}\cos^2\theta\, d\theta-\frac{k}{2\pi}\frac{\left (\int_{0}^{2\pi} \frac{\rho}{p'(\rho)}\cos\theta\, d\theta\right )^2}{\int_{0}^{2\pi} \frac{\rho}{p'(\rho)}\, d\theta}.\nonumber\\
\label{inho9}
\end{eqnarray}
Therefore, if we restrict ourselves to even perturbations ($b_y=0$ and
$b_x\neq 0$) the condition of marginal stability is given by
Eq. (\ref{inho8}) and if we restrict ourselves to odd perturbations
($b_x=0$ and $b_y\neq 0$) the condition of marginal stability is given
by Eq. (\ref{inho9}). For general perturbations, we must satisfy both
Eqs. (\ref{inho8}) and (\ref{inho9}). Note that using
Eq. (\ref{fo20}), which becomes for the HMF model
\begin{eqnarray}
 \frac{\rho}{p'(\rho)}=-\int \frac{\frac{\partial f}{\partial v}(\theta,v)}{v}\, dv,
\label{inho9b}
\end{eqnarray}
we can express the criteria of marginal stability (\ref{inho8}) and
(\ref{inho9}) directly in terms of the distribution function
$f(\theta,v)=f(\epsilon)$.

For homogeneous distributions $\rho={\rm const.}$, we immediately find
that Eqs. (\ref{inho8}) and (\ref{inho9}) both reduce to the condition
of marginal stability 
\begin{eqnarray}
c_s^2=\frac{kM}{4\pi}, \quad {\rm or}\quad 
1+\frac{k}{2}\int_{-\infty}^{+\infty}\frac{f'(v)}{v}dv=0.
\label{inho9c}
\end{eqnarray}

For inhomogeneous isothermal distributions (see Eq. (20) of
\cite{cvb}) for which $p=\rho T$, using Eqs. (21) and (26) of \cite{cvb} and the relation $I_2(t)=I_0(t)-\frac{2}{t}I_1(t)$, we find that Eq. (\ref{inho8}) is always
satisfied. This implies that an inhomogeneous isothermal distribution
is always marginally stable with respect to even perturbations
($b_y=0$ and $b_x\neq 0$) of the form (\ref{inho6}), for any
temperature $T$. This can be checked explicitly. For an isothermal
distribution, these perturbations can be written
\begin{eqnarray}
q(\theta)=e^{-\beta B\cos\theta}-1,
\label{inho10}
\end{eqnarray}
and a direct substitution shows that they are solution of Eq. (\ref{inho2}) with $\lambda=0$. Finally, for inhomogeneous isothermal distributions, using Eqs. (20), (21) and (26) of \cite{cvb},  we find that Eq. (\ref{inho9}) is equivalent to
\begin{eqnarray}
\left (\frac{I_1(x)}{I_0(x)}\right )^2+\frac{2}{x}\frac{I_1(x)}{I_0(x)}=1,
\label{inho11}
\end{eqnarray}
with $x=\beta B$. The only solution of this equation is $x=0$ leading
to $T=\frac{kM}{4\pi}=T_c$. Indeed, it is shown in \cite{cvb} that the
inhomogeneous isothermal distributions are always stable ($\lambda\ge
0$) so that the point of marginal stability corresponds to the point
$x=0$ where the branch of inhomogeneous states begins. At the
temperature $T=T_{c}$, the branch of homogeneous states becomes
unstable and bifurcates to the branch of inhomogeneous states (second
order phase transition). Therefore, in that case, the marginal
stability condition for homogeneous and inhomogeneous states
coincides.  However, this is not necessarily the case for more general
distributions like for example the Lynden-Bell (or Fermi-Dirac)
distribution where a first order phase transition takes place
\cite{antoPRL,reentrant}. In that case, the new stability criteria
(\ref{inho8}) and (\ref{inho9}) can give information about the
stability/instability of the inhomogeneous states.

\end{document}